\documentclass[preprint, 11pt, times]{elsarticle}

\usepackage{amssymb}
\usepackage{amsmath}
\usepackage{bm}

\usepackage{lipsum}  
\usepackage[margin=1in]{geometry}
\usepackage{pdflscape}
\usepackage{tabularx}
\usepackage{booktabs}
\usepackage{enumitem}
\usepackage{soul}        

\journal{Advances in Colloid and Interface Science}

\begin{document}
	
	\begin{frontmatter}
		
		
		
		
		\title{From Interface Dynamics to Darcy Scale Description of Multiphase Flow in Porous Media}
		
		
		\author[label1,label2,label3]{Steffen Berg} 
		
		\affiliation[label1]{organization={Shell Global Solutions International B.V.},
			addressline={Grasweg 31}, 
			city={Amsterdam},
			postcode={1031WG}, 
			state={},
			country={The Netherlands}}

        \affiliation[label2]{organization={Porelab, Norwegian University of Science and Technology}, Department of Physics,
			addressline={S.P. Andersens vei 15B}, 
			city={Trondheim},
			postcode={N-7491}, 
			country={Norway}}  

        \affiliation[label3]{organization={Imperial College London}, Department of Earth Science and Engineering,
			city={London},
			postcode={SW7 2BP}, 
			country={United Kingdom}}

		\author[label4]{Ryan T. Armstrong} 
		
		\affiliation[label4]{organization={School of Civil and Environmental Engineering, University of New South Wales},
			addressline={}, 
			city={Sydney},
			postcode={NSW 2052}, 
			state={},
			country={Australia}}

        \author[label5]{Maja R{\"u}cker}
		\affiliation[label5]{organization={Eindhoven University of Technology},
			addressline={PO Box 513}, 
			city={Eindhoven},
			postcode={5600 MB}, 
			state={},
			country={The Netherlands}}

		\author[label2]{Alex Hansen} 
		

		\author[label6]{Signe Kjelstrup} 
		\author[label6]{Dick Bedeaux} 
		
		\affiliation[label6]{organization={Porelab, Norwegian University of Science and Technology}, Department of Chemistry,
			addressline={S.P. Andersens vei 15B}, 
			city={Trondheim},
			postcode={N-7491}, 
			country={Norway}}

		
		\begin{abstract}
            An outstanding characteristic of porous media, desired in many applications, is the large surface area, which facilitates solid-fluid interactions, making porous media an extreme case in colloid and interface science. 
            In two-fluid systems, wetting and the balance of capillary and viscous forces control fluid displacement processes, leading to a wide range of complex flow regimes with rich spatio-temporal dynamics. Macroscopic two-phase flow is historically described through the phenomenological extensions of Darcy's law. Besides many other shortcomings and inconsistencies, it covers only connected pathway flow in the capillary-dominated flow regime in a rigorous manner while other flow regimes with moving interfaces and associated topological changes are entirely implicit. 
            
            Given the lack of adequate descriptions, upscaling multiphase flow from pore to Darcy scale represents a long-standing challenge paving into the fields of thermodynamics, statistical mechanics and integral geometry. In this review, we compare novel concepts which have been largely motivated by experimental insights, enabled by significant advances in pore-scale imaging and modeling over the last decade. 
            They cover the three dominant flow regimes in a rigorous manner: (I) the capillary-dominated regime, consisting mainly of connected pathway flow with capillary fluctuations covered by the space-time averaging approach and by the extended nonequilibrium thermodynamic theory (NET), resulting in linear laws; (II) the nonlinear flow regime, where capillary states become increasingly accessible by viscous mobilization leading to ganglion dynamics and intermittency, which is described by the statistical thermodynamics approach; and (III) the viscous limit consisting mainly of drop-traffic, which leads again to a linear law described by the NET approach, which utilizes the fluctuation-dissipation theorem and Onsager reciprocal relationships. Most applications are in regime I which is the most complex and least intuitive to understand because it is a ''frozen state''. It is more intuitive to start with regime III being from the perspective of dynamics the most comprehensive, and then approach successively regime II and I.

            Spatio-temporal fluctuations are an inherent part of the novel approaches and thereby avoid previous limiting assumptions, which in key aspects are inconsistent with experimental observations of fluctuations from pore to Darcy scales. Currently only a minimum set of state variables is used to provide a proof of concept which can be extended to the four Minkowski functionals,  representing the geometric state variables for capillarity. 

            We conclude with open questions and invite to contribute to steer the theoretical advances towards application. The most immediate being the use of the fourth new concept, the co-moving velocity, which utilizes inherent symmetries in the 2-phase Darcy equations, to constrain the functional form of relative permeability-saturation relationships for either validating the relative permeability data or simplifying the relative permeability measurements. 
            In the future, the choice of state variables and the statistical thermodynamics approach that establishes relationships between them can be used to replace empirical hysteresis models. More generally, the more rigorous grounding of transport laws in thermodynamic concepts opens new possibilities for the study of coupled transport phenomena, including phase behavior/phase changes and other more complex processes, e.g., in electrochemical devices. 
            
		\end{abstract}
		
		
		
		
	\end{frontmatter}
	
		

		
	  \section{Introduction}\label{sec:introduction}
  Porous media systems (particularly those involving multiple fluid and solid phases) are distinct from other surface science applications owing to their exceptionally high surface area, which scales linearly with the system size. As a result, interfacial effects remain significant across a wide range of length scales, unlike in other open systems where the influence of surface tension diminishes with scale, e.g., transitioning from droplet-based flows to open-channel regimes. In porous media, surface forces continue to govern multiphase flow over relatively long distances, making these systems a unique subset of transport problems in which conventional assumptions about scale separation break down. In addition, the complex interplay of viscous and capillary forces coupled with pore scale processes that change fluid topology introduces spatio-temporal dynamics that create dynamic structures in fluid-fluid interfaces and associated flow regimes at complexity and associated length scales beyond the morphology of the confining pore space. 
  This lack of a clear separation of scales poses a fundamental challenge that is not typically encountered in other fields of engineering or surface science, where scalability often simplifies the theoretical treatment. 
  As such, the upscaling of multiphase flow in porous media (from the pore scale to the Darcy scale) has long been a central unresolved problem~\cite{hassanizadehUpscalingMultiphaseFlow2005}. Despite decades of research from various perspectives, the development of fully predictive physics-based upscaling models remains elusive.
  For a few selected flow regimes, such as connected pathway flow, analytical upscaling methods exist~\cite{tullerHydraulicConductivityVariably2001}. However, significant assumptions have been made, and in general, we lack a coherent framework that would allow the prediction of all flow regimes, including ganglion dynamics~\cite{avraamFlowRegimesRelative1995} and intermittency~\cite{gaoPoreOccupancyRelative2019,raizadaDynamicModeDecomposition2025} and the respective hydraulic transport coefficient in porous media, that is, relative permeability. In such flow regimes, the flow of disconnected non-wetting phase clusters~\cite{ramstadClusterEvolutionSteadystate2006} can have a significant contribution to the overall flux~\cite{Armstrong2016} which is not captured in the traditional upscaling methods.

While reviews for previous upscaling approaches exist~\cite{battiatoTheoryApplicationsMacroscale2019}, in this review, we provide an update on recent advances in the theoretical and experimental understanding of multiphase flow in porous media, with a particular focus on the upscaling problem in relation to the wide range of flow regimes. Motivated by breakthroughs in pore-scale imaging, modeling, and data interpretation over the past decade, we present four emerging frameworks that challenge conventional phenomenological descriptions. These include thermodynamic and statistical mechanical approaches that explicitly incorporate spatiotemporal fluctuations, topological changes, and nonequilibrium effects inherent to multiphase displacement processes. Central to these developments is the recognition that fluid configurations evolve through discrete, dissipative events, such as ganglion dynamics and Haines jumps~\cite{Haines1930} that persist across scales and are associated with topological changes of fluid pathways. The review also explores how these new perspectives enable more useful definitions of state variables (e.g., via Minkowski functionals~\cite{Armstrong2019} or Hill's system variables \cite{Bedeaux2024_nano}), provide theoretical constraints on relative permeability through symmetry arguments, offer space-time averaging strategies that replace ad hoc representative elementary volume (REV) postulates, define the configuration entropy of heterogeneous systems on the level of pore scale occupancy, and develop the application of the fluctuation dissipation theorem \cite{Bedeaux2021,Bedeaux2022}. By unifying these approaches, this review sets the foundation for a new and truly novel generation of Darcy-scale transport equations grounded in first principles and aligned with observable physics.

The recent advances covered include the following.
        \begin{itemize}
            \item Combination of topological concepts and state variables that arise from them which sets the foundations to identify flow regimes. 
            \item Extension of nonequilibrium thermodynamics to porous media flow. Formulation of fluctuation-dissipation theorems to determine transport coefficients.
            \item Development of a classical statistical mechanics framework with Boltzmann-type statistics applied to the capillary energy scale that would open the door to predict flow regimes,
            
            \item Identification of intrinsic symmetry relationships of two-phase Darcy type flow equations to constrain the range of possible relative permeability functions.
        \end{itemize}
        
Although several approaches are currently in progress, we aim to establish this review as a starting point for future research. Therefore, this review provides milestones in theoretical developments and shows how experimental insights from primarily pore-scale imaging techniques have influenced concept ideation and development of theory. 
It provides an assessment of the current status, lists the solved problems, and questions that are still open. Finally, we provide an outlook. 

 In the following chapters, we start with a brief summary of historical developments within the context of key questions in the upscaling of multiphase flow in porous media relevant to the flow regimes and their key characteristics in terms of requirement for theoretical descriptions. We then focus on developments over the past decade, where significant progress has been made in the thermodynamic description of multiphase flow in porous media. Subsequently, we provide an overview of the developing approach with a comparison between the main similarities and differences. Finally, we provide a perspective for further research. 

        \subsection{Porous media applications and scales}
\begin{figure}[ht]
			\includegraphics[width=1.0\linewidth]{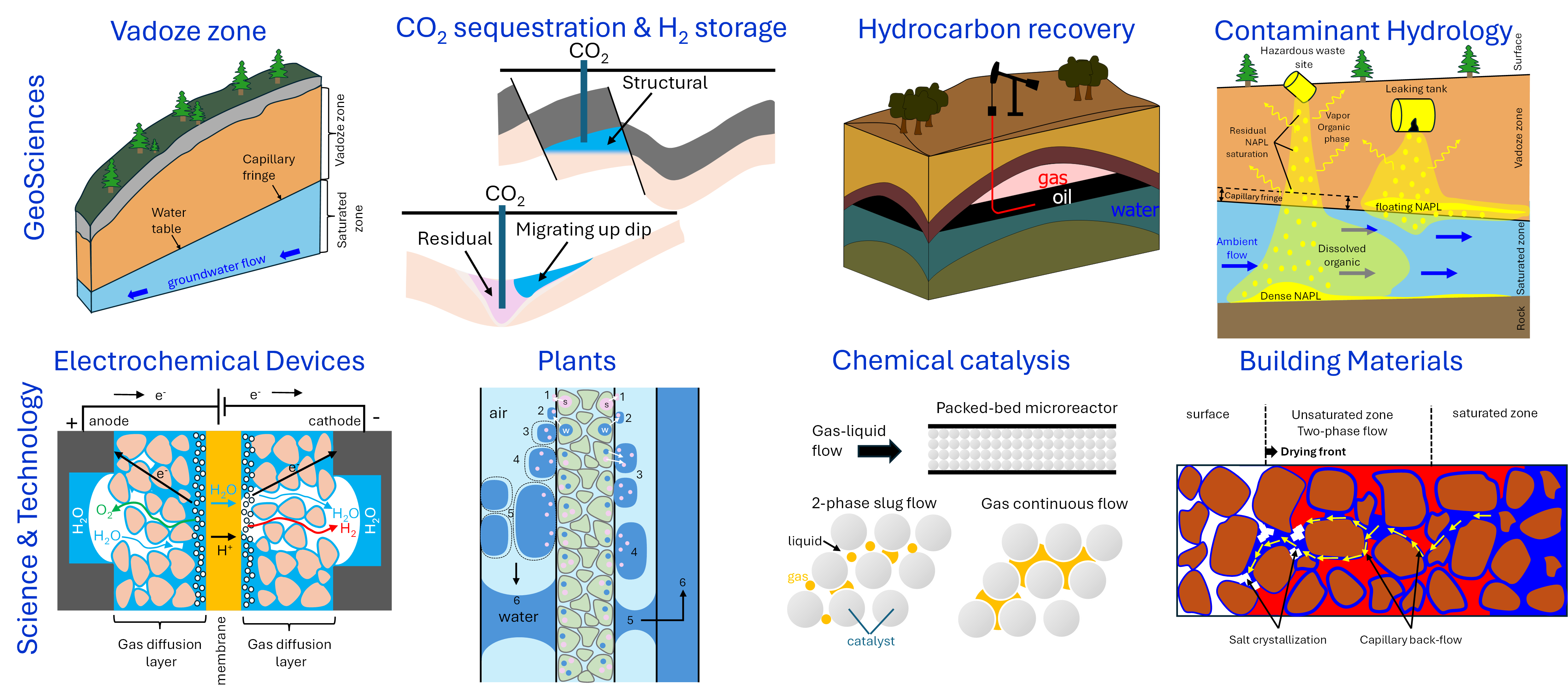}
			\caption{Overview of selected examples in science and technology where multiphase flow in porous media is relevant~\cite{maasViscousFingeringCCS2024,maasViscousFingeringCCS2024,simonInfluenceGasDiffusion2017,zhangSimilaritiesDifferencesGas2024,Hochberg2019,yueMultiphaseFlowProcessing2018,miriNewInsightsPhysics2015}.}
			\label{fig:OverviewMultiphaseFlowPorousMedia}
		\end{figure}

        Many aspects of human life depend centrally on transport through porous media~\cite{bearDynamicsofFluidsinPorousMedia1972,wangPredictionsEffectivePhysical2008,sahimiFlowTransportPorous2011,feder2022physics}. Examples include the production of drinking water from underground aquifers~\cite{katsanouSurfaceWaterGroundwater2017, helmigMultiphaseFlowTransport1997,schijvenRemovalVirusesSoil2000,torkzabanVirusTransportSaturated2006}, water and moisture retention in soil for agriculture~\cite{kuhlmannInfluenceSoilStructure2012,luGeneralizedSoilWater2016,vereeckenModelingSoilProcesses2016} and water transport in / melting of snow~\cite{colbeckPhysicalAspectsWater1978,illangasekareModelingMeltwaterInfiltration1990,waldnerEffectSnowStructure2004,krolRapidMRIProfiling2025}.  Biological processes in the human body, such as nutrient transportation in tissues and bones~\cite{Goldsztein2005,vaughanPoroelasticModelDescribing2013}, are central. A significant sub-class of relevant problems involves the flow of multiple fluid phases in porous media~\cite{classMultiphaseProcessesPorous2006}, such as  flow of water in soil in the Vadoze zone~\cite{geeRecentStudiesFlow1991, rybakMultirateTimeIntegration2015} and its impact on ecosystems~\cite{wankmullerGlobalInfluenceSoil2024}, recovery of oil and gas~\cite{dakeFundamentalsReservoirEngineering2010,lakeFundamentalsEnhancedOil2014,mohantyPhysicsOilEntrapment1987,stegemeierRelationshipTrappedOil1974,morrowWettabilityItsEffect1990,tangInfluenceBrineComposition1999,sorbiePolymerImprovedOilRecovery1991,hirasakiRecentAdvancesSurfactant2011,morrowRecoveryOilSpontaneous2001}, underground storage of carbon dioxide~\cite{kumarReservoirSimulationCO22005,nordbottenInjectionStorageCO22005,juanesImpactRelativePermeability2006,koppInvestigationsCO2Storage2009,classBenchmarkStudyProblems2009,Perrin2010,szulczewskiLifetimeCarbonCapture2012,tuckerCarbonCaptureStorage2018,buiCarbonCaptureStorage2018,wangPoreScaleModelingMultiphase2025} and hydrogen~\cite{heinemannEnablingLargescaleHydrogen2021,Higgs2022,Jha2021,lysyyHydrogenRelativePermeability2022,Gao2023,alzaabiWettabilityPoreOccupancy2025,gomezmendezInsightsUndergroundHydrogen2024,dokhonPressureDeclineGas2024}, nutrient uptake~\cite{rooseMathematicalModelPlant2001} and embolism in plants~\cite{Hochberg2019,Hochberg2019,brodersenDynamicsEmbolismRepair2010}, salt precipitation~\cite{ottMicroscaleSoluteTransport2014,ottSaltPrecipitationDue2015,ottSaltPrecipitationDue2021} during drying~\cite{pratRecentAdvancesPorescale2002} of soil~\cite{orAdvancesSoilEvaporation2013,lehmannCharacteristicLengthsAffecting2008,zhangNumericalStudyEvaporationinduced2014,hassaniPredictingLongtermDynamics2020,hassaniGlobalPredictionsPrimary2021} and building materials~\cite{Desarnaud2015,lavegliaCircularDesignMaterial2024,grossogiordanoSurfaceCouplingWater2024,madridControllingSaltWeathering2025,madridEffectCementContent2025}, removing contaminants from underground aquifers~\cite{abriolaMultiphaseApproachModeling1985,pennellInfluenceViscousBuoyancy1996,ahmadiLargescalePropertiesTwophase1996,sogaReviewNAPLSource2004,palmerPrinciplesContaminantHydrogeology2019,Pak2020}, but also technological processes such as chemical reactors~\cite{dudukovicMultiphaseReactorsRevisited1999,gladdenRecentAdvancesMRI2003,sauerschellMethanationPilotPlant2022}, fuel cells and electrolysis where multiphase flow occurs in gas diffusion layers~\cite{Owejan2009,Alink2011,Doerenkamp2021,Shrestha2020,Eller2017,Bazylak2009,Shafaque2020,Lee2020}. An overview of situations in which multiphase flow in porous media is relevant is shown in Fig. ~\ref{fig:OverviewMultiphaseFlowPorousMedia}.

\begin{figure}[ht]
			\includegraphics[width=0.9\linewidth]{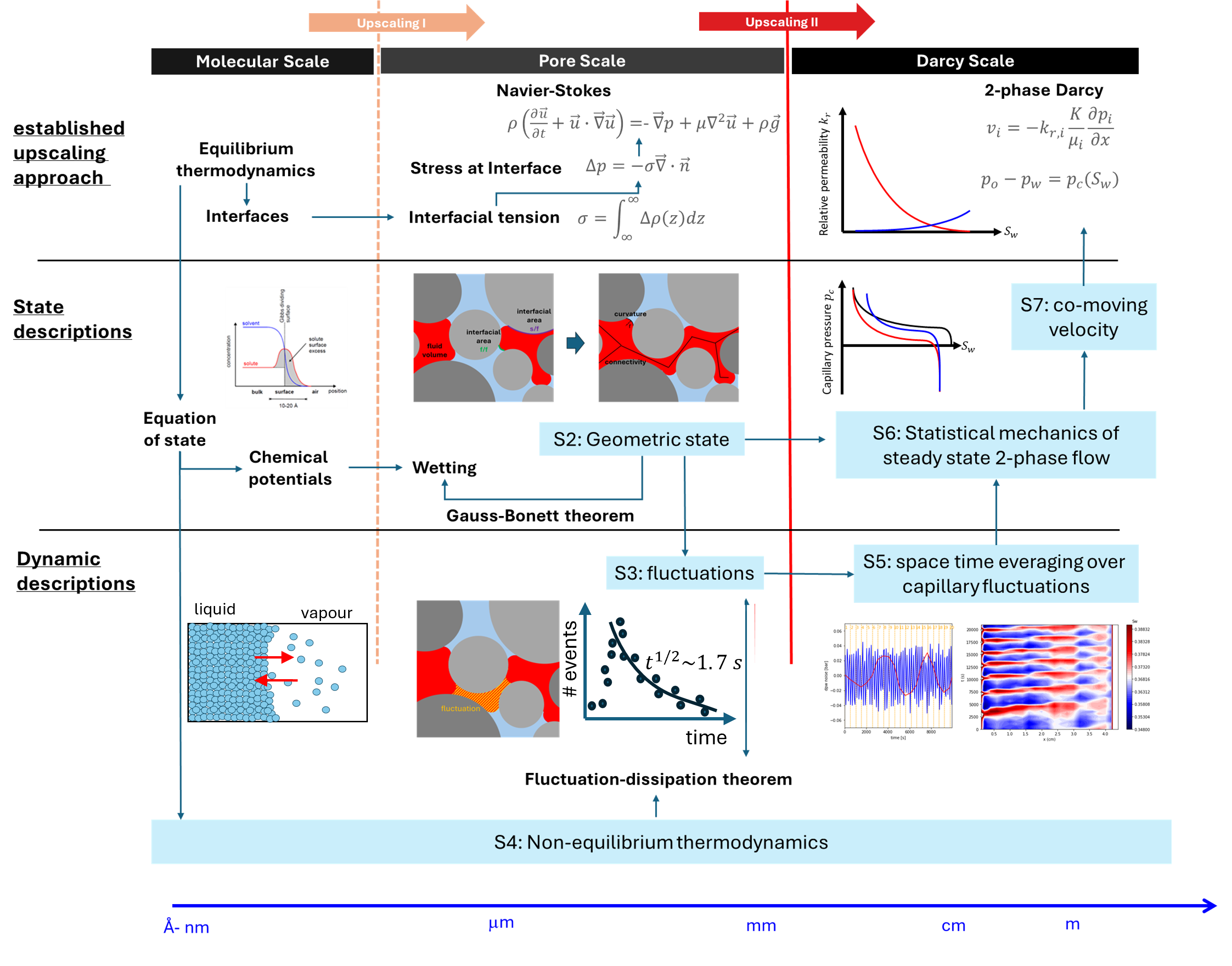}
			\caption{Multiphase flow in porous media from molecular to Darcy scale. While the length scales associated with the structure of the porous medium itself also generate a multi-scale problem, there are two upscaling steps where physical concepts change: (1) the upscaling of the fluids from molecular to continuum hydrodynamic scale and (2) the upscaling from the pore scale where pores are discrete to a continuum mechanics (Darcy scale) description of the porous medium. This review focuses on (2), i.e. the upscaling of the multiphase flow from the pore to the Darcy scale.  
            Four novel approaches are introduced, which are indicated in the light-blue boxes, including the  respective sections (S5-S7) \cite{Ruecker2021}. 
            The key novelty are the consideration of the geometric (capillary) state as starting point (S2) and the explicit consideration of fluctuations at the capillary energy scale (S3) \cite{armstrongSubsecondPorescaleDisplacement2014}.}
			\label{fig:cartoonupscaling}
		\end{figure}

        Many applications involving multiphase flow in porous media are inherently multiscale~\cite{hassanizadehUpscalingMultiphaseFlow2005} and cover a wide range of flow regimes in terms of fluid velocity and associated force balance. While transport in aquifers and in the vadose zone is slow with typical velocities of about 1 foot per day, and associated flow regimes are dominated by capillary forces, in other applications such injection of gasses in underground geological formations near the injection point or transport in gas diffusion layers can be fast and viscous-dominated. 
        An overview of the length scales relevant for multiphase flow in porous media is shown in Fig. ~\ref{fig:cartoonupscaling} from the molecular scale to Darcy scale.
        We have included the molecular scale in the overview in Fig. ~\ref{fig:cartoonupscaling} because, on the one hand, the concept of the Gibbs dividing surface introduces discrete phases, but also because fundamental concepts such as nonequilibrium thermodynamics and the fluctuation-dissipation theorem have been originally developed at the molecular scale, but are also valid at larger scales.
        Fig.~\ref{fig:cartoonupscaling} also serves as an outline of this review paper, where key concepts are linked with the respective section numbers.
        
        The fundamental challenge in describing these multiphase flow processes in porous media is the complexity at the pore scale, where the interplay of capillary and viscous forces leads to a complex pore-scale fluid configuration at the microscopic level (which is also difficult to monitor because most relevant porous media are not transparent to visible light), whereas the scale relevant for applications typically involves thousands to millions of pores or more. Therefore, for most relevant applications, a description of transport in porous media at the continuum level (Darcy scale) is required. Traditionally, Darcy-scale constitutive relationships and transport equations have been introduced either empirically or as phenomenological extensions~\cite{adlerMultiphaseFlowPorous1988}. Although in many cases fit-for-purpose, this is from a fundamental perspective unsatisfactory and also comes along with a number of consequences that cause practical problems. Phenomenologically introduced constitutive relationships and transport equations do not fully specify state variables and therefore are operated with an insufficient set of state variables, which may cause apparent or perceived hysteresis. They contain empirical parameters to describe material behavior that cannot be predicted within the phenomenological framework itself, but need to be either measured experimentally or predicted by numerical simulations from the scale below~\cite{zhaoComprehensiveComparisonPorescale2019,ramstadPoreScaleSimulationsSingle2019a}. 

\begin{figure}[ht]
			\includegraphics[width=1.0\linewidth]{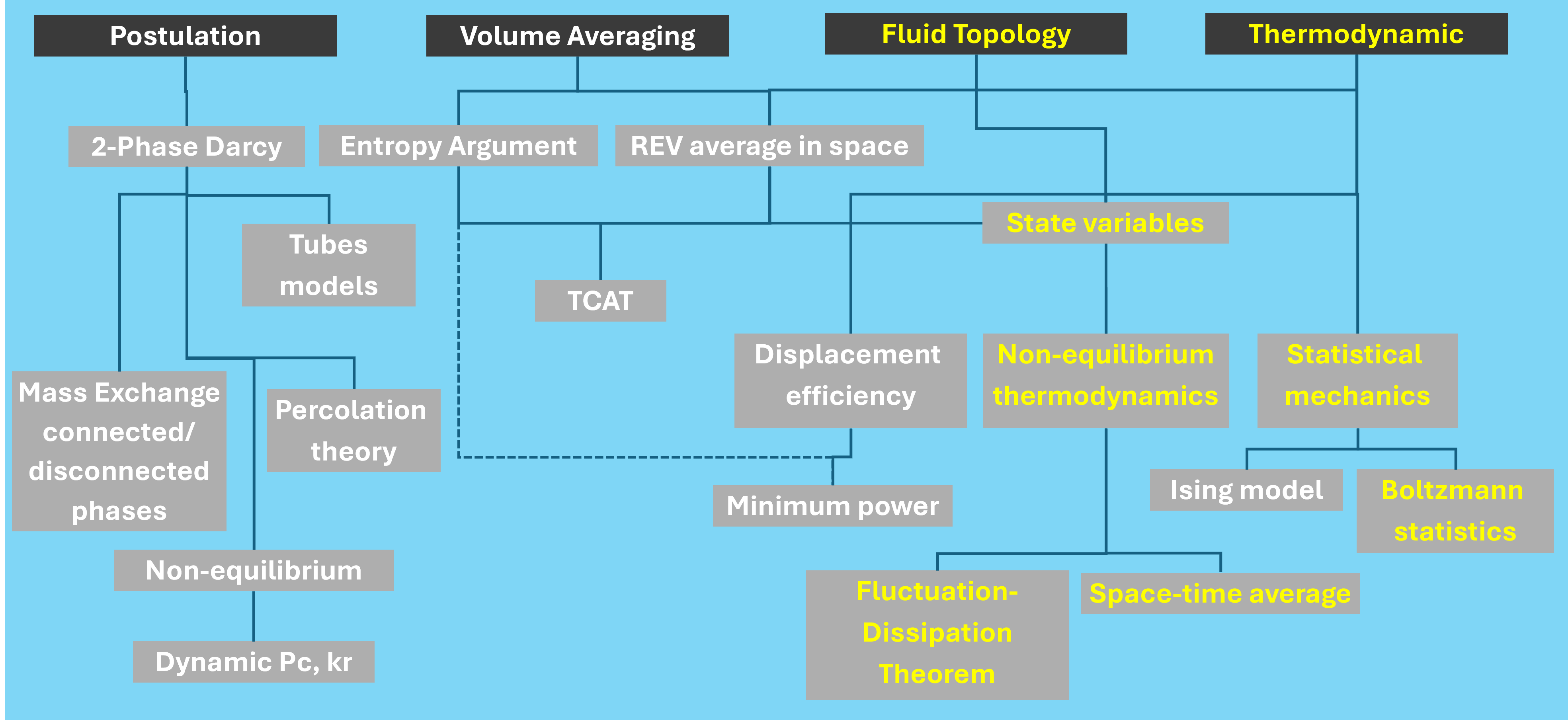}
			\caption{Overview of approaches for developing descriptions for Darcy-scale multiphase flow in porous media. While the approaches on the left have the postulated 2-phase Darcy equations as the starting point, approaches towards the right often involve upscaling from the pore to the Darcy scale. The focus of this review is on the paths indicated in yellow. }
			\label{fig:OverviewCartoon}
		\end{figure}

        While the experimental measurements of, for instance, relative permeability have their own challenges, the upscaling problem is a conceptually relevant subject, as it implicitly provides a route to predict parameters in Darcy-scale models and the constitutive relationships and transport equations themselves, such that we no longer have to rely on phenomenological relationships. For multiphase flow in porous media, the upscaling problem is significantly more complex compared with single-phase flow because flow dynamics generates a spatio-temporal structure of immiscible fluid phases. A wide range of upscaling approaches has been proposed. Fig.~\ref{fig:OverviewCartoon} provides an (incomplete) overview of upscaling approaches for multiphase flow in porous media from the pore to the Darcy scale and attempts to categorize them into intuitive classes, while recognizing a very large degree of interconnectivity and the multidimensional nature of the clustering.

        Although significant research has been conducted on the subject of upscaling multiphase flow from the pore to the Darcy scale, notable progress has also been made in the past decade, largely driven by advances in imaging technology/methods and computational methods/performance/capacity~\cite{bluntPorescaleImagingModelling2013}. Experimental insights obtained from synchrotron-based micro-CT flow experiments in porous materials in combination with extensive use of Lattice-Boltzmann simulations have led to  significantly improved insights and ultimately the identification of the geometric state function for capillarity and the derivation of the 2-phase Darcy transport equations. Furthermore, novel thermodynamic approaches provide a consistent and complementary derivation of the transport equations.

        Note that the term ''upscaling'' is also used in the context of heterogeneity scales~\cite{nordahlIdentifyingRepresentativeElementary2008}, which is not considered in this review. This review focuses on the conceptual upscaling where at the step from one to another scale, the physical concept changes, i.e. going from pore scale with the 2nd order Navier-Stokes equation to the Darcy scale with a 1st order equation, implying that the description level has changed. The conceptual steps are summarized as follows:
        \begin{itemize}
            \item from molecular scale to continuum hydrodynamics~\cite{dengHilbertsSixthProblem2025} but discrete with pores
            \item from discrete pores to a porous medium, i.e. continuum description for both the fluid and the porous medium
        \end{itemize}

        \subsection{Multiphase flow regarded as a mass and momentum balance problem}\label{sect:massmomentumbalance}
        Historically, approaches to address these challenges are closely linked to understanding the type of problem of multiphase flow.
        Initially, the problem was considered a transport problem, i.e. subject to mass and momentum balance. 
        This was probably motivated by the fact that upscaling single-phase flow from the pore to the Darcy scale is a transport problem, e.g., upscaling Stokes flow from the pore to the Darcy scale and the derivation of the Darcy law \cite{neumanTheoreticalDerivationDarcys1977, whitakerFlowPorousMedia1986}.
        However, such a derivation was not possible for multiphase flow because of the mobile liquid-liquid interfaces. In some sense, the root of the issue of pore-scale fluid topology is the division of regions of different compositions into fluid phases and the introduction of interfaces using the Gibbs dividing surface concept~\cite{AdamsonGast1997,Hassanizadeh2024}. Historically, this complication was addressed by considering only the Darcy scale transport with effective transport coefficients that would account for the complex fluid topology and respective dynamics at the pore scale. 
        The origin of capillary action in porous media date back to the mid 1800s~\cite{Gauss1830,Einstein_1901}. 
        Buckingham first attempted to develop a mathematical description of the two-phase flow in porous soil in the early 1900s~\cite{Buckingham1907, narasimhanBuckingham1907Appreciation2005}.
        The transport equations for two-phase flow in porous media were postulated by Richards~\cite{richardsCAPILLARYCONDUCTIONLIQUIDS1931}, Muskat \& Meres~\cite{muskatFlowHeterogeneousFluids1936}, and Wyckoff \& Botset~\cite{wyckoffFlowGasLiquidMixtures1936} in the 1930s and complemented by Leverett with a description of capillary pressure~\cite{leverettCapillaryBehaviorPorous1941} in the 1940s. In their work Darcy's law for single-phase flow is generalized in a phenomenological extension for 2-phase flow where the flux $v_{\alpha}$  

        \begin{equation}
            v_{\alpha} = - \frac{k_{r,\alpha}}{\mu_{\alpha}} K \cdot \nabla p_{\alpha}
            \label{eqn:twophasedarcy}
        \end{equation}

        \noindent is linearly proportional to the negative pressure gradient $\nabla p_{\alpha}$ as only driving force, and the proportionality factors absolute permeability $K$, the viscosity $\mu_{\alpha}$ of phase $\alpha$ and the relative permeability $k_{r,\alpha}$ of phase $\alpha$. The relative permeability factor $k_{r,\alpha}$ was introduced to account for the interaction between multiphase phases co-occupying the pore space. The phase pressure difference between wetting and non-wetting fluids $p_{nw} - p_w$ by the quasi-static capillary pressure $p_c$. In the absence of external pressure gradients, gradients in capillary pressure can drive flow, which gives rise to spontaneous imbibition~\cite{hammondSpontaneousImbibitionSurfactant2011,caiDiscussionEffectTortuosity2011,masonDevelopmentsSpontaneousImbibition2013}. 

        While this approach can be justified from an engineering perspective, it comes with highly unsatisfactory aspects and shortcomings:
        \begin{itemize}
            \item the correctness of the flow equations is not guaranteed and alternative equations have been proposed which have additional terms~\cite{HassanizadehGray1993}

            \item pore-scale dynamics and associated high Reynolds numbers $Re \approx 1$ ~\cite{armstrongModelingVelocityField2015b} are not creeping flow ($Re<1$) anymore, 
            which is the underlying assumption of the Darcy flow~\cite{hubbertDarcysLawField1956, bearFundamentalsTransportPhenomena1984,kingEffectivePropertiesFlow1995} and would thereby violate the requirements for the underlying single-phase Darcy equation from which the phenomenological extension was made.
            
            \item the 2-phase Darcy equations are generic linear transport models and have no physical contents before relative permeability is given physical contents that allows to constrain them.
            
            \item state variables are not specified which implies that the extent of the parameter space is unknown, and potentially very large

            \item this leads to apparent hysteresis in many flow parameters and functions and additional phenomenological models (''scanning curves''~\cite{killoughReservoirSimulationHistoryDependent1976,kjosavikRelativePermeabilityCorrelation2002,masalmehImprovedCharacterizationModeling2007}) have to be invoked to describe specific saturation paths 

            \item and additional phenomenological correlations are required to describe the dependency of irreducible saturation as a function of initial saturation~\cite{landCalculationImbibitionRelative1968,juanesImpactRelativePermeability2006} and capillary number~\cite{dullienPorousMediaFluid1979,lakeFundamentalsEnhancedOil2014,armstrongCriticalCapillaryNumber2014a,hilferCapillarySaturationDesaturation2015}.
            
            \item without specifying state variables it is practically impossible to falsify the transport equations (the effect of missing terms would be lumped into relative permeability~\cite{niessnerComparisonTwoPhaseDarcys2011}) which therefore cannot be a theory within the philosophical framework of science by Karl Popper~\cite{Popper34}

            \item only a small fraction of pore scale flow regimes is therefore covered in a rigorous manner while flow regimes with complex spatio-temporal dynamics and associated toplogical changes of interfaces are only covered implicitly and in a phenomenological manner, while the non-linear, viscous-dominated flow regimes are not covered at all. 
            
        \end{itemize}

\begin{figure}[ht]
		  \includegraphics[width=1.0\linewidth]{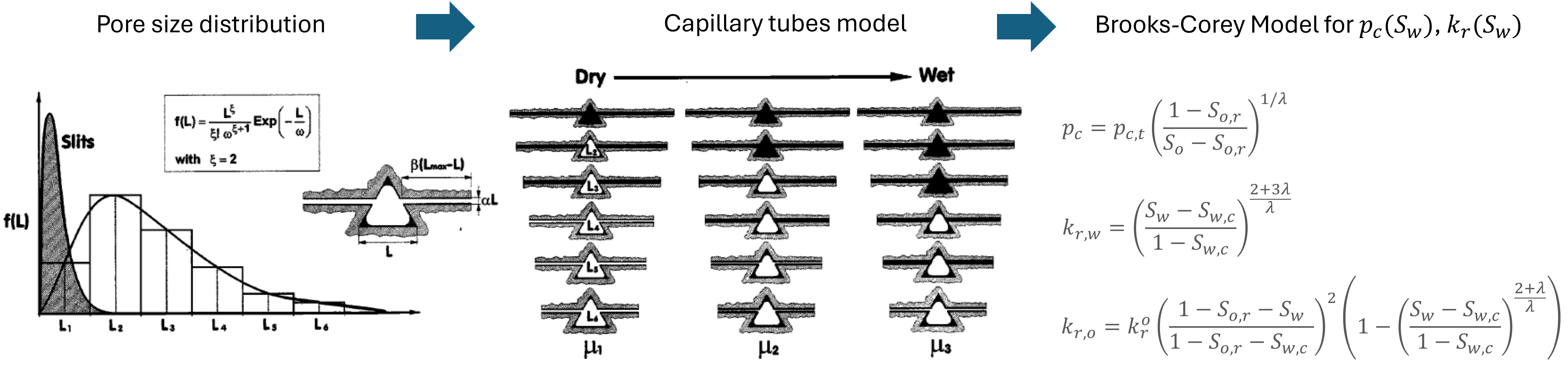}
			\caption{Assuming a connected pathway flow regime, upscaling from pore to Darcy scale for multiphase flow can be conceptually achieved by assuming a capillary tubes model with the size distribution linked to the pore size distribution~\cite{tullerHydraulicConductivityVariably2001}. Assuming a log-normal pore size distribution, the exponent of the log-normal pore size distribution is linked to the exponent $\lambda$ in the Brooks-Corey model~\cite{brooksPropertiesPorousMedia1966} for the relative permeability and capillary pressure-saturation functions (for drainage). Adapted from~\cite{tullerHydraulicConductivityVariably2001}}
			\label{fig:BrooksCorey}
 	\end{figure}

        From an experimental evidence perspective it is ''surprising'' that the formulation of Eq. ~\ref{eqn:twophasedarcy} covers next to connected pathway flow (for which eq. ~\ref{eqn:twophasedarcy}) can be analytically derived, it is essentially a network of tubes subject to Poiseuille flow~\cite{krolLocalHydraulicResistance2021}, see Fig.~\ref{fig:BrooksCorey}) also linear (flux-force balance) flow regimes which are subject to moving interfaces and ganglion dynamics. Why exactly that is so from a fundamental physics perspective will be a key discussion point of this review paper. 
        
        This description remained the status quo until the 1990s, although much attention has been paid to determining or predicting the relative permeability $k_r$ and capillary pressure $p_c = p_n - p_w$ functions, which were treated as explicit functions of wetting phase saturation $S_w$ only; that is, $S_w$ was considered the only state variable. However, it has been recognized that there are more parameters that influence $k_r$ and $p_c$, such as wettability~\cite{andersonWettabilityLiteratureSurvey1987a,morrowWettabilityItsEffect1990,mccaughanMolecularDynamicsSimulation2013,zhengSurrogateModelsStudying2020}, and that $k_r$ and $p_c$ depend on the displacement process, e.g. drainage and imbibition had different $k_r(S_w)$ and $p_c(Sw)$, which was interpreted as hysteresis~\cite{andersonWettabilityLiteratureSurvey1987a, lenhardModelHystereticConstitutive1987,jerauldEffectPorestructureHysteresis1990}. 
        Several attempts have been made to derive Eq. ~\ref{eqn:twophasedarcy} by volume averaging methods~\cite{battiatoTheoryApplicationsMacroscale2019}, with the most prominent approaches being by Marle~\cite{marleMacroscopicEquationsGoverning1982}, Whitaker~\cite{whitakerFlowPorousMedia1986,whitakerFlowPorousMedia1986a} and Quintard \cite{quintardTwophaseFlowHeterogeneous1988,quintardTransportOrderedDisordered1993,quintardTransportOrderedDisordered1994}. The constitutive relationships between the state variables are then formulated as closure relationships. Examples include the capillary pressure-saturation $p_c - S_w$ relationship in which the phase pressure difference between the wetting phase pressure $p_w$ and non-wetting phase pressure $p_n$ is approximated by the quasi-static capillary pressure, e.i., the Laplace pressure related to the curvature of the microscopic liquid-liquid meniscus at equilibrium. This is simplistic and does not necessarily hold under dynamic flow conditions where interfaces and contact lines move~\cite{latva-kokkoScalingDynamicContact2007,snoeijerMovingContactLines2013,mcclureTrackingInterfaceCommon2016,akaiWettingBoundaryCondition2018,qiuPhasefieldModelingTwophase2025}, as will be discussed later.
        A review of classical upscaling methods is provided by \cite{arceArtScienceUpscaling2005} in the review paper by Battiato \textit{et al.} \cite{battiatoTheoryApplicationsMacroscale2019}.
        Despite such approximations and other shortcomings, these approaches have played a very important role in the development of theory and continue to be further developed~\cite{Fadili2004,Lasseux2022,lasseux2023upscaled,zhangUpscalingNavierStokesCahnHilliardModel2025,briones-carrilloUpscaledCoefficientsImmiscible2025}. 
        Another class of approaches used the postulated 2-phase Darcy equations as the starting point, but used pore-scale models such as tube models~\cite{PermeabilityPorousMaterials1950,burdineRelativePermeabilityCalculations1953,mualemNewModelPredicting1976,tullerHydraulicConductivityVariably2001,lenhardParametricModelPredicting,genuchtenClosedformEquationPredicting1980,petersPredictionAbsoluteUnsaturated2023} (see for example, Fig. ~\ref{fig:BrooksCorey}) or their numerical counterpart pore network modeling~\cite{patzekShapeFactorHydraulic2001,patzekShapeFactorCorrelations2001,silinMicrotomographyPoreScaleModeling2011,bluntPorescaleImagingModelling2013} and percolation theory~\cite{heibaPercolationTheoryTwoPhase1992,ramstadFluxdependentPercolationTransition2009} to derive relative permeability-saturation functions and more complex processes, such as reactive transport~\cite{raoofPoreFlowComplexPorenetwork2013}. 

        Starting in the mid-1990s, new conceptual approaches to describe multiphase flow were developed. Sahimi and co-workers started developing connections between classical continuum models, percolation and fractal theory to cellular automata and simulated annealing~\cite{sahimiFlowPhenomenaRocks1993}.    
        Hilfer et al. used percolation concepts combined with mechanistic approaches for describing ganglion dynamics~\cite{hilferMacroscopicEquationsMotion1998, hilferMacroscopicTwophaseFlow2000,hilferPercolationBasicConcept2010}, where connected and disconnected phases were treated as separate entities with mass exchange between them. To this point, the problem of multiphase flow in porous media has been classified as a flow problem, i.e., subject to mass and momentum balance, which are either postulated in the end or based on volume averaging methods that postulate a representative elementary volume. 

        In some sense, the situation is not very different from the way other transport laws have been introduced, such as Ohm's law. Commonly, transport laws are introduced empirically on the basis of limited experimental observation with a limited set of variables and parameters, and used in practical applications before they are rigorously derived from first principles. Such rigorous derivations are surprisingly complex. For instance, the derivation of Ohm's law requires stepping outside the boundaries of Maxwell's equations, involving linear response theory, operating in Fourier space with current-current correlation functions using a Kubo formulation~\cite{ferryKuboFormulaLinear1992}.

        \subsection{Multiphase flow regarded as a thermodynamic problem}\label{sect:thermodynamicproblem}
        The thermodynamic perspective of multiphase flow started in the 1970s with the landmark paper by Morrow \cite{morrow1970physics} and possibly earlier work. But it took until the 1990s when 
        Hassanizadeh \& Gray~\cite{hassanizadehMechanicsThermodynamicsMultiphase1990, HassanizadehGray1993} treated the transport problem as a thermodynamic problem to derive Darcy scale transport equations. Mass-, momentum-, and energy balances are formulated at the pore scale and then upscaled to the Darcy scale by volume averages, whereby interfaces are treated as separate entities. The system is closed using a thermodynamic argument~\cite{grayThermodynamicallyConstrainedAveraging2005} such as the second law of thermodynamics with non-negative entropy production rate~\cite{HassanizadehGray1993}. 
        This approach is a nonequilibrium thermodynamics approach, given the fact that there is flow and movement of interfaces and that the system is not in equilibrium, such as a static mixture of molecules that would lead to a static phase equilibrium. Complementary/alternative formulations of the thermodynamic closure argument such as global energy ~\cite{grayClosureConditionsTwoFluid2002,jongschaapEquilibriumThermodynamicsCallens2001} or entropy minimum~\cite{niessnerComparisonTwoPhaseDarcys2011} or minimum (non-negative) entropy production rate~\cite{hassanizadehMechanicsThermodynamicsMultiphase1990} are considered as well in particular when it comes to determining specific transport coefficients.  However, specific arguments with respect to entropy generation rate beyond the non-negative requirement by the second law of thermodynamics such as minimum or maximum entropy generate rate are generally controversial and related to the length scale of the system relative to a diffusion length scale relevant for equilibration~\cite{veveakisReviewExtremumPostulates2015}.
        In addition to saturation, the interfacial area also became a state variable, and in addition to the pressure gradient, other driving forces, such as saturation gradients and interfacial area gradients, also became driving forces~\cite{niessnerComparisonTwoPhaseDarcys2011}. 
        This approach was developed further by Gray \& Miller into the thermodynamically constrained averaging theory (TCAT)~\cite{grayThermodynamicallyConstrainedAveraging2005,millerThermodynamicallyConstrainedAveraging2005a,grayThermodynamicallyConstrainedAveraging2006,millerThermodynamicallyConstrainedAveraging2008,grayThermodynamicallyConstrainedAveraging2009,jacksonThermodynamicallyConstrainedAveraging2009,grayThermodynamicallyConstrainedAveraging2009a,grayThermodynamicallyConstrainedAveraging2010,jacksonThermodynamicallyConstrainedAveraging2012a,grayIntroductionThermodynamicallyConstrained2014,millerPedagogicalApproachThermodynamically2017,battiatoTheoryApplicationsMacroscale2019,millerThermodynamicallyConstrainedAveraging2018}. Ultimately these approaches are based on Callen's postulational approach on thermodynamics~\cite{jongschaapEquilibriumThermodynamicsCallens2001}.
        The work by Hassanizadeh \& Gray and TCAT has influenced a new generation of experimental work and new developments~\cite{Culligan2004,porterLatticeBoltzmannSimulationsCapillary2009,meisenheimer2020exploring,Herring2013,Schlueter2016}. Even approaches that are not directly based on the groundbreaking work of Hassanizadeh \& Gray or follow different routes or take different assumptions are still very much influenced by it. Their work has significantly influenced experimental work~\cite{Culligan2004} and numerical modeling ~\cite{porterLatticeBoltzmannSimulationsCapillary2009,dyeCapillaryPressureDynamics2013,McClure2018} and was the starting point to determine relationships between state variables. An example of this is the interfacial area as a function of saturation. 
        More generally, this family of approaches performs upscaling from pore to Darcy scale by assuming the existence of a representative elementary volume (REV) for multiphase flow where fluctuations of individual parameters such as the saturation or the pressure average out in space. At the Darcy scale, the underlying pore-scale flow regimes are no longer considered and implicitly captured in Darcy-scale state variables, which are within the framework of saturation, interfacial area, and capillary pressure. 
        
       The experimental results of Avraam and Payatakes~\cite{avraamFlowRegimesRelative1995} and Constatinides and Payatakes~\cite{constantinidesNetworkSimulationSteadystate1996} enabled these authors to develop a mechanistic model for steady-state two-phase flow. Note that different terminologies exist such as static, stationary and steady-state. Here we adopt the terminology "steady-state" which is accepted in the porous media (i.e. "steady-state method"), physics and chemical engineering commmunities.
       Their dominant flow regime was ganglion dynamics. The porous medium was modeled as a network of randomly sized unit cells of the constricted-tube type. A set of linear equations gave same-time values for instantaneous pressures at all network nodes and the corresponding flow rates through the cells. The model was subsequently used to examine the effect of network dimensionality and wettability, and to investigate whether optimum operating conditions appeared in steady-state two-phase flow in pore networks~\cite{valavanidesSteadyStateTwoPhaseFlow2012, valavanidesTruetomechanismModelSteadystate, valavanidesReviewSteadyStateTwoPhase2018a, valavanidesFlowRateDependency2023}. It has been reported that the efficiency of the flow process depends on its spontaneity, which is measurable by the rate of global entropy production. In the DeProF model, the latter is the sum of two contributions: the rate of mechanical energy dissipation at a constant temperature and the conformational entropy production, which is directly related to the number of internal flow arrangements. 

        \begin{figure}[ht]
        \includegraphics[width=0.6\linewidth]{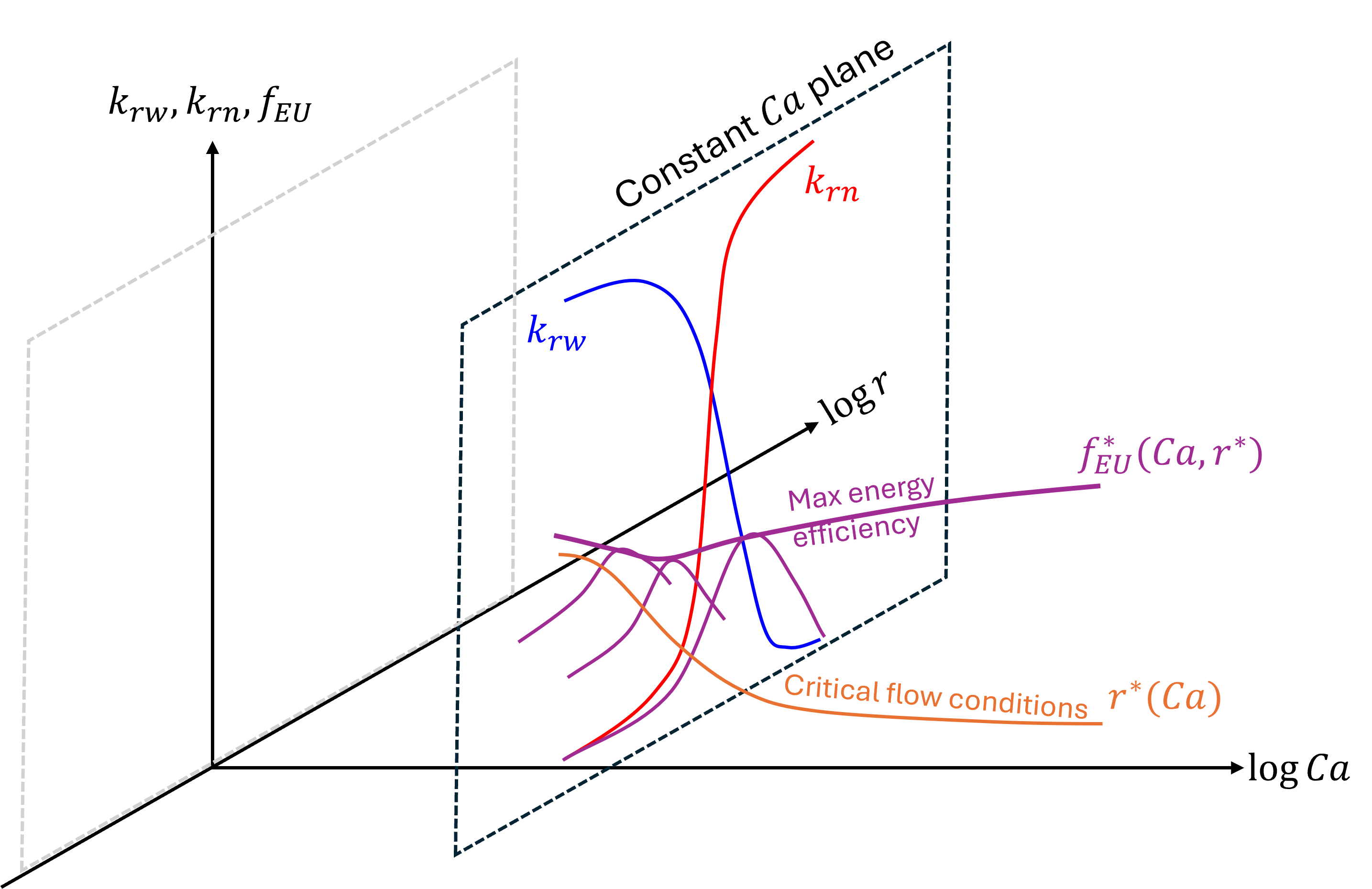}
			\caption{Illustration of the key results of the DeProF theory by Valavanides~\cite{valavanidesReviewSteadyStateTwoPhase2018a}:             
            Energy efficiency map relating steady-state relative permeability $k_r$ as a function of flow rate ratio of non-wetting and wetting fluid phases $r=q_n/q_w$ and capillary number $Ca$ to energy efficiency expressed by $f^*_{EU}$. For explanations of the figure in detail we refer to the original work by Valavanides~\cite{valavanidesReviewSteadyStateTwoPhase2018a}.}
			\label{fig:Valavanides2018}
	\end{figure}

       The DeProF algorithm simulations indicate that for every oil-water-pore network system, Optimum Operating Conditions exist for the flow rate, $r^*(Ca)$, as displayed in Fig.~\ref{fig:Valavanides2018}, for which the rate of global entropy production becomes maximum, i.e. when the process is as spontaneous as physically possible.  The mechanistic model then predicts the relative permeabilities from the concept of decomposition in prototype flows,  accounting for the pore-scale mechanisms and network-wide cooperative effects. DeProF is considered sufficiently simple and fast to be of practical use.
       In this manner, entropy production governs the flows of the DeProF model. The model invokes the maximum entropy principle to search for constitutive equations for flow, suggesting that two-phase flows arise and accordingly take the path of least resistance. No symmetry principle was invoked in this context, meaning that the assumption of time-reversal invariance (microscopic reversibility) was also not applied. The strength of the model is that it does not require the validity of these assumptions. A somewhat related concept, from a pore scale perspective, is the principle of minimum power, which determines the pore scale flow regimes and, more specifically, the configuration of liquid-liquid interfaces~\cite{gaoNewInsightsInterface2025}. For the capillary-dominated flow regime, the system adapts pathways with a minimal surface energy change rate. 
        
        Another class of approaches worth mentioning involves upscaling to the Darcy scale from the molecular scale. 
        Standnes derived Darcy's law for single-phase flow by upscaling from the molecular scale~\cite{dengHilbertsSixthProblem2025} using the Langevin equation~\cite{standnesImplicationsMolecularThermal2018, standnesDissipationMechanismsFluids2021, standnesDerivationConventionalGeneralized2022}, which results in a temperature dependency of the absolute permeability of the porous medium. As a premise for his derivation of temperature-dependent permeability~\cite{standnesDissipationMechanismsFluids2021}, Standnes hypothesized that there are two types of energy dissipation in a system: viscous and thermal dissipation. However, this is incorrect. An isothermal system exposed to a pressure difference has only viscous dissipation. The fact that this dissipation leads to heat transfer to the surroundings is another issue~\cite{Berg2013}. The relative and absolute permeabilities may be functions of the temperature; for example, any Onsager coefficient may be a function of relevant state variables.
                
        The gradient in the total chemical potential can be a more general driving force for transport than the pressure-dependent part, as observed by Standnes and coworkers~\cite{standnesUsingTotalChemical2024}. This driving force was used to generalize the gradient in the capillary pressure to explain the drainage/imbibition hysteresis. It may be interesting to see such an analysis anchored in the entropy production of an REV, as defined below in Section~\ref{sect:constructionofREV}. Overall the approaches discussed are increasingly leading into the direction of treating multiphase flow in porous media as a non-equilibrium thermodynamics problem in soft matter~\cite{bresmeTopicalIssueThermal2019}.

        \subsection{Nonequilibrium dynamics}
        In the framework of the family of theories developed by Hassanizadeh \& Gray, some aspects of pore scale dynamics are captured in nonequilibrium concepts such as dynamic capillary pressure ~\cite{hassanizadehThermodynamicBasisCapillary1993a,nieberDynamicCapillaryPressure2005,botteroNonequilibriumCapillarityEffects2011} which were further developed into a full nonequilibrium description for multiphase flow by Barenblatt \cite{barenblattMathematicalModelNonequilibrium2003} which started in the 1970s. In these descriptions, the explicit rate dependence of transport parameters was introduced, which led to a much more complex solution such as traveling waves and saturation overshoots, which were observed experimentally. However, these concepts are still extensions of phenomenological descriptions and are not rigorously derived from the pore scale, and fluctuations~\cite{dattaSpatialFluctuationsFluid2013,dattaFluidBreakupSimultaneous2014,masalmehLowSalinityFlooding2014,reynoldsCharacterizingFlowBehavior2015,soropRelativePermeabilityMeasurements2015,gaoXrayMicrotomographyIntermittency2017,gaoPoreOccupancyRelative2019,gao2020pore,linImagingMeasurementPoreScale2018,alcornCorescaleSensitivityStudy2020,spurinIntermittentFluidConnectivity2019,wangObtainingHighQuality2019,menkeUsingNanoXRMHighContrast2022} have not been addressed.

        \subsection{Topological principles}
        Topological concepts were introduced to the problem of 2-phase flow in porous media in the 1990s and the 2000s by research teams led by Mecke~\cite{Mecke_98,meckeEulerCharacteristicRelated1991}, Arns \cite{Arns2001} and Hilfer~\cite{hilferMacroscopicCapillarityHysteresis2006}. Vogel~\cite{Vogel2010} used topological descriptions, including Minkowski functionals~\cite{hadwiger1957vorlesungen,Klain_95} to parameterize the morphology of the pore space~\cite{chernyavskiyQuantitativeDescriptionInternal2025}.
        for \textit{single-phase flow} and to determine the respective representative elementary volume (REV)~\cite{sadeghnejadMinkowskiFunctionalEvaluation2023}. 
        This work was then extended by Wildenschild, Herring, Schl\"uter and others to experimentally study the role of interfacial area~\cite{Culligan2004} and topology~\cite{Arns2001, Vogel2010,  Herring2013, Herring2015, Schlueter2016} in \textit{multiphase flow}. Most of the interfacial area works were motivated by the work of Hassanizadeh \& Gray~\cite{HassanizadehGray1993} who articulated three of the four Minkowski functionals as state variables from thermodynamic principles. The study by Porter~\cite{porter2010image} is an important milestone, as it was one of the first studies to demonstrate that we can measure a specific interfacial area with reasonable accuracy using X-ray computed micro-tomography. 
        The detailed understanding of fluid topology on the pore scale was driven by the tremendous progress in the direct visualization of pore-scale displacement processes that became accessible by synchrotron-based X-ray computed micro-tomography~\cite{Berg2013, Pak2015, Bultreys2024} first applied to multiphase flow in natural rock, but later also applied to multiphase flow processes in gas diffusion layers in fuel cells~\cite{Zenyuk2016}. 
        The understanding that fluid topology evolves in discrete jumps and asymmetries between drainage and imbibition in terms of fluid topology is the main cause of perceived hysteresis~\cite{Schlueter2016,holtzmanOriginHysteresisMemory2020,nepalMechanismsInterfaceJumps2025} has then led to the identification of the Euler characteristic as the last missing state variable~\cite{Ruecker2015} and the identification of the geometric state function~\cite{McClure2018, Armstrong2019, McClure2020} which represents a complete set of state variables for capillarity. There were also strong indications that these topological concepts were useful for the description/parameterization of transport, such as relative permeability~\cite{Liu2017, Khorsandi2017, Purswani2021, AlZubaidi2023}. 
        The research group led by Armstrong discovered that topology is also a key concept for the characterization of wetting~\cite{armstrong2021multiscale} which was an important step since now fluid configurations caused by intermediate and mixed-wet solid surfaces could be integrated into the geometric state variable concept. 
        
        \subsection{Energy dissipation of pore scale displacement events, non-ergodicity, space-time averaging REV and derivation of the 2-phase Darcy equations without a postulated REV}
        The insights gained from experiments with \textit{in situ} imaging conducted over the past decade paint a more complex picture, which suggests a transition from the pore scale to the Darcy scale.
        The first key observation was that pore scale displacement processes are very dissipative and can consume up to 85\% of the invested pressure-volume work, i.e., depending on the morphology of the pore space and associated pore scale flow regime only a minority of the invested work causes actual flow~\cite{sethEfficiencyConversionWork2007, Berg2013}.
        The (closely related) second key observation was that pore-scale processes cause non-ergodicity~\cite{McClure2021} which consequently requires space-time averaging to compute Darcy-scale averages~\cite{mcclureThermodynamicsFluctuationsBased2021}. This has important consequences for the concept of a representative elementary volume (REV), which could not be a concept in space alone. 
        It was also observed experimentally that fluctuations of individual parameters, such as saturation and pressure (largely caused by the very dissipative pore-scale displacement events), do not average out in space~\cite{Ruecker2021} on the transition between the pore and Darcy scale. This challenges all volume-averaging-based upscaling concepts and implies that a theory that would derive the Darcy scale transport equations for multiphase flow would need to consider and handle fluctuations in individual state variables.
        Using space-time averaging concepts~\cite{McClure2021}, the 2-phase Darcy equations were then finally derived on the basis that the collective energy dynamics of fluctuations in state variables (which are spatio-temporally correlated) would average out for steady-state flow conditions~\cite{McClure2022}.
        The respective representative elementary volume (REV) is not postulated as in previous approaches, such as homogenization~\cite{arbogastDerivationDoublePorosity1990,Fadili2004} and volume averaging~\cite{quintardTwophaseFlowHeterogeneous1988}, but is defined through a space-time average. 
        For a homogeneous sandstone rock under co-injection, such a space-time REV can be reached for domain sizes of 2-4 mm~\cite{McClure2025} which is an important observation demonstrating that Darcy-scale transport can be reached at typical sample sizes of a few centimeters, consistent with the current experience of most researchers.
        
        The derivation of the 2-phase Darcy equations for steady-state conditions is settled in the sense that we now know that the historically postulated transport equations are indeed correct, but it still does not solve the problem of how to derive the transport coefficients such as relative permeability.
        It is also an open question of what happens under non-steady-state conditions.

        \subsection{Effective rheology concept}\label{sect:effectiverheology}
        An alternative formulation for the 2-phase Darcy equations (eq.~\ref{eqn:twophasedarcy}) is the effective rheology model by Hansen and co-workers~\cite{fyhn2021rheology,fyhn2023effective}. The fundamental argument is that the 2-phase Darcy equations are inconsequential in the sense that saturation $S_w$ is a mixing concept at Darcy scale (where pore scale fluid configurations are not identifiable anymore) while phase pressures $p_w$ and $p_{nw}$ are chosen to be identifiable in order to have a transport equation for each phase with respective relative permeability $k_r$.
        Capillary pressure $p_c = p_{nw} - p_w$ is then used as closure relationship~\cite{quintardTransportOrderedDisordered1993} which has the consequence that capillary pressure $p_c$ enters as a state variable. Respective capillary coupling terms (resulting from normal-stress boundary condition at pore scale fluid configurations~\cite{lealAdvancedTransportPhenomena}) are then lumped into relative permeability $k_r$. 
        However, phase pressures of wetting $p_w$ and non-wetting phases $p_{nw}$ are inherently a pore scale concept and often not accessible at Darcy scale e.g. in core flooding experiments. Instead, only an effective pressure drop can be measured. Therefore, if only an effective pressure drop is accessible, and we pursue the mixing concent at Darcy scale with more consistency, the pore-scale interaction between wetting and non-wetting phases would be described by an effective rheology. Instead of operating with 2 equations (one for wetting and one for non-wetting phases), 2-phase flow is described by only one transport equation with an effective phase flux which is related to an effective pressure gradients and an effective rheology~\cite{fyhn2021rheology,fyhn2023effective}.
        Ultimately, the Darcy scale description level is a choice i.e. exactly which parameters are chosen to be identifiable. However, with the choice of each parameter e.g. phase pressures we inherit an additional state variable (here capillary pressure $p_c$). 
                
        \subsection{New approaches}
        More recently, new concepts that complement and extend the status quo have been introduced.  One concept is the co-moving velocity~\cite{royCoMovingVelocityImmiscible2022,alzubaidiImpactWettabilityComoving2024,hansenLinearityComovingVelocity2024}.  This occurs naturally in the recently introduced statistical thermodynamics approach for immiscible two-phase flow in porous media \cite{hansenStatisticalMechanicsFramework2023}.  This is a general framework for immiscible two-phase flow in porous media, which contains the 2-phase Darcy equations as a special case.  Now, assuming the 2-phase Darcy equations with the relative permeabilities containing only saturation as a variable, the co-moving velocity introduces a relation between them. This constrains the possible choices of relative permeability pairs. 
        One of the traditional challenges associated with the postulated 2-phase Darcy equations is that there is no theoretical functional form for the relative permeability-saturation functions, and several functional forms are equally permissible~\cite{picchiRelativePermeabilityScaling2019,bergSensitivityUncertaintyAnalysis2021}. This wide range of possible choices introduces significant uncertainty when determining relative permeability in experiments~\cite{bergSensitivityUncertaintyAnalysis2021}. The constraints imposed by the concept of co-moving velocity have the potential to reduce the experimental effort and simplify experimental protocols~\cite{hansenLinearityComovingVelocity2024}.  
        
        Bedeaux, Kjelstrup, and co-workers formulated nonequilibrium thermodynamics for time-dependent REV variables of multiphase flow \cite{Kjelstrup2018,Kjelstrup2019,Bedeaux2021,Bedeaux2022,Bedeaux2024_nano,Bedeaux2025}. The fluctuating, independent and coarse-grained variables of the REV were derived from the entropy production. The variables were constructed by  adding volume-, area-, and line- variables. The coarse-grained variables therefore contain Minkowski variables. Fluctuation-dissipation theorems  and  Onsager's reciprocal relationships were formulated ~\cite{Bedeaux2021,bedeauxFluctuationdissipationTheoremsMultiphase2025}. This is a novel approach to the upscaling of multiphase flow in porous media from the pore to the Darcy scale, which explicitly honors the fluctuations of the system. Although independently derived, the approach is intellectually close to the work of McClure \textit{ et al.}~\cite{McClure2022, McClure2025} which indicates that there is a new common view of how multiphase flow is examined. This has the potential to provide deeper insight into the mechanism behind the relative permeability functions.

        Lastly, Hansen and coworkers introduced a statistical thermodynamics framework~\cite{hansenThermodynamicsImmiscibleTwophase2009} for immiscible and incompressible two-phase flow in porous media~\cite{hansenStatisticalMechanicsFramework2023} in order to obtain macroscopic properties~\cite{sahimiFlowPhenomenaRocks1993}.  The framework essentially postulates Boltzmann occupancy statistics at the capillary energy scale by applying the Jaynes maximum entropy principle, whereas a previous bottom-up approach had difficulties caused by the non-differentiability of the Euler characteristic in the geometric state function~\cite{mcclureGeometricEvolutionSource2019}. The newly introduced concepts of the agiture and the flow derivative are analogs of temperature and chemical potential in standard (thermal) statistical mechanics. This new formalism opens another route to gaining deeper insight into the simplistic relative permeability theory, while still keeping the number of variables tractable.  As previously mentioned, the co-moving velocity occurs naturally in this approach.
		
		\cleardoublepage
		\newpage
		\section{The Geometric State}\label{sec:geometricstate}

        \subsection{State variables of the 2-phase Darcy equations}
        Since the 2-phase Darcy equations have historically been introduced in a phenomenological manner, there has been no clear specification of the state variables. The only state variables chosen were saturation $S_w$ and to allow us to close the set of equations (energy and mass conservation) also capillary pressure $p_c$. This led to the implicit assumption that the relative permeability and capillary pressure are functions of saturation only, i.e., $k_r(S_w)$ and $p_c(S_w)$. Although always presented as a function of saturation $S_w$, in the original publications on the two-phase Darcy equations~\cite{richardsCAPILLARYCONDUCTIONLIQUIDS1931, wyckoffFlowGasLiquidMixtures1936, muskatFlowHeterogeneousFluids1936, leverettCapillaryBehaviorPorous1941} there is no clear restriction to the saturation. Many experimental observations then suggested dependency on the direction of saturation change~\cite{carlsonSimulationRelativePermeability1981}, wettability~\cite{andersonWettabilityLiteratureSurvey1987a,al-futaisiImpactWettabilityAlteration2003,abdallahFundamentalsWettability2007}, capillary pressure interfacial tension, viscosity, flow rate~\cite{masalmehImpactRelativePermeability2010a} and potentially other parameters. This leaves, in principle, a large set of parameters on which relative permeability and capillary pressure depend. 
        
        Thus, it is important to distinguish between the parameters and state variables. The (quasistatic) state is, in principle, defined by the pore scale fluid configuration, while parameters of the porous medium and fluids such as tortuosity~\cite{ghanbarianTortuosityPorousMedia2013}, wettability, and process parameters such as the displacement scenario (drainage, imbibition~\cite{caiGeneralizedModelingSpontaneous2014}, steady-state, unsteady-state, gas coming out of solution) influence this fluid configuration. Therefore, the primary objective is to identify the variables that characterize the geometric state of the pore-scale fluid configuration. This can be achieved in principle by a wide range of concepts, for instance, by using spherical harmonics. However, this method has the disadvantage of arriving at an infinite series, implying an infinite number of state variables. Minkowski functionals, which also provide a decomposition of the geometry of quasi-static pore-scale fluid distributions, have several advantages, particularly being a complete set of geometrical measures \cite{hadwiger1957vorlesungen}. 
        
        \subsection{Interfacial area as an additional state variable}
        Historically, the discovery that the capillary state can be characterized by the four Minkowski functionals has evolved in parallel between the two communities. The first was concerned with thermodynamic upscaling methods of multiphase flow \cite{HassanizadehGray1993} and introduced the interfacial area (which is one of the Minkowski functionals) as a new state variable \cite{joekar-niasarTrappingHysteresisTwophase2013}.

 \begin{figure}[ht]
			\includegraphics[width=0.8\linewidth]{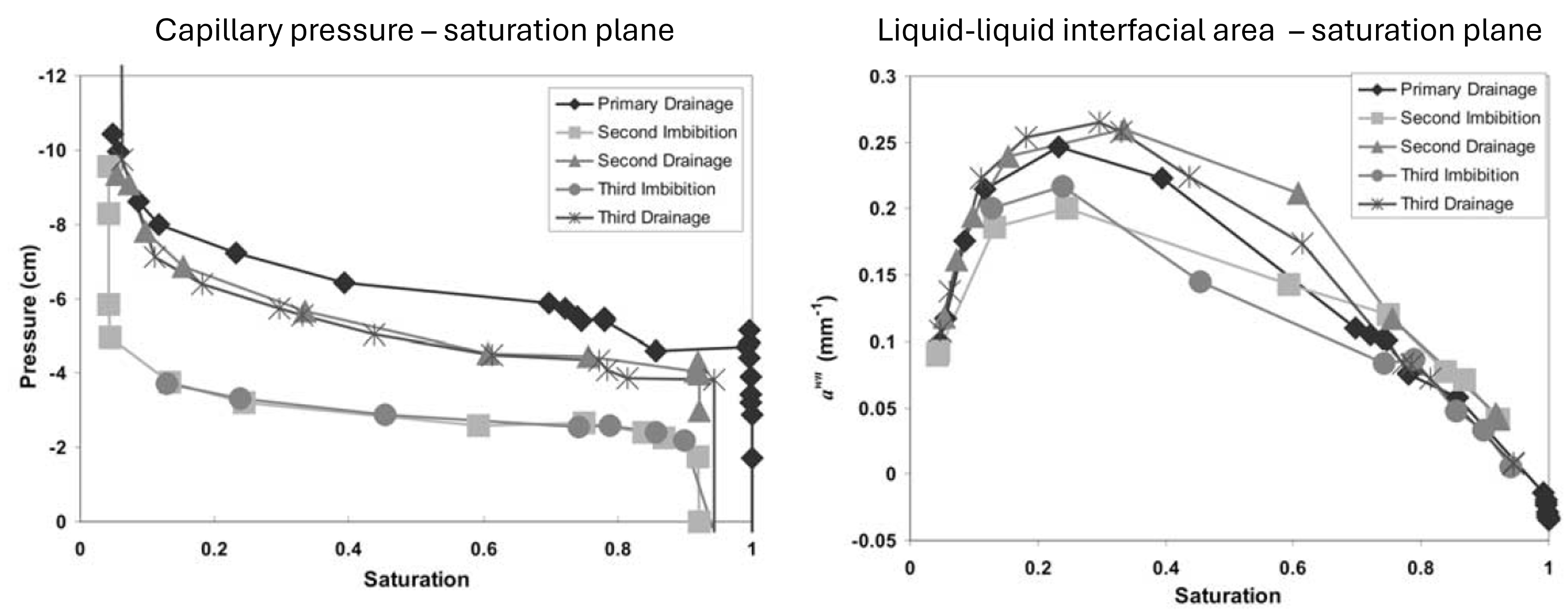}
			\caption{Capillary pressure-saturation plane (left) and liquid-liquid interfacial area - saturation plane ~\cite{Culligan2004}. }
			\label{fig:Culligan2004}
		\end{figure}
        
        The thermodynamics-based derivation of the Darcy scale two-phase flow transport equations~\cite{HassanizadehGray1993} considers next to the pressure-volume work also the interfacial energy as driving forces and thereby introduces the interfacial area as a state variable, thus considering three state variables capillary pressure $p_c$, saturation $S_w$ and liquid-liquid interfacial area $a_{n,w}$, that is, $p_c - S_w - a_{n,w}$. Culligan and co-workers~\cite{Culligan2004,porterLatticeBoltzmannSimulationsCapillary2009} experimentally measured this relationship, as shown in Fig. ~\ref{fig:Culligan2004}. The interfacial area typically has a maximum at $S_w<0.5$. 
        However, with saturation and interfacial area as state variables, it has not been possible to completely close the capillary pressure hysteresis~\cite{porterLatticeBoltzmannSimulationsCapillary2009, joekar-niasarNonequilibriumEffectsCapillarity2010a, dyeCapillaryPressureDynamics2013, meisenheimer2020exploring}.

         \subsection{Bi-continuous interfaces and (negative) Gaussian curvature}
\begin{figure}[ht]
			\includegraphics[width=0.8\linewidth]{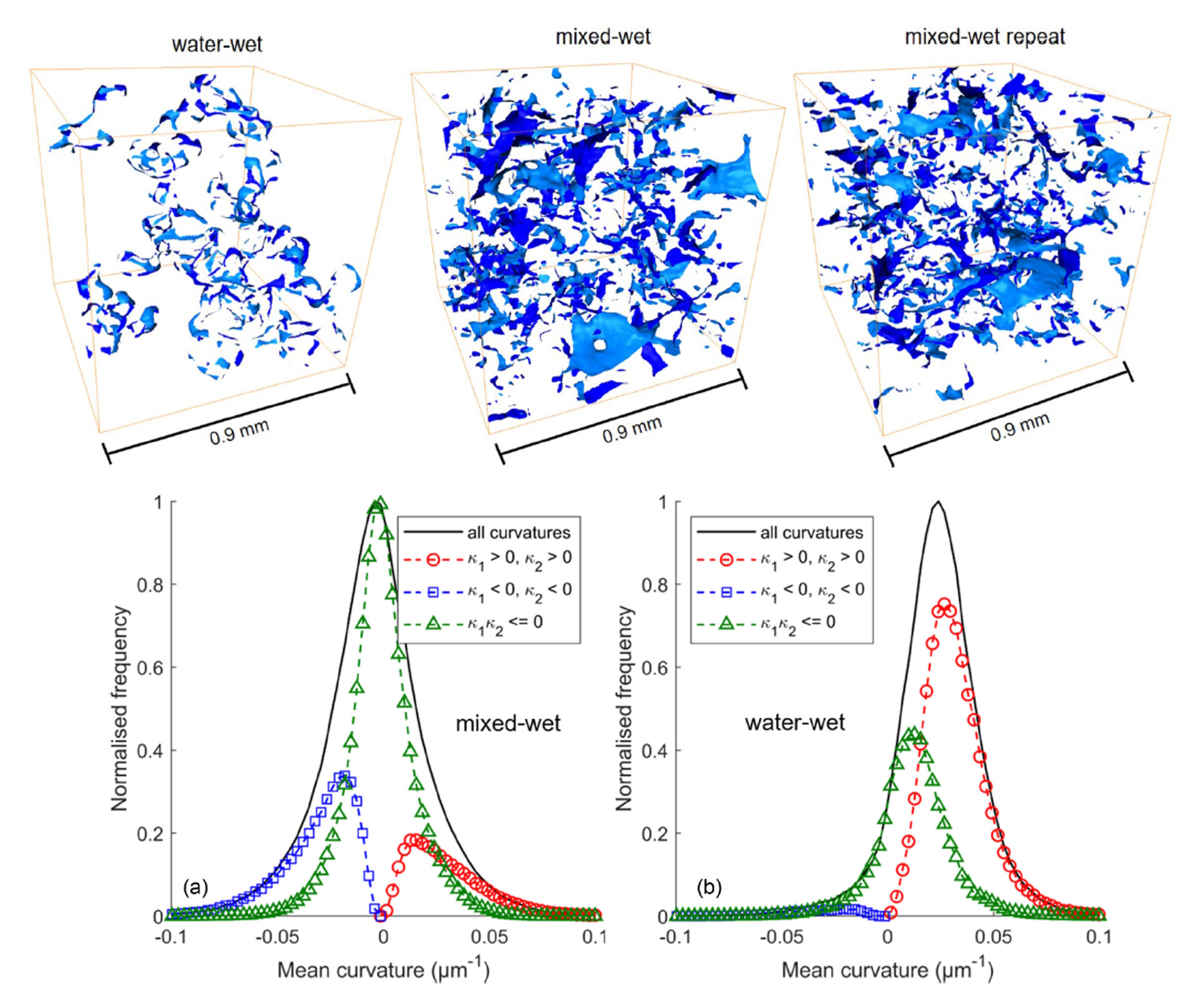}
			\caption{In mixed-wet porous media, minimal surfaces with bi-continuous, saddle-point like structures are observed (top). These are characterized by considering a negative Gaussian curvature $\kappa_1 \cdot \kappa_2 \leq 0$. Adapted from Lin \textit{et al.}~\cite{linMinimalSurfacesPorous2019}}
			\label{fig:minimalsurfaces}
		\end{figure}

         Imaging the pore-scale fluid configurations during multiphase flow experiments in porous media such as mixed wet rock~\cite{linMinimalSurfacesPorous2019} and gas diffusion layers consisting of hydrophilic fibers and hydrophobic binders in electrochemical devices \cite{shojaeiMinimalSurfacesPorous2022} by micro-CT shows bicontinuous, minimal surfaces with saddle-point structures that provide simultaneous connectivity for wetting and non-wetting phases. An example is shown in Fig. ~\ref{fig:minimalsurfaces}.         

        These are characterized by a negative Gaussian curvature $\kappa_1 \cdot \kappa_2 \leq 0$. This is a clear demonstration that for the characterization of the capillary state, it is not sufficient to consider only the mean curvature, which is a common practice when, for instance, probing the morphology of pore space (with spheres, which have bi-convex menisci)~\cite{silinPoreSpaceMorphology2006}. 

        \begin{equation}
            \bar \kappa = \frac{1}{2} \left( \kappa_1 + \kappa_2  \right) = \frac{1}{2} \left( \frac{1}{r_1} + \frac{1}{r_2}  \right)
        \end{equation}\label{eqn:meancurvature}

        \noindent where $\kappa_i=1 / r_i$ are the 2 principal curvatures of the interfaces (with the associated radii of curvature $r_i$). 
        Instead, it is necessary to consider the Gaussian curvature $\kappa_1 \cdot \kappa_2$.

        It is very possible that the formation of bi-continuous interfaces that favor connected pathway flow regimes is a consequence of the principle of minimum power~\cite{gaoNewInsightsInterface2025,valavanidesReviewSteadyStateTwoPhase2018a}, as connected pathway flow is less dissipative than pore-scale displacement events~\cite{Berg2013}, \textit{i.e.} the flow follows the path of least resistance~\cite{bejanadrianShapeStructureEngineering2000}. This implies that bi-continuous interfaces are enabled by the wetting boundary condition, but form as a consequence of flow boundary conditions.


        \subsection{Topological changes, Minkowski functionals and state variables}\label{sect:Minkowskifunctionals}
\begin{figure}[ht]
			\includegraphics[width=1.0\linewidth]{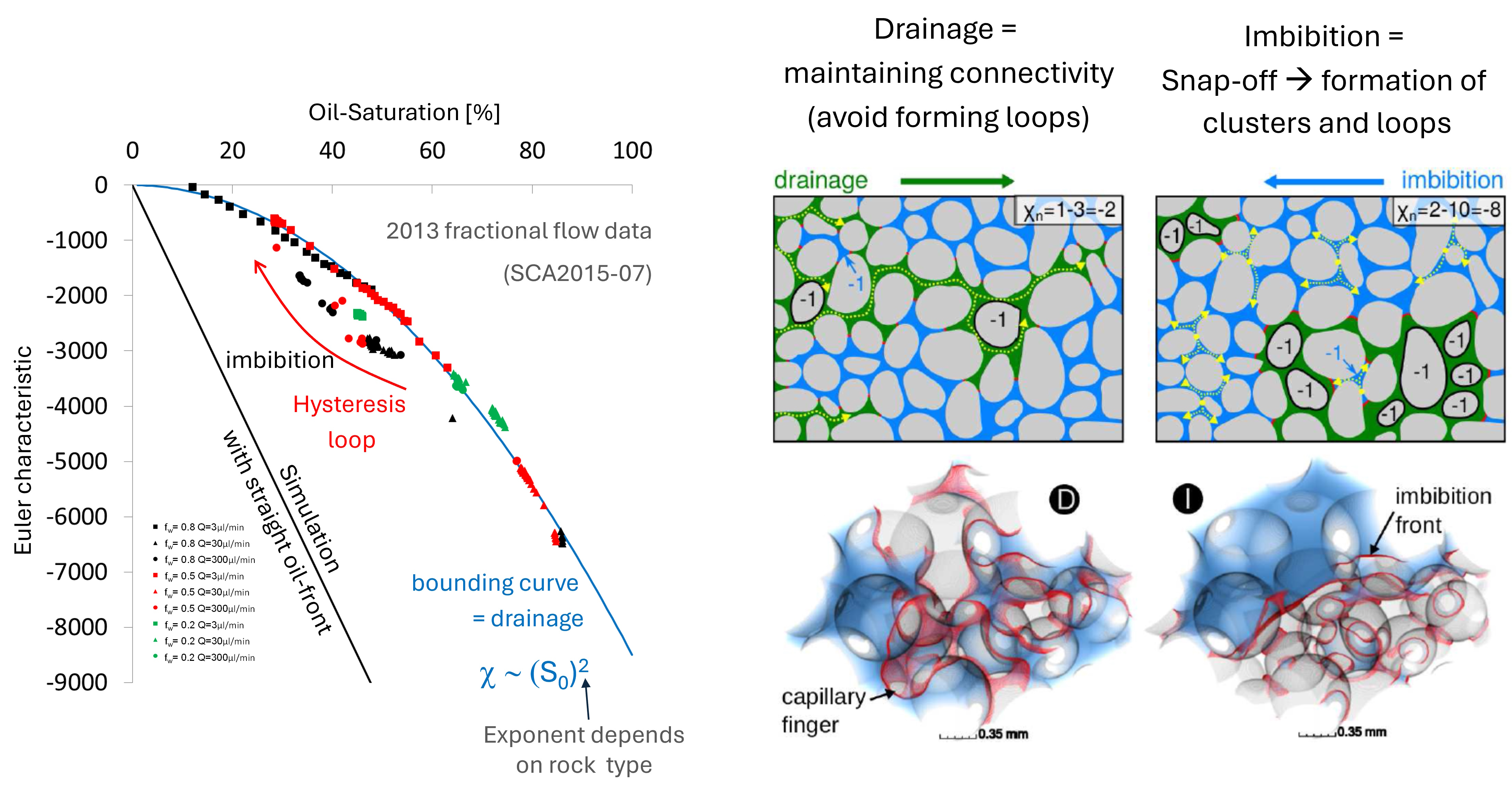}
			\caption{Euler characteristic $\xi$ as function of saturation for a sequence of drainage and imbibition cycles in a multiphase flow experiment imaged by micro-CT~\cite{Ruecker2015} (left). The hysteresis loop in $\xi$ was attributed to an asymmetry of pore-scale displacement mechanisms where Haines jumps in drainage preserve connectivity, while snap-off events in imbibition preserve loops~\cite{Schlueter2016}. }
			\label{fig:EulercharacteristicvsSaturationRueckerSchlueter}
		\end{figure}

        The other community was concerned with characterizing the morphology of the pore space through Minkowski functionals and parameterizing properties such as permeability~\cite{Vogel2010,Schlueter_Vogel_11}. The description was then extended to characterize fluid distributions in micro-CT flow experiments~\cite{Herring2013, Herring2015}. It was realized that pore-scale displacement events were the main source of capillary pressure hysteresis~\cite{Schlueter2016} because irreversibility is introduced through interface jumps~\cite{nepalMechanismsInterfaceJumps2025,yanWettabilitydrivenPorefillingInstabilities2025}.
        The breakthrough occurred when both communities were connected, and hysteresis in the Euler characteristic $\xi$ was observed in drainage-imbibition cycles~\cite{Ruecker2015} (Fig. ~\ref{fig:EulercharacteristicvsSaturationRueckerSchlueter}). This was attributed to an asymmetry of pore-scale displacement mechanisms in which Haines jumps~\cite{Haines1930} in drainage preserve connectivity, while snap-off events in imbibition steadily reduce loops~\cite{Schlueter2016}.

\begin{figure}[ht]
			\includegraphics[width=1.0\linewidth]{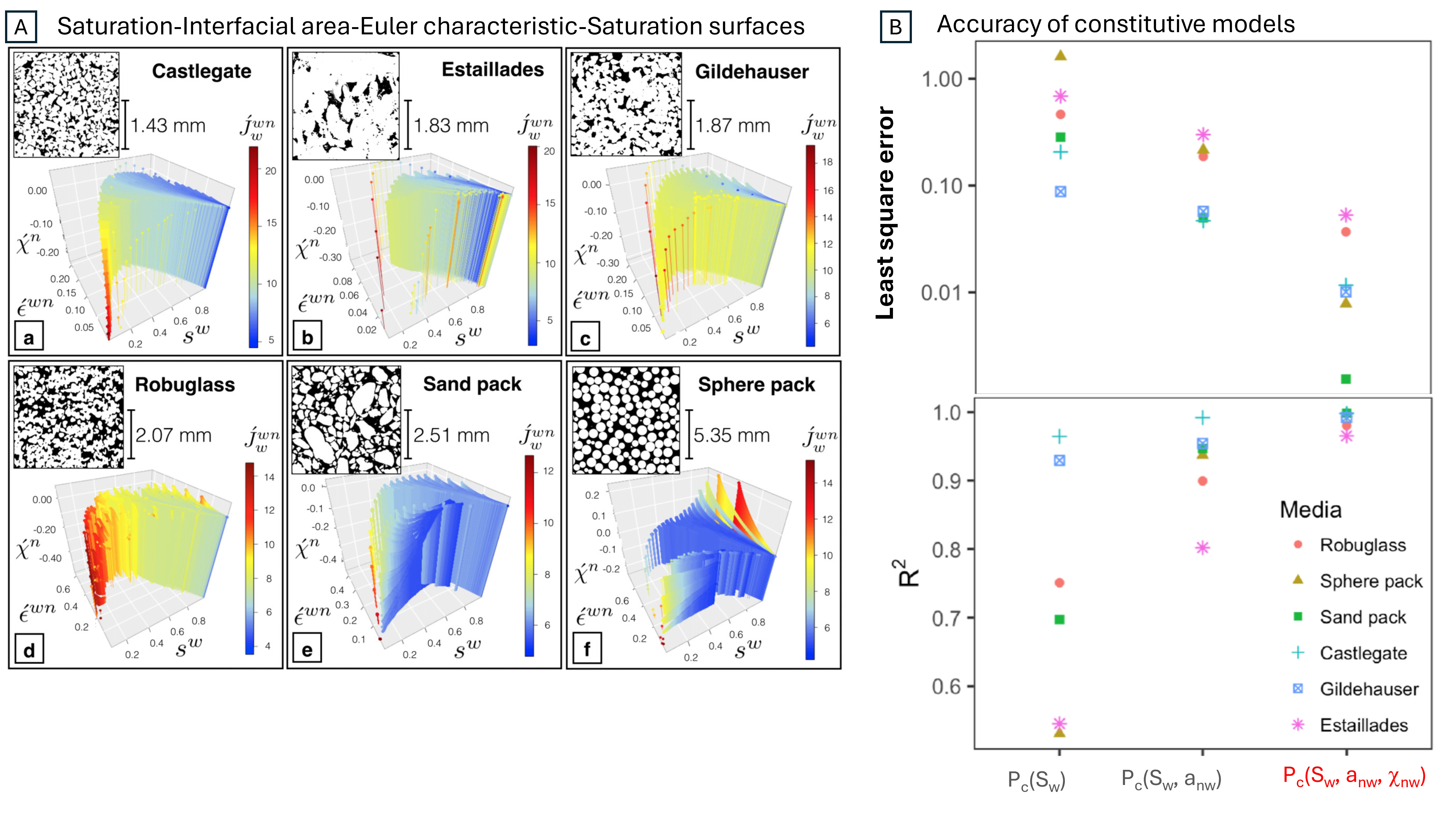}
			\caption{Saturation-Interfacial area-Euler characteristic-Saturation surfaces obtained from 260000 LBM simulations (A) and Accuracy of respective constitutive models (B), demonstrating that the 4 Minkowski functionals represent a capillary state-function that eliminates capillary pressure hysteresis (taken from \cite{McClure2018}). }
			\label{fig:capillarystatefunction}
		\end{figure}

        Although in previous work based on thermodynamic arguments it has already been clear that the state variables of capillarity involve saturation $S_w$ and capillary pressure $p_c$ (pressure volume work) and interfacial area $a_{nw}$ (representing together with interfacial tension the interfacial energy), McClure and co-workers have shown that together with the Euler characteristic $\chi_{nw}$, which complements the set of 4 Minkowski functionals~\cite{hadwiger1957vorlesungen,Klain_95} the capillary pressure hysteresis can be closed. 
        The 4 Minkowski functionals, which are named after the German/Polish/Lithuanian-German/Russian mathematician Hermann Minkowski, for the non-wetting phase $n$ for a closed manifold $\Omega_n$ with surface $\Gamma_n$ are 

        \begin{eqnarray}
            M_0^n & = & \lambda \left( \Omega_n \right) = \int_{\Omega_n}dr \label{eqn:minkowski1} \\
            M_1^n & = & \lambda \left( \Gamma_n \right) = \int_{\Gamma_n}dr \label{eqn:minkowski2} \\
            M_2^n & = & \int_{\Gamma_n} \left( \frac{1}{R_1} + \frac{1}{R_2}\right) dr \label{eqn:minkowski3}\\
            M_3^n & = & \int_{\Gamma_n} \frac{1}{R_1 R_2}  dr = 2 \pi \chi_{nw} \label{eqn:minkowski4}
        \end{eqnarray}
        
        \noindent where $M_0^n$ is the volume of phase $n$, which is related to saturation $S_w=1-S_n = 1-M_0^n/V_{pore}$, $M_1^n$ the interfacial area, $M_2^n$ the mean curvature which is related to the capillary pressure by interfacial tension $\sigma_{nw}$ through $p_c = \sigma_{nw} \cdot M_0^n$, and $M_3^n$ is related to the Euler characteristic $\xi_{nw}$ for closed manifolds by $M_3^n = 2\pi \chi_{nw}$. 
        In essence, the four Minkowski functionals represent a complete set of state variables for capillary pressure. In other words, there is no hysteresis of capillary pressure for inert systems, that is, systems where the geometry of the pore space and wetting properties remain constant during saturation changes. For such situations, the capillary pressure hysteresis reported in the literature is only a perceived hysteresis caused by an insufficient number of state variables. Fig.~\ref{fig:capillarystatefunction} shows how the '' capillary pressure hysteresis'' is closed. 
        That is, in principle already mathematically proven via Hadwiger's theorem~\cite{hadwiger1957vorlesungen,Klain_95}, which proves that all rigid motion-invariant, continuous valuations (meaning additive properties) on the space of convex bodies (compact convex sets) are linear combinations of intrinsic volumes, i.e. Minkowski functionals. This implies that the static capillary pressure, which is related to the mean curvature, which is the third Minkowski functional from Eq. ~\ref{eqn:minkowski3} can be expressed using the linear combinations of the other three Minkowski functionals. This was demonstrated using 260000 LBM simulations, in which the saturation-interfacial area-Euler characteristic-saturation surfaces for capillary pressure were calculated and then fitted with a piecewise polynomial function (Fig. ~\ref{fig:capillarystatefunction}A)] ~\cite{McClure2018}. 
        
        Mean squared error of individual simulation conditions from the fitted surface and the correlation coefficient ~\ref{fig:capillarystatefunction}B) show that saturation, as the only state variable, leaves a large hysteresis. In addition, including the interfacial area did not completely close the hysteresis. Including also the Euler characteristic, i.e. the fourth Minkowski functional, closes the capillary pressure hysteresis to the level of accuracy of the numerical simulations. 

        The advantages of using the four Minkowski functionals as capillary state functions can be summarized as follows:

        \begin{itemize}
			\item capillary pressure hysteresis is closed (in essence, there is no hysteresis, the hysteresis is only apparent when considering an incomplete set of state variables, such as $p_c - S_w$).
            \item only limited number of state variables required (in 3 dimensions there are 4 Minkowski functionals~\cite{meckeEulerCharacteristicRelated1991, McClure2018, Armstrong2019})
            
            \item with in general $D+1$ Minkowksi functional in $D$ dimensions (which provides a natural explanation for the impact of dimensionality on multiphase flow in porous media~\cite{lei3DgeometrytriggeredTransitionMonotonic2025})
           
            \item following a mathematical theorem (Hadwiger's theorem )~\cite{hadwiger1957vorlesungen,Klain_95,Bedeaux2024_nano}
            \item providing a link between pore scale flow regime and topological changes~\cite{Schlueter2016}
			\item providing a link to wetting~\cite{Wang2024}
              \item while the concept is geometric, there is a link to thermodynamics~\cite{ederRoleMinkowskiFunctionals2018, simeskiModelingAdsorptionSilica2020} i.e. the Minkowski functionals allow us to decompose the energy of the system into a geometric function and thermodynamic parameters such as interfacial tension. 
              \item Geometric variables are essential for nanoscale systems and may also play a role on the micrometer scale by changing the value of state variables \cite{Hill1994,Bedeaux2024_nano,Galteland2019}. 
		\end{itemize}


        \subsection{Geometric evolution: prediction of Euler characteristic}
        When introducing additional state variables, such as interfacial area and Euler characteristic, it is then necessary to measure, model, or predict their evolution. The goal is to use inherent dependencies, symmetries, and other properties to predict their evolution with as few parameters as necessary. 
        Using the geometric state variables (4 Minkowski functionals) is then an advantage over other approaches since McClure \textit{et al.} \cite{McClure2020} showed that geometric evolution is hierarchical in nature, with a topological source term that constrains how structure can evolve with time. This places the Euler characteristic $\chi$ in a central position. 
        Because the four Minkowski functionals are not independent of each other, the degrees of freedom were then reduced from four to three using nondimensional groups identified by a Buckingham Pi-theorem approach. Hence, the geometric state is described by only 3 non-dimensional groups as  

        \begin{equation}
           F(\phi_i,W_i,X_i) = 0\;.
           \label{eq:geom-state}
         \end{equation}

         \noindent where $\phi_i$, $W_i$ and $X_i$ are functions of the 4 Minkowski functionals from eq.~\ref{eqn:minkowski1}–\ref{eqn:minkowski4}. 
        
\begin{figure}[ht]
			\includegraphics[width=1.0\linewidth]{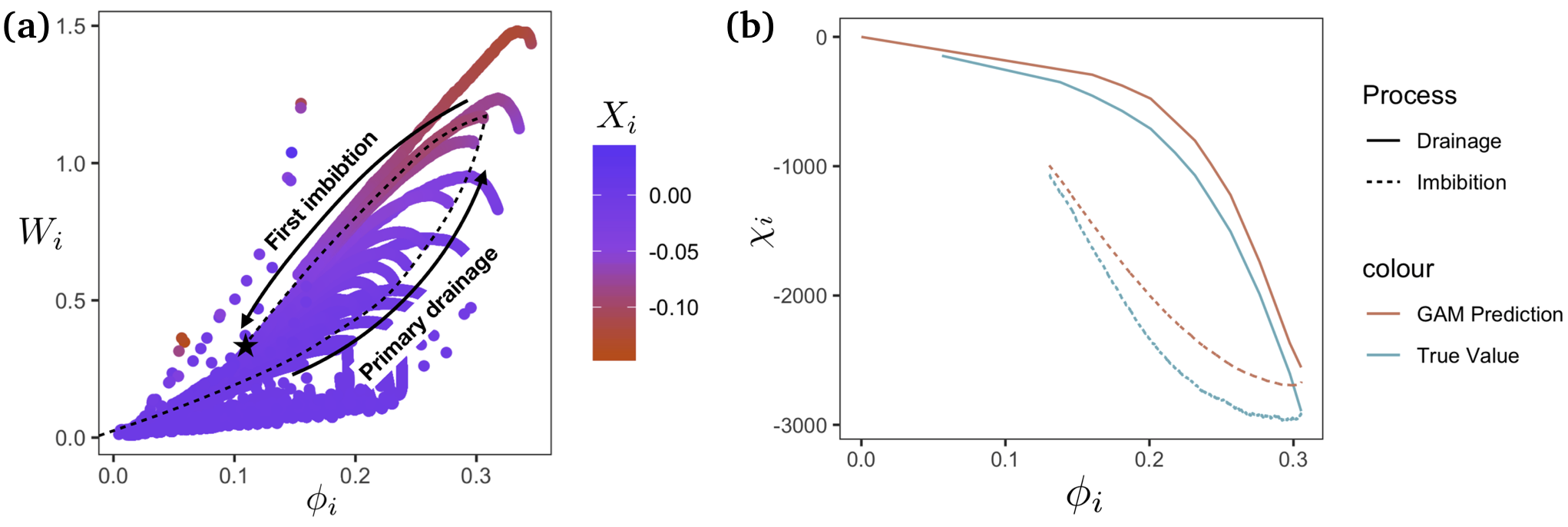}
			\caption{Evaluation of the non-dimensional form to predict Euler characteristic: (a) Hysteresis loop for simulated drainage-imbibition sequence within a sand pack based on the non-dimensional representation obtained. The star represents the residual non-wetting fluid that remains trapped within the pore structure following the first imbibition; (b) Prediction of Euler characteristic for a simulated drainage-imbibition sequence within a sand pack with the GAM approximation for $f(\phi_i,W_i)$.  Taken from~\cite{McClure2020}}
			\label{fig:Hysteresis-predict}
		\end{figure}

        Further reduction to two degrees of freedom can be achieved by using assumptions on the evolution of the Euler characteristic. An example of predicting the Euler characteristic for drainage and imbibition is shown in Fig.~\ref{fig:Hysteresis-predict}. 
                
\begin{figure}[ht]
			\includegraphics[width=1.0\linewidth]{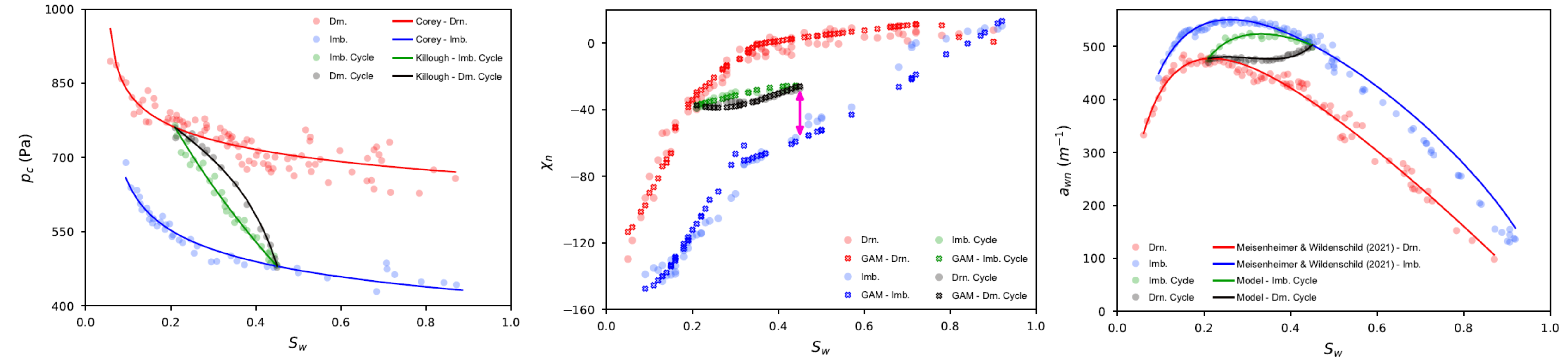}
			\caption{Prediction of capillary pressure, the Euler characteristic of the non-wetting phase $\chi_n$ and the interfacial area $a_{wn}$ for drainage, imbibition and scanning curves~\cite{kjosavikRelativePermeabilityCorrelation2002,masalmehImprovedCharacterizationModeling2007}.  Taken from~\cite{ebadiTopologicalEvolutionUnexplored2024}}
			\label{fig:EbadiTopologicalEvolution}
		\end{figure}

        Another example is shown in Fig. ~\ref{fig:EbadiTopologicalEvolution} where in addition to drainage and imbibition cycles, scanning curves~\cite{kjosavikRelativePermeabilityCorrelation2002,masalmehImprovedCharacterizationModeling2007} were also predicted.
        
        More generally speaking, the geometric evolution of, for instance, non-wetting phase clusters by pore-scale displacement events is described by a so-called Minkowski sum~\cite{mcclureGeometricEvolutionSource2019} where the change in volume is 

        \begin{equation}\label{eqn:Minkowskisum}
            V(\Omega_i \, \oplus \, \delta \zeta) - V(\Omega_i) = \alpha_1 A_i \delta + \alpha_2 H_i \delta^2 + \alpha_e \chi_i \delta^3.
        \end{equation}

        Because the Euler characteristic $\chi$ is discrete, pore-scale displacement events that involve topological changes cause differential volume changes to become discontinuous and non-differentiable. 
        Note that this is not a consequence of natural laws, but rather relates back to the choice of introducing discrete phases in discrete regions in space separated by the Gibbs dividing surface~\cite{AdamsonGast1997}, while in reality, concentration fields vary continuously with continuous density gradients.

        These discontinuities (which are introduced by choice) propagate through the Minkowski sum from Eq. ~\ref{eqn:Minkowskisum} into several other quantities. 
        This affects many fundamental relationships in physics, such as Noether's theorem~\cite{noether1918,noetherInvariantVariationProblems1971} which are derived based on the assumption of differentiability. Therefore geometric discontinuities, such as those experienced through the Minkowski sum from Eq. ~\ref{eqn:Minkowskisum}, result in symmetry breaking~\cite{viscardyViscosityNewtonModern2010} when singular points are considered. 
        
\begin{figure}[ht]
			\includegraphics[width=1.0\linewidth]{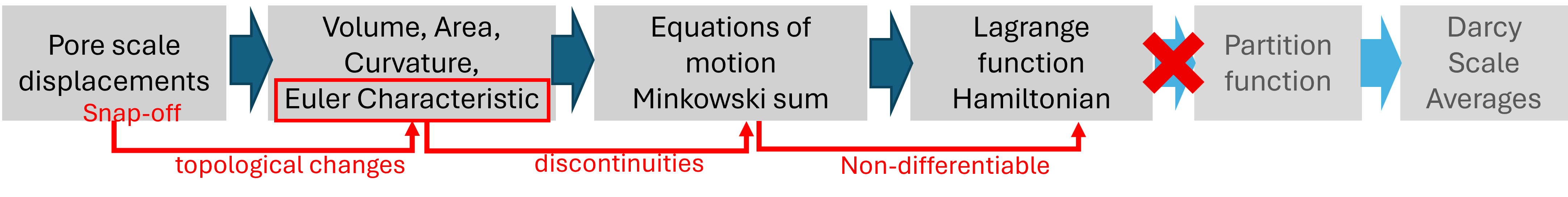}
			\caption{Workflow from pore scale displacements towards a partition function. Due to the Euler characteristic $\chi$ being discrete, the differential volume in the Minkowski sum from eq.~\ref{eqn:Minkowskisum} becomes non-differentiable which becomes a major challenge for computing the partition function and from that Darcy scale averages~\cite{mcclureGeometricEvolutionSource2019}.}
			\label{fig:workflowpartitionfunction}
		\end{figure}

        This also applies to classical thermodynamics, which links geometric invariants to standard thermodynamic quantities. Therefore, discontinuous geometric effects can propagate to the associated physical variables.
        The discontinuities in Eq. ~\ref{eqn:Minkowskisum} has consequences~\cite{mcclureGeometricEvolutionSource2019} for deriving relationships between state variables and transport equations by using the conventional formalism with the classical statistical mechanics approach, which begins with constructing the Euler-Lagrange function and Hamiltonian, and then the partition function, as sketched in Fig. ~\ref{fig:workflowpartitionfunction}. This problem was circumvented by the approach taken by Hansen and co-workers, described in Section ~\ref{sect:statisticalmechanics}. 
        
        \subsection{Relative permeability as a function of the geometric state}
\begin{figure}[ht]
			\includegraphics[width=0.6\linewidth]{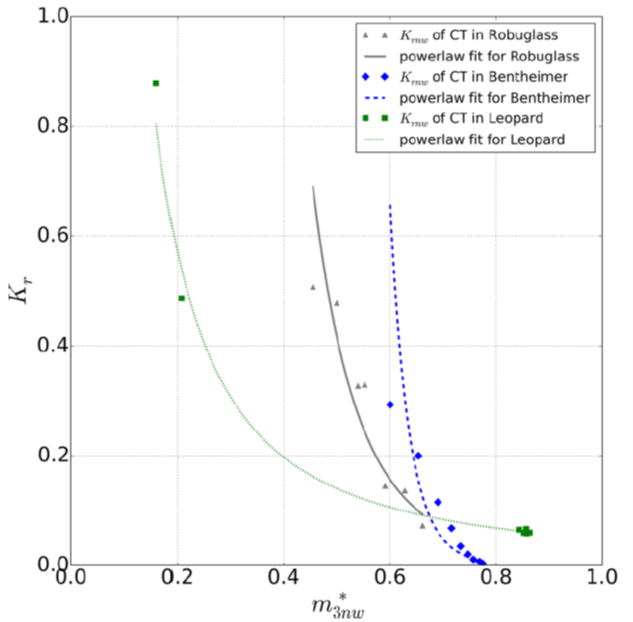}
			\caption{Non-wetting phase relative permeability $k_{r,n}$ as a function of the 4th Minkowksi functional $m^*_{3nw}$ (which is related by $2 \pi$ to the Euler characteristic $\chi$, see eq.~\ref{eqn:minkowski4}) for 3 different porous media. (taken from~\cite{Liu2017})}
			\label{fig:relpermeulercharacteristic}
		\end{figure}

       Relative permeability can be parameterized as a function of the Euler characteristic $\chi$ as shown by Armstrong and colleagues~\cite{Liu2017} which is shown in Fig. ~\ref{fig:relpermeulercharacteristic}. In addition, electrical resistivity~\cite{liuInfluenceWettabilityPhase2018} can be parameterized using the (scaled) Euler characteristic.
       This makes sense for flow regimes where the flux is dominated by connected pathway flow \cite{Armstrong2016} i.e. where the (relative) flux is then simply a function of connectedness \cite{gloverConnectednessTheoryRelative2025} which is represented by the Euler characteristic. 
       A more detailed follow-up study demonstrated generally how effective permeability in multiphase flow in porous media is determined by its geometric state~\cite{AlZubaidi2023} as defined by the Minkowski functionals from Eq. ~\ref{eqn:minkowski1}–\ref{eqn:minkowski4}.

        The group by Johns has developed in parallel semi-empirical saturation-Euler characteristic relationships $S_{w,nw} - \chi_{w,nw}$ to describe relative permeability hysteresis \cite{Khorsandi2017,Purswani2021,mukherjeeModelingRelativePermeability2025}. The key concept is to formulate the relative permeability as a total differential analogous to an equation of state. 

        \begin{equation}
            dk_r = \underbrace{\frac{\partial k_r}{\partial S}dS + \frac{\partial k_r}{\partial \hat{\chi}}d\hat{\chi}}_\textrm{Phase distribution} 
            + \underbrace{\frac{\partial k_r}{\partial I}dI}_\textrm{Wettability}
            + \underbrace{\frac{\partial k_r}{\partial N_{Ca}}dN_{Ca}}_\textrm{Capillary number}
            + \underbrace{\frac{\partial k_r}{\partial \lambda}d\lambda}_\textrm{Rock structure}
        \end{equation}
        
\begin{figure}[ht]
			\includegraphics[width=1.0\linewidth]{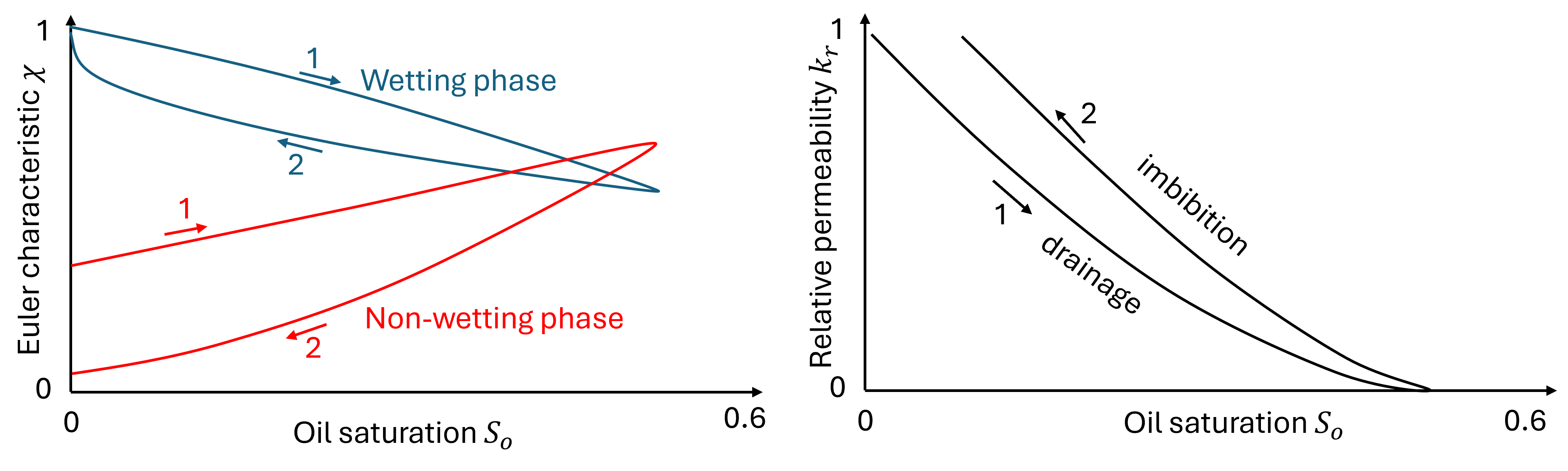}
			\caption{Relationship between Euler characteristic $\chi$ for wetting and non-wetting phases as a function of saturation (left) and the resulting relative permeability-saturation curves for drainage and imbibition (right)~\cite{Khorsandi2017}.}
			\label{fig:JohnsRelpermEulercharacteristic}
		\end{figure}

        An early example is shown in Fig.~\ref{fig:JohnsRelpermEulercharacteristic} where the relationship between the normalized Euler characteristic $\hat{\chi}$ and saturation is parameterized as 

        \begin{equation}
             \hat{\chi} =
            \begin{cases}
               (\hat{\chi}_0 - 1) \left(\frac{S-1}{S_0-1}  \right)^{\alpha_{\chi}}+1 & \frac{\partial S}{\partial t} > 0\\
               \left[\frac{n+1}{C(1-n)} \left(\frac{1}{S^{n+1}} - \frac{1}{S_0^{n+1}}  \right)+\hat{\chi}_0^{n+1}  \right]^{\frac{1}{n+1}} & \frac{\partial S}{\partial t} < 0
            \end{cases}       
         \end{equation}
                 
        In follow-up work, the method was further refined~\cite{Purswani2021} and the tuning of the constitutive relationship was improved using artificial neural networks~\cite{mukherjeeModelingRelativePermeability2025}.

        \cleardoublepage
        \newpage
        \subsection{The Gauss-Bonnet theorem, deficit curvature and contact angle}

\begin{figure}[ht]
			\includegraphics[width=1.0\linewidth]{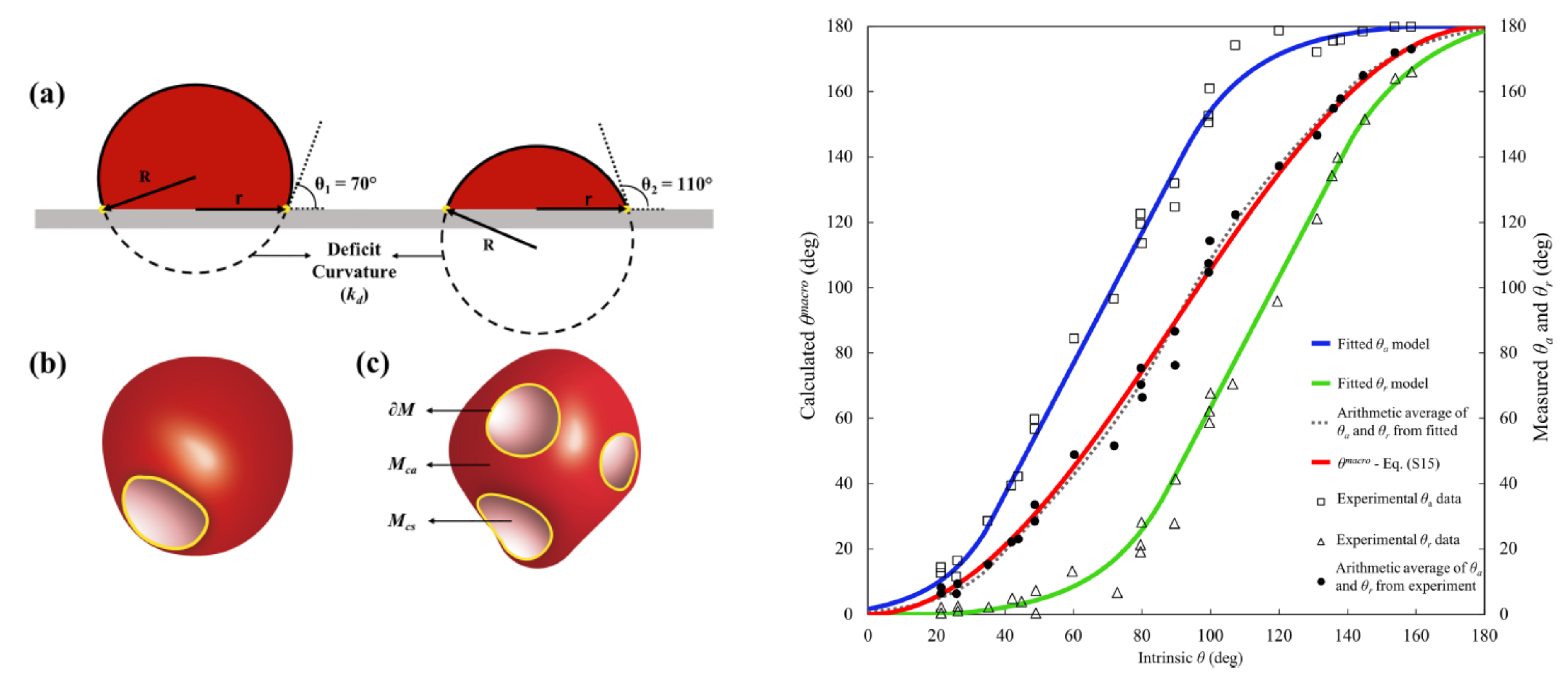}
			\caption{Left - (a) A 2-D schematic diagram of fluid clusters on a solid surface. Larger deficit curvature corresponds with
             a larger contact angle; see the definition of deficit curvature. The 3-D analog fluid cluster in the system
             which has (b) only a single contact line loop and (c) multiple contact line loops. The three-phase contact line $\partial M$
             subdivides the boundary surface for a cluster into the surface that separates the cluster and fluid $M_{ca}$, and the surface
             between the cluster and solid $M_{cs}$. The contact lines (yellow lines) can be decomposed into a discrete number of loops
            for each cluster. Right - Comparison of experimental and modeled hysteretic contact angle data from the work of Morrow ~\cite{morrowEffectsSurfaceRoughness1975} to the theoretical results of the deficit curvature model. The theoretical results are for a droplet on a solid surface, which are in strong agreement with the arithmetic average of $\theta_a$ and $\theta_r$ for both the experimental data and hysteresis model. Hence, it is evident that $\theta_{macro}$ involves not only an average of the microscopic contact angles to describe wetting at the macroscale, but also a global geometrical measure that accounts for contact angle hysteresis.
            Taken from Sun \textit{et al.}~\cite{sunProbingEffectiveWetting2020i}}
			\label{fig:deficitcurvature}
		\end{figure}

        The relation between the integral of the Gaussian curvature $\kappa_T$ and Euler characteristic $\chi$ holds for closed manifolds $M$. In the context of multiphase flow in porous media, this implies a completely non-wetting situation where the non-wetting phase is completely surrounded by the wetting phase, for instance, oil or gas in a strongly water-wet system being separated by water (films) from the solid surface. However, in most realistic situations, there is partial wetting where a three-phase contact line (which might be associated with a line tension~\cite{lawLineTensionIts2017}) is formed and the non-wetting phase no longer has a closed surface, as shown in Fig. ~\ref{fig:deficitcurvature}. In this case, the relationship between the Gaussian curvature and Euler characteristic is described by the Gauss-Bonnet theorem \cite{sunProbingEffectiveWetting2020i}. 

        \begin{equation}
            2 \pi \chi(M) = \int_{M}{\kappa_T}dS + \int_{\partial M}{\kappa_g}dC
            \label{eqn:GaussBonnetTheorem}
        \end{equation}

        The Gauss-Bonnet theorem can be regarded as the conservation equation for Gaussian curvature, which adds up to $2 \pi$ times the Euler characteristic $\chi$. 
        The first term on the right-hand side describes the surface integral of the Gaussian curvature of the droplet surface. However, this surface integral cannot be fully executed because of the intersection of the droplet with the solid. This ''deficit curvature'' (dashed line in Fig. ~\ref{fig:deficitcurvature}a) is captured in the second term on the right-hand side, which is the line integral of the geodesic curvature $\kappa_g$. 

        Sun et al. associated the deficit curvature, i.e., the line integral of the geodesic curvature, with contact angle $\theta$. It is important to note that the contact angle is a geometric concept. 
        In the next step, they showed that the Young-Laplace equation can be derived from the deficit curvature by applying equilibrium thermodynamics conditions with the additional assumption that the droplet has a hemispherical geometry and sitting on a flat surface~\cite{sunCharacterizationWettingUsing2020}. In a follow-up study, they demonstrated that more complex wetting states such as the Cassie-Baxter and Wenzel states can also be derived from deficit curvature, and thereby this concept becomes a universal and most general description of wetting~\cite{sunUniversalDescriptionWetting2022a}.  
        
\begin{figure}[ht]
			\includegraphics[width=0.7\linewidth]{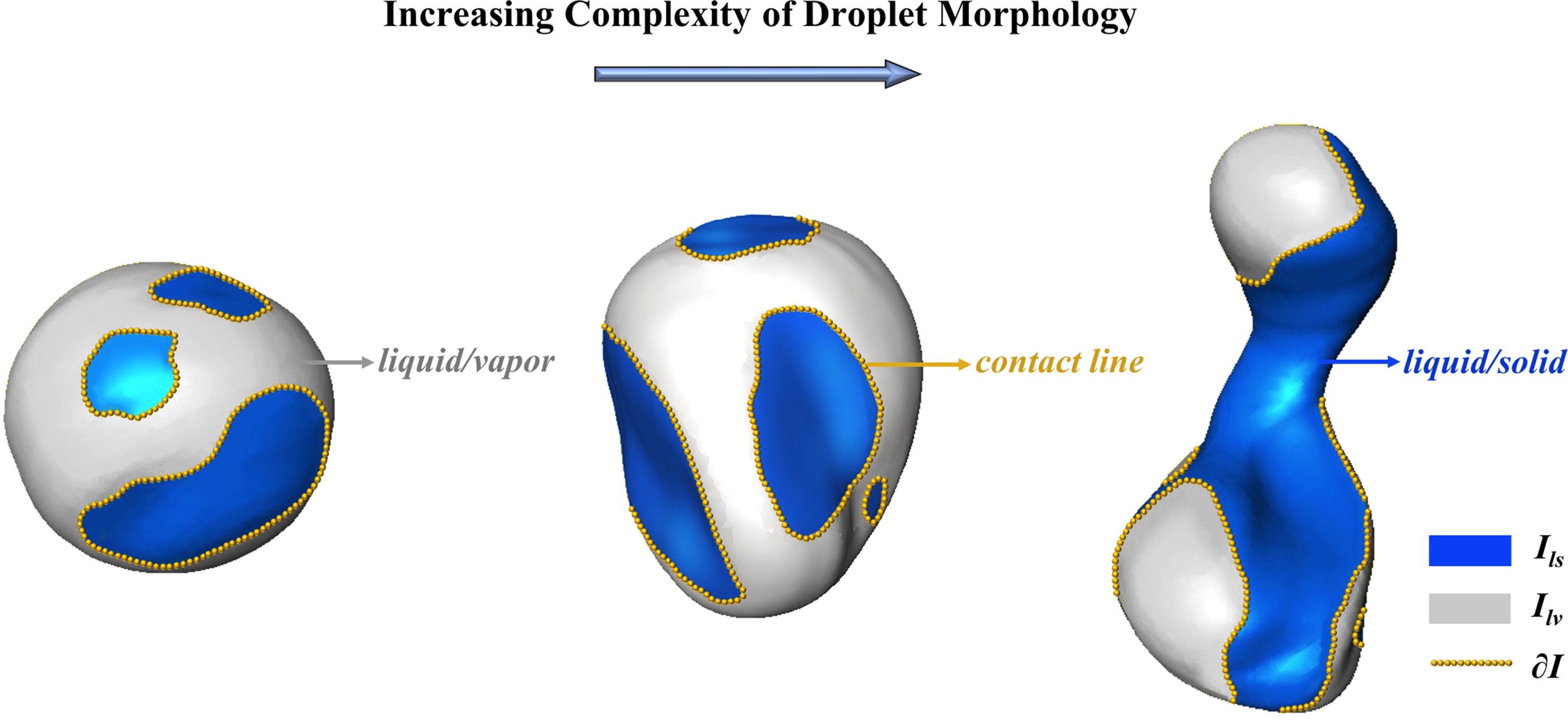}
			\caption{Contact loop for increasingly complex morphology.}
			\label{fig:contactloop}
		\end{figure}

        A common misconception (due to how historically the contact angle was measured~\cite{haghaniReviewWettabilityCharacterization2025}) is that the common loop is often called a point or line. This is not the case; it is a loop and has curvature, as illustrated in Fig. ~\ref{fig:contactloop}. Calling it a line is not helpful when trying to understand complex morphologies due to different wetting conditions.

		\section{Nonequilibrium effects and flow regimes}\label{sec:nonequilibriumeffects}

        \subsection{Pore scale displacements, flow patterns and flow regimes}\label{sec:flowregimes}

\begin{figure}[htbp]
			\includegraphics[width=0.8\linewidth]{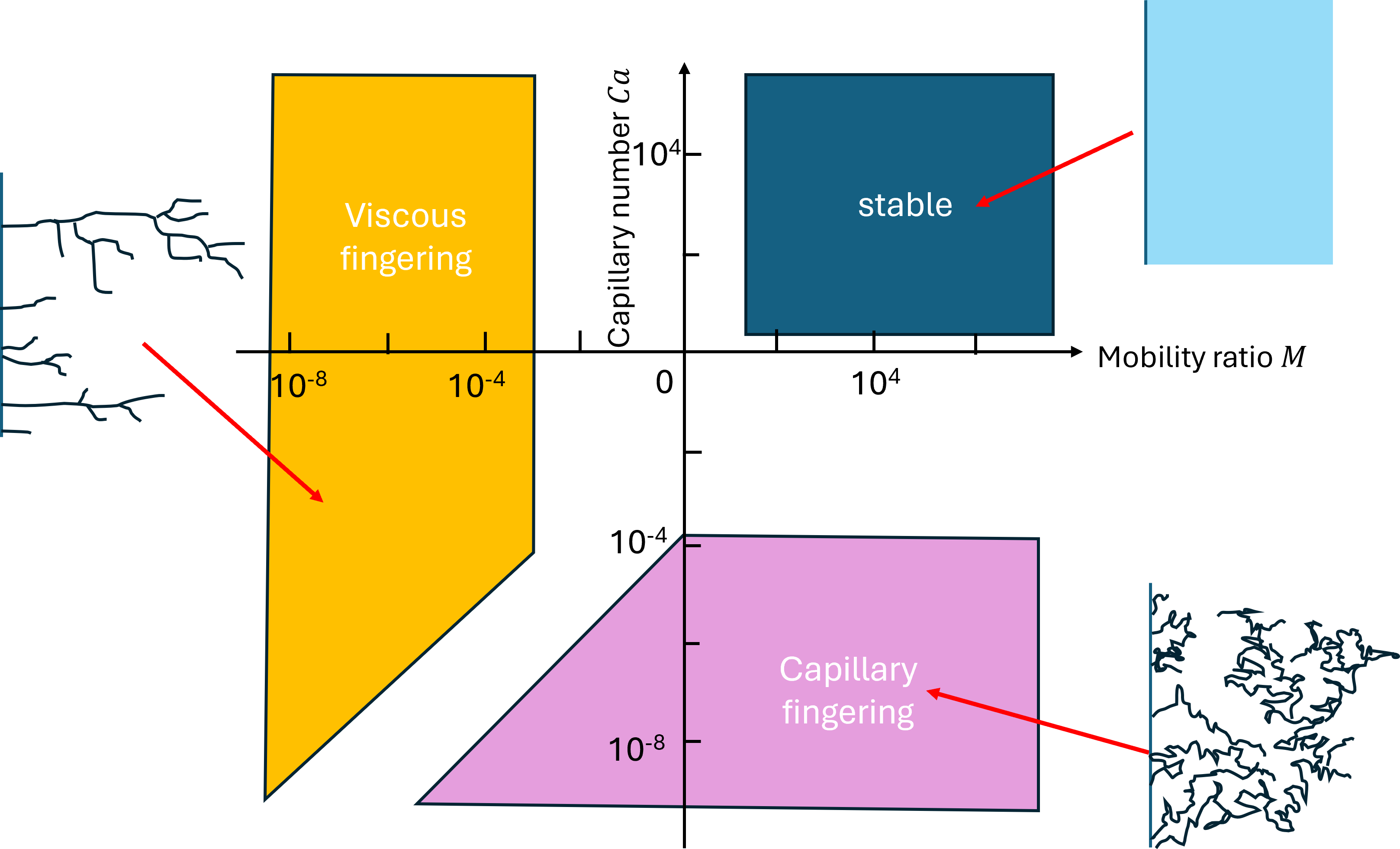}
			\caption{Pore scale flow regimes categorized into stable, capillary fingering and viscous fingering based on capillary number $Ca$ and viscosity ratio $M$ ~\cite{lenormandNumericalModelsExperiments1988}}
			\label{fig:Lenormandregimes}
		\end{figure}

        Pore-scale flow regimes have traditionally been categorized based on capillary number and viscosity ratio into stable, capillary, and viscous fingering~\cite{lenormandNumericalModelsExperiments1988} as shown in Fig. ~\ref{fig:Lenormandregimes}.

\begin{figure}[htbp]
			\includegraphics[width=0.7\linewidth]{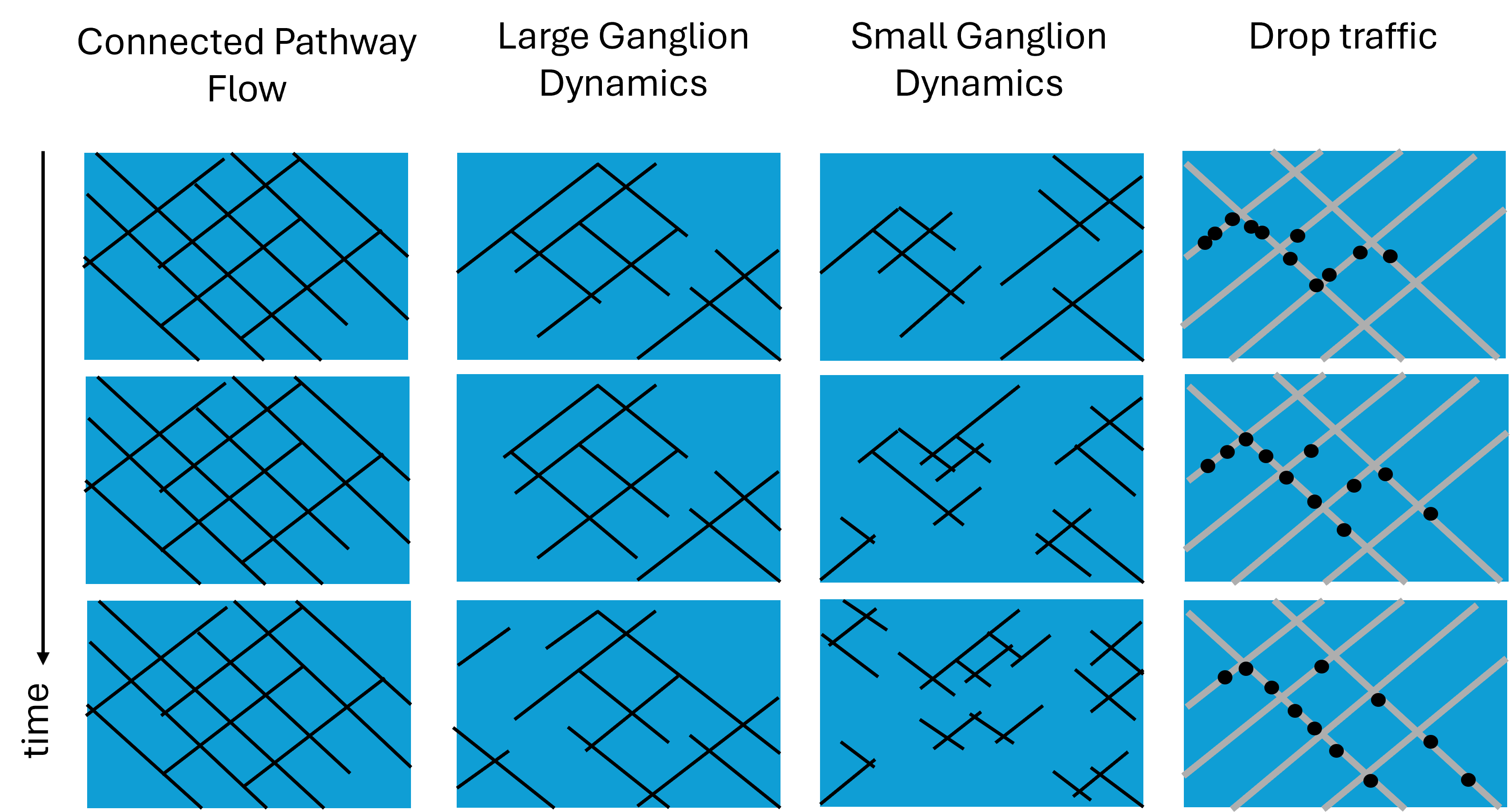}
			\caption{Pore scale flow regimes include next to connected pathway flow also large and small ganglion dynamics and drop traffic~\cite{avraamFlowRegimesRelative1995} in which the flowing phases are not permanently connected from inlet to outlet.}
			\label{fig:AvraamPayatakesJFM1995}
		\end{figure}

        This description, like many other models for 2-phase flow, such as capillary tube models~\cite{tullerHydraulicConductivityVariably2001} (from which the Brooks-Corey capillary pressure and relative permeability model~\cite{brooksPropertiesPorousMedia1966} can be analytically derived) and quasistatic pore network models~\cite{bluntPorescaleImagingModelling2013,bluntMultiphaseFlowPermeable2017} assume implicitly that flux occurs only through connected pathways.  However, Avraam \& Payatakes~\cite{payatakesDynamicsOilGanglia1982, avraamFlowRegimesRelative1995, avraamGeneralizedRelativePermeability1995} demonstrated in micromodels that there are more flow regimes, which can be part of ganglion dynamics and drop traffic (Fig. ~\ref{fig:AvraamPayatakesJFM1995}). 

\begin{figure}[htbp]
			\includegraphics[width=0.3\linewidth]{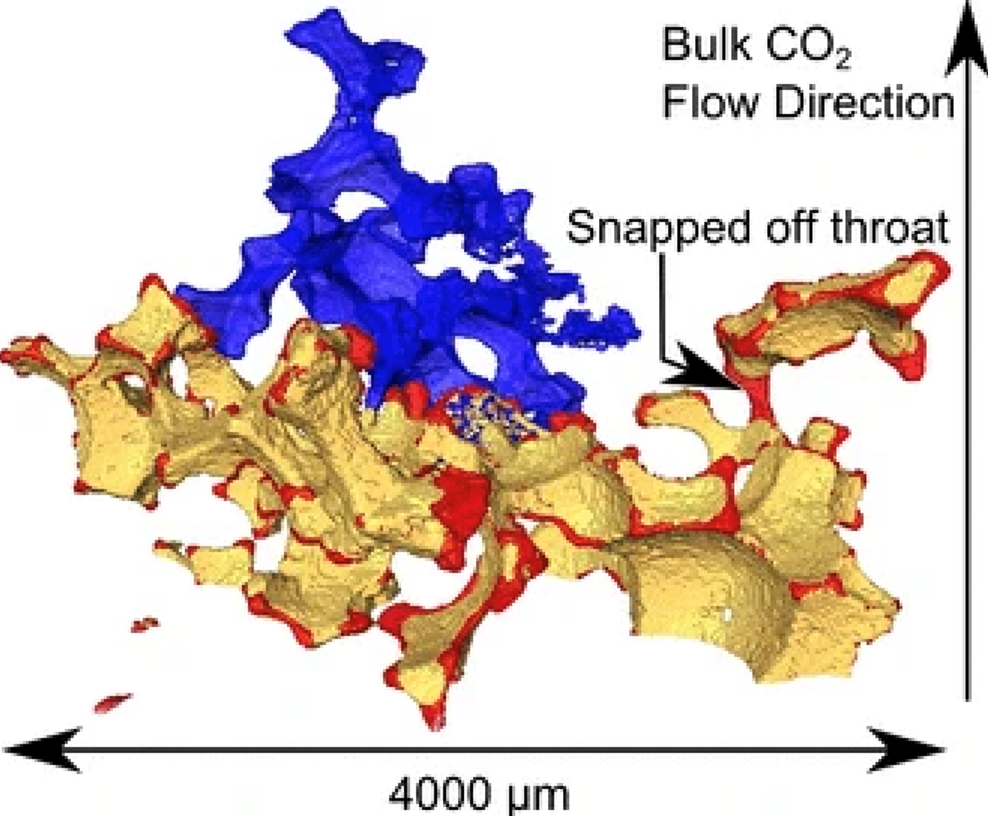}
			\caption{Cooperative pore filling event where the newly filled pore space (blue) leads to a retraction of menisci (red) in the regions already filled with non-wetting phase, and even a snap-off event. Taken from~\cite{andrewImagingDynamicMultiphase2015}.}
			\label{fig:AndrewCooperativeFilling}
		\end{figure}

       Ganglion-dynamic flow regimes also occur in 3D porous media~\cite{Ruecker2015} and can transport flux~\cite{Armstrong2016}. Ganglion dynamics are  complex phenomena consisting of cooperative dynamics at the pore scale. This can already occur during individual pore filling events such as Haines jumps~\cite{Haines1930} (Fig. ~\ref{fig:relaxationphenomena} on the right), where the filling event of a pore with a non-wetting phase leads to a retraction of menisci in adjacent pores. This has also been observed in 3D porous media, as shown in Fig. ~\ref{fig:AndrewCooperativeFilling}. In essence, the fluid volume for the Haines jump is drawn from nearby menisci, leading to much higher local flow rates than those supplied, for instance, by an injection pump. This can lead to more complex dynamics, such as cooperative filling events and burst dynamics~\cite{Berg2013, bergMultiphaseFlowPorous2014a, andrewImagingDynamicMultiphase2015} where filling events extend to multiple pores, as shown in Fig. ~\ref{fig:AndrewCooperativeFilling}. 
       
\begin{figure}[htbp]
			\includegraphics[width=1.0\linewidth]{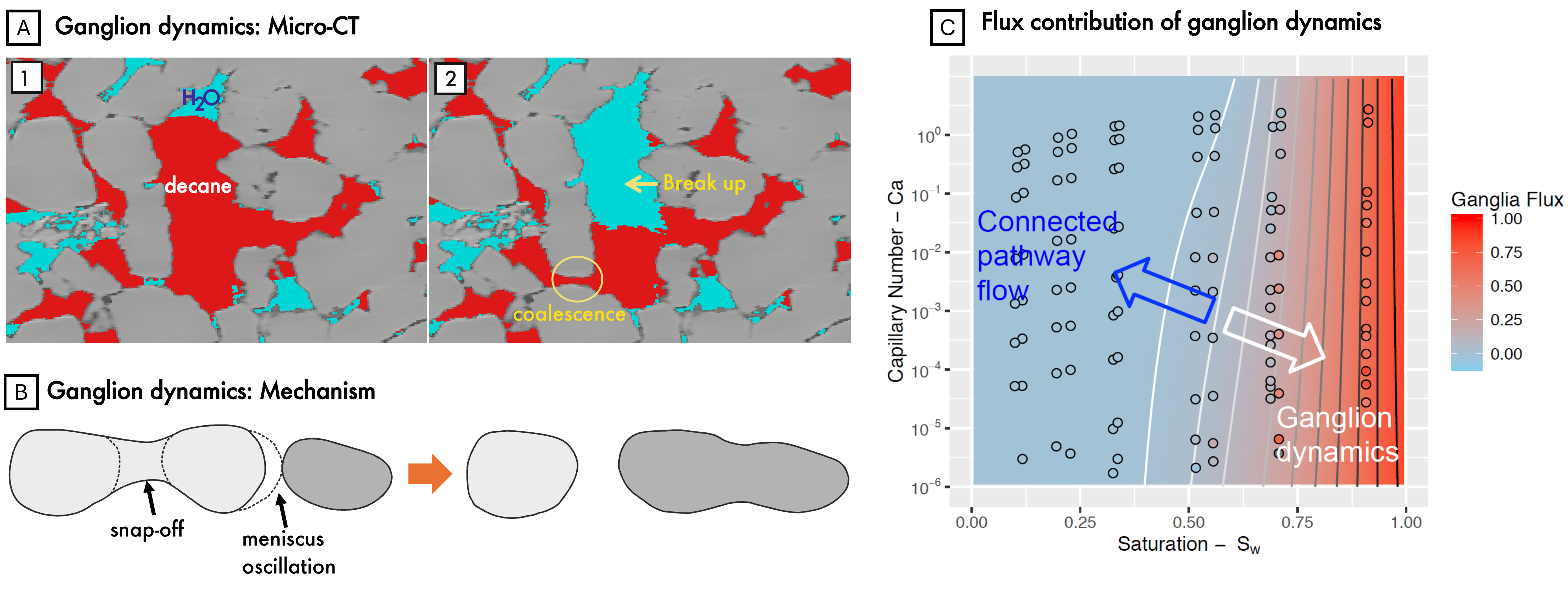}
			\caption{(A) Ganglion dynamics in imbibition imaged in-situ by micro-CT where a snap-off event causes a meniscus oscillation which leads to coalescence and re-connection (B) Taken from~\cite{Ruecker2015}. Such coalescence events are key mechanisms to restore reversibility across pore-scale displacement events with topological changes in the pore-scale fluid distribution, while over most of the mobile saturation range, the contribution of moving ganglia to the total flux is small (C). For most of the mobile saturation range, the dominant flux contribution is connected pathway flow. Only close to the residual non-wetting phase saturation there is significant flux contribution of moving clusters (taken from~\cite{Armstrong2016}).}
			\label{fig:RueckerGanglionDynamics}
		\end{figure}

       This can also lead to disconnection-reconnection processes in which the phases remain mobile even though they are not permanently connected~\cite{Ruecker2015}. 
       An example is shown in Fig. ~\ref{fig:RueckerGanglionDynamics} where during imbibition, a snap-off event causes a meniscus oscillation that leads to coalescence and reconnection. This implies that phases that are not permanently connected can remain mobile, which constitutes the flow regime. However, the flux contribution of moving ganglia remains very small compared to connected pathway flow, except for a small saturation range close to the residual non-wetting phase saturation, as shown by the Lattice Boltzmann simulations in Fig. ~\ref{fig:RueckerGanglionDynamics}C \cite{Armstrong2016}.

\begin{figure}[ht]
			\includegraphics[width=1.0\linewidth]{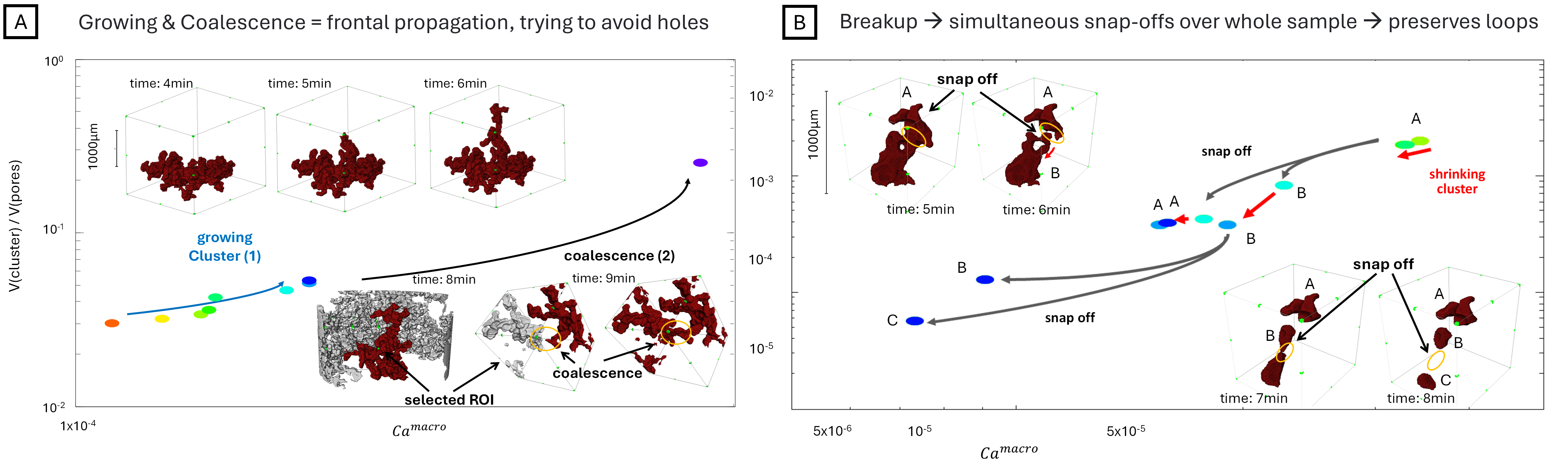}
			\caption{Growth and coalescence (A) and cooperative breakup (B) regimes observed in drainage and imbibition cycles. In a representation of relative cluster volume (i.e. non-wetting phase saturation) vs. cluster-based capillary number~\cite{armstrongCriticalCapillaryNumber2014a} these regimes are characterized by trajectories because the cluster-based capillary number contains the length of the cluster~\cite{Ruecker2015,Ruecker2021}. }
			\label{fig:flowregimestrajectoriesclustercapillarynumber}
		\end{figure}

        This observation is significant because it demonstrates that in typical capillary-dominated flow regimes, microscopic reversibility is restored beyond the topological change by capillary fluctuations triggered by capillary events even several pores away~\cite{armstrongModelingVelocityField2015b}. 
        For isolated snap-off events, in absence of such capillary fluctuations,    
        microscopic reversibility is only given up to topological change and not beyond~\cite{Steijn2009}. This can be also seen by following the fate of clusters over longer times during coalescence and breakup as displayed in Fig.~\ref{fig:flowregimestrajectoriesclustercapillarynumber} which ultimately restores reversibility in the capillary-dominated flow regimes where capillary states become accessible via sequences of coalescence and breakup, which is associated with trajectories in saturation. 

        The insights gained from \textit{in situ} imaging experiments conducted during the past decade paint a more complex picture that suggests a transition from the pore to the Darcy scale, where fluctuations of individual parameters such as saturation and pressure never average out and are an integral part of flow regimes~\cite{Ruecker2021}. Additionally, pore-scale displacement events are dissipative. Depending on the flow regime, between 10 and 90\% of the invested energy is dissipated in the pore-scale displacement event and hence is not available locally to drive flow~\cite{sethEfficiencyConversionWork2007,Berg2013}. 
        Therefore, a theory must handle fluctuations in individual state variables and consider their energy dynamics.

          \subsection{Flow regimes and dependent and independent state variables}
\begin{figure}[ht]
			\includegraphics[width=0.7\linewidth]{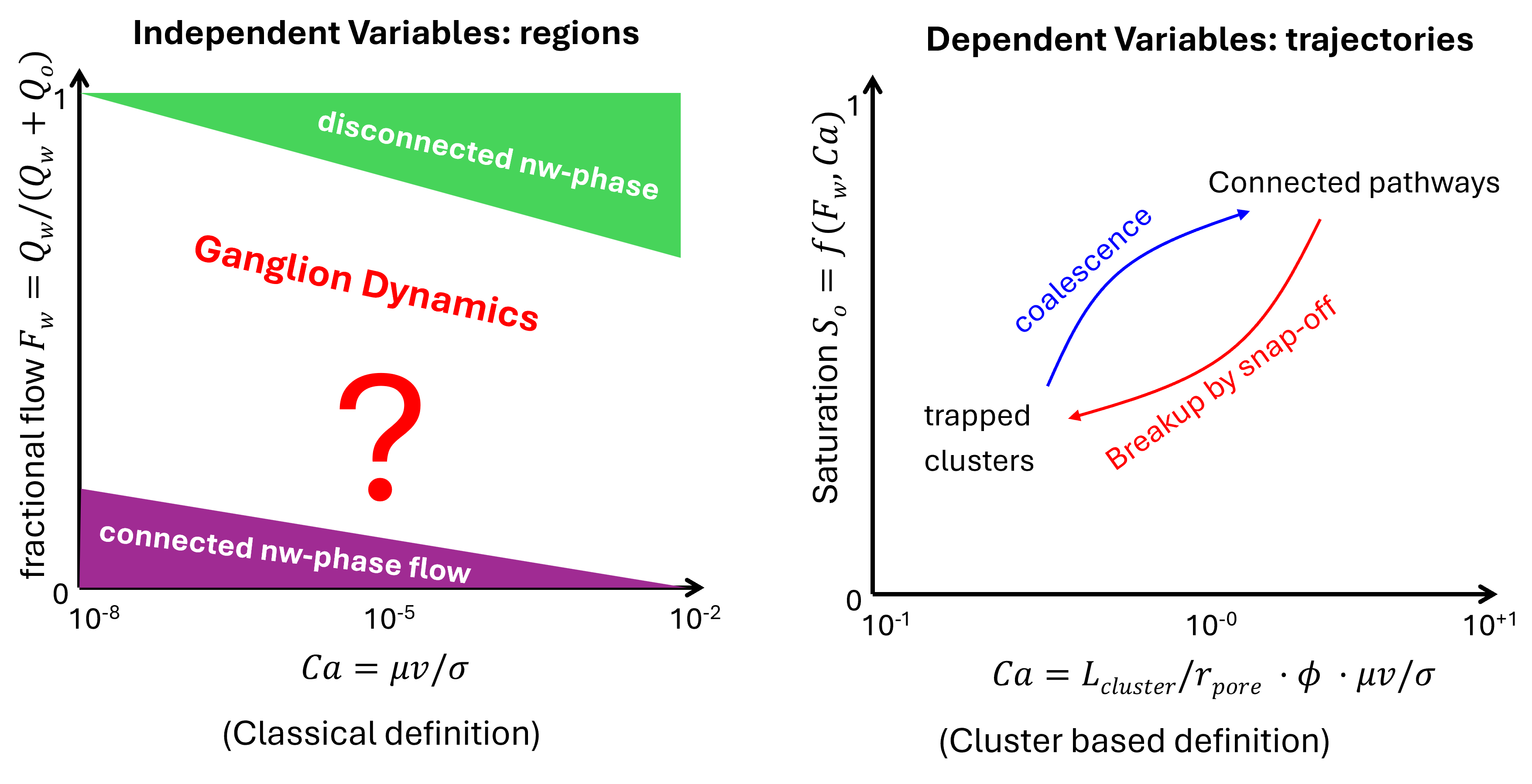}
			\caption{Sketch of the complications when constructing ''phase diagrams'' for flow regimes with respect to the choice of state variables. Independent parameters such as fractional flow $f_w$ and the traditionally defined capillary number $Ca=\mu v /\sigma$ (eq.~\ref{eqn:capillarynumber}) span a phase diagram with \textit{independent}, where flow regimes are static regions (left). However, for dependent variables such as saturation $S_w$ or Minkowski functionals (which are in this context all \textit{dependent variables}) and the cluster-based capillary number from eq.~\ref{eqn:clusterbasedcapillarynumber}~\cite{armstrongCriticalCapillaryNumber2014a}, flow regimes manifest themselves as trajectories~\cite{Ruecker2015,Ruecker2021} as observed in experiments shown in Fig.~\ref{fig:flowregimestrajectoriesclustercapillarynumber}. (after \cite{Ruecker2015}) }
			\label{fig:flowregimes}
		\end{figure}

        The Minkowski functionals which represent the geometric state are applicable for quasi-static situations as they do not contain time. For flow new variables and parameters become important such as viscosity ratios, capillary number~\cite{Armstrong2016, zhang2022nonlinear,suwandiRelativePermeabilityVariation2022} (visco-capillary balance),  fractional flow $f_w$,  Ohnesorge number $Oh$~\cite{chenInertialEffectsProcess2019,zacharoudiouPoreScaleModelingDrainage2020} (capturing inertial effects), etc. However, there are potential pitfalls. A ''phase diagram'' represents the Darcy scale flow regime, and therefore, the respective parameterization needs to be Darcy scale as well. However, some concepts, such as the traditionally defined capillary number 
        
        \begin{equation}\label{eqn:capillarynumber}
            Ca=\frac{\mu_w v}{\sigma }
        \end{equation}
        
        \noindent are pore-scale definitions. One of the consequences is that the transition from capillary-dominated to viscous dominated flow occurs at (this) capillary number $10^{-5}$ which is misleading (flow regimes should be defined by the dimensionless number changes from $<1$ to $>1$). The cluster-based capillary number concept~\cite{armstrongCriticalCapillaryNumber2014a}
        \begin{equation}\label{eqn:clusterbasedcapillarynumber}
              Ca_{cluster}= \frac{L_{cl} \mu_w V_{Darcy}}{k_{r,w} P_c} \approx \frac{L_{cl}}{r_{pore}} Ca
        \end{equation}
        \noindent which takes the length scale of non-wetting phase clusters $L_{cl}$ into account correctly marks in capillary de-saturation experiments the transition between capillary and viscous regimes around $1$ but leads to a \textit{dependent} variable. However, as shown in Fig. ~\ref{fig:flowregimes}, only \textit{independent} variables define regimes in static regions, whereas for more meaningful variables, such as the cluster-based capillary number~\cite{armstrongCriticalCapillaryNumber2014a} the regimes manifest as trajectories.  

        \subsection{Fluctuations from pore to Darcy scale}      
\begin{figure}[htbp]
			\includegraphics[width=1.0\linewidth]{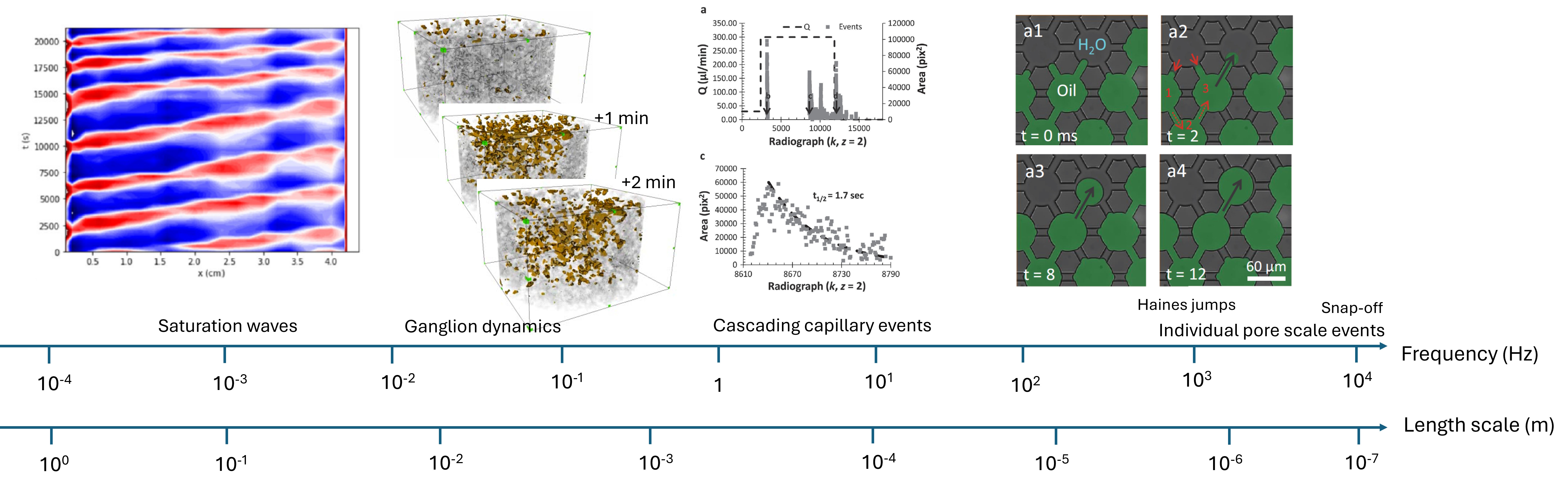}
			\caption{Relaxation phenomena in multi-phase flow in porous media within typical frequency ranges~\cite{bedeauxFluctuationdissipationTheoremsMultiphase2025}. This overview shows that there is a continuous transition from phenomena in individual pores (high frequency, right~\cite{armstrongInterfacialVelocitiesCapillary2013}) over cooperative displacement events ~\cite{armstrongSubsecondPorescaleDisplacement2014,ruckerConnectedPathwayFlow2015} to the Darcy scale (traveling wave solutions that follow fractional flow physics on the left~\cite{Ruecker2021}) without a length scale at which fluctuations have averaged out.}
			\label{fig:relaxationphenomena}
		\end{figure}

        Fluctuations and relaxation phenomena in multiphase flow~\cite{dattaFluidBreakupSimultaneous2014,masalmehLowSalinityFlooding2014,reynoldsCharacterizingFlowBehavior2015,soropRelativePermeabilityMeasurements2015,gaoXrayMicrotomographyIntermittency2017,gaoPoreOccupancyRelative2019,gao2020pore,linImagingMeasurementPoreScale2018,alcornCorescaleSensitivityStudy2020,spurinIntermittentFluidConnectivity2019,wangObtainingHighQuality2019,menkeUsingNanoXRMHighContrast2022} range from (sub)millisecond time scales associated with pore scale displacement events~\cite{dicarloAcousticMeasurementsPorescale2003,armstrongInterfacialVelocitiesCapillary2013} to cascading and cooperative phenomena ~\cite{armstrongSubsecondPorescaleDisplacement2014} on the second time scale to ganglion dynamics~\cite{Ruecker2015}        
        at the minute scale, and traveling wave solutions~\cite{Ruecker2021} at the scale of tens of minutes. An overview of the different relaxation phenomena and the associated time/frequency scale is shown in Fig. ~\ref{fig:relaxationphenomena}. 

        At the pore scale, displacement events are cooperative; for instance, during Haines jumps, the advancing meniscus in one pore is connected via viscous pressure gradients to several receding menisci in adjacent pores~\cite{armstrongInterfacialVelocitiesCapillary2013}. Furthermore, at the mesoscale, which consists of many pores, Haines jumps manifest themselves as cascading, avalanche, or bust-like events followed by a relaxation time of a few seconds ~\cite{armstrongSubsecondPorescaleDisplacement2014}.
        While the frequency of small cooperative events that involve up to 10 pores follows the predictions of simple percolation models, larger burst events that involve hundreds or thousands of pores are significantly more  frequent than the prediction by percolation theory~\cite{bergMultiphaseFlowPorous2014a} and provide the most significant contribution to saturation changes at the Darcy scale. 

\begin{figure}[htbp]
			\includegraphics[width=1.0\linewidth]{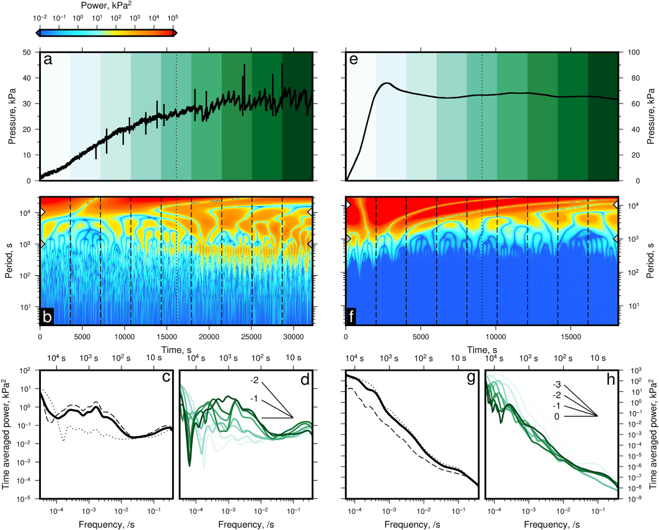}
			\caption{Spectral analysis of pressure time series from pore-scale experiments. (a) Black curve = pressure from the gas/water experiment. Green strips and dotted line = time intervals indicated in panels b–d. (b) Power spectrum calculated by transforming the black curve in panel (a). The dashed and dotted lines correspond to the time intervals indicated in panel (a). 
            Gray/white arrowheads indicate limits on low-pass filters (at periods, $P \approx 10^3$ and $\approx 10^4$ s). (c) Thick black curve = time-averaged, rectified, power spectrum for the entire series. Dotted and dashed curves = time-averaged power for the first and second halves of the time series, respectively (separated by dotted lines in panels a and b). 
            (taken from~\cite{spurin2023pore})}
			\label{fig:Spurinfluctuations}
		\end{figure}

        Wavelet analysis of the pressure response during multiphase flow experiments reveals a wide spectral range of fluctuations~\cite{spurinRedNoiseSteadyState2022, spurin2023pore}. The example displayed in Fig.~\ref{fig:Spurinfluctuations} reveals the fluctuation dynamics for intermittent gas-liquid displacements ranging from below $10^{-4}$~Hz to above $10^{-1}$~Hz.

\begin{figure}[htbp]
        \begin{center}
        \includegraphics[width=1.0\textwidth]{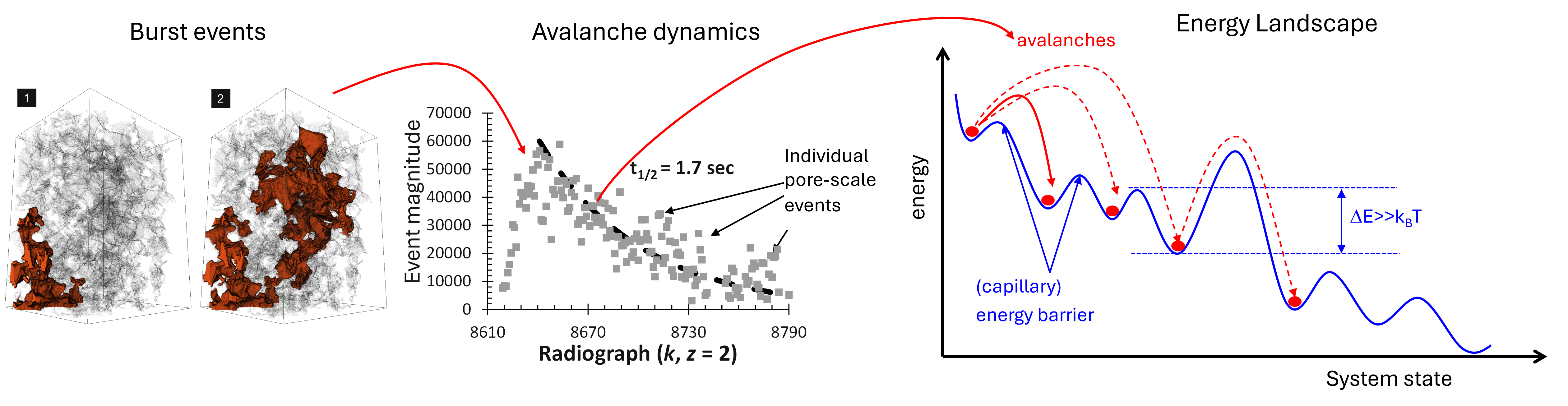}
        \end{center}
        \caption{Cartoon of the energy landscape~\cite{onuchicTHEORYPROTEINFOLDING1997,bhandarEnergyLandscapesBistability2004,debenedettiSupercooledLiquidsGlass2001,debenedettiEquationStateEnergy1999,cueto-felguerosoDiscretedomainDescriptionMultiphase2016,wangTrappedLiquidDrop2013} which is motivated by the experimental observation of avalanche-like dynamics and burst events~\cite{crandallDistributionOccurrenceLocalizedbursts2009} and relaxation phenomena~\cite{armstrongSubsecondPorescaleDisplacement2014}.  \label{fig:energylandscape}}
        \end{figure}

        Some of the experimental observations such as burst dynamics and subsequent avalanche-like dynamics suggest~\cite{armstrongSubsecondPorescaleDisplacement2014} a complex energy landscape with meta-stable states~\cite{onuchicTHEORYPROTEINFOLDING1997,bhandarEnergyLandscapesBistability2004,debenedettiSupercooledLiquidsGlass2001,debenedettiEquationStateEnergy1999,cueto-felguerosoDiscretedomainDescriptionMultiphase2016,wangTrappedLiquidDrop2013} as indicated in the cartoon in Fig.~\ref{fig:energylandscape}. However, the exact form of the energy landscape is an open question.

        These examples show that there is a continuous transition from phenomena in individual pores (high frequency, right~\cite{armstrongInterfacialVelocitiesCapillary2013}) over cooperative displacement events ~\cite{armstrongSubsecondPorescaleDisplacement2014,ruckerConnectedPathwayFlow2015} to the Darcy scale (traveling wave solutions that follow fractional flow physics on the left~\cite{Ruecker2021}) without a length scale at which fluctuations have averaged out. This implies that
        \begin{itemize}
            \item Fluctuations and related complex spatio-temporal dynamics are an inherent part of the system and need to be an integral part of the upscaling strategy. Simple volume averaging or homogenization will not suffice.
            \item a representative elementary volume for Darcy scale multiphase flow cannot be defined on a spatial dimension alone. Instead, it is defined based on space-time averaging~\cite{McClure2025}.
        \end{itemize}

        \cleardoublepage
        \newpage
        \subsection{Relaxation phenomena impact state variables}
\begin{figure}[ht]
			\includegraphics[width=0.6\linewidth]{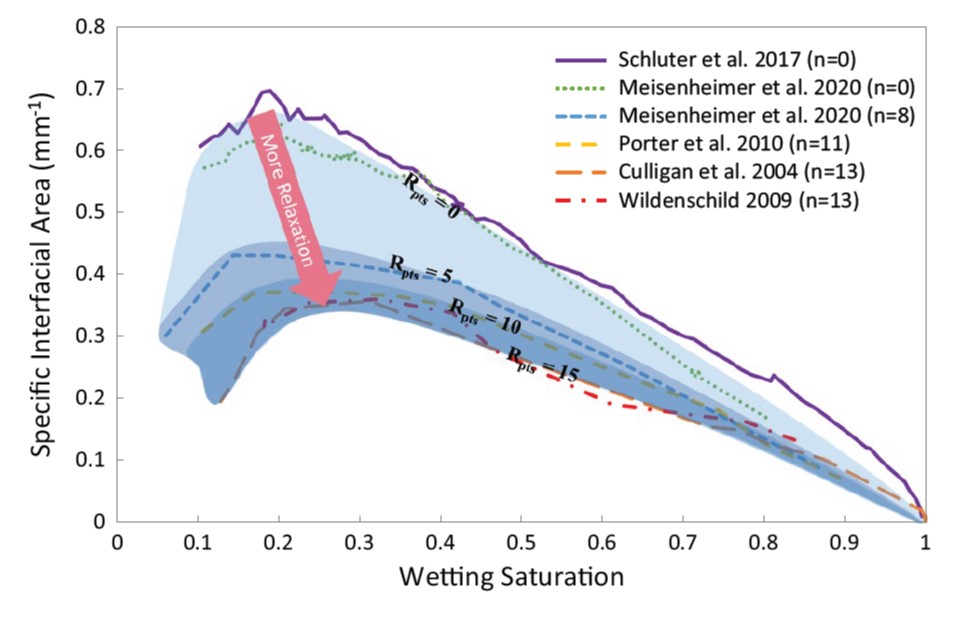}
			\caption{Specific interfacial area $a_{nw}$ vs. saturation $S_w$ for various primary drainage experiments with glass beads where the number of equilibrium steps $R_{pts}$ in each experiment was decreased from 15 to 0, i.e. less and less equilibration was allowed to happen. The key finding is that the more equilibration steps, the more relaxation and less interfacial area generated. This underlines the impact of nonequilibrium effects on state variables, such as interfacial area. Taken from \cite{meisenheimerPredictingEffectRelaxation2021}}
			\label{fig:Meisenheimeranwrelaxation}
		\end{figure}

        These relaxation phenomena can lead to nonequilibrium effects, depending on whether the time scales of external forcing are faster or slower than the intrinsic relaxation times associated with these phenomena. This has a direct impact on the behavior of state variables, such as the interfacial area $a_{nw}(S_w)$, as shown in Fig. ~\ref{fig:Meisenheimeranwrelaxation}. The more the system is allowed to equilibrate, i.e., the more relaxation, the less interfacial area $a_{nw}$ is generated~\cite{meisenheimerPredictingEffectRelaxation2021}. This demonstrates the impact of nonequilibrium effects on state variables, such as interfacial area. 

\begin{figure}[htbp]
			\includegraphics[width=0.8\linewidth]{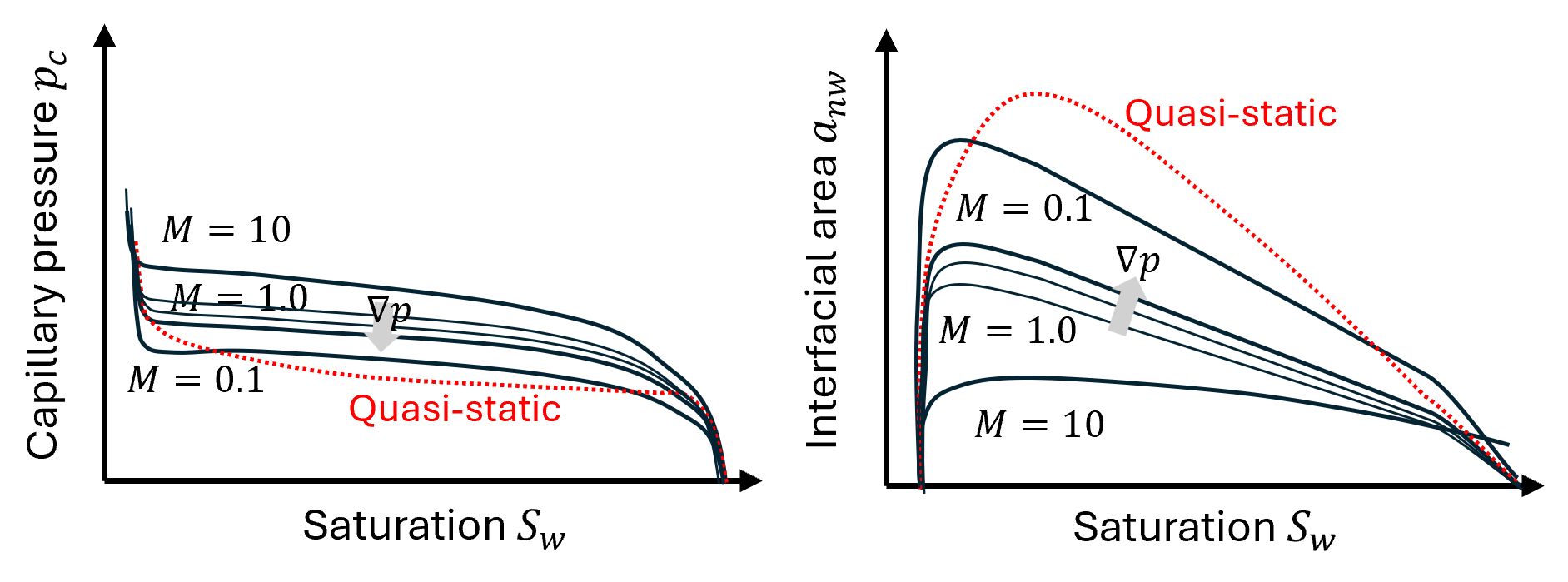}
			\caption{Nonequilibrium effects on capillary pressure $p_c(S_w)$ and interfacial area $a_{nw}(S_w)$. Dynamic pore network modeling simulations were performed for a range of pressure gradients $\nabla p$ and viscosity ratios $M$. Non-equilibrium effects consistently impact the $p_c - S_w - a_{nw}$ surface~\cite{joekar-niasarNonequilibriumEffectsCapillarity2010a}.}
			\label{fig:JoekarNiasarJFM2010}
		\end{figure}

        Similar effects were also observed for the dynamic pore network modeling~\cite{joekar-niasarNonequilibriumEffectsCapillarity2010a} of capillary pressure $p_c(S_w)$ and interfacial area $a_{nw}(S_w)$ as shown in Fig. ~\ref{fig:JoekarNiasarJFM2010}. The higher the pressure gradient during flow, i.e., the less relaxation, the higher the interfacial area, which is consistent with the experimental findings~\cite{meisenheimerPredictingEffectRelaxation2021} from Fig.~\ref{fig:Meisenheimeranwrelaxation}. 

        The relaxation behavior in the dynamic capillary pressure was then described by introducing a relaxation term $-\tau \partial S_w / \partial t$ into the phase pressure difference $p_n - p_w$ \cite{hassanizadehThermodynamicBasisCapillary1993a,hassanizadehDynamicEffectCapillary2002}

        \begin{equation}
            p_n - p_w = p_c(S_w) - \tau \frac{\partial S_w}{\partial t}
            \label{eqn:dynamicpc}
        \end{equation}

\begin{figure}[htbp]
			\includegraphics[width=0.8\linewidth]{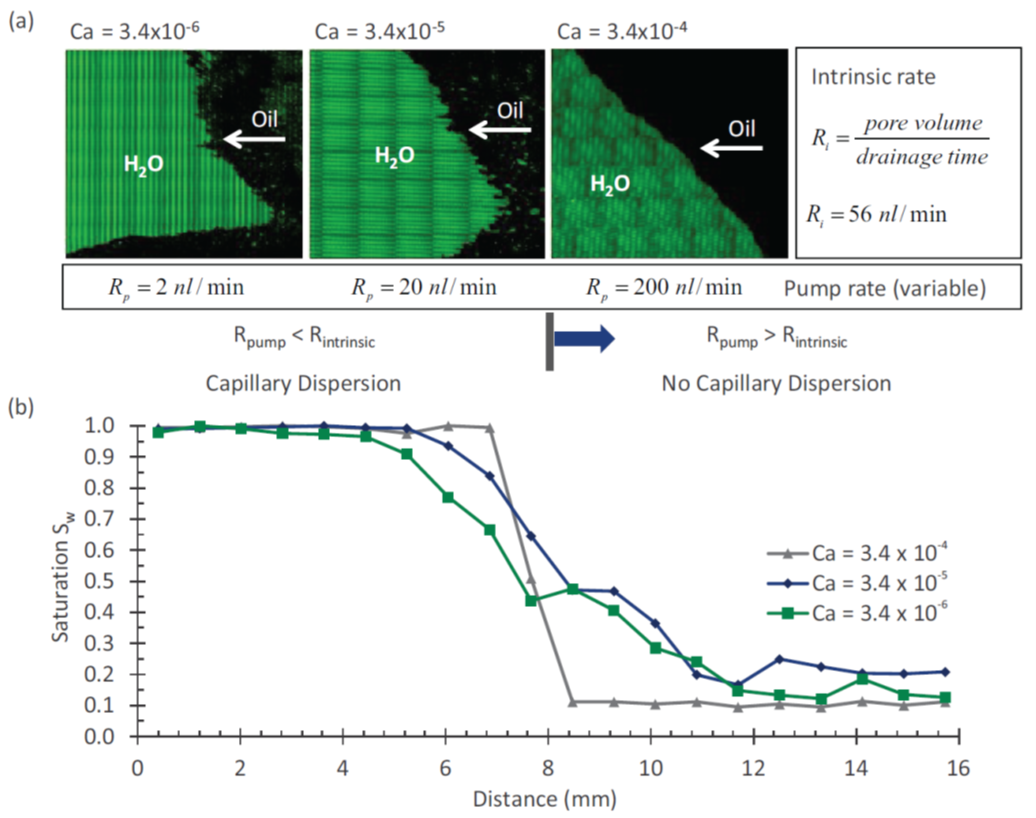}
			\caption{Displacement front in a micromodel as a function of injection rate $R_p$.  As long as the injection rate $R_p$ is smaller than the intrinsic rate of Haines jumps $R_i$ the front is dispersed, which is in line with Lenormand's capillary fingering. For $R_p > R_i$ the front becomes sharp. This is consistent with an interpretation of capillary-dominated vs. viscous dominated flow, but is seen from the perspective of intrinsic time scales vs. time scales of external forcing.  Taken from \cite{armstrongInterfacialVelocitiesCapillary2013}}
			\label{fig:ArmstrongInterfacialVelocity}
		\end{figure}

        For $\tau \rightarrow 0$ the closed relationship of the classical phase pressure difference $p_n - p_w = p_c(S_w)$ with the equilibrium capillary pressure $p_c(S_w)$ is obtained. 
        The relaxation times $\tau$ are related to the range of physical phenomena depicted in Fig. ~\ref{fig:relaxationphenomena} and the associated time scales. These range from milliseconds for individual pore-scale events, such as Haines jumps~\cite{Haines1930} and seconds for cascading events~\cite{armstrongSubsecondPorescaleDisplacement2014} to hours~\cite{schluterTimeScalesRelaxation2017}. The relevant relaxation time scale is not always entirely clear.    
        It is clear that when a system is externally driven, it matters whether the rate of the external driving is comparable to the relaxation timescale of the intrinsic processes. Examples are polymer rheology, where the material behavior is viscous when the forcing is slower than the relaxation time and elastic when the forcing is faster than the intrinsic relaxation time. For multiphase flow in porous media, the work of Armstrong~\cite{armstrongInterfacialVelocitiesCapillary2013} demonstrated that a disperse displacement front was observed when the external forcing was slower than the intrinsic relaxation time and sharp when the external forcing was faster than the intrinsic relaxation time (see Fig. ~\ref{fig:ArmstrongInterfacialVelocity}, 
        
        \subsection{Non-monotonic saturation profiles highlight importance of relaxation phenomena at macroscopic scales}
\begin{figure}[htbp]
			\includegraphics[width=0.4\linewidth]{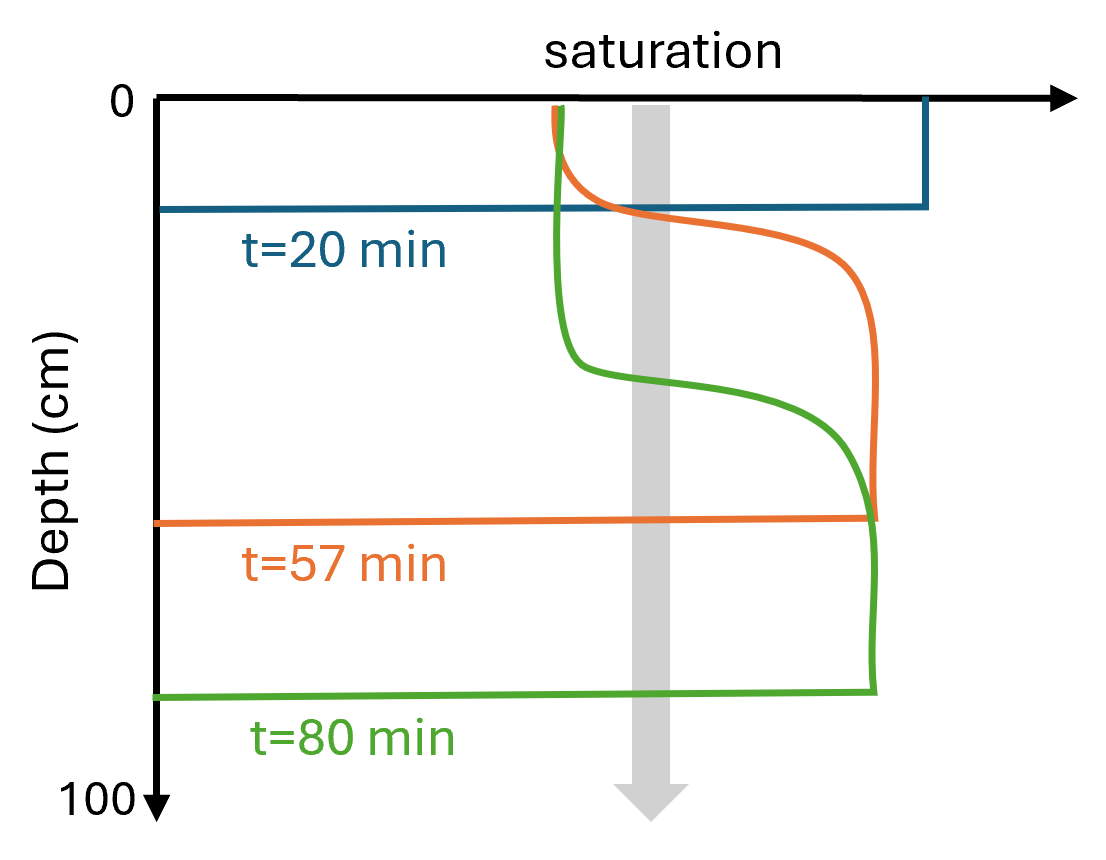}
			\caption{Water saturation and water pressure profiles during the infiltration of water in initially dry glass beads. A saturation overshoot was observed for which a range of models, including nonequilibrium models, were considered~\cite{shiozawaUnexpectedWaterContent2004}.}
			\label{fig:shiozawa2004}
		\end{figure}

        Eq.~\ref{eqn:dynamicpc} has been the starting point for a class of nonequilibrium models for Darcy-scale multiphase flow in porous media to explain saturation overshoot~\cite{shiozawaUnexpectedWaterContent2004,dicarloExperimentalMeasurementsSaturation2004}, which is observed when infiltrating water under specific conditions in dry soil (Fig. ~\ref{fig:shiozawa2004}). Non-monotonic saturation profiles were observed and could not be explained within the classical 2-phase Darcy description with static capillary pressure and relative permeability, which predict only monotonic solutions. 
        
        Several models have been proposed to explain this effect, ranging from fingering models~\cite{cueto-felguerosoNonlocalInterfaceDynamics2008} to nonequilibrium models~\cite{botteroNonequilibriumCapillarityEffects2011, steinleInfluenceInitialConditions2016}. The latter involves a combination of hysteresis and relaxation of the relative permeability and capillary pressure saturation functions, although Steinle \& Hilfer demonstrated that hysteresis is not a necessary model parameter, but the effect can be completely explained by a dynamic nonequilibrium effect~\cite{steinleInfluenceInitialConditions2016}. 

        These concepts have been further developed into non-equilibrium suction models in soil~\cite{einavHydrodynamicsNonEquilibriumSoil2023} which demonstrate a significantly different water retention than equilibrium models, that is, generate hysteresis. A key ingredient in the non-equilibrium model is meta-stable states, which are indicative of a non-trivial energy landscape with multiple local minima as sketched in Fig.~\ref{fig:energylandscape}.

        \subsection{Traveling saturation waves and nonlinear dynamics}
\begin{figure}[htbp]
			\includegraphics[width=1.0\linewidth]{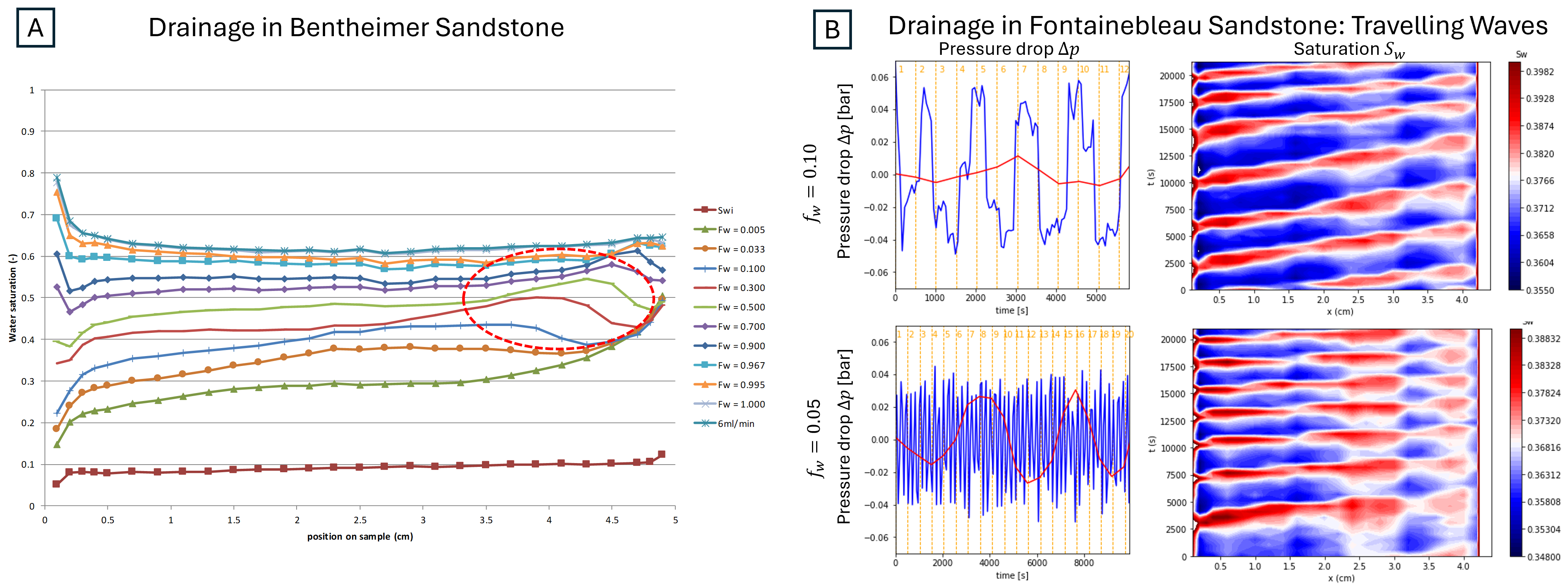}
			\caption{Non-monotonic saturation profiles in steady-state 2-phase flow special core analysis (SCAL) experiments with the purpose to determine relative permeability, conducted in strongly water-wet sandstone rock. (A) Drainage in the Bentheimer sandstone and (B) drainage in the Fontainebleau sandstone. In both cases, non-monotonic saturation profiles were observed. The space-time contour plot in (B) clearly shows traveling waves, and the corresponding pressure drop shows a very periodic behavior \cite{Ruecker2021}. Taken from \cite{Ruecker2021}}
			\label{fig:travellingwaves}
		\end{figure}

        Non-monotonic saturation profiles have also been observed in conventional steady-state core flooding experiments in special core analysis (SCAL) intended to determine the relative permeability (Fig. ~\ref{fig:travellingwaves}). The detailed space-time in Fig. ~\ref{fig:travellingwaves}B, and the pressure drop signal shows a traveling wave. 

\begin{figure}[htbp]
			\includegraphics[width=1.0\linewidth]{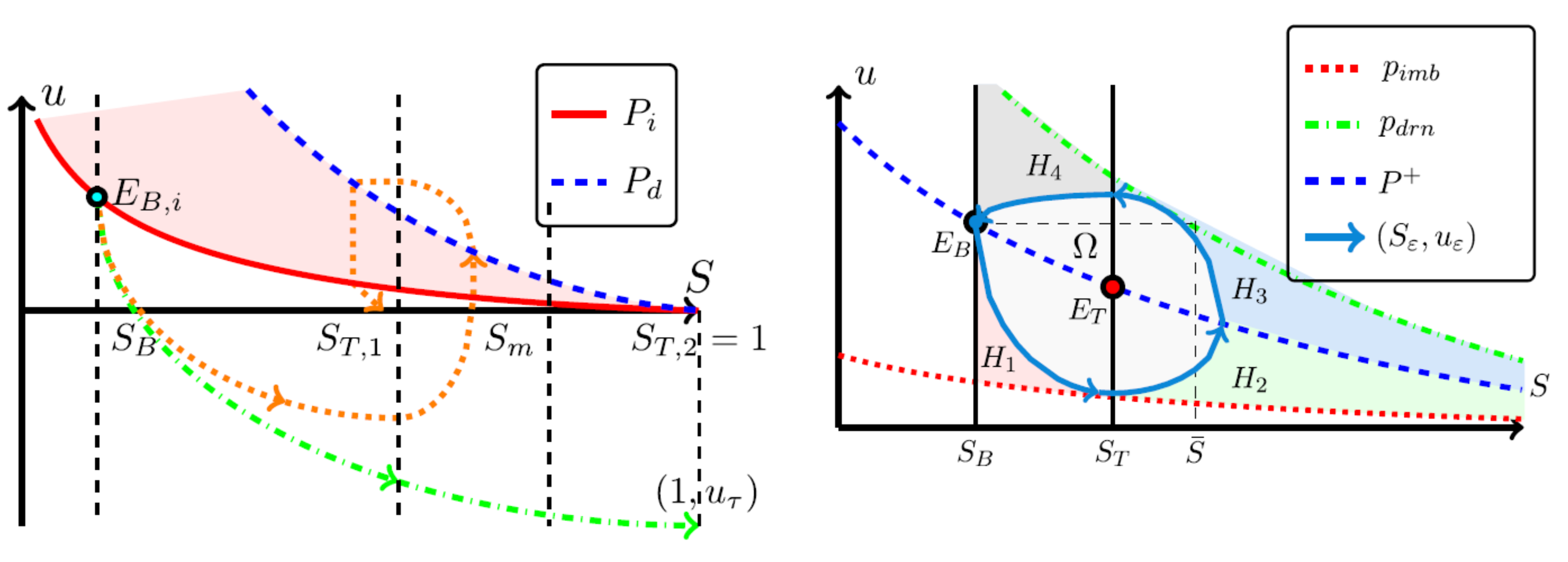}
			\caption{The analytical solutions for traveling waves with nonequilibrium models show orbits in the suction $u=-p$ -- saturation $S$ plane (left, taken from \cite{mitraWettingFrontsUnsaturated2019}). Under special conditions, closed orbits are possible ( taken from \cite{vanduijnTravellingWaveSolutions2018}). }
			\label{fig:MitraOrbits}
		\end{figure}

        R{\"u}cker and coauthors demonstrated that traveling waves follow fractional flow physics and are thus Darcy-scale phenomena~\cite{Ruecker2021}. However, the magnitudes of the pressure oscillations are similar to pore-scale displacement events, and the magnitude of the saturation change is similar to that in ganglion dynamics. Overall, the significance of this observation is that there is a continuous transition of nonequilibrium phenomena from the pore to the Darcy scale, as shown in Fig. ~\ref{fig:relaxationphenomena}.
        Such traveling waves have been described by nonequilibrium models that exhibit orbits in the pressure-saturation plane, as shown in Fig. ~\ref{fig:MitraOrbits} \cite{vanduijnTravellingWaveSolutions2018,mitraWettingFrontsUnsaturated2019,mitraFrontsTwophasePorous2020a}. The closed orbits shown in Fig. ~\ref{fig:MitraOrbits} on the right is a possible explanation for the periodic pressure and saturation behavior observed in Fig. ~\ref{fig:travellingwaves} (B)). 

        Because such solutions are, in principle, permissible and also transport more flux than a saturation profile constant in time~\cite{Ruecker2021}, traveling waves will form when triggered. Triggers can be capillary instabilities at the inlet and outlet, such as bubble pinch-off~    \cite{unsalBubbleSnapoffCapillaryBack2009, unsalCoCountercurrentSpontaneous2007, unsalCocurrentCountercurrentImbibition2007} but can in principle also be pore-scale displacement events inside the porous medium itself. Therefore, even if inlet instability is the trigger, the effect is still intrinsic to the porous medium itself. 

        While the Barenblatt nonequilibrium model for two-phase flow in porous media~\cite{barenblattMathematicalModelNonequilibrium2003} considers the dynamic effects as a nonequilibrium extension to the two-phase Darcy equations, which then leads to corrections, the more recent work that embraces the dynamics effects more in the context of nonlinear dynamics~\cite{vanduijnTravellingWaveSolutions2018,mitraWettingFrontsUnsaturated2019,mitraFrontsTwophasePorous2020a} shows the emergence of dynamic spatiotemporal patterns at the Darcy scale that are also observed experimentally~\cite{Ruecker2021} implying that new dynamics emerge rather than being a correction. 
        The wide spatio-temporal spectrum of nonequilibrium effects observed from pore to Darcy scale paints a picture where spatio-temporal dynamics (''fluctuations'')  -- which previously has often been mistaken as noise -- continuously transitions from pore to Darcy scales, thereby questioning traditional REV concepts~\cite{bearDynamicsofFluidsinPorousMedia1972}. 

\begin{figure}[htbp]
			\includegraphics[width=0.8\linewidth]{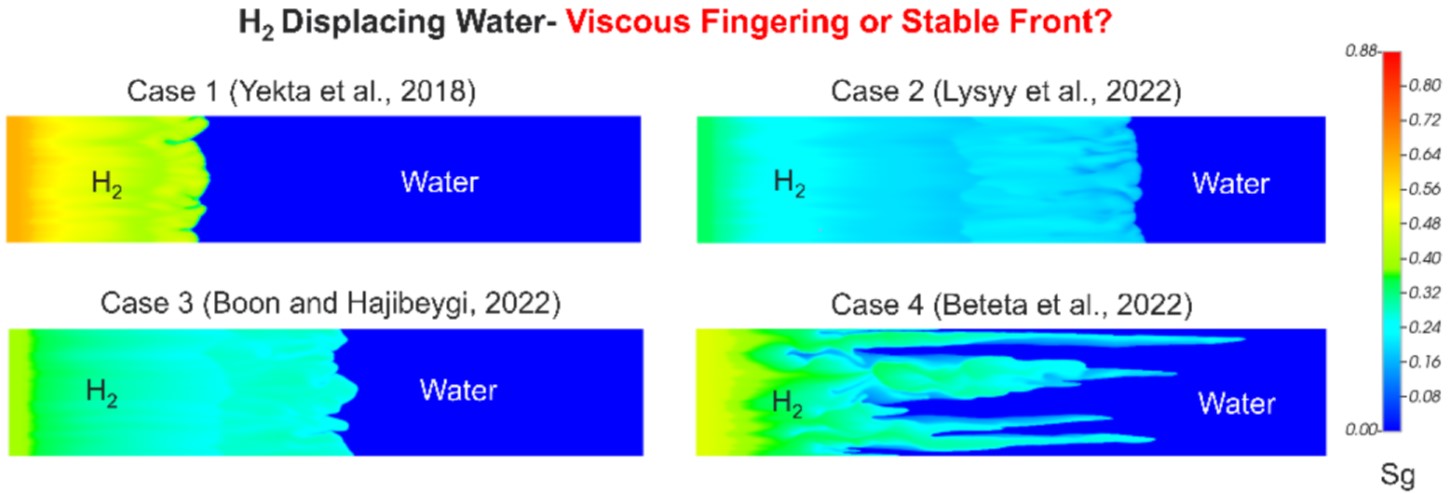}
			\caption{Numerical modeling of viscous fingering for underground storage of $\rm{H_2}$. The authors argue that conventionally measured, that is, equilibrium relative permeability, does not predict the expected viscous fingering patterns, whereas nonequilibrium relative permeability-saturation functions~\cite{betetaRoleImmiscibleFingering2022} do (taken from \cite{wangInherentErrorsCurrent2025}). }
			\label{fig:betetaviscousfingering}
		\end{figure}

        Skauge, Sorbie, and co-workers raised the provocative question of whether relative permeability determined in laboratory experiments under stable displacement conditions is applicable for unstable, that is, nonequilibrium displacement processes such as viscous fingering~\cite{wangInherentErrorsCurrent2025}. The key point of the argument is that the relative permeability-saturation functions determined from conventional core flooding experiments do not reveal the shape/pattern of viscous fingering observed experimentally~\cite{betetaRoleImmiscibleFingering2022}. This argument was then applied to the underground storage of $\rm{CO_2}$~\cite{larkiViscousFingeringDynamics2025} and hydrogen~\cite{wangHowUsefulAre2025}.  As shown in Fig.~\ref{fig:betetaviscousfingering} it is demonstrated that the nonequilibrium relative permeability function from~\cite{betetaRoleImmiscibleFingering2022} correctly predicts the expected viscous fingering pattern. 
        
        There is certainly the question of whether the arguments of the aforementioned nonequilibrium approaches are also applicable to viscous fingering, an unstable and therefore nonequilibrium process. The arguments from Beteta, Sorbie, Skauge, and co-workers may need to be investigated in more detail. One question is that in their experimental work, the Saffman-Taylor equation~\cite{scotsonXrayComputedTomography2021} is used to explain the finger wavelength, which is clearly pore-scale fingering, while for field applications, we are interested in Darcy-scale viscous fingering, which exhibits a very different scaling relationship for the finger wavelength as a function of viscosity ratio, interfacial tension, etc., see \cite{maasViscousFingeringCCS2024, ottWavelengthViscousUnstableCO2Brine2025, bergStabilityCO2brineImmiscible2012} and references therein. 

\begin{figure}[htbp]
			\includegraphics[width=1.0\linewidth]{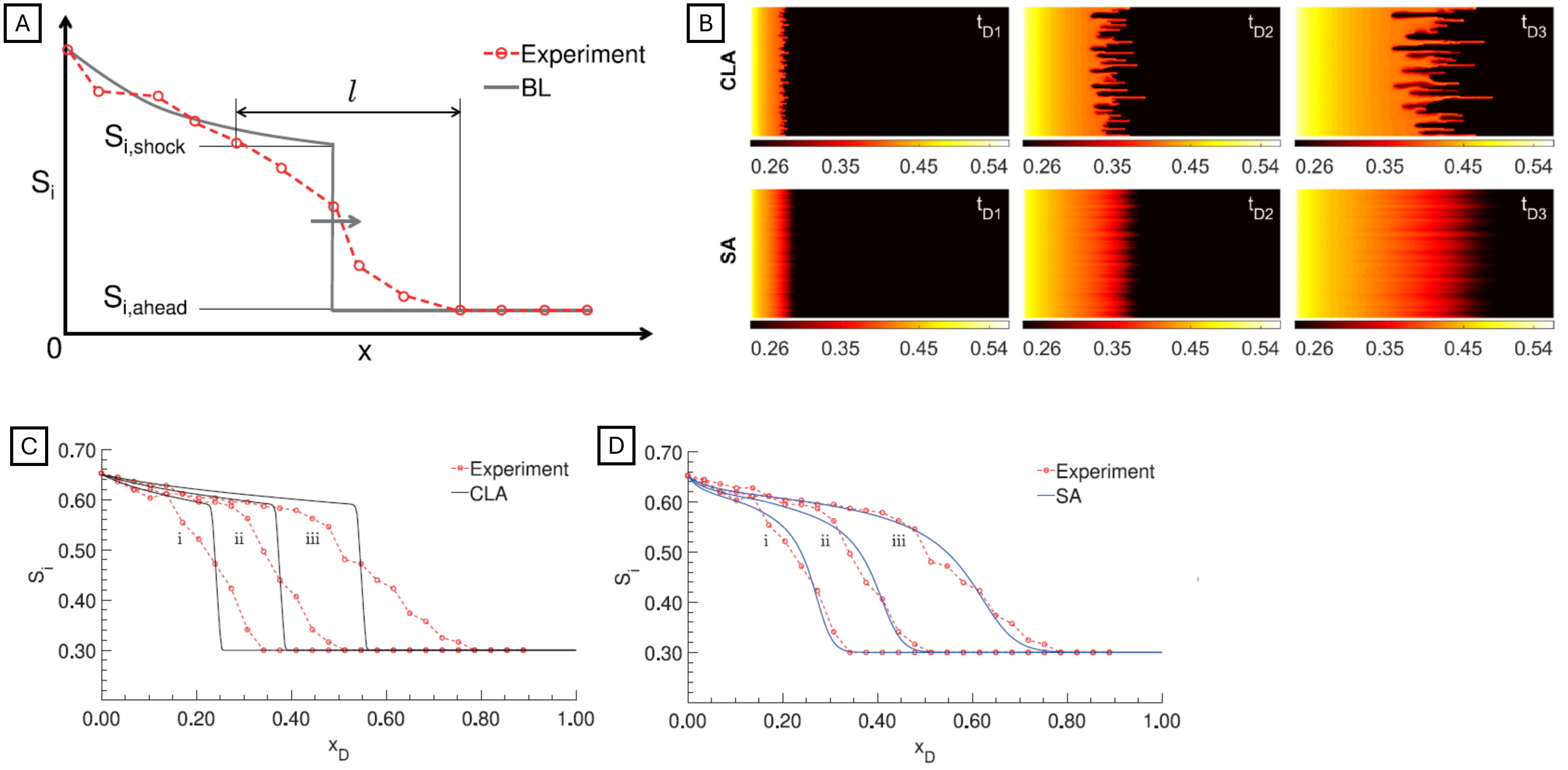}
			\caption{nonequilibrium model from Aryana and co-workers~\cite{aryanaNonequilibriumEffectsMultiphase2013,wangExtensionDarcysLaw2019} explains the experimentally measured saturation profiles in unstable / viscous fingering immiscible flow. The underlying concept is that in unstable displacement, the saturation front is not as sharp as the classical Buckley-Leverett profile but extended (A). Therefore, spatial averaging over a dynamic length scale $l$ is performed, which then influences the shape of fingers (B), which provides a much closer match with experimental observation (D) than the sharp-interface model (C) (adapted from \cite{wangExtensionDarcysLaw2019}). }
			\label{fig:AryanaKovscek}
		\end{figure}

        Such a nonequilibrium approach has been explored by Aryana et al. ~\cite{aryanaNonequilibriumEffectsMultiphase2013,wangExtensionDarcysLaw2019} who explained the experimentally measured saturation profiles in unstable multiphase flow, that is, subject to viscous fingering with a nonequilibrium model that involves spatial averaging over a dynamic length scale. As shown in Fig.~\ref{fig:AryanaKovscek} this provides a much closer match with experimental observations than the traditional sharp-interface model for unstable displacements~\cite{renBayesianModelSelection2017}.
        
        \cleardoublepage
		\newpage
		\subsection{Darcy and non-Darcy flow regimes}
        \label{sec:non-linear}

Darcy found a linear relationship between the flow rate and pressure drop. The generalized Darcy equations (\ref{eqn:twophasedarcy}), on the other hand, contain the option that they are nonlinear in the relation between the flow rate and pressure drop.  This would occur if the relative permeabilities depended on the capillary number.  In fact, it is very easy to imagine this to be so because the relative permeability reflects the ganglion structure, which in turn depends on the flow rate. It was then a surprise when Tallakstad et al.\ in 2009 investigated steady-state flow in a Hele-Shaw cell filled with fixed glass beads 
, thus constituting a (quasi-) two-dimensional model porous medium, that not only was the flow $Q$ rate vs.\ pressure drop $\Delta p$ non-linear, but it was even a power law, \cite{tallakstad2009steady,tallakstad2009steadyb} 

        \begin{equation}\label{eqn:forcefluxnonlinear}
            \Delta p\propto Q^{\alpha}\;,
        \end{equation}
with $\alpha=0.54\pm 0.08$. In the experiment, the total flow rate $Q$ and fractional flow were the control parameters.  


Tallakstad et al.\ in 2009 proposed a heuristic argument for why the constitutive equation should be a power law, which also pinned the exponent $\alpha$ down to 1/2 \cite{tallakstad2009steadyb}. A simplified version of this argument is as follows. Assuming that the average pressure gradient across the Hele-Shaw cell is $\Delta p/L$ and the typical capillary pressure barrier across the fluid-fluid interfaces is $p_c$, groups smaller than a length scale $l=L p_c/\Delta p$ will not move.  Suppose that the width of the Hele-Shaw cell is $W$.  The effective mobility is then proportional to $m\propto W/l$; therefore, the flow rate is given by $Q=m\Delta p\propto \Delta p^2$, giving $\alpha=1/2$.  

Subsequently, confusion arose when Rassi et al. \cite{rassi2011nuclear} found a strong dependence of the exponent $\alpha$ on the fractional flow rate.  

Sinha and Hansen \cite{sinha2012effective} found the same value for $\alpha$ using a homogenization technique borrowed from percolation theory.  They supported this result through numerical simulations using a dynamic pore-scale network model \cite{sinha2021fluid}. Notably, the model was run using two different boundary conditions. The first closely mimicked those used in the experiments: a fixed flow rate $Q$ and a fractional flow rate $f$ were fixed, and the pressure difference across the model, $\Delta p$ was measured.  In the second mode, the two-dimensional network was deformed to form a torus, i.e., introducing bi-periodic boundary conditions.  The pressure drop across the model $\Delta p$ and the saturation $S_w$ were fixed, and the flow rate $Q$ was measured.  In both modes, a law somewhat different from (\ref{eqn:forcefluxnonlinear}) was found,
\begin{equation}\label{eqn:forcefluxnonlinear-2}
            \Delta p-\Delta p_m \propto Q^{\alpha}\;,
        \end{equation}
where $\Delta p_m$ is the threshold pressure in the sense that the power law regime must end when the flow rate is zero.  In practice, when the flow rate $Q$ and fraction flow rate $f$ are the control variables, one will see a transition to a different regime before $Q=0$ is reached. The threshold pressure $\Delta p_m$ is caused by the interfaces blocking the flow.  Feder et al.\ \cite{feder2022physics} showed that the threshold pressure is a finite-size effect. More precisely, whereas the pressure drop increases as the system size $L$ keeps the flow rate constant, the threshold pressure $\Delta p_m$ increases as the square root of $L$, such that $\Delta p_m/\Delta p \sim 1/\sqrt{L} \to 0$ as $L\to \infty$.  This work was followed by \cite{roy2024immiscible}.

\begin{figure}[ht]
            \begin{center}
			\includegraphics[width=0.5\linewidth]{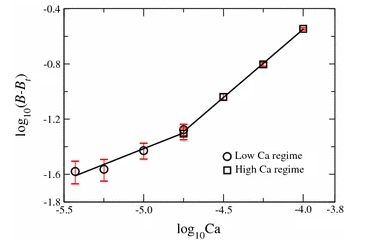}
            \end{center}
			\caption{Dimensionless excess pressure drop $B-B_t$ as a function of flow rate $Q$ as recorded in the experiments of \cite{sinha2012effective}. There are two regimes: one low-capillary number regime characterized by an exponent $\alpha=0.46\pm 0.05$ and a high-capillary number regime characterized by $\alpha=0.99\pm 0.02$ (from \cite{sinhaEffectiveRheologyTwoPhase2017}).}
			\label{fig:sinha-codd}
		\end{figure}

New experiments by the Codd-Seymour group, who published Rassi et al.\ \cite{rassi2011nuclear}, in collaboration with Sinha and Hansen, assuming a non-zero $\Delta p_m$ in the analysis, gave $\alpha=0.46$ in the non-linear regime \cite{sinhaEffectiveRheologyTwoPhase2017}.  Furthermore, they showed that there is a transition from the nonlinear to the linear Darcy regime at high capillary numbers (see Fig. \ \ref{fig:sinha-codd}).

The transition from the nonlinear to the linear regime, where the flow rate $Q$ is proportional to the pressure drop $\Delta p$ was expected because the viscous forces overwhelm the capillary forces, so that the flow is essentially that of a single Newtonian fluid.  As seen in Fig.\ \ref{fig:sinha-codd}, there is a well-defined capillary number --- or equivalently, pressure drop --- for which the transition from the power law to Darcy flow occurs. Roy et al.\ studied how this pressure drop would evolve with system size $L$ \cite{roy2024immiscible} analytically using capillary fiber bundles and numerically using a dynamic pore-scale network model \cite{sinha2021fluid}, finding that the transition pressure gradient shrinks with increasing $L$.

Determining the value of exponent $\alpha$ is highly sensitive to the threshold pressure $\Delta p_m$.  Fyhn et al.\  \cite{fyhn2023effective} constructed a mixed wet model, where $\Delta p_m$ would be zero by construction.  Adapting the dynamic pore network model to this model yielded a value $\alpha=0.39+\pm 0.01$.

The use of the capillary fiber bundle has proven very useful in understanding the mechanisms behind the nonlinear regime; see \cite{sinhaEffectiveRheologyBubbles2013,xu2014non,roy2019effectiveb,fyhn2021rheology,lanza2022non,cheon2023steady} in addition to Roy et al.\ \cite{roy2024immiscible}.  

The work we have quoted thus far has not involved film flow which is a general challenge to be captured by pore scale imaging and modelling~\cite{zhaoComprehensiveComparisonPorescale2019} due to finite resolution (except for pore network models which have ''infinite'' resolution with respect to films~\cite{bluntMultiphaseFlowPermeable2017}).  Aursj{\o} et al.\ \cite{aursjo2014film} used rapeseed/canola oil and a water-glycerol mixture as immiscible fluids in a Hele-Shaw cell, similar to that used by Tallakstad et al.\ \cite{tallakstad2009steady}.  They used three fractional flows, $f_{oil}=1/3$, 1/2, and 2/3. Surprisingly, they found $\alpha=0.74\pm 0.05$ for $f_{oil}=1/3$ and $\alpha=0.67\pm 0.05$ for $f_{oil}=1/2$ and 2/3. No theory has been presented to explain this result.

\begin{figure}[ht]
            \begin{center}
			\includegraphics[width=0.45\linewidth]{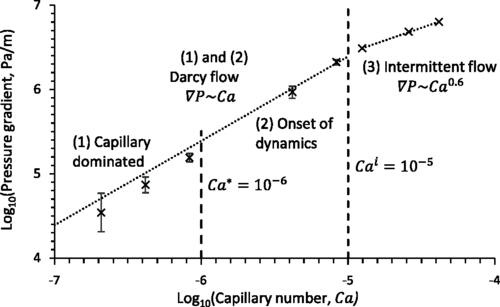}
            \end{center}
			\caption{Pressure drop as a function of capillary number, i.e., $\Delta p$ vs.\ $Q$ as measured by Gao et al.\ \cite{gao2020pore}. (From \cite{gao2020pore}.)
            }
			\label{fig:gao2020}
		\end{figure}

The Bijeljic \& Blunt group studied the other end of the pressure drop versus flow rate diagram.  When $Q$ and fractional flow are the control parameters, double percolation of the two immiscible fluids must occur when the flow rate is so low that the viscous forces are unable to overcome the capillary forces holding the fluid-fluid interfaces in place \cite{gao2020pore,zhang2021quantification,zhang2022nonlinear}.  In this regime, we expect a Darcian flow, that is, the pressure drop $\Delta p$  is proportional to the flow rate $Q$.  This is indeed what was observed (see Fig. \ \ref{fig:gao2020}.   

They divided the pressure drop vs. flow rate curve into three regimes.  The low-flow-rate regime is capillary-dominated, and there is no motion at the fluid-fluid interfaces.  In this regime, the relation between pressure drop and flow rate is  
linear, i.e., Darcian.  At higher flow rates, a second regime characterized by strong pressure fluctuations appears.  Then, the third nonlinear regime is reached, characterized by $\alpha=0.60\pm 0.01$.  This transition is not related to the threshold pressure $\Delta p_m$ reported in other studies.  

\begin{figure}[ht]
            \begin{center}
			\includegraphics[width=0.4\linewidth]{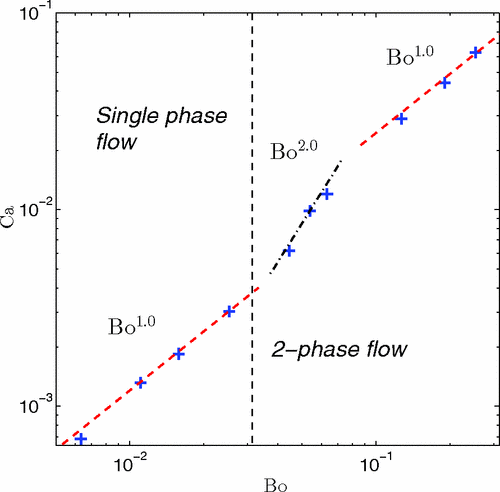}
            \end{center}
			\caption{Capillary number vs.\ bond number $Bo$ for a lattice Boltzmann simulation of immiscible two-phase flow in porous media showing all three flow regimes --- linear-non-linear-linear.  (From \cite{yiotis2013blob}.)
            }
			\label{fig:talon-1}
		\end{figure}

\begin{figure}[ht]
            \begin{center}
			\includegraphics[width=0.60\linewidth]{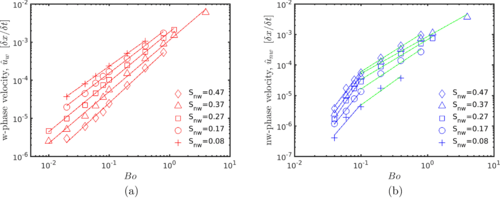}
            \end{center}
			\caption{The Darcy velocity of (a) the wetting fluid and (b) the non-wetting fluid as a function of the bond number $Bo$ (From \cite{yiotis2019nonlinear}).  
            }
			\label{fig:yiotis}
		\end{figure}

Yiotis et al.\ \cite{yiotis2019nonlinear} studied the nonlinear regime in terms of ganglia dynamics, and the three regimes correspond approximately to the different flow regimes displayed in Fig. ~\ref{fig:AvraamPayatakesJFM1995}. They used the Lattice Boltzmann method with a constant body force as the driving force. Hence, rather than using the capillary number as a variable, the bond number $Bo$ (gravity over capillary force) is the proper control variable, but that is rather a technicality associated with the specific simulation technique, and this is ultimately equivalent to the capillary number $Ca$. They also measured the Darcy velocities of the wetting and non-wetting fluids as a function of the bond number $Bo$ (see Fig. \ \ref{fig:yiotis}.  They found that the wetting fluid followed a power law with an exponent close to unity, whereas the non-wetting fluid showed a power-law regime (with an exponent of approximately $2/3$ which has been recently confirmed \cite{botticiniOriginPressureflowNonlinearity2025}) followed by a linear regime.  This is in accordance with the hypothesis that nonlinearities are caused by the dynamics of non-wetting ganglia.      

Can the nonlinear regime influence the shape of the viscous fingers?  L{\o}voll et al.\ \cite{lovoll2004growth} and Toussaint et al.\ \cite{toussaint2005influence} asked the following question: what is the link between the pressure gradient on the surface of a viscous finger and the growth probability of the viscous finger?  They measured this relation experimentally using a Hele-Shaw cell, finding that the growth probability would be proportional to the {\it square\/} of the pressure gradient. This is a surprising result, breaking with the belief that the diffusion-limited aggregation (DLA) model would model the proper dynamics \cite{barabasi1995fractal}, which assumes a linear relation. This work has been followed up computationally using a dynamic pore-scale network model \cite{sinha2024disorder}, essentially confirming the original experimental measurements.  A consequence of this is that the shape of the viscous fingers will be controlled by whether the flow is locally in the linear or nonlinear regime.

We may summarize our discussion so far: There are in general three different flow regimes encountered with respect to the visco-capillary balance (expressed through the capillary number $Ca$):
        \begin{enumerate}
            \item a linear regime I at small capillary numbers with $\alpha=1$
            (which sub-divides into regime Ia with connected pathway flow only and Ib where ganglion dynamics is present but the overall flux is still dominated by connected pathway flow)
            \item a non-linear regime II at intermediate capillary numbers with $0.4 < \alpha < 0.75$ \cite{roy2019effectiveb,yiotis2019nonlinear} where for mobilization of ganglia $\alpha=0.5$ can be derived analytically~\cite{sinhaEffectiveRheologyBubbles2013}
            \item a linear regime III at high capillary numbers with $\alpha=1$
        \end{enumerate}

\begin{figure}[ht]
            \begin{center}
			\includegraphics[width=1.0\linewidth]{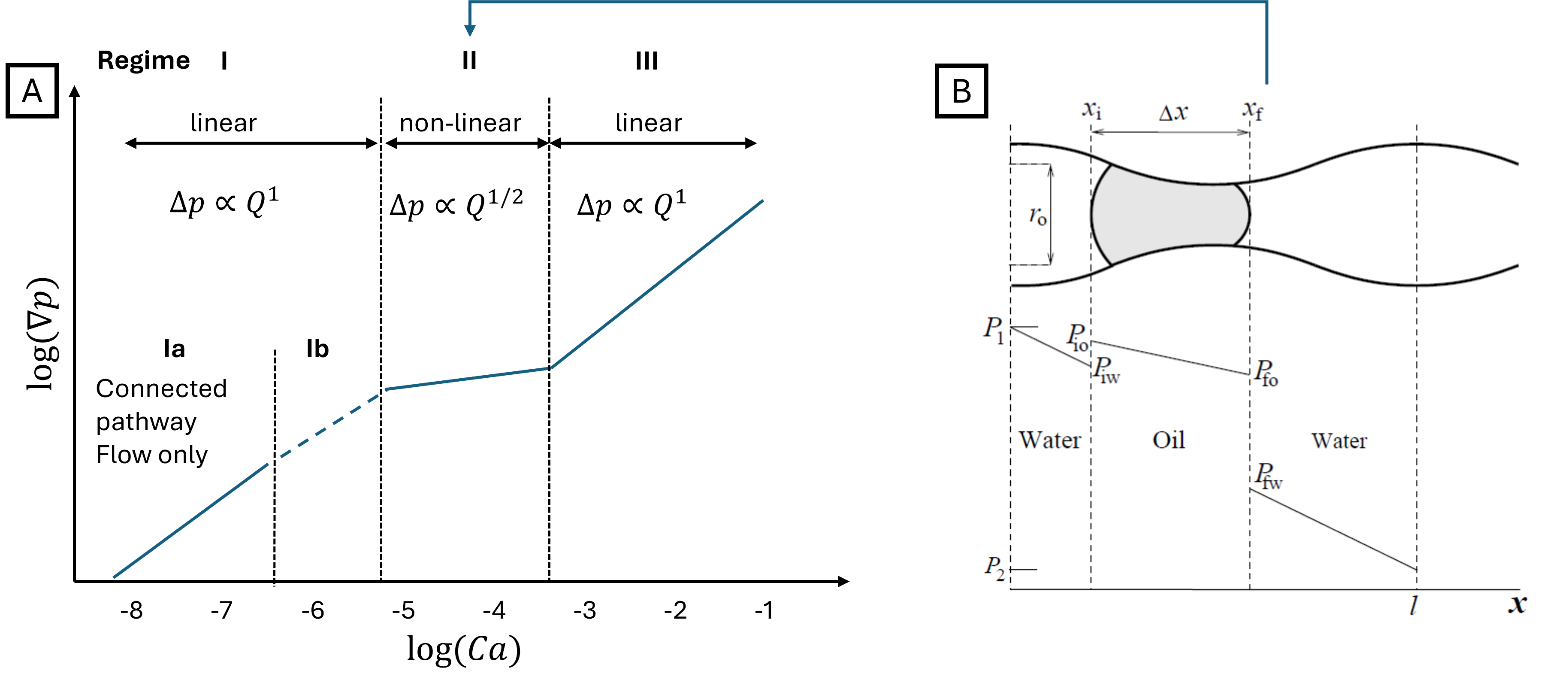}
            \end{center}
			\caption{(A) Cartoon summarizing the flow regimes from Fig.~\ref{fig:sinha-codd}-\ref{fig:yiotis} and (B) the mobilization of individual ganglia for which in regime II the exponent $\alpha = 0.5$ in the pressure gradient-flux relationship was calculated analytically (taken from~\cite{sinhaEffectiveRheologyBubbles2013}).)
            }
			\label{fig:flowregimeslinearnonlinear}
		\end{figure}

This is illustrated in Fig. ~\ref{fig:flowregimeslinearnonlinear} where the three flow regimes are sketched conceptually, and non-linear regime II consists of ganglion dynamics, where non-wetting phase clusters are mobilized by viscous forces to the extent that it dominates the flux. For simple geometries, the power law exponent can be derived analytically. 

The second two regimes are illustrated in Fig. ~\ref{fig:sinha-codd} and the first two regimes are shown in Fig. \ \ref{fig:gao2020}. Fig.\ \ref{fig:talon-1} shows a Lattice Boltzmann simulation displaying all three regimes simultaneously \cite{yiotis2013blob}.

In summary, the important insight from the linear - nonlinear–linear behavior reflects and, to some extent, also characterizes flow regimes. 
		
		
		\cleardoublepage
		\newpage 
        
		\section{The two-phase Darcy equations described by non-equilibrium thermodynamics (NET)}\label{sec:Derivation2PhaseDarcyEquationsEntropyGeneration}

In linear flow regimes, one may hope that non-equilibrium thermodynamics (NET) can be applied. Kjelstrup, Bedeaux, and coworkers have fronted a new development of classical nonequilibrium thermodynamics (NET) in order to be able to deal with transport in porous media \cite{Kjelstrup2018,Kjelstrup2019,Bedeaux2022,Bedeaux2024_nano}. 
In classical NET, the temporal evolution of a system is governed by the rate of entropy production~\cite{burelbachUnifiedDescriptionColloidal2018}. Local equilibrium is assumed, meaning that Gibbs equation is assumed valid in any volume element. This assumption is in fact used in all thermodynamic modeling of homogeneous systems, without normally being stated. It was found to be true for surfaces which are described using Gibbs excess variables \cite{Kjelstrup2020}. 

        The standard NET procedure for homogeneous systems was adapted to deal with two-phase flow in porous media. A REV was first defined, and local equilibrium was assumed for the REV. The variables were coarse-grained by integrating over REV space and suitable time. An example of such a variable is the effective pressure computed from the grand potential. The grand potential contains the products of 1) all bulk volumes of the REV multiplied by their appropriate bulk pressure, 2) all surface areas times their appropriate surface tension, and 3) all three-phase contact lines times their appropriate line tension. These are all well-defined measurable variables, including geometric ones. The products of the grand potential are assumed to be additive (weakly interacting). Hill's nano-thermodynamics can then be used for each (confined) subsystem \cite{Bedeaux2024_nano}.  The weakly coupled additive contributions to the REV can be used to account for small system effects.  This coarse-graining procedure brings us directly from the pore scale to the Darcy scale since it includes integration over space and time. This procedure is unlike that of Hassanizadeh \& Grey \cite{HassanizadehGray1993} who wrote macroscopic equations for each and every phase and interface, providing a formidable set of governing equations to be reduced later.

\begin{figure}[ht]
			\includegraphics[width=1.0\linewidth]{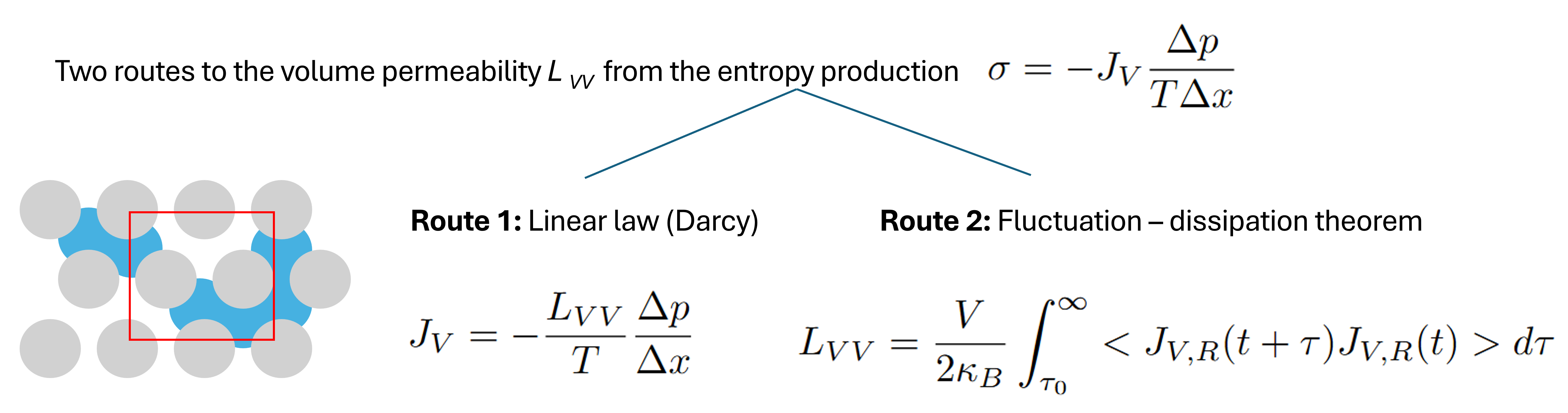}
			\caption{Two routes to the coefficient $L_{VV}$ derived from the entropy production. In Route 1, the coefficient is computed from the linear law. This law states that the volume flow $J_V$ is proportional to the driving force $-\Delta p/(T \Delta x)$. In Route 2, $L_{VV}$ is obtained from the autocorrelation function of the volume flow integrated from $\tau_0$ to $\infty$, where $\tau_0$ is the lower limit for detection of fluctuations, $V$ is the volume of the REV, $\kappa_B$ plays the role of an effective Boltzmann constant and $J_{V,R}$ is the fluctuation random contribution to the average flow $J_V$.} 
			\label{fig:relpermfromentropygenerationroutes}
		\end{figure}

          The entropy production, $\sigma$, is described by the independent fluxes $J_i$ and their  conjugate driving forces. There are also fluctuation-dissipation theorems (FDT) connected with the entropy production \cite{Green1954,Kubo1966,Kjelstrup2020}.  In order to apply FDT to porous media transport, we assume that the REV is in local equilibrium \cite{Bedeaux2022, Bedeaux2024_nano}. The idea is that fluxes which fluctuate around their  average can provide  information on the REV permeability. The FDT was formulated for porous media in analogy with the way FDTs are formulated for homogeneous systems. This way to obtain the permeability is alternative to the procedure using linear laws.  The two routes to the permeability of porous media are illustrated in Fig. ~\ref{fig:relpermfromentropygenerationroutes}. Symbols are further defined in the text below.
       
In classical NET, the entropy production of the volume element is obtained from the Gibbs equation for the homogeneous phase. When written for porous media, the entropy production was obtained from a Gibbs equation written for the REV \cite{Kjelstrup2018, Kjelstrup2020, Bedeaux2022}. By introducing balance equations for mass, energy, and momentum into the Gibbs equation for coarse-grained variables, expressions for the entropy production as well as the entropy flux could be identified. The thermodynamic flux-force relations can then be obtained directly for the Darcy scale. The symmetry of the macro-scale coefficient matrix is linked to the fluctuation-dissipation theorem \cite{Kjelstrup2020}, written here for fluctuating fluxes of the REV. Using Hill's thermodynamics for small systems, one can deal with fluids that are confined in the REV. This additional possibility was elaborated, i.e., for slit pores and matrices of spheres \cite{Bedeaux2021,Bedeaux2024_nano, Galteland2019, Rauter2020}. Hill's method connects the capillary pressure to the effective pressure of the REV and helps define the driving force conjugate to the volume flow. It contains a contribution which is proportional to the curvature of the fluid-fluid interface in the definition of the (so-called) \textit{integral} pressure.

The variables of the REV are coarse-grained because they contain a sum of contributions from the single phases, surface areas, and contact lines. A typical REV showing such contributions is illustrated by the red square in Fig.\ref{fig:relpermfromentropygenerationroutes}.

In this NET approach to porous media transport, the Gibbs equation is written for the entire REV, with variables that are already coarse-grained. This trick makes it possible to move directly from contributions at the pore scale to the Darcy scale of transport. 
As the method \cite{Bedeaux2021,Bedeaux2024_nano} adds contributions from volumes, surfaces, and contact lines in order to define REV properties, it contains contributions from Minkowski functionals~\cite{hadwiger1957vorlesungen,Klain_95}. The many variables introduced by Hassanizadeh \& Gray \cite{hassanizadehMechanicsThermodynamicsMultiphase1990,HassanizadehGray1993,Gray1998} are also included. The set of variables can be said to represent an equation of state, cf. \ref{eq:geom-state}. Such an equation can be used to define the assumption of local equilibrium for the REV.

NET in combination with Hill's systematic method for confined fluids \cite{Hill1994,Bedeaux2024_nano} can thus be used to define a REV using only a small number of variables. The environment plays a crucial role in the theoretical setup, as it enables the control of selected variables.  Chemical potentials and temperature can, for instance, be controlled by reservoirs in contact with the system. The grand potential
is a relevant thermodynamic property of the porous media. In small systems, the pressure depends on controlled variables such as the volume, leading to the definition of the integral pressure.

The additivity of the REV variables makes it possible to express Gibbs equation using lumped
variables. Additions are possible when 
the REV is in local equilibrium. It is likely that such a system is  ergodic and that fluctuations are time-reversal invariant. Under such conditions, it is also likely that the nonequilibrium thermodynamic theory can be extended to porous media. We can thus benefit from a systematic theory for the determination of  constitutive relations for the fluxes, including Onsager symmetry relations and
fluctuation-dissipation theorems. 

\subsection{Construction of REV variables. An example}\label{sect:constructionofREV}

To see how the coarse-graining of REV variables is performed in practice,
considered a common example of a porous rock with two fluid flows. The REV has volume $V^{\text{REV}}=V$. First, we choose the set of variables that will be controlled by the reservoirs. Here, we use the temperature, $T$, and
chemical potentials and $\mu ^{n}$, $\mu ^{w}$, $\mu ^{r}$ as control variables. Therefore, the abbreviation GC refers to the grand potential of the system. The grand potential $X^{\text{GC,REV}}$ has a statistical basis in the grand canonical partition function $\Xi$, as follows:
\begin{equation}
X^{\text{GC,REV}}(T,V^{\text{REV}},\mu ^{n},\mu ^{w},\mu ^{r})=-k_{\text{B}%
}T\ln \Xi (T,V^{\text{REV}},\mu ^{n},\mu ^{w},\mu ^{r})  \label{R3.1}
\end{equation}%
 We assume next that 
contributions to the grand potential are \text{additive}. This is true when the states of the
wetting and non-wetting fluids, the rock, the surfaces, and the contact line are weakly coupled. The probability distribution in phase space is then
equal to the product of the probability distributions of the subsystems. It
follows that the grand canonical partition function of the REV is the
product of the partition functions of the subsystems, here the \textit{
volumes, surface areas, and contact lines}. The grand potential, $X$, can be written as 
\begin{eqnarray}
 X^{\text{GC,REV}}(T,V^{\text{REV}},\mu ^{n},\mu ^{w},\mu ^{r})    \notag
 & = & X^{n,\text{%
GC},\text{REV}}(T,V^{n,\text{REV}},\mu ^{n})  \nonumber \\
& + &X^{w,\text{GC},\text{REV}}(T,V^{w,\text{REV}},\mu ^{w})+X^{r,\text{GC},%
\text{REV}}(T,V^{r,\text{REV}},\mu ^{r})  \nonumber \\
&+ &X^{nr,\text{GC},\text{REV}}(T,\Omega ^{nr,\text{REV}},\mu ^{n},\mu
^{r})+X^{wr,\text{GC},\text{REV}}(T,\Omega ^{wr,\text{REV}},\mu ^{w},\mu
^{r})  \nonumber \\
&+&X^{nw,\text{GC},\text{REV}}(T,\Omega ^{nw,\text{REV}},\mu ^{n},\mu
^{w})+X^{nwr,\text{GC},\text{REV}}(T,\Lambda ^{nwr,\text{REV}},\mu ^{n},\mu
^{w},\mu ^{r})  \nonumber \\
&+&X^{\text{conf}}(T,V^{\text{REV}})  \label{R3.3}
\end{eqnarray}%
The origin of a variable is shown by superscripts. 

The different bulk phases are immiscible. 
The origin of the last configurational contribution \cite{Bedeaux2024_nano} is a probability distribution 
in phase space, which contains a factor that gives the probability
distribution of fluids over all possible choices of subvolumes. By integrating over the entire phase space, we also integrate over these distributions. 
The Minkowski functionals in Eq.8 can be seen as embedded in this description of the REV. These are variables that contribute to the REV variables; see below the computation of the effective pressure $\widehat{p}$, the integral pressure in Hill's theory \cite{Galteland2019,Rauter2020}.

The grand potential is now specified for our example. It is equal to minus the
integral pressures times the volume for each bulk phase, plus the integral surface- (or line-)
tension times the surface area of each surface (or line length of each line):
\begin{eqnarray}
&&\widehat{p}V^{\text{REV}}=\left( \widehat{p}^{\text{mat}}+\widehat{p}^{%
\text{conf}}\right) V^{\text{REV}}=\widehat{p}^{n}V^{n,\text{REV}}+\widehat{p%
}^{w}V^{w,\text{REV}}+\widehat{p}^{r}V^{r,\text{REV}}  \nonumber \\
&&-\widehat{\gamma }^{nr}\Omega ^{nr,\text{REV}}-\widehat{\gamma }%
^{wr}\Omega ^{wr,\text{REV}}-\widehat{\gamma }^{nw}\Omega ^{nw,\text{REV}}-%
\widehat{\gamma }^{nwr}\Lambda ^{nwr,\text{REV}}+\widehat{p}^{\text{conf}}V^{%
\text{REV}}  \label{R2.14}
\end{eqnarray}
We see how the additive terms lead to  the integral pressure $\widehat{p}$ when we divide by $V^{\text{REV}}$. The gradient in the effective pressure or driving force $-\nabla \widehat{p}/T$ has several separate contributions. Gradients in capillary pressure can therefore drive the flow in the absence of an external pressure gradient ~\cite{hammondSpontaneousImbibitionSurfactant2011,caiDiscussionEffectTortuosity2011,masonDevelopmentsSpontaneousImbibition2013}.
The volume, masses of the components, and Gibbs energy have material contributions only. All other thermodynamic energies, like the internal
energy, the Helmholtz energy, and the enthalpy, have configurational
contributions. 

For the volume of the REV we have 
\begin{equation}
V^{\text{REV}}=V^{r,\mathrm{REV}}+V^{n,\text{REV}}+V^{w,\mathrm{REV}}+V^{nwr,%
\text{REV}}\equiv V^{r,\mathrm{REV}}+V^{p,\text{REV}}+V^{nwr,\text{REV}}
\label{R1.2}
\end{equation}%
The contact line has volume because the dividing surfaces do not cross each other along the same line. However, this contribution can typically be neglected. 

For the masses we have%
\begin{eqnarray}
M_{n}^{\text{REV}} &=&M_{n}^{n,\text{REV}}+M_{n}^{nr,\text{REV}}+M_{n}^{nw,%
\text{REV}}+M_{n}^{nwr,\text{REV}}  \nonumber \\
M_{w}^{\text{REV}} &=&M_{w}^{w,\text{REV}}+M_{w}^{wr,\text{REV}}+M_{w}^{nwr,%
\text{REV}}  \nonumber \\
M_{r}^{\text{REV}} &=&M_{r}^{r,\text{REV}}  \label{R2.3}
\end{eqnarray}%
The equimolar (or equimass) surface of the rock is a convenient 
fluid-rock dividing surface. Likewise, the wetting fluid dividing surface is useful as a
fluid-fluid dividing surface. Furthermore, the position of the contact lines
can be such that $M_{r}^{nwr,\text{REV}}=0$. The total mass of each component in the REV,
however, is \emph{independent} of the location of the dividing surfaces and contact line. 
%
%
The entropy of the REV is the sum of bulk entropies, excess interfacial entropies, excess line
entropies and configurational entropy contribution \cite{Bedeaux2022}: 
\begin{eqnarray}
S^{\text{REV}} &=&S^{\text{mat}}+S^{\text{conf}}=S^{n,\text{REV}}+S^{w,\text{%
REV}}+S^{r,\text{REV}}+S^{nr,\text{REV}}  \nonumber \\
&&+S^{wr,\text{REV}}+S^{nw,\text{REV}}+S^{nwr,\text{REV}}+S^{\text{conf}}
\label{R2.7a}
\end{eqnarray}%
The sum of bulk, excess interfacial, excess line entropies is a \textit{%
material} \textit{contribution}, $S^{\text{mat}}$, to entropy. There are
many ways to distribute the phases, and the origin of the \textit{%
configurational contribution} comes from the many configurations that have
the same component volumes, surface areas and contact line lengths in the
REV. The integral of the logarithm of the corresponding probability
distribution times $k_{\text{B}}$\ gives the configurational contribution to
the entropy. Both the material and the configurational contributions are
additive.

\subsection{Fluctuation-dissipation theorems}
      
Fluctuations on the molecular scale have been used to derive transport coefficients on the macroscale in homogeneous systems, since the formulation of the Green-Kubo formulas in the fifties of the last
centuries \cite{Callen1951, Green1954,Kubo1966}. Diffusion coefficients and thermal conductivities are two examples of homogeneous solutions \cite{Liu2011,Bresme2014}.
Current-current correlation functions have been used to formulate electric conductivities ~\cite{ferryKuboFormulaLinear1992}.

FDT-formulas constitute part of the foundation of NET for homogeneous media, and it has been suggested that this is also true for porous media. Fluctuations in fluxes of extensive variables are then central; this means volume flow fluctuations \cite{Bedeaux2022}. However, this method has barely been investigated for porous media \cite{Winkler2020,Alfazazi2024}. 
Alfazazi et al. \cite{Alfazazi2024} computed the ratio between the permeability obtained
from the linear (Darcy) law and FDT. The ratio (called $F$) of the two
permeabilities was not unity, but turned out to be independent of the phase saturation, as predicted \cite{Umar2025}.  Winkler et al.\cite{Winkler2020} made a first try to apply 
FDT to networks and found a symmetric matrix of coefficients in a 2-dimensional, honeycomb network, containing two 
immiscible and incompressible fluids. These important and surprising results indicate that Onsager relations \cite{Onsager1931a,Onsager1931b} may apply. Moura \textit{et al.} \cite{Moura2024}  used the time rate of change of fluctuations in phase saturation, and found that it is possible to define a REV in a size-independent manner in a 2D-Hele-Shaw cell. Therefore, there may exist a foundation for the application of FDTs to porous media. 
The pressure fluctuations observed by Rücker et al. \cite{Ruecker2021} provide a strong indication of the existence of nonthermal fluctuations.

\subsubsection{The linear laws. Route 1 to $L_{ij}$. }
To find the full set of Onsager coefficients of a REV, we return to NET for two free variables. The entropy production is%
\begin{equation}
\sigma =J_{w}(-\frac{1}{T}\frac{\Delta \mu _{w}}{\Delta x})+J_{n}(-%
\frac{1}{T}\frac{\Delta \mu _{n}}{\Delta x}
)  \label{FD1}
\end{equation}%
where $J_{w}$ and $J_{n}$ are independent particle flows in m$^{-2}$ s$^{-1}$, of the wetting and non-wetting fluids, respectively. The components can be regarded as mixed in the REV at a meso-level. The two-flux, two-force variable set allows for a description of the mass movement caused by gradients in pressure and phase saturation. The flux-force relations are in general: 
\begin{align}
J_{w}= & \Omega _{ww}(-\frac{1}{T}\frac{\Delta \mu_{w}}{\Delta x})+\Omega
_{wn}(-\frac{1}{T}\frac{\Delta \mu_{n}}{\Delta x})  \notag \\
J_{n}= & \Omega _{nw}(-\frac{1}{T}\frac{\Delta \mu_{w}}{\Delta x})+\Omega
_{nn}(-\frac{1}{T}\frac{\Delta \mu_{n}}{\Delta x})  \label{FD2}
\end{align}%
The dimension of the Onsager coefficient, $\Omega_{ij} $, is K m$^{-1}$ s$^{-1}$ J$%
^{-1}$.  The equations are reduced in the case that the phase saturation in constant, and $\Delta \mu_i = \Delta \mu_i^c +  V_i \Delta p =  V_i \Delta p$. 
\begin{align}
J_{w}& =-(\Omega _{ww}V_{w}+\Omega _{wn}V_{n})\frac{1}{T}\frac{\Delta p}{\Delta x}  \notag \\
J_{n}& =-(\Omega _{nw}V_{w}+\Omega_{nn}V_{n})\frac{1}{T}\frac{\Delta p}{\Delta x}
\label{FD4}
\end{align}%
For the total volume flow $J_V$, Eq.\ref{FD4} gives one linear law for the REV:
\begin{eqnarray}
J_{V} &=&J_{w}V_{w}+J_{n}V_{n}=-(L_{ww}+2L _{wn}+L _{nn})\frac{1}{T}%
\frac{\Delta p}{\Delta x}  \notag \\
&=&-L_{VV}\frac{1}{T}\frac{\Delta p}{\Delta x}  \label{FD5}
\end{eqnarray}%
where we assumed that the Onsager symmetry relation applies, $\Omega _{nw}=\Omega _{wn}$ and $L_{ij} = V_w \Omega_{ij} V_n$. There is only one independent flux in the porous medium. 

\begin{figure}
    \centering
    \includegraphics[width=0.5\linewidth]{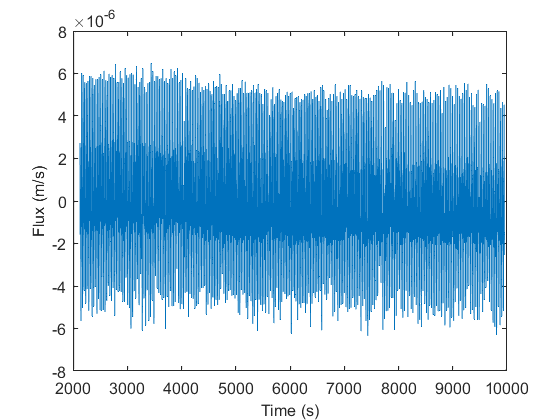}
    \caption{Fluctuations in the steady state total volume flow of two-phases in sintered glass. These fluctuations contain information on flow permeability. Courtesy of Physics of Fluids \cite{Alfazazi2024}}
    \label{fig:FDT-J_V}
\end{figure}

Around each average value of a steady-state flow, $J_V$, there is a fluctuating contribution, $J_{i,R}$. This random part of the total flux is central in the fluctuation-dissipation theorem (FDT). The fluctuating part reflects a broad range of phenomena or mechanisms of transport (Fig. ~\ref{fig:relaxationphenomena}).  
An example of flow fluctuations  is shown in Fig.\ \ref{fig:FDT-J_V}. The average of the fluctuating contributions to the volume flux is zero by definition.

\subsubsection{Fluctuations at steady state.  Route 2 to $L_{ij}$ via FDT.}
The fluctuating contributions to the  (particle) flux are correlated. The correlation was formulated for independent fluxes in homogeneous systems in terms of fluctuation-dissipation theorems (FDT) \cite{Callen1951,Green1954,Kubo1966} 
\begin{equation}
C_{ij}(\mathbf{r}-\mathbf{r}^{\prime },t-t^{\prime })\equiv \left\langle
J_{i,R}(\mathbf{r},t)J_{j,R}(\mathbf{r}^{\prime },t^{\prime })\right\rangle
=2k_{B}\Omega _{ij}\delta (\mathbf{r}-\mathbf{r}^{\prime })\delta
(t-t^{\prime })  \label{FD6}
\end{equation}
All fluxes in this definition are local-valued, and not integrated over the cross-section of the porous medium normal to the direction of flow \cite{bedeauxFluctuationdissipationTheoremsMultiphase2025}. They apply to one REV, and the averages of their random contributions are always zero. The average of the total flux is the average of the collective effort to move in one direction. The equation applies for steady as well as unsteady states, which makes it very versatile. On the REV time-scale, the random contributions will appear as Gaussian white noise. 
The central term expresses the flux correlations at position $\mathbf{r}$ and time $t$, after the same event took place at $\mathbf{r}^{\prime },t^{\prime }$. 

The range, over which the fluctuating contributions are correlated, is small compared to the range of change in the average fluxes.  However, the REV of a porous medium is significantly different from that of a homogeneous fluid. The fluctuations in the REV are no longer only molecular. Mixing is also possible on the meso-level, \textit{e.g.} as in ganglion dynamics. This may lead to other types of fluctuations. A meso-level description was therefore postulated based on  the presence of pores.  In the case of porous media, the idea is to include pore-level fluctuations at the mesolevel~\cite{dattaFluidBreakupSimultaneous2014,masalmehLowSalinityFlooding2014,reynoldsCharacterizingFlowBehavior2015,soropRelativePermeabilityMeasurements2015,gaoXrayMicrotomographyIntermittency2017,gaoPoreOccupancyRelative2019,gao2020pore,linImagingMeasurementPoreScale2018,alcornCorescaleSensitivityStudy2020,spurinIntermittentFluidConnectivity2019,wangObtainingHighQuality2019,menkeUsingNanoXRMHighContrast2022}. This amounts to examination of particle flux correlations in terms of the above equation. 

This view of the nature of the fluctuations adopts the bold hypothesis of microscopic reversibility, or time-reversal invariance, in a new context. Some inspiration for this hypothesis is obtained from electrochemistry \cite{Rubi}, where equilibrium exchange rates are much larger than unidirectional rates and the activation energies for the two-way traffic are large. Support can also be obtained from the results of Steijn et al. \cite{Steijn2009}. They have shown 
that there is full microscopic reversibility in capillary events right up to the point when a topological change occurs. After this point, the event can be dealt with as rare, going spontaneously back to the original state.

The correlations in the REV may best be described by delta functions. The pre-factor on the right-hand side of the above equation arise from $
\Omega _{ij}+\Omega _{ji}=2\Omega _{ij}$ (Onsager relations are used). 
The fluctuations in the total volume flow are given by $J_{V,R}=V_wJ_{w,R}+V_nJ_{n,R}$. 
The FDT for the total volume flow fluctuations is%
\begin{equation}
C_{VV}(\mathbf{r}-\mathbf{r}^{\prime },t-t^{\prime })\equiv \left\langle
J_{V,R}(\mathbf{r},t)J_{V,R}(\mathbf{r}^{\prime },t^{\prime })\right\rangle
=2k_{B}L_{VV}\delta (\mathbf{r}-\mathbf{r}^{\prime })\delta
(t-t^{\prime })  \label{FD7}
\end{equation}%

An Onsager coefficient is a measure of the correlations' strengths. Noise is a characteristic of the underlying events. It reflects snap-offs, Haines' jumps, etc.
that occur on the REV scale. Boltzmann's constant enters because the 
integration, in principle, includes all scales, including the molecular scale. In the experiment of Alfazazi et al. \cite{Alfazazi2024}
sampling occurred at a frequency of 10$^{-5}$ s. It was therefore proposed
to integrate from $\tau _{0}=10^{-5}$ to $\infty $ \cite{Alfazazi2024}. 
Measurement and simulation techniques are normally limited, leaving the experimenter with a sampling window. We may therefore expect that there is an
effective Boltzmann constant different from the normal constant $k_{B}$ that
appears in the FDT of homogeneous systems. So far, experiments indicate \cite{Alfazazi2024} that we must replace $k_{B}$ 
by $\kappa _{B}$ when we are dealing with porous media. The factor is likely to be particular for each porous medium and the experimental setup.

From experimental results, we find the fluctuating volume flux $J_{V,R}$. We average this over the volume of the REV,  
volume $V^{\text{REV}}$, and obtain 
\begin{equation}
C_{VV}\equiv \left\langle
J_{V,R}(t)J_{V,R}(t^{\prime })\right\rangle 
\end{equation}
where $C_{VV}$ is the correlation function of the volume. The function depends  on the time only. We next integrated over the appropriate time difference, and obtain
\begin{equation}
    L_{VV} = \frac{V^{\text{REV}}}{2 \kappa_B}\int_{\tau_0}^{\infty}C_{VV} dt 
\end{equation}


 FDT was applied to steady-state two-phase flow in the linear flux-force regime, and the ratio of the coefficients obtained from the FDT and from the experiment was formulated \cite{Alfazazi2024}:
\begin{equation}
F\equiv \frac{L_{VV}/T}{C_{VV}}=\frac{V^{\text{REV}}}{2\kappa _{B}T}
\label{FD12}
\end{equation}
The ratio should be a constant of the experiment.
Ideally, FDT should give the same permeability as the linear law, as formulated with the conjugate flux-force pair. However, the heterogeneous nature of the sample, and the restrictions on the sampling time may lead to deviations. 
 
\subsubsection{Memory effects. Zero frequency component.}

The correlation functions $C_{ij}$ may be time-dependent (contain memory effects). The total volume flow becomes
\begin{eqnarray}
J_{V}(t) =
-\int_{t}^{\infty }dt^{\prime }L_{VV}(t-t^{\prime })(\frac{1}{T}%
\frac{\Delta p}{\Delta x})(t^{\prime })  \label{FD15}
\end{eqnarray}%
FDT applied to
the volume flow fluctuations gives now
\begin{equation}
C_{VV}(\mathbf{r}-\mathbf{r}^{\prime },t-t^{\prime })\equiv \left\langle
J_{V,R}(\mathbf{r},t)J_{V,R}(\mathbf{r}^{\prime },t^{\prime })\right\rangle
=2\kappa _{B}L _{VV}(t-t^{\prime })\delta (\mathbf{r}-\mathbf{r}%
^{\prime })  \label{FD17}
\end{equation}%
The average integral of the fluctuating flux over the REV
volume $V^{\text{REV}}$ can be obtained. The correlation function is next integrated over 
time. 
Clearly, the time integrals must be of sufficient length. In the simulations \cite{Winkler2020} and experiments \cite{Alfazazi2024,Moura2024}, $C_{VV}$ was directly determined. The factor $F$ can be obtained from the zero frequency expression of FDT.  
The zero-frequency part of the FDT plays the same role as the steady-state value in the determination of the Onsager coefficients, and therefore also for the value of $F$.  
   

\subsection{Validity of NET}
According to the discussion in Sections 4.1-4.3, we may expect NET to be applicable if the REV is in local equilibrium (is described by the Gibbs equation). Local equilibrium can also be defined in terms of a valid equation of state. The condition applies at steady and non-steady states. The REV that is ergodic and can be expected to obey microscopic reversibility \cite{Onsager1931a,Onsager1931b} on the time scale of the investigation. 

The stated properties can be tested. A system with ganglion dynamics may be relevant for such tests \cite{Alfazazi2024}. The two immiscible phases will then mix on the REV scale during transport by overtaking each other's positions. Also testable is the way to obtain permeabilities. In a valid application of FDT, the permeabilities from FDT are the same as those from the linear laws. This property has not yet been observed in porous media. The lack of agreement between methods on this point promoted the definition of an effective Boltzmann constant, $\kappa_B$, dependent on the measuring technique \cite{Alfazazi2024}. More support is required before this ideas can come to real use. 

Obviously, systems exist that are not ergodic, where the flux-force relations are nonlinear, or where microscopic reversibility does not apply, cf. Section 3. These classes of systems mostly fall outside the region of NET validity. Systems with channel flow may belong to this category. It is, however, possible to describe chemical reactions on the meso-level \cite{Rubi}. In NET, the permeability is a function of state variables. In the present context, such variables are the saturation, temperature, interface area, and curvature or line lengths. Hill's thermodynamics for small systems can take these into account. Its use is only in its infancy. 

However, the transport coefficients in NET cannot be functions of the driving forces or fluxes. They cannot be derived from the FDT unless $i$ and $j$ refer to independent events.


\subsection{Open questions}

If the conditions for validity of NET apply, Darcy's law and its extensions will follow. NET then provides the transport law with a strong theoretical basis, lending it open to extensions. A major extension would be to allow for other driving forces. The product of the heat flux and its thermal driving force can be added to the entropy production in Eq. \ref{FD2} \cite{Kjelstrup2018}. This provided a symmetric matrix of transport coefficients to deal with diffusion, thermal diffusion, and/or thermal osmosis. For an example applied to oil reservoirs, see ~\cite{hafskjoldSoretSeparationThermoosmosis2022}.  The meaning of thermo-diffusion or thermal osmosis is well described by NET for  two-component homogeneous systems, but the methods reviewed here  have not been used for porous media \cite{hafskjoldSoretSeparationThermoosmosis2022}.  

Completely new possibilities arise from the application of NET to porous media transport because the new definition of the REV enables us to describe fluid confinements, and thereby transport in such systems. 
The grand potential of the REV formulated using Eq. \ref{R2.14} is a sum of products which contain geometric variables. We can use it to replace \textit{i.e.} the capillary pressure gradient as a driving force contribution. The overall driving force is the gradient of the integral pressure, $-\nabla \widehat {p}$. This pressure corrects the normal pressure with area and line-dependent terms, see Eq. \ref{R2.14}. There is no correction when the capillary pressure is zero. In the classical limit, when $\nabla \widehat{p} = \nabla p$, we do not need Hill's thermodynamics. Bulk pressures differ typically between the phases.  Galteland et al. \cite{Galteland2019,Rauter2020} found that the integral pressure was constant across a two-phase boundary in equilibrium, and that this observation was consistent with the validity of Young's and Young-Laplace's law. 

The formulation of FDT for porous media is only in its infancy. More work is needed to determine how the FDT route can be viable and the results can be understood. 
The fluctuation-dissipation theorems (FDT) for homogeneous systems cover the complete range of fluctuations, from the molecular scale and up. This has made the method invaluable for the determination of transport properties. The same also applies to FDT for porous media. However, this application is hardly investigated. The fluctuations under consideration are again from the molecular scale and up. A peculiar situation now is that a meso-level of fluctuations can become important. A typical example is ganglion dynamics, a phenomenon on the capillary energy scale. Such dynamics are also contained in FDT. Much work needs to be done to fully explore this path to transport properties, understand, and classify phenomena. For instance, for porous media measurements, there is a lower boundary for integration posed by the measurement sampling technique \cite{Alfazazi2024}. Considerable work is required to clarify this integration limit. 

Diffusion takes place in the presence of gradients in chemical potential (composition), and  diffusion is the key element in many processes that involve gases such as ripening~           \cite{dechalendarPorescaleModellingOstwald2018,xuEgalitarianismBubblesPorous2017,BergDeterminationCriticalGas2020,gaoCapillarityPhasemobilityHydrocarbon2021,wangCapillaryEquilibriumBubbles2021,dokhonPressureDeclineGas2024,goodarziTrappingHysteresisOstwald2024}. These phenomena can be described within NET for porous media using the chemical potential as a variable. Their role in hysteresis should be investigated. For the typical low flow rates / small capillary numbers, the diffusive flux can be equally large as the advective flux (i.e. a P\'eclet number around 1)~\cite{gaoCapillarityPhasemobilityHydrocarbon2021}.

		
		\cleardoublepage
		\newpage
		\section{Derivation of the 2-phase Darcy equations by space-time averaging over capillary fluctuations}\label{sect:spacetimeaveraging}

        The starting point is the understanding that relaxation phenomena are an integral part of multiphase flow and exist over a very wide range of length and time scales, as shown in Fig. ~\ref{fig:relaxationphenomena}.
        Fluctuations, sometimes also referred to as intermittency~\cite{gaoPoreOccupancyRelative2019,zhang2021quantification,spurinRedNoiseSteadyState2022,wangTimeAndSpaceAveragingApplied2024b} (although there might be subtle differences) are observed from the pore scale to length scales where fractional flow physics is applicable, as shown in the work of Rücker \textit{et al.} \cite{Ruecker2021} (see also Fig. ~\ref{fig:travellingwaves}. Hence, there is no single length scale between the pore and Darcy scales where these fluctuations average out in space-average only. Therefore, a credible upscaling approach of multiphase flow from the pore to the Darcy scale must consider capillary fluctuations. 
        
\begin{figure}[ht]
			\includegraphics[width=0.6\linewidth]{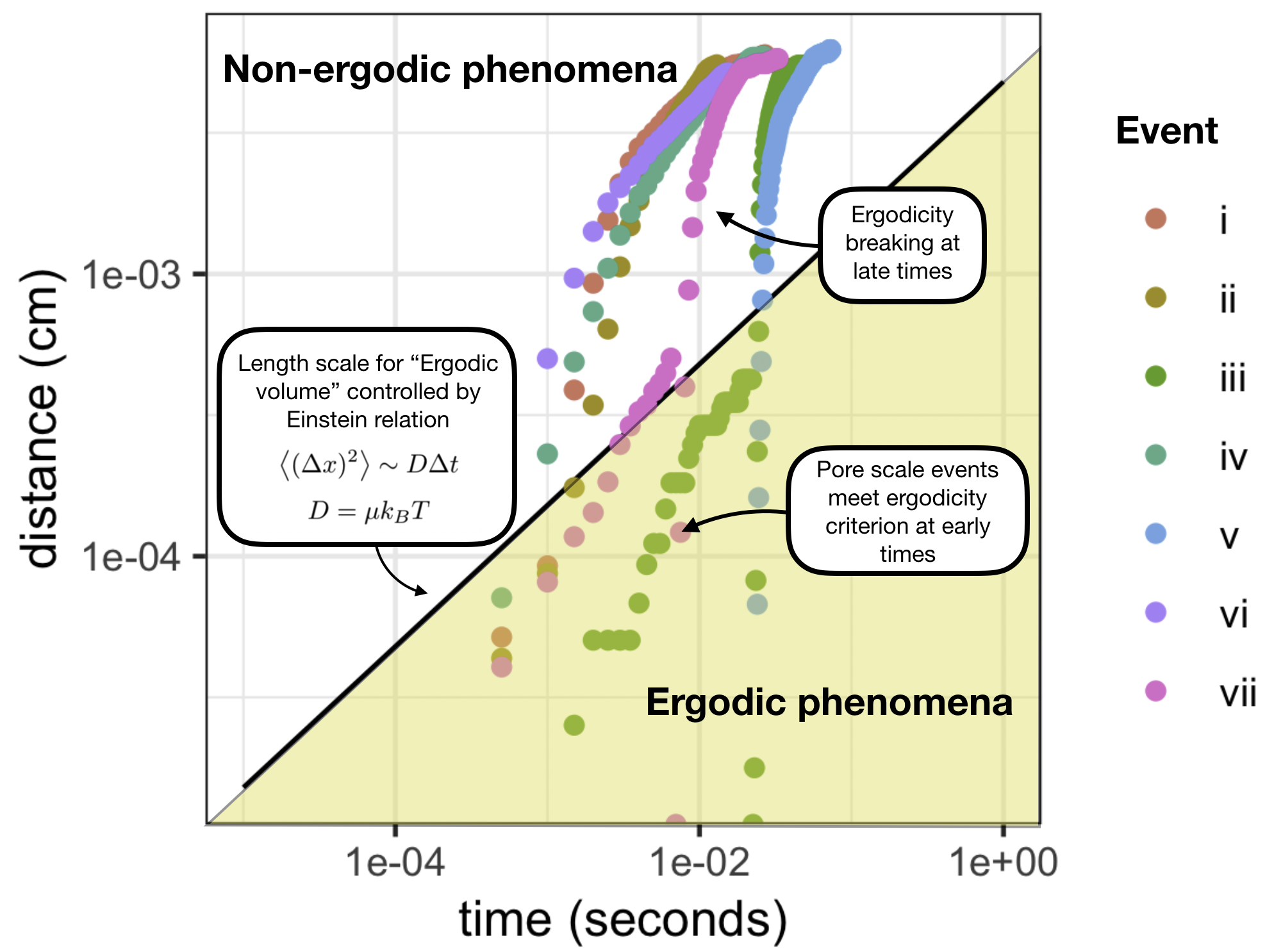}
			\caption{Space-time trajectory of Haines jumps~\cite{armstrongInterfacialVelocitiesCapillary2013} which scale beyond the diffusive-mixing scale that would allow thermal equilibration, and hence break ergodicity at late times (taken from \cite{McClure2021}).}
			\label{fig:porescale-ergodicity}
		\end{figure}

        Capillary fluctuations are caused by pore-scale displacement events~\cite{lenormandMechanismsDisplacementOne1983}. Individual pore-scale displacement events such as Haines jumps~\cite{Haines1930,Berg2013} have space-time trajectories that scale beyond the diffusive-mixing scale, as shown in Fig. ~\ref{fig:porescale-ergodicity} and become effectively nonergodic ~\cite{McClure2021}. Furthermore, pore-filling events can also occur in cyclic sequences that lead to intermittent connections or so-called dynamic connectivity \cite{reynolds2017dynamic}. As explored by \cite{spurin2023pore}, the spectral signature based on pressure data during co-current flow depends on the frequency and pore volumes injected (time), see Fig.\ \ref{fig:porescale-ergodicity}, demonstrating the difficulty of upscaling when only space averages for a specific time are considered.

        Therefore, to compute the averages of the Darcy scale, \textit{ space and time averaging} must be performed~\cite{mcclureThermodynamicsFluctuationsBased2021, wangTimeAndSpaceAveragingApplied2024b} instead of volume-based homogenization. A sequence of publications by McClure~\textit{et al.} establishes the basic problem of non-ergodic capillary displacements, and the necessity of space-time averaging was proposed as a general formalism for upscaling systems with fluctuations~\cite{mcclureThermodynamicsFluctuationsBased2021}. 

        The flux–force relation for multiphase flow in porous media is derived by applying nonequilibrium thermodynamics with explicit time-and-space averaging to account for capillary fluctuations and pore-scale dynamics \cite{mcclureThermodynamicsFluctuationsBased2021}. Starting from a thermodynamic description of the internal energy, an entropy inequality was constructed, incorporating reversible work, surface energy changes, and fluctuation terms. By introducing geometric constraints and eliminating non-independent terms, the inequality is rearranged into a flux–force form. This enables the derivation of phenomenological relations, such as Darcy’s law and dynamic capillary pressure, which remain valid provided that the condition of steady-state is satisfied, ensuring that fluctuations are averaged over the selected domain and time scale.
        
\begin{figure}[ht]
		   \includegraphics[width=0.8\linewidth]{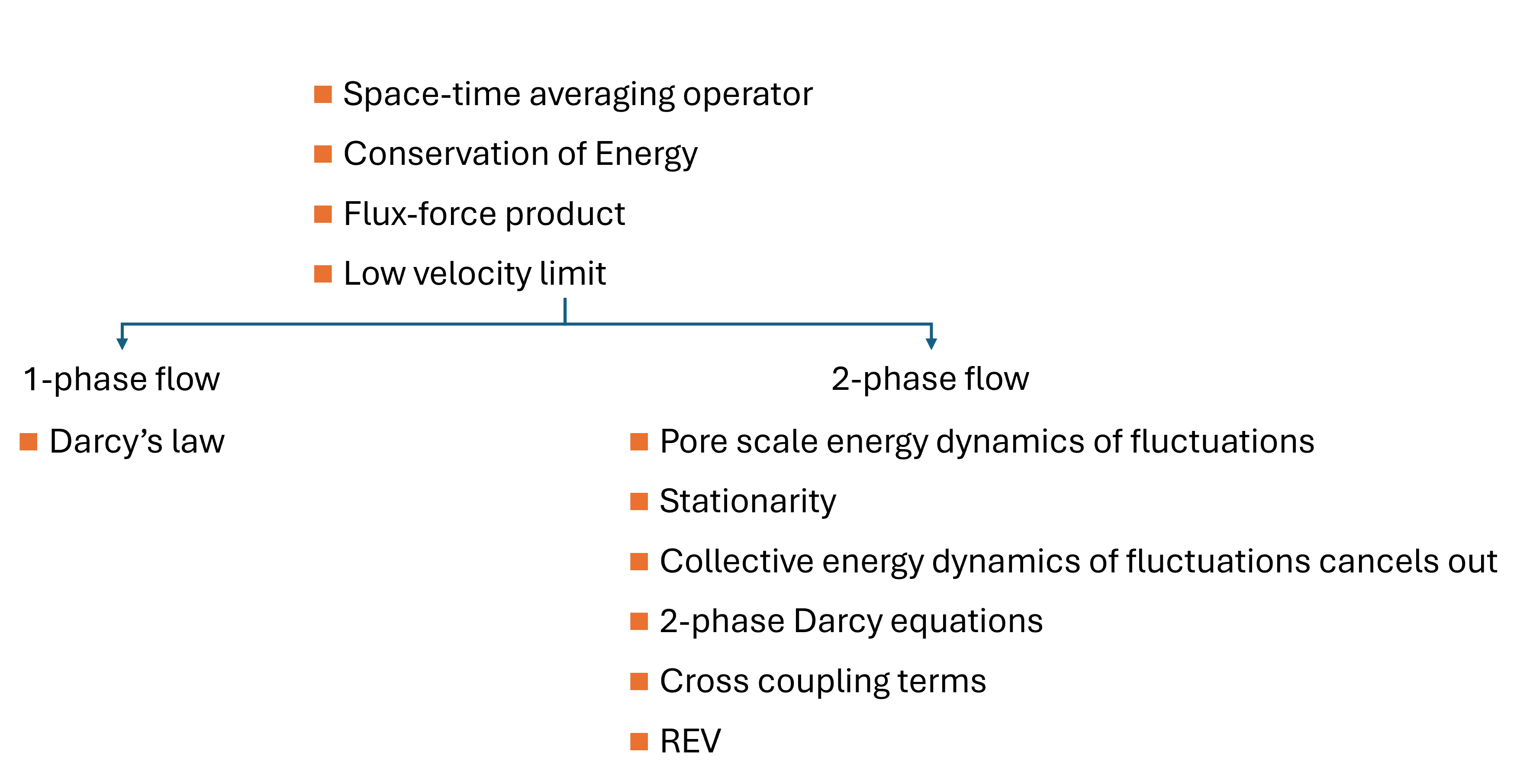}
		     \caption{Sketch of the derivation of the 2-phase Darcy equation from space-time averages of the pore scale capillary fluctuations~\cite{McClure2022}.}
		\label{fig:workflowspacetimeaverage}
		\end{figure}

        A high-level overview of the derivation of the 2-phase Darcy formulation under the condition of steady-state is sketched in  Fig. ~\ref{fig:workflowspacetimeaverage}. A complete derivation is provided in ~\cite{McClure2022}.
        
        In this context, steady-state refers to a condition in which the average thermodynamic state of the system does not change with time, although microscopic fluctuations (e.g., capillary bursts, Haines jumps~\cite{Haines1930}, or intermittency) may still occur. Specifically, it means that over a chosen time interval and spatial domain, all relevant extensive quantities (e.g., energy, entropy, fluid volumes) and intensive quantities (e.g., pressure, chemical potential) are statistically constant in time when averaged. This assumption was validated with 2-phase LBM simulations~\cite{McClure2022}, which demonstrated that the space-time averaging of the collective energy dynamics of fluctuations vanishes when a sufficiently long time scale is considered. Steady-state is a critical aspect to ensure that the fluctuation terms do not contribute net energy, which may occur under unsteady-state conditions. The latter is a condition that has yet to be explored.

        \subsection{Key steps in the derivation of phenomenological equations}
        In the following, we show the key steps in the derivation of the 1-phase and 2-phase Darcy equations. For capillary fluctuations in multiphase flow in porous media specifically, space-time averages of all thermodynamic variables have been defined via the space-time averaging operator

        \begin{equation}
             \big<f\big > \equiv \frac{1}{\lambda \mathcal{V} }\int_{\Lambda} \int_\Omega fdVdt
             \label{eqn:spacetimeaveragingoperator}
        \end{equation}
        
        \noindent where $\Omega$ is the spatial domain of extent $\mathcal{V}$ and $\Lambda$ is the temporal domain with time interval $\lambda$. 
        The starting point is the conservation of energy, which is derived from standard conservation principles.

        \begin{equation}
              \frac{\partial }{\partial t} \frac{U}{\mathcal{V}}
               + \nabla \cdot \Big(\frac{\mathbf{u} U}{\mathcal{V}} \Big)
               -\bm{\sigma} : \nabla \mathbf{u} - \nabla \cdot \mathbf{q}_h - {q}_s  = 0\;,
               \label{eq:energy-i}
        \end{equation}
        where $\mathbf{u}$ denotes the flow velocity, $\bm{\sigma}$ denotes the stress tensor, and $\mathbf{q}_h$ denotes the heat flux. The stress tensor $\bm{\sigma}$ can be decomposed into the mean and the deviatoric components ${\bm{\sigma}} = -{p} \bm I + {\bm{\tau}}$ which then leads to 

        \begin{equation}
            \frac{\partial }{\partial t} \frac{U}{\mathcal{V}}
             +\nabla \cdot \Big[\mathbf{u} \Big( \frac{U}{\mathcal{V}}+ p \Big)- \mathbf{q}_h\Big] 
             -\mathbf{u} \cdot \nabla p
             -\bm{\tau} : \nabla \mathbf{u} = 0
             \label{eq:energy-ii}
        \end{equation}
        
        In the next step, the space-time averaging is performed. Conservation of energy for the region $\Omega$ as a whole for time interval $\lambda$ is given by the integral of Equation \ref{eq:energy-ii}
        \begin{equation}
                \int_{\Lambda}\int_{\Omega} \Bigg(
                \frac{\partial }{\partial t} \frac{U}{\mathcal{V}}
                +\nabla \cdot \Big[\mathbf{u} \Big( \frac{U}{\mathcal{V}} + p \Big)- \mathbf{q}_h\Big]
                - \mathbf{u} \cdot \nabla p
                -\bm{\tau} : \nabla \mathbf{u} \Bigg) dV dt = 0
                \label{eq:energy-iii}
        \end{equation}
        
        After several transformations this equation is re-written in form of the flux-force product

        \begin{equation}
              -\bar{\mathbf{u}} \cdot \nabla \big( \bar{\phi} \bar{p} \big) 
               = \frac{1}{\lambda V}\int_{\Lambda}\int_{\Omega}\bm{\tau} : \nabla \mathbf{u}\;dV dt  
               \label{eq:energy-v}
        \end{equation}

        \noindent where the work done due to the fluctuations cancels

        \begin{equation}
               \frac{1}{\lambda V}\int_{\Lambda}\int_{\Omega}
               \Bigg(  
               \mathbf{u}^\prime\cdot \nabla p
               + \frac{S}{\mathcal{V}} \frac{\partial T^\prime}{\partial t}
               + \frac{N_k}{\mathcal{V}}\frac{\partial \mu_k^\prime}{\partial t} 
               \Bigg) dV dt = 0 \;. 
               \label{eq:fluctuation}
        \end{equation}

        For single-phase flow of a Newtonian fluid with viscosity $\mu$ we then obtain 

        \begin{equation}
               \int_{\Lambda}\int_{\Omega}
               \bm{\tau} : \nabla \mathbf{u}\; dV dt  = 
               \mu \int_{\Lambda} \int_{\Omega}
               \mathbf{E}  : \nabla \mathbf{u}\; dV dt  
               \label{eq:Darcy-viscosity}
        \end{equation}

        \subsubsection{1-phase flow: Darcy's law}
        Darcy's law is then obtained for a sufficiently small driving force from a linear expansion 
        
        \begin{equation}
              \bar{\mathbf{u}} = -\frac{\mathbf{\mathsf{L}}}{\mu} \cdot
              \nabla \big( \bar{\phi} \bar{p} \big)
              \label{eq:darcy-ii}
        \end{equation}
        where the tensor $\mathbf{\mathsf{L}}$ contains linear 
        phenomenological coefficients, consistent with both the approach of Onsager \cite{Onsager1931a, Onsager1931b} and volume-averaging theory \cite{whitakerFlowPorousMedia1986}.
        For constant porosity $\phi$, Darcy's law is obtained in the familiar form
        \begin{equation}
              \bar{\mathbf{q}} = -\frac{\mathbf{\mathsf{K}}}{\mu} \cdot
              \nabla  \bar{p}  \;,
              \label{eq:darcy-standard}
        \end{equation}
        where the flux $\bar{\mathbf{q}} = \bar{\phi}\bar{\mathbf{u}}$.

       \subsubsection{2-phase flow}\label{sect:spacetimeaveraging2phaseflow}
       For multiphase flow, the collective energy dynamics of pore-scale fluctuations is expressed over their degrees of freedom defined in the geometric state variables (Minkowski functionals, see section~\ref{sect:Minkowskifunctionals}). 
       2-phase LBM simulations were used to assess the extent to which the collective energy dynamics of pore-scale displacement events averages out \cite{McClure2021}, i.e.

       \begin{equation}
              \Big<\mathbf{u}^\prime\cdot \nabla p + {\frac{A_s}{V} \left( \frac{\partial \gamma_s^\prime }{\partial t} - h_s \frac{\partial \Pi_s^\prime }{\partial t} \right)   } - \phi_w \frac{\partial p_w^\prime }{\partial t} - \phi_n \frac{\partial p_n^\prime }{\partial t} \Big> = 0
             \label{eq:2phase-darcy-fluctuation}
       \end{equation}
       where $p_i$ is the fluid pressure for $i \in \{w, n\} $, $\gamma_s$ is the fluid-solid surface energy along the solid material, $\Pi_s $ is the disjoining pressure, $\phi_i$ is the fluid volume fraction for $i \in \{w, n\} $, $A_s$ is the surface area of the solid material, and $h_s$ is the film thickness along the solid material.      
       
\begin{figure}[ht]
			\includegraphics[width=1.0\linewidth]{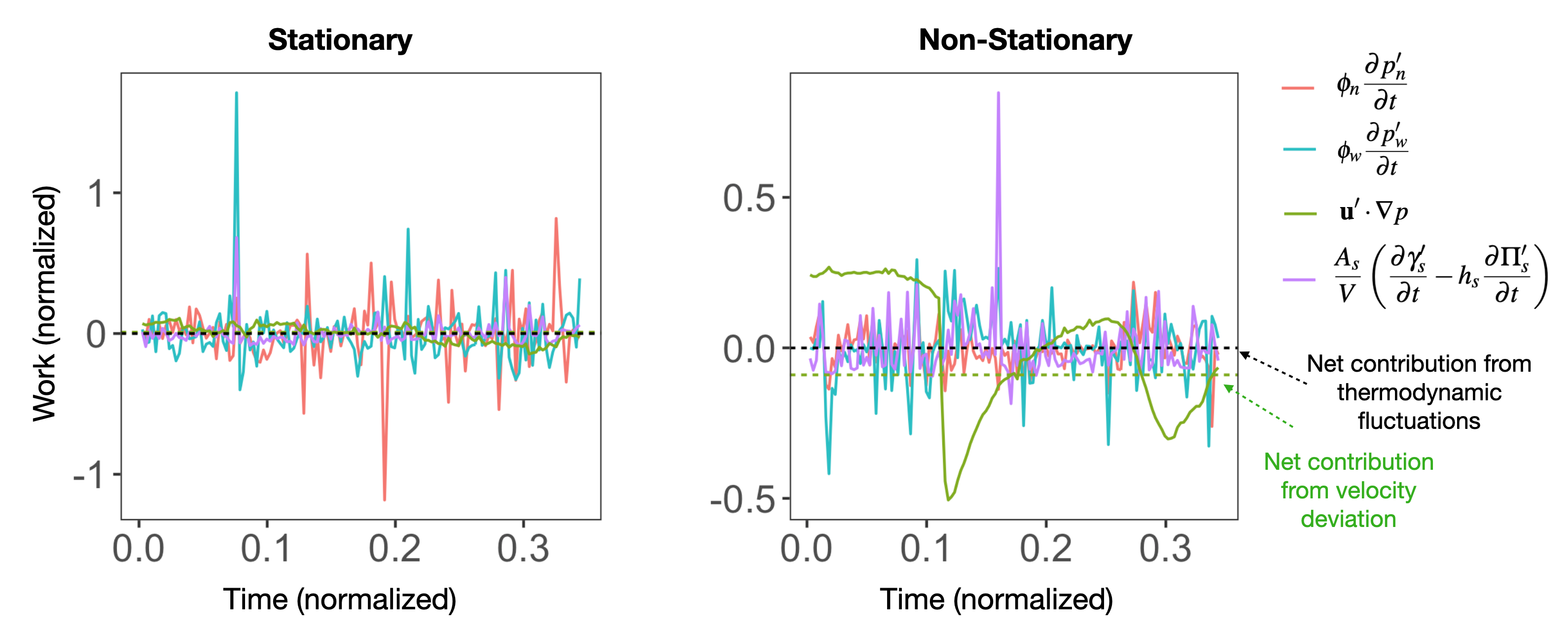}
			\caption{Contribution of all energy fluctuation terms for steady-state and non-steady-state displacements. For the steady-state case (left), the net contribution of work based on the sum of all fluctuations is zero. For the non-steady-state case (right), the net work due to velocity deviation is significant compared with the rate of dissipation in the system. (taken from \cite{McClure2022}).}
			\label{fig:steady-state-vs-nonsteady-state}
		\end{figure}

       To account for intermittent connectivity, which can be assessed via the Euler characteristic $\chi$, the system must be averaged over a sufficiently long time interval $\lambda$ \cite{McClure2022}. As shown in Fig.~\ref{fig:steady-state-vs-nonsteady-state} which has been obtained from LBM simulations for steady-state conditions, the collective energy dynamics averages out, but not for non-steady-state conditions. 
       
        The condition that the collective energy dynamics of fluctuations averages out in space-time averages results in the 2-phase Darcy equations. With eq.~\ref{eq:energy-v} we obtain for 2 fluid phases 
        \begin{equation}
              -\big[{ \bar{\omega}_w \bar{\mathbf{u}}_w + \bar{\omega}_n \bar{\mathbf{u}}_n  } \big]
              \cdot  \nabla \big(\bar{\phi} \bar{p}\big)
               = \frac{1}{\lambda V}  \int_{\Lambda}\int_{\Omega}
              \bm{\tau} : \nabla \mathbf{u}\;dV dt \;.
             \label{eq:energy-vi}
        \end{equation}

        \noindent where $\omega_i$ are the mass fractions of the two fluid phases. 
        Similar as the considerations for single-phase flow we then obtain 
        \begin{equation}
             {\omega}_i \bar{\mathbf{u}}_i  
              = - \frac{ \mathbf{\mathsf{L}}_i}{\mu_i} \cdot
             \nabla \big( \bar{\phi} \bar{p} \big) \;,
             \label{eq:darcy-iii}
        \end{equation}

        \noindent which has the functional form of the 2-phase Darcy equation, identifying 

        \begin{equation}
              \bar{s}_i \approx \omega_i \;, \quad
              \bar{\mathbf{q}}_i \approx  \bar{\phi} \bar{s}_i \bar{\mathbf{u}}_i
        \end{equation}

        Relative permeability is then expressed as 
        \begin{equation}
             \mathbf{\mathsf{L}}_i  =  k^r_i \mathbf{\mathsf{K}} \;.
        \end{equation}

        Note that cross-coupling terms between fluid phases~\cite{standnesNovelRelativePermeability2017} are obtained when the pressure gradients of the reach fluids become accessible. 
        
        In essence, this derivation demonstrates why, in the presence of capillary-driven, complex spatio-temporal dynamics of fluid-fluid interfaces with associated topological changes of connected pathways, the linear flux-force relationship still holds and the 2-phase Darcy equation remains applicable. In a nutshell, the reason is that in this still capillary-dominated flow regime, the collective energy dynamics of all interfacial dynamics averages out (in space-time averages), which keeps the viscous dissipation in the connected pathway flow as the dominant factor, for which a linear flux-force relationship holds. All the complex spatio-temporal dynamics do \textit{in this flow regime} change the configuration of the connected pathway flow.

        \subsection{Definition of a multiphase REV}\label{sect:multiphaseREV}
        The space–time averaging approach can be used to define a representative elementary volume for multiphase systems at the Darcy scale~\cite{McClure2025}. The spectrum of the relaxation phenomena in Fig. ~\ref{fig:relaxationphenomena} and the associated length scales suggests that there is no length scale, at least until the scale of many centimeters, where multiphase flow variables such as saturation and pressure would average out in space. Consequently, a multiphase flow REV must be larger than the length scale of the traveling waves \cite{Ruecker2021}. Because the length of the largest nonwetting phase cluster for typical flow conditions in sandstone rock is on the order of several centimeters~\cite{armstrongCriticalCapillaryNumber2014a}, this would imply that a multiphase REV would have to be several times this length to provide a robust average. Consequently, it would not be possible to compute multiphase flow parameters such as capillary pressure and relative permeability from pore-scale simulations, where typical domain sizes are only a few millimeters. Even the relative permeability from traditional special core analysis experiments performed on rock samples of a few centimeters in length would be questionable.  
        
\begin{figure}[ht]
			\includegraphics[width=1.0\linewidth]{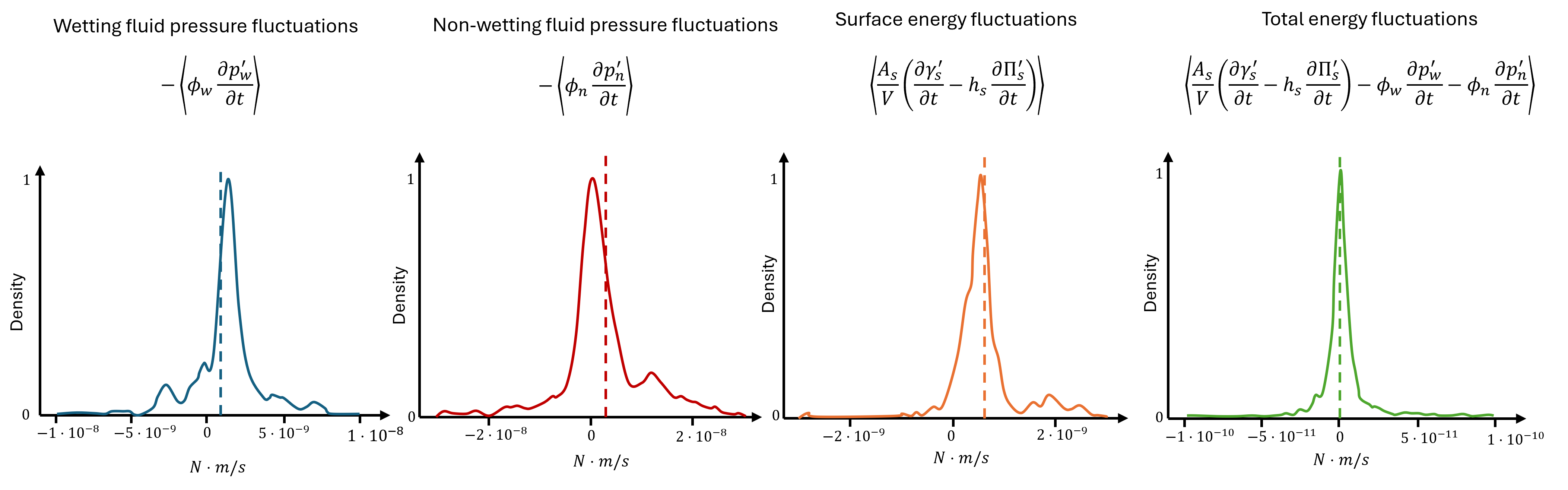}
			\caption{Histograms for fluctuation terms for steady-state flow in a Bentheimer sandstone for a water-wet situation at $S_w=0.43$ computed with Lattice Boltzmann simulations~\cite{mcclureLBPMSoftwarePackage2021}. While fluctuations for wetting phase pressure, non-wetting phase pressure and surface energy do not average out (non-zero mean, dashed lines), the fluctuations of the total energy dynamics averages out (zero mean) ~\cite{McClure2025}.}
			\label{fig:multiphaserev}
		\end{figure}

        However, space-time averaging brings a new perspective to the definition of an REV.  Fig.~\ref{fig:multiphaserev} shows histograms of the fluctuation terms for individual terms (such as the pressure-volume terms, but also the interfacial energy terms) and phases, and the combined fluctuations of the energy dynamics of all terms.        
               
        Fig.~\ref{fig:multiphaserev} shows that already at the length scale of 1-2 mm the histogram of combined fluctuations has zero mean, which suggests a multiphase REV of corresponding length scales. Note that the single-phase REV, i.e. porosity and permeability REV, for this particular rock is of similar size. 
        This implies that already few millimeter-sized domains can provide meaningful multiphase flow parameters when applying the space-time averaging. Similar results have also been observed for experiments conducted on a mini-core plug imaged with synchrotron-based micro-CT under co-current steady-state flow \cite{wangTimeAndSpaceAveragingApplied2024b}.

        %

        \subsection{Open questions}

        \begin{itemize}
            \item In the space-time averaging approach, molecular fluctuations are not considered explicitly.  Their dissipation is lumped into viscous dissipation terms under steady-state conditions so that space-time averages of fluctuations at the capillary energy scale are averaged out. 
            The question is what the consequences are actually. 
            This restriction was not imposed in the NET approach.

             \item Fluctuations are decomposed into contributions per degree of freedom as expressed by fluctuations in state variables for which Minkowski functionals are selected, which is intuitive. However, the Minkowski functionals are not independent from each other, which implies that the degrees of freedom represented by Minkowski functionals are not independent. The question is, what are the consequences? It is of course possible to formulate a reduced, independent set of non-dimensional groups~\cite{McClure2020} but these are less intuitive than the Minkowski functionals. 
            
        \end{itemize}

		
		\cleardoublepage
		\newpage
		\section{Statistical thermodynamics approach}\label{sect:statisticalmechanics}

        The central theme of this review is the scale-up problem, i.e., how to connect the physics of flow in porous media at the pore scale with a description of the same process at the Darcy scale.  The scale-up problem is of course not unique to flow in porous media.  The best description of the general problem was given by Anderson in his essay {\it More is Different\/} from 1972 \cite{anderson1972more}.  Anderson was a theoretical condensed matter physicist who won the Nobel Prize in 1977.  The essay {\it More is Different\/} was his reaction to what he saw as the arrogance of the high-energy physics community when they proclaimed that they were only ones doing fundamental physics, since everything consists of elementary particles (such as electrons, quarks, etc.). Anderson writes ``...the reductionist hypothesis does not by any means imply a ``constructionist" one: The ability to reduce everything to simple fundamental laws does not imply the ability to start from those laws and reconstruct the universe." He goes on: ``The constructionist hypothesis breaks down when confronted with twin difficulties of scale and complexity. The behavior of large and complex aggregates of elementary particles, it turns out, is not to be understood in terms of a simple extrapolation of the properties of a few particles. Instead, at each level of complexity entirely new properties appear, and the understanding of the new behaviors requires research which I think is as fundamental in its nature as any other."  We see here the introduction of the concept of {\it emergent properties.\/}   When scaling up from one level of description to another, the concepts involved change character and contents.  Take neuroscience.  At the cellular level, the brain is a large axion switching board. However, such a description is not useful for a psychiatrist with a patient on the couch.  The concepts used here are completely different, yet they describe the same system, the brain, but at different levels of scale. 

        This is precisely the point of view we take with respect to upscaling immiscible two-phase flow from the pore scale to the Darcy scale, which in this section we treat as a multi-body problem (i.e. the individual pores and their occupancy with wetting or non-wetting phases) for which its macroscopic i.e. upscaled behavior is described out of the statistics of pore level occupancy. The approach is different from those described so far, but they are --- as it were --- written on a palimpsest, namely statistical mechanics. The roots of statistical mechanics can be traced back to the last half of the nineteenth century~\cite{viscardyViscosityNewtonModern2010}.  In the middle of that century, equilibrium thermodynamics was largely in place. At the same time, the atomistic hypothesis was increasingly believed.  Clearly, atoms are very small, while at the same time thermodynamics was a theory developed in the wake of the steam engine, machinery that is in comparison huge.  How can we find a connection between small atoms and these huge machines?  This is a scale-up problem in this context: Find a description of the behavior of atoms at their scale, which leads to thermodynamics when scaled up. This is what statistical mechanics does, and it does this with tremendous success.     

        In order to avoid the in Fig.~\ref{fig:workflowpartitionfunction} illustrated complications with non-differentiable Euler-Lagrange function and Hamiltonian, out of which partition functions could be constructed, here a different route is taken.
        When Shannon introduced in 1948 his function for measuring ignorance in information theory, i.e., information entropy \cite{shannon1948mathematical}, he also opened the door for a generalization of statistical mechanics to information theory.  In 1957, Jaynes implemented this generalization of statistical mechanics from being a scale-up theory for molecular matter to a theory focused on information \cite{jaynes1957information}. This was done by replacing Boltzmann's statistical interpretation of thermodynamic entropy with Shannon's generalized information entropy, which is a quantitative measure of what is {\it not\/} known about a system. Let us consider a stochastic process that can produce $N$ possible outcomes, numbered from 1 to $N$, assuming a probability $p_i$ for the $i$th outcome $x_i$.  If we know nothing about the process, its information entropy must be at a maximum because our ignorance is maximal.  To utilize this, Shannon generalized to $N$ outcomes the Laplace Principle of Insufficient Reason \cite{de1995philosophical}, which states that {\it the optimal choice of probabilities for a stochastic process with two possible outcomes is to assign them equal probabilities.\/} Hence, the optimal choice when there are $N$ outcomes is to assign all probabilities the same value, $p_i=1/N$.  Assuming a number of symmetries that the function of ignorance must obey, Shannon ended up with the form  
        \begin{equation}
        \label{eq:eq4-1}
        S_I=-\sum_{i=1}^N p_i \log p_i\;,
        \end{equation}
        which is the information or Shannon entropy.  This function has a maximum $\log N$ when $p_i=1/N$. 
        
        What happens to information entropy if we {\it do\/} know something about the system?  Suppose we know the average over the possible outcomes, i.e., 
        \begin{equation}
        \label{eq:eq4-2}
        \langle x\rangle =\frac{1}{N}\sum_{i=1}^N p_i x_i\;.
        \end{equation}
        This is the question that Jaynes posed and then answered by generalizing the principle of insufficient reason further: the probabilities should be assigned so that the information entropy, equation (\ref{eq:eq4-1}), is maximized, given the constraint (\ref{eq:eq4-2}). The result is a probability $p_i$ given by 
        \begin{equation}
        \label{eq:jaynes-1}
        p_i=\frac{e^{-\lambda x_i}}{Z}\;,
        \end{equation}
        where 
        \begin{equation}
        \label{eq:jaynes-2}
        Z=\sum_{k=1}^N e^{-\lambda x_k}\;,
        \end{equation}
        is the normalization factor, also known as the {\it partition function,\/}
        and where $\lambda$ is a new variable that is determined by the equation
        \begin{equation}
        \label{eq:jaynes-3}
        \langle x\rangle=-\frac{\partial }{\partial \lambda} \log{Z}\;.
        \end{equation}
        We can see in Equation (\ref{eq:jaynes-1}) that an exponential probability appears.  This is for the same reason that the Boltzmann distribution appears in molecular systems~\cite{viscardyViscosityNewtonModern2010}. 

        There is one complication that we need to address before letting this approach loose on the immiscible two-phase flow in porous media problem: We will be dealing with {\it continuous distributions.\/} That is, rather than having a discrete set of $N$ possible outcomes, we have a continuous set of outcomes on the interval $x_{\min} \le x \le x_{\max}$ with probability density $p(x)$.  What then is information entropy?  Suppose that the discrete distribution $p_k$ may be approximated by a probability density $p(x_k)$ such that $p(x_k) \Delta x_k\approx p_k$ where $\Delta x_k=(x_{k+1}-x_{k})$. We let $k\to\infty$, finding that the information entropy (\ref{eq:eq4-1}) becomes
        \begin{equation}
        \label{eq:jaynes-4}
        S_I=-\int_{x_{\min}}^{x_{\max}} p(x)dx\  \log[p(x) dx]\;.
        \end{equation}
        This is worrisome because the integral is infinite. However, it is infinite in an interesting way.  To see this, we introduce the cumulative probability
        \begin{equation}
        \label{eq:jaynes-5}
        P(x)=\int_{x_{\min}}^x p(x')dx'\;,
        \end{equation}
        which is the probability of finding a value of the variable that is $x$ or less. We have $P(x_{\min})=0$ and $P(x_{\max})=1$.  We write the entropy in terms of the cumulative probability, finding
        \begin{equation}
        \label{eq:jaynes-6}
        S_I=-\int_{x_{\min}}^{x_{\max}} p(x)dx\  \log[p(x) dx]=-\int_0^1 dP\  \log[dP]\;,
        \end{equation}
        since $dP=p(x)dx$.  We may perform the integral to the right by discretizing the interval 0 to 1 into $N$ intervals $\Delta P=1/N$. The integral is therefore 
        \begin{equation}
        \label{eq:jaynes-7}
        S_I=-\lim_{N\to\infty}\sum_{k=1}^N\frac{1}{N}\ \log\left[\frac{1}{N}\right]=\lim_{N\to\infty} \log[N]\to \infty \;.
        \end{equation}
        Here is an important observation: Any probability distribution $p(x)$ may be expressed through a corresponding cumulative probability $P(x)$, which means that the corresponding information entropy may be expressed as the right-hand side of equation (\ref{eq:jaynes-6}).  Information entropy is {\it invariant\/} with respect to the probability distribution!  In other words, the information entropy defined in equation (\ref{eq:jaynes-4}) is useless as a basis for the Jaynes technique, as it cannot be maximized. However, if we split the expression inside the logarithm in equation (\ref{eq:jaynes-4}) into 
        \begin{equation}
        \label{eq:jaynes-8}
        S_I=-\int_{x_{\min}}^{x_{\max}} p(x)dx\  \log[p(x)]-
        \int_{x_{\min}}^{x_{\max}} p(x)dx\  \log[dx]\;,
        \end{equation}
        the second term is easy to calculate,
        \begin{equation}
        \label{eq:jaynes-9}
        -\int_{x_{\min}}^{x_{\max}} p(x)dx\  \log[dx]=\lim_{N_x\to\infty} \log[N_x]\;,
        \end{equation}
        where $N_x$ is the number of intervals we divided the range of $x$ into.
        Combining equations (\ref{eq:jaynes-7}), (\ref{eq:jaynes-8}) and (\ref{eq:jaynes-9}), we find
        \begin{equation}
        \label{eq:jaynes-10}
        \Delta S_I=-\int_{x_{\min}}^{x_{\max}} p(x)dx\  \log[p(x)]=\lim_{N,N_x\to\infty} \log\left[\frac{N}{N_x}\right]\;,
        \end{equation}
        which is a finite expression. This {\it differential\/} information entropy is what we will use, as it depends on the probability distribution $p(x)$ and may therefore be maximized. 
       
\begin{figure}
\begin{center}
\includegraphics[width=0.5\textwidth]{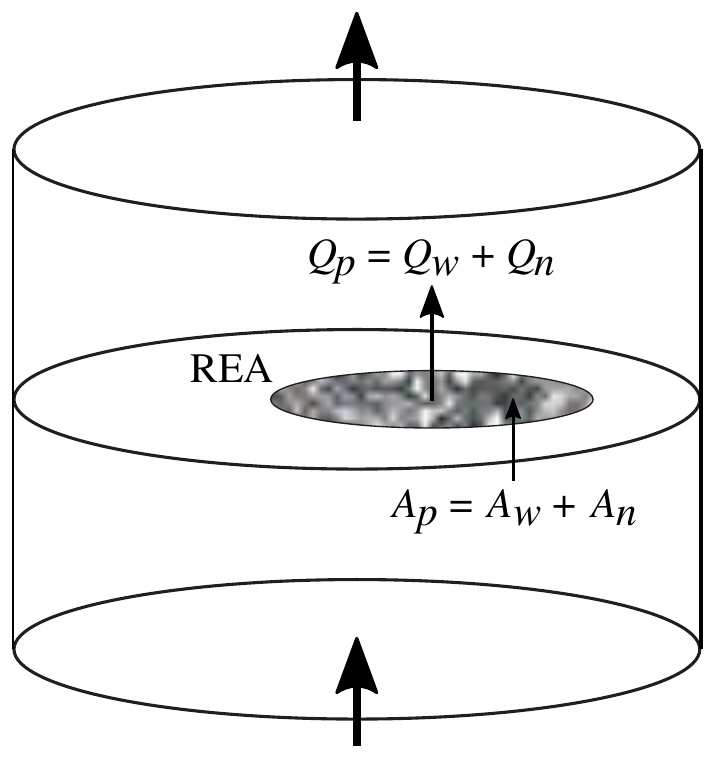}
\end{center}
\caption{We see here an imaginary cut through a porous medium sample orthogonal to the average flow direction. We choose a point in this cut and draw a circle around it with area $A$. This is a {\it Representative Elementary Area\/} (REA). We partition the area $A$ into a matrix area $A_m$ and a pore area $A_p$.  The pore area
$A_p$ may, in turn, be partitioned into an area cutting through the wetting fluid $A_w$ and an area cutting through the non-wetting area $A_n$, so that $A_p=A_w+A_n$.  Furthermore, there is a volumetric flow rate $Q_p$ passing through the REA.  This may be split into a wetting volumetric flow rate $Q_w$ and a non-wetting volumetric flow rate $Q_n$ such that $Q_p=Q_w+Q_n$. (From \cite{sinha2025thermodynamics}.)}\label{entropy-fig1}
\end{figure}          

\subsection{Defining the system}\label{sect:jaynes}

We now set the stage for using the Jaynes technique for the immiscible two-phase flow problem. The narrative here follow that of Hansen et al.\ \cite{hansenStatisticalMechanicsFramework2023} and Sinha and Hansen \cite{sinha2025thermodynamics}.  

We assume a cylindrical pore sample, as shown in Fig.\ \ref{entropy-fig1}.  Two immiscible fluids enter at the bottom and leave at the top.  The sidewalls are sealed.  We assume that the sample is statistically uniform throughout.  At some distance from the lower and upper edges, we assume the flow to be in a steady state --- that is, averages over the flow fluctuate around well-defined values.  We introduce a coordinate system $(x,y,z)$, with the $z$ axis oriented along the average flow direction. 

\begin{figure}
\begin{center}
\includegraphics[width=0.5\textwidth]{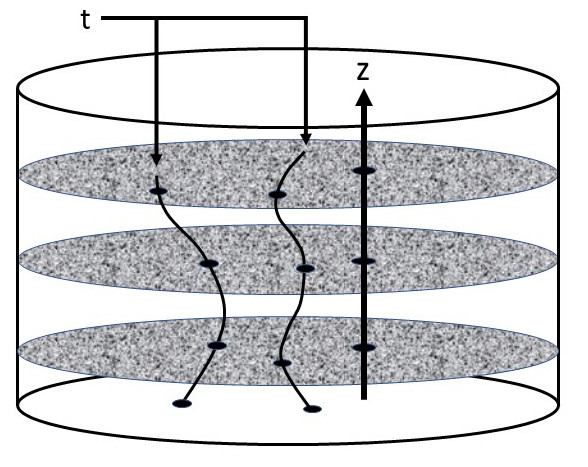}
\end{center}
\caption{A stack of three cuts through the porous sample. We also show the paths of two Lagrangian fluid elements that end at a time $t$. (From \cite{hansenStatisticalMechanicsFramework2023}.)}\label{stack-fig}
\end{figure}

Let us now imagine making $N$ imaginary cuts orthogonal to the average flow direction in this region, each a distance $dz$ apart from the others, so that the position of cut number $n$ along the $z$-axis is $z_n=z_0+(n-1)dz$, with $n=1,\cdots N$.  Fig.\ \ref{stack-fig} shows $N=3$ such cuts. We pick the cut at $z=z_n$.  It 
cuts through the matrix and pore space. As the pore space is fully fluid filled, the plane formed by the cut will cut through either the wetting or the non-wetting fluid. The matrix does not move, but there will be a velocity field associated with the fluids filling pore space.  We may formalize this in the following way:  Each point in the cut is characterized by two numbers, the first one is 0 if the point is in the matrix, 1 if it is in the wetting fluid, and 2 if it is in the non-wetting fluid.  The second number is the velocity at that point, which is zero if it is in the matrix or if the fluid at that point is stagnant. These numbers form a field over the cut, ${\cal X}={\cal X}(x,y;z_n,t)$, where $(x,y)$ are coordinates mapping the cut and $t$ is time. We will refer to ${\cal X}$ as the flow configuration --- even though it also contains the geometry of the matrix. Let us now imagine a moving cut in the following.  

Time keeps track of the motion of each Lagrangian fluid element moving through the porous medium. We show the motion of two such fluid elements through the porous sample in Fig.\ \ref{stack-fig}.  The position of the two fluid elements are shown at a given time $t$.  We note that the two end points, defining the position of the fluid elements, are at different values of $z$. 

We may parametrize time in a different way by basing ourselves on the {\it injected pore volumes\/} as time unit. If the transversal area of the porous sample is $A$, the porosity is $\phi$, and the volumetric flow rate of the fluids is $Q_p(t)$, the injected pore volume $V_I$ is given by
\begin{equation}
\label{eq-injected-volume}
V_I=\int_0^t Q_p(t')dt'\;.
\end{equation}
We may define a position along the $z$-axis associated with the injected pore volume,
\begin{equation}
\label{eq-injected-z}
z_I=\frac{V_I}{A\phi}\;.
\end{equation}
We may now imagine following a cut at $z_I$ moving through the porous sample at a speed $Q_p/A\phi$, thus letting the $z$ axis act as a time axis. If we record the changing fluid and matrix configurations $X$ in the moving cut at fixed interval $dz$, we may define a a probability density for the fluid and matrix configurations along the moving cut, ${\tilde p}({\cal X})$ from the sampling of the cuts. 

If we on the other hand choose a given cut $z_n$ and observe the changing fluid configurations as clock time $t$ passes, the matrix structure will be fixed. Hence, sampling over this cut will not be representative.  We define a configurational probability for this cut as $p_n({\cal X})$. 

\subsection{The REA ensemble}
\label{REA}

Due to the sealed sides of the porous media sample shown in Fig.\ \ref{entropy-fig1}, 
the average flow direction is along the $z$-axis. A plane cuts through the sample orthogonally to
the $z$-axis.  In this plane, we choose a point. Around this point, we choose 
an area $A$, e.g., bounded by a circle, as shown in Fig.\ \ref{entropy-fig1}. We assume that the area 
is large enough for averages of variables characterizing the flow are well-defined, but 
not larger.  Furthermore, we assume the linear size of the area to be larger than
any relevant correlation length in the system.  This defines the {\it Representative 
Elementary Area\/} (REA) \cite{bear2012introduction} at the chosen point. We may do the same at 
any point in the plane, and we may do this at any other such plane. In the same way as we defined 
$\cal X$ as the fluid and matrix configuration, we may define $X$ as the fluid and matrix configuration
for the REA.  Clearly, $X$ is a subset of $\cal X$. We also define the fluid configuration in 
the rest of the plane that is not part of the REA, $X_r$.  We will refer to this part of the plane as 
the ``reservoir."
Hence, we have that 
\begin{equation}
\label{eq:xxr}
{\cal X}=X\cup X_r\;.
\end{equation}
A central question is now, how independent are the configurations $X$ and $X_r$? If they are
independent, we may focus entirely on the REA configurations $X$ as we may write the 
configurational probability for the entire plane as
\begin{equation}
\label{eq:tildeppp}
\tilde p({\cal X})=p(X)p_r(X_r)\;,
\end{equation}
where $p(X)$ is the configurational probability for the REA and $p_r(X_r)$ is the 
configurational probability for the reservoir.

Fyhn et al.\ \cite{fyhnLocalStatisticsImmiscible2023} studied the validity of equation (\ref{eq:tildeppp})
in a two-dimensional dynamic pore network model \cite{sinha2021fluid}. By changing the size of the two-dimensional 
sample, while keeping the size of the REA fixed, they checked whether the statistical distributions 
of $Q_p$ and $A_w$ were dependent on the size of the sample.  Only a very weak dependency was found
which decrease with size.  It is therefore realistic to assume Equation (\ref{eq:tildeppp}) to be valid 
for large enough samples and REAs.

We now consider the REA shown in Fig.\ \ref{entropy-fig1}. We use the Jaynes maximum entropy principle to
construct a statistical mechanics where the REA defines the choice of ensemble.  

We assume a volumetric flow rate $Q_p$ passing through the REA. This may be split into a volumetric flow rate of the wetting fluid $Q_w$ and the volumetric flow rate of the non-wetting fluid $Q_n$, so that
\begin{equation}
\label{eq:eq2-1}
Q_p=Q_w+Q_n\;.
\end{equation} 
        The REA has an area $A$. An area $A_p<A$ cuts through pore space.  This area may be divided into the area cutting through the pore space filled with wetting fluid, $A_w$ and the area cutting though the pore space filled with the non-wetting fluid, $A_n$, so that 
\begin{equation}
\label{eq:eq3-1}
A_p=A_w+A_n\;.
\end{equation}

        The quantities that describe the REA depend on $X$, so that we have $Q_p(X)$, $Q_w(X)$, $Q_n(X)=Q_p(X)-Q_w(X)$, $A_p(X)$, $A_w(X)$, and $A_n(X)=A_p(X)-A_w(X)$. The pore area $A_p$ is special because it does not depend on the velocity field or the distribution of the two fluids, but only on the matrix. This will become important later. 
        
        The probability density for configuration $X$ is $p(X)$. We define a {\it configurational differential entropy,\/}
        \begin{equation}
        \label{eq:eq4-3}
        \Delta S_I=-\int dX\ p(X)\ \ln p(X)\;.
        \end{equation}

We assume that we know the averages in time of the variables just described:
\begin{eqnarray}
Q_u&=&\int dX\ p(X)\ Q_u(X)\;,\label{eq:eq4-4}\\
A_w&=&\int dX\ p(X)\ A_w(X)\;.\label{eq:eq4-5}
\end{eqnarray}

The variable $Q_u$ is related to the volumetric flow rate $Q_p$, but we defer its definition until Equation (\ref{eq:eq4-32}). 

We now follow the Jaynes prescription identifying the probability distribution $p(x)$ that maximizes the differential information entropy $\Delta S_I$ in (\ref{eq:eq4-3}), given constraints (\ref{eq:eq4-4}) and (\ref{eq:eq4-5}).  The result is
\begin{equation}
\label{eq:eq4-7}
p(X;\lambda_u,\lambda_w)=\frac{1}{Z(\lambda_u,\lambda_w}\ \exp\left[-\lambda_u Q_u(X)-\lambda_w A_w(X)\right]\;,
\end{equation}
where 
\begin{equation}
\label{eq:eq4-8}
Z(\lambda_u,\lambda_w)=\int\ dX\ \exp\left[-\lambda_u Q_u(X)-\lambda_w A_w(X)\right]
\end{equation} 
is the partition function. Two new {\it emergent variables\/} have appeared: $\lambda_u$ and $\lambda_w$. 

We may write the partition function as
\begin{equation}
\label{eq:eq4-13}
Z(\lambda_u,\lambda_w)=\exp\left[-\lambda_u Q_z(\lambda_u,\lambda_w)\right]\;,
\end{equation}
where $Q_z(\lambda_u,\lambda_w)$ is another new variable that will get its
interpretation in due time. We note that the pore area $A_p$ and pore geometry are specific for the chosen REA.  At a later point, we will also average over the position of the REA.  

The partition function is a generating function for the probability distribution $p(x)$. This means that knowing it, the statistics of the system are fully known. However, calculating the partition function is exceedingly difficult and can only be performed for simple model systems
\cite{baxter1985exactly}.  This means that a direct calculation of the immiscible two-phase flow problem is out of reach.  However, a byproduct of the statistical mechanics approach is the emergence of a set of thermodynamics-like relations between the REA-scale variables.  This framework is in place to determine whether it is possible to calculate the partition function explicitly or not.  This framework is the end product of the statistical thermodynamics approach to this problem.    

We start by calculating the differential information entropy $\Delta S_I$ by inserting Equation (\ref{eq:eq4-7}) in Equation (\ref{eq:eq4-3}) and using 
Equation (\ref{eq:eq4-13}).  We obtain
\begin{equation}
\label{eq:eq4-13-1}
\Delta S_I(Q_z,A_w)=-\lambda_u Q_z(\lambda_u,\lambda_w)+\lambda_u Q_u+\lambda_w A_w\;,
\end{equation}
which we rewrite as
\begin{equation}
\label{eq:eq4-13-2}
Q_z=Q_u-\frac{1}{\lambda_u}\ \Delta S_I+\frac{\lambda_w}{\lambda_u}\ A_w\;.
\end{equation}

It is convenient to define the two new variables
\begin{eqnarray}
\theta&=&+\frac{1}{\lambda_u}\;,\label{eq:eq4-15}\\
\mu&=&-\frac{\lambda_w}{\lambda_u}\;.\label{eq:eq4-16}
\end{eqnarray}

We may then write Equation (\ref{eq:eq4-13-2}) as
\begin{equation}
\label{eq:eq4-13-3}
Q_u(\Delta S_I,A_w,)=Q_z(\theta,\mu)+\theta \Delta S_I+\mu\ A_w \;,
\end{equation}
where we have used that this is a Legendre transform assuming convexity of the involved functions, since the configurational differential entropy
\begin{eqnarray}
\Delta S_I&=&-\left(\frac{\partial Q_z}{\partial \theta}\right)_{\mu}\;,\label{eq:eq4-9}\\
A_w&=&-\left(\frac{\partial Q_z}{\partial \mu}\right)_{\theta}\;.\label{eq:eq4-16-1}
\end{eqnarray}

We have finally arrived at defining the volumetric flow rate $Q_p$ in Equation (\ref{eq:eq2-1}),
\begin{equation}
\label{eq:eq4-31}
Q_p(\theta,A_w)=Q_z(\theta,\mu)+\mu A_w\;,
\end{equation}
and the relation between $Q_u$ and $Q_p$ is then
\begin{equation}
\label{eq:eq4-32}
Q_p(\theta,A_w)=Q_u(S,A_w)-\Delta S_I\theta\;.
\end{equation}
We summarize these results in Table \ref{table:flow_rates}.

We see that $\theta$ defined in (\ref{eq:eq4-15}) acts in a similar function as the temperature in ordinary thermodynamics.  Hansen et al.\ \cite{hansenStatisticalMechanicsFramework2023} named it {\it agiture\/}, which stands for {\it agitation temperature\/}.  They also called $\mu$ defined in (\ref{eq:eq4-16}), the {\it flow derivative\/}. It is analogous to the chemical potential in ordinary thermodynamics. We note that the unit of the agiture is flow rate, whereas the unit of the flow derivative and the flow pressure is velocity.

\begin{table}
\begin{center}
\begin{tabular}{ |c|c| } 
\hline
Control variables & Volumetric flow rate \\
\hline
$\theta,\mu$& $Q_z=-\theta \log Z$   \\ 
$\theta, A_w$ & $Q_p=Q_z+\mu A_w$  \\ 
$\Delta S_I, A_w$ & $Q_u=Q_p+\theta \Delta S_I$  \\ 
\hline
\end{tabular}
\end{center}
\caption{As in ordinary thermodynamics where we operate with different free energies,
we have here different volumetric flow rates depending on what the control variables are, see equations (\ref{eq:eq4-13}), (\ref{eq:eq4-31}), and (\ref{eq:eq4-32}).
}\label{table:flow_rates}
\end{table}

It should be noted that the flow rate $Q_p=Q_p(\theta,A_w)$ is measured directly in the laboratory.  This will become clear once we have identified the
agiture $\theta$ in terms of more common variables. The relation between $Q_p$ and the control variables $\theta$ and $A_w$ is a constitutive relation.  As we have already remarked, in principle, this may be found by calculating the partition function $Z$.  However, in practice, this is difficult.  If we find $Q_p$ via other means, e.g., by measuring it or computing it, which is the aim of the two other approaches we review here, we then have access to full thermodynamics with all its variables and relations.  This is completely analogous to ordinary thermodynamics, which we may see as a vast machine that requires an equation of state inserted to crank out results concerning that particular system.  

\subsection{Averaging over the position of the REAs}
\label{sec:pi}

So far, we have considered time averages over a given REA.  We now need to consider averaging over different REAs, corresponding to move from one cut to $N$ as described in Section \ref{sect:jaynes}.

Each REA presents its own static pore structure and accompanying pore area $A_p$.  In terms of the standard terminology of statistical mechanics, we would refer to this kind of disorder as {\it quenched\/}, as it is time independent. The disorder from the changing fluid configurations is called {\it annealed.\/}     

The flow rate averaged over fluid configurations but not over pore structure at fixed $\theta$ and $\mu$ (and $A_p$), is found from Equation (\ref{eq:eq4-13}),  
\begin{equation}
\label{eq:eq4-130}
Q_z(\theta,\mu)=-\theta \log Z(\theta,\mu)\;.
\end{equation}
We now need to average this quantity over the pore structure, i.e., over different REAs,
\begin{equation}
\label{eq:eq4-131}
\langle Q_z(\theta,\mu)\rangle =-\theta \langle \log Z(\theta,\mu)\rangle \;,
\end{equation}
where $\langle\cdots\rangle$ is the average over the pore structure in an REA over the direction perpendicular to the REA, which according to Fig.~\ref{stack-fig} is also the time coordinate, i.e. the average is then essentially a space-time average. 
The method to first perform the averaging over the annealed disorder and then over the quenched disorder was developed in the seventies and eighties in connection with {\it spin glass theory\/} \cite{binder1986spin}.  It relies on a mathematical trick that --- to be honest --- looks crazy. The amazing thing is that it works!  The {\it replica trick\/} is based on the mathematical relation
\begin{equation}
\label{eq:replica}
\log[x]=\lim_{n\to 0} \frac{x^n-1}{n}\;.
\end{equation}

We consider $n$ REAs --- called replicas --- and construct
\begin{equation}
\label{eq:replica2}
\langle \prod_{i=1}^n Z_i(\theta,\mu) \rangle\;,
\end{equation}
where $Z_i$ is the partition function for REA number $i$. When the calculation is completed, $n$ is a variable in the expression.  This is then analytically continued to non-integer values, and the limit $n\to 0$ is taken as in Equation (\ref{eq:replica}).

\begin{figure}
\begin{center}
\includegraphics[width=0.7\textwidth]{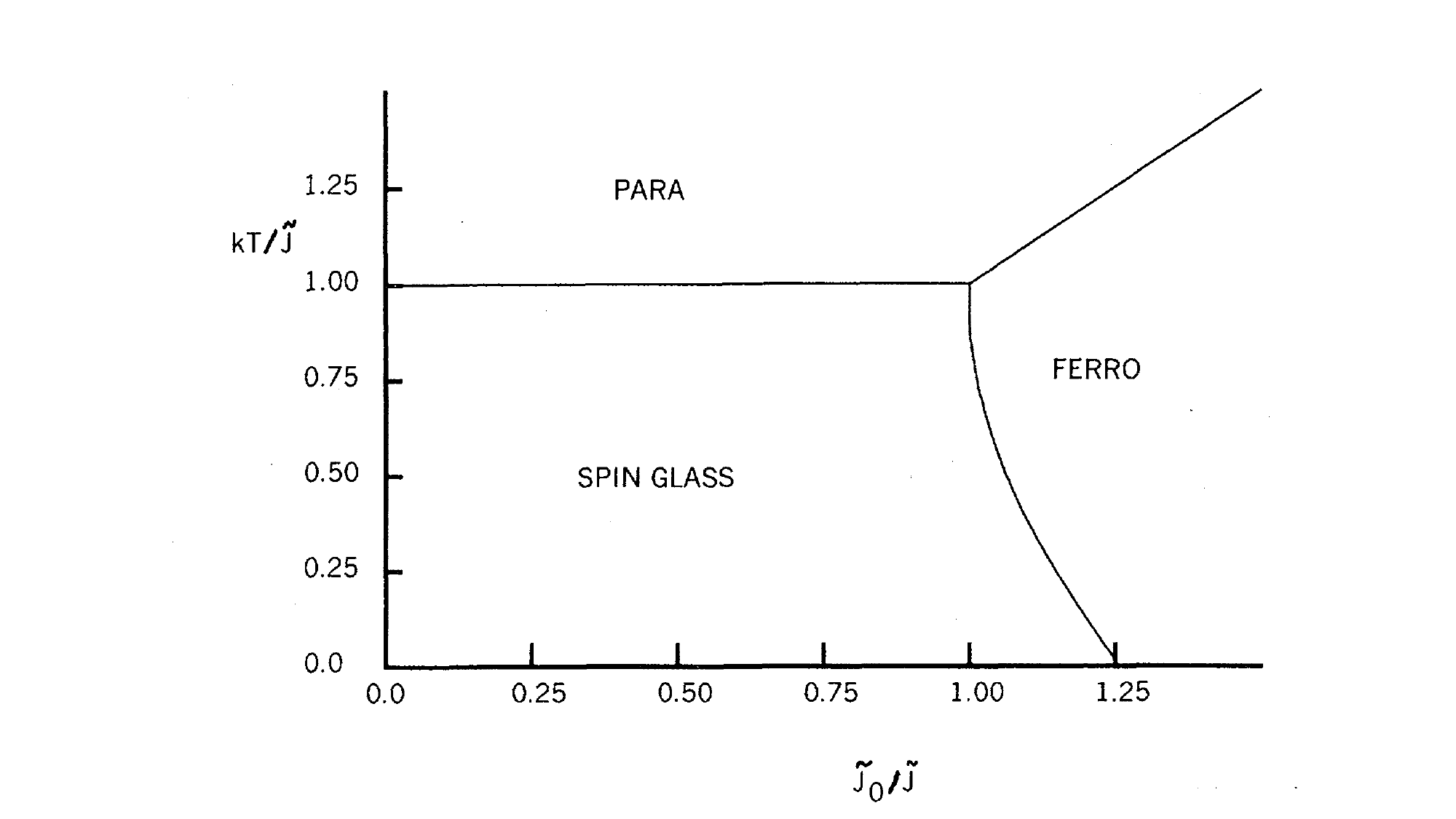}
\end{center}
\caption{\label{fig:sk-phase diagram} The phase diagram of the Sherrington-Kirkpatrick spin glass.  The two axes are $\tilde{J}_0/\tilde{J}$ and $kT/\tilde{J}$, where $\tilde{J}_0=J_0 N$ and $\tilde{J}=J N^{1/2}$. There are three phases: a paramagnetic phase --- nor magnetization in other words --- for high temperature $kT$ ($k$ is the Boltzmann constant and $T$ is the temperature).  When the average coupling constant $J_0$ is large compared to the variance of the coupling constants, $J$, we have the ferromagnetic phase, i.e., it is an ordinary magnet.  Lastly, when $J_0$ and $kT$ are small, we are dealing with a spin glass.   
(From \cite{sherrington1975solvable}.)}
\end{figure}

Let us now make a detour into spin glasses. Spin systems model magnets.  Each magnetic atom in the magnet has a magnetic moment that points in a specific direction.  When they all point in the same direction, the system has an overall magnetic moment making it into a ferromagnet.  When they point in arbitrary directions, the system is not magnetic and is a paramagnet.  The magnetic moments of the atoms are proportional to a quantum variable called the spin.  The quantum nature of the system only allows the spins to point in certain directions, the simplest being ``up" or ``down". Let us now consider $N$ spins, $S^i$, where $i=1,\cdots, N$. Each spin takes the value $S^i=\pm 1$. The spins interact with each other. Let us consider two spins, $S^i$ and $S^j$, where $i\neq j$. The two spins contribute $-J_{i,j}S^i S^j$ to the internal energy of the system.  $J_{i,j}$ is the coupling constant. We sum over all the spins to get the total internal magnetic energy, 
\begin{equation}
\label{eq:sk-ham}
H=-\sum_{i\neq j}^N J_{i,j} S^i S^j\;,
\end{equation}
and the probability to find a certain spin configuration $\{S^i\}$ is proportional to the Boltzmann factor
\begin{equation}
\label{eq:sk-boltzmann}
p\left(\{S^i\}\right)= \exp\left[\frac{1}{T}\sum_{i\neq j}^N J_{i,j} S^i S^j\right]\;,
\end{equation}
where $T$ denotes temperature. Now, assume that the coupling constants are distributed according to a Gaussian distribution around an average $J_0$ and with a variance $J$. 
This is the simplest spin glass, the {\it Sherrington-Kirkpatrick model\/} \cite{sherrington1975solvable}.  The resulting phase diagram is shown in Fig.\ \ref{fig:sk-phase diagram}.  There are three phases: a paramagnetic phase --- no magnetization in other words --- for high temperature $kT$ ($k$ is the Boltzmann constant).  When the average coupling constant $J_0$ is large compared to the variance of the coupling constants, $J$, we have the ferromagnetic phase, i.e., it is an ordinary magnet.  Finally, when $J_0$ and $kT$ are small, we are dealing with spin glass. This occurs when the coupling constants $J_{ij}$ have a mixture of signs. 
This phase is characterized by the average spin,
\begin{equation}
\label{eq:para-definition}
\frac{1}{N}\sum_{i=1}^N \overline{S^i}
\end{equation}
is zero, but the Edwards-Anderson order parameter 
\begin{equation}
\label{eq:EA-definition}
\frac{1}{N}\sum_{i=1}^N \overline{S^i}^2
\end{equation}
is larger than zero.
Here, $\overline{\cdots}$ is an average over time. 
This means that the local magnetization is non-zero, but averaged over the entire sample, it is zero.  The system is {\it frustrated,\/}, which manifests itself in that the energy function has large numbers of local minima.  This leads to {\it hysteresis.\/}

As a side remark, we point out that the phase diagram in Fig.\ \ref{fig:sk-phase diagram} turned out not to be the whole story.  Parisi \cite{binder1986spin} found another class of solutions that leads to a slightly different phase diagram.  

\begin{figure}
\begin{center}
\includegraphics[width=0.5\textwidth]{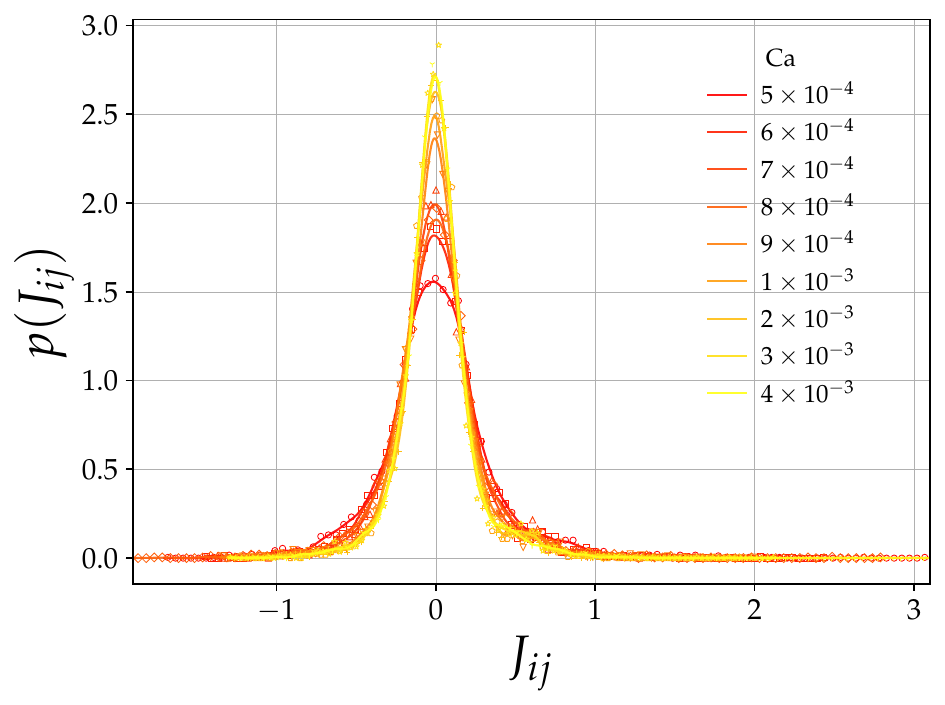}
\end{center}
\caption{Probability distribution $p(J_{ij})$ of the coupling constants $J_{ij}$, generated using the mapping described in the text from a pore network model for immiscible two-phase flow run at different capillary numbers $Ca$ and average saturation $S_w=1/2$.   
The distribution has a Gaussian maximum, centered somewhat on the positive side so that the coupling constants consist of a mixture of positive and negative values. It was generated by the mapping described in the text from the pore network model to the spin model. The mixture of positive and negative coupling constants signals that the mapped system is a spin glass.
(Courtesy Santanu Sinha.)
\label{fig:porous-sk} }
\end{figure}

Sinha et al.\ \cite{sinha2025sk} mapped the dynamics of immiscible two-phase flow in porous media under steady-state flow conditions on a spin model.  Each link in the pore network model has a saturation $s_i$ associated with it.  They then imagine a spin $S^i$ associated with the link, and the value of the spin is then assigned according to rule, 
\begin{equation}
\label{eq:spin_glass1}
S^i=
\begin{cases}
-1\ \text{for } s_i<1/2\;,\\
+1\ \text{for } s_i\ge 1/2\;,
\end{cases}
\end{equation}
i.e., a majority rule.  They run the model and measure the spin correlations --- i.e., transform the saturations to spins according to Equation (\ref{eq:spin_glass1}) and then measure the correlation functions
\begin{equation}
\label{eq:singles}
M_i=\overline{S^i}\;,
\end{equation}
and 
\begin{equation}
\label{eq:doubles}
C_{ij}=\overline{S^iS^j}\;.
\end{equation}
Now, the question is the following: Can one construct a series of coupling constants $J_{ij}$ in the spin model Hamiltonian (\ref{eq:sk-ham}) so that the model produces the same correlations as the pore network model.  This is achieved by invoking the Jaynes maximum entropy principle.  That is, the Shannon entropy $S_I$ is maximized by treating $M_i$ and $C_{ij}$ as constraints, leading to the determination of the coupling constant. The probability density for $J_{ij}$ for an average saturation $S_w=1/2$ and different capillary numbers $Ca$ is shown in Fig.\ \ref{fig:porous-sk}. The resulting mixture of positive and negative coupling constants $J_{ij}$ in Fig.\ \ref{fig:porous-sk} signals that the spin system is in the spin glass phase.  Sinha et al.\ then reason in the following way: if the spin system that the flow problem is mapped onto is in a glassy phase, so must also the flow problem itself. Such a {\it glassy flow phase\/} would be characterized by hysteresis, broken ergodicity (Fig.~\ref{fig:porescale-ergodicity}), and dynamics over a wide range of time scales, which is essentially observed in regime I in Fig.~\ref{fig:flowregimeslinearnonlinear}.    

        \subsubsection{Statistical mechanics approach to water retention}
        \label{subsubsec:Louge}

        We use the opportunity to review another use of spin models in connection with immiscible two-phase flows in porous media.  Xu and Louge \cite{xuStatisticalMechanicsUnsaturated2015} use statistical mechanics to study the fluid retention curve in a porous medium where a fluid (e.g.\ water) competes with a gas (e.g.\ air), assuming that the fluid is the wetting phase and the gas is the non-wetting phase. The fluid retention curve relates the macroscopic capillary pressure $P_c=P_n-P_w$ to saturation $S_w$, i.e., $S_w=S_w(P_c)$. This is the inverse of the capillary pressure curve $P_c=P_c(S_w)$ introduced by Leverett \cite{leverettCapillaryBehaviorPorous1941}. Similarly, the capillary pressure curve is highly hysteretic, as is the fluid retention curve.  This means that it follows a different path depending on whether the fluid is imbibed into the porous medium or drained from it, that is,\ $S_w=S_w^i(P_c)$ for imbibition and $S_w=S_w^d(P_c)$ for drainage. What is the origin of hysteresis? This is the question that Xu and Louge answer by considering nearest-neighbor interactions which is the interfacial energy between fluids in adjacent pores.  

        They see the porous medium as a collection of pores that communicate through pore throats.  In practice, each pore may be partially filled with fluid and gas, but they approximate that each pore is either completely filled with water or completely empty.  This allows them to assign a binary variable to each pore $i$, $S^i$ which takes the value $+1$ if the pore is gas filled and $-1$ if filled with water. 
        
        Suppose now that pore $i$ which shares pore throats with $n$ neighbors, changes from being gas filled to fluid filled or vice versa, i.e., $S^i \to -S^i$.  What is the energy $\Delta E(S^i\to -S^i)$  that needs to be supplied to accomplish this?  There is the work done to fill or drain the pore,
        \begin{equation}
        \label{eq:Louge1}
        \Delta E_v=-S^i\int_{v_p}^0 (P_w-P_n)dv=-P_c v_pS^i\;,
        \end{equation}
        where $v_p$ is the pore volume.  Then there is the work associated with the surface energies between the wetting and non-wetting fluids. $\gamma_{wn}$ and between the wetting fluid and the matrix walls, $\gamma_{wm}$,
        \begin{equation}
        \label{eq:louge2}
        \Delta E_p=(\gamma_{wn}-\gamma_{wm})a_pS^i\;,
        \end{equation}
        where $a_p$ is the area of the pore. Both of these work terms only involve a single pore. Lastly, there is work associated with the emptying or filling of the adjacent pores with fluid.  This is due to the fluid-gas interfacial energy associated with the interfaces passing through the pore throats.  Suppose there are $N$ adjacent pores to pore $i$.  The work is
        \begin{equation}
        \label{eq:louge3}
        \Delta E_n=-\sum_{j=1}^N S^j \gamma_{wn} a_{i,j}S^i\;,
        \end{equation}
        where $a_{i,j}$ is the transversal area of pore throat $i$ and $j$. The total energy change $\Delta E(S^i\to -S^i)$ is given by
        \begin{equation}
        \label{eq:louge4}
        \Delta E(S^i\to -S^i)=\Delta E_v+\Delta E_p+\Delta E_n\;.
        \end{equation}
        The energy of a single pore $i$ is then given by 
        \begin{equation}
        \label{eq:louge5}
        E_i=\frac{1}{2}\Delta E(S^i\to -S^i)=\frac{1}{2}\left[-P_c v_p+(\gamma_{wn}-\gamma_{wm})a_p-\sum_{j=1}^n S^j \gamma_{wn} a_{i,j}\right]S^i\;,
        \end{equation}
        and the total energy is 
        \begin{equation}
        \label{eq:louge6}
        H=\sum_{i=1}^N E_i\;,
        \end{equation}
        where $N$ is the total number of pores.  This is a spin model Hamiltonian of the same type we have already met in Equation (\ref{eq:sk-ham}) --- but with one large difference: Only spins that are neighbors are connected.  In Section \ref{sec:pi}, all spins were connected irrespective of distance between them.  

        Spin models governed by Hamiltonians as in Equation (\ref{eq:louge6}) are known as {\it Ising models.\/} This model, first introduced in Ernest Ising's doctoral thesis in 1924 \cite{ising1925beitrag}, has since been central in statistical mechanics.  Lars Onsager managed to calculate the partition function for the Ising model implemented on a square lattice in 1944 \cite{onsager1944crystal}. A quote in this connection is in order: ``In 1945, Casimir complained that in Dutch isolation during the war, five years of hectic American activity in physics had passed him by. To that the Swiss Nobel prize winner Pauli quipped that the only development during those five years worth mentioning was Onsager’s solution of the Ising model" \cite{hauge2005lars}.
        Kenneth G.\ Wilson won the 1982 physics Nobel Prize for inventing an {\it approximate\/} method to calculate the Ising partition function near critical points.  

        An important property of Ising models is that they contain phase transitions.  They may be of first order or second order (also known as critical points or lines) between ferromagnetic or paramagnetic states.  The first order transitions are hysteretic.  Xu and Louge proceed to identify the hysteretic behavior of the water retention curve with such hysteretic first order transitions of the Ising model defined in Equation (\ref{eq:louge6}).  

\subsection{From REA to points}
\label{sect:intensive}

We now return to the statistical thermodynamics approach after this detour. The variables $Q_u$, $Q_z$, $Q_p$, $A_w$, $A_p$, and $\Delta S_I$ are extensive in the area of the REA: double its area $A$, and all these variables double. On the other hand, the variables $\theta$ and $\mu$ are intensive.  They do not change when $A$ changes.  
We rescale $Q_p$, $A_w$, and $A_p$ by $A$, defining
\begin{eqnarray}
\phi&=&\frac{A_p}{A}\;,\label{eq:eq5-1}\\
S_w\phi&=&\frac{A_w}{A}=\frac{A_w}{A_p}\phi\;,\label{eq:eq5-2}\\
v_d&=&v_p\phi=\frac{Q_p}{A}=\frac{Q_p}{A_p}\phi\;,\label{eq:eq5-3}\\
s&=&\frac{\Delta S_I}{A}\;,\label{eq:eq5-6}
\end{eqnarray}
where $\phi$ is the porosity, $S_w$ the wetting saturation, $v_d$ the Darcy velocity, $v_p$ the pore or seepage velocity, and $s$ the differential information entropy area density. 

In order to keep track of what happens to the variables in, e.g., $Q_p(\theta,A_w,A_p,A)$, we express the extensivity through the scaling relation 
\begin{equation}
\label{eq:eq5-4}
\lambda Q_p(\theta,A_w,A_p,A)=Q_p(\theta,\lambda A_w,\lambda A_p,\lambda A)\;.
\end{equation}

We set the scale factor $\lambda=1/A$, finding
\begin{equation}
\label{eq:eq5-5}
v_d(\theta,S_w,\phi)=v_p(\theta,S_w,\phi)\phi=\frac{1}{A}Q_p(\theta,A_w,A_p,A)=Q_p(\theta,S_w\phi,\phi,1)\;.
\end{equation}

Using the same arguments, we find that
\begin{equation}
\label{eq:eq5-7}
s=s(\theta,S_w,\phi)\;.
\end{equation}

The variables defined in equations (\ref{eq:eq5-1}) to (\ref{eq:eq5-6}) and equation (\ref{eq:eq5-7}) are variables that may be localized to a point in space and time, $(\vec x,t)$ (hence the title of the section).  Therefore, they are fields at the Darcy scale. 

 The velocities of the two fluids can be defined in many different ways.  We have the Darcy velocities $v_{di}$ (m/s), which is the volumetric flow rate of fluid $i$ per area. We may {\it non-dimensionalize the Darcy velocities\/} by using the parameters at hand, permeability $K$ (m$^2$), the pressure gradient $p'=\partial P/\partial x$ (Pa/m), and the viscosity $\mu_i$ (Pa s), finding
    \begin{equation}
    \label{eq:non-dim-kr}
    k_{ri}=-\frac{\mu_i v_{di}}{K p'}\;,
    \end{equation}
    where we have added a minus sign as the pressure gradient $p'$
    is negative in the flow direction. It is of course no accident that we use the symbol for the relative permeabilities for the nondimensionalized Darcy velocities, as that it what they are.  If we invert equation (\ref{eq:non-dim-kr}), we find
    \begin{equation}
    \label{eq:Darcy-double}
    v_{di}=-\frac{K}{\mu_i}\ k_{ri}\ p'\;,
    \end{equation}
    which are the two-fluid Darcy equations.  Therefore, these equations are generic. Their only physical contents as they stand is to relate the Darcy velocity $v_{di}$ to its non-dimensional version $k_{ri}$.  

    It turns out that for the discussion to follow, the Darcy velocities --- with or without dimensions --- are not the most convenient to use.  Rather, we use the {\it seepage\/} --- or {\it average pore\/} --- velocities $v_i$. The relation between seepage velocity and Darcy velocity is
    \begin{equation}
    \label{eq:darcy-seepage}
     v_i=\frac{v_{di}}{\phi S_i}\;.
    \end{equation}

\subsection{Emergent variables}
\label{sect:emergent}

The two variables, $\theta$ and $\mu$ emerge because of the scale-up process.  These are the conjugate variables to the entropy $\Delta S_I$ and wetting area $A_w$, respectively.  How can we interpret them? 

\begin{figure}
\begin{center}
\includegraphics[width=0.15\textwidth]{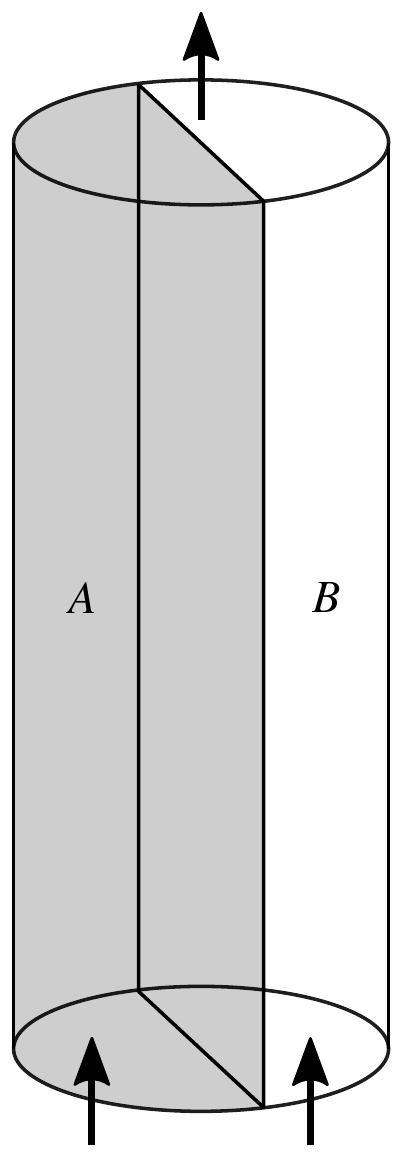}
\end{center}
\caption{{A} cylindrical porous sample consisting of two halves, A and B, having different properties. Two immiscible fluids are injected at the bottom edge. The two halves are in direct contact, and fluids may pass unrestricted between them. (From \cite{sinha2025thermodynamics}.)
\label{entropy-fig2}}
\end{figure}

We follow Hansen et al. \ \cite{hansenStatisticalMechanicsFramework2023} and consider a cylindrical porous medium (see Fig.\ \ref{entropy-fig2}).  We imagine a plane running through the cylinder axis parallel to the average flow direction.  On one side of the plane, the porous medium has one set of properties; on the other side, it has another set of properties.  The difference may be in the chemical composition of the matrix, or it may consist of different porosities or topological structures.  We name the two halves ``A'' and ``B'', respectively.  We assume that the two halves are statistically homogeneous along the flow axis.  

We now assume that two immiscible and incompressible fluids are simultaneously pushed through the porous cylinder.  Away from the edge where the two fluids are injected, the flow is in a steady state.  This means that the flow statistics are invariant along the axis.  However, this does not mean that the fluid--fluid interfaces do not move.  At sufficient flow rates, they do, and as a result, fluid clusters merge and break up. 

As the statistical distributions describing the flow are invariant along the flow axis, the configurational entropy $\Delta S_I$ is invariant along this axis.  As shown by Hansen et al.\ \cite{hansenStatisticalMechanicsFramework2023}, this means that the agiture in sectors A and B, $\theta^A$ and $\theta^B$, are equal,
\begin{equation}
\label{eq:eq6-1}
\theta^A=\theta^B\;.
\end{equation}

In ordinary thermodynamics, the conjugate of a conserved quantity is constant in a heterogeneous system at equilibrium. This is a generalization of the argument for the temperature being the same everywhere in a system at equilibrium, as the entropy is conserved and the temperature is its conjugate. This reasoning may be repeated for flow problems. 

The wetting area $A_w$ is not conserved along the flow axis at the pore scale.  This implies that $\mu^A\neq \mu^B$. However, $A_w$ is conserved {\it at the continuum scale,\/}, as will be apparent later.  This is an emergent law of conservation. We therefore conjecture that \cite{hansenStatisticalMechanicsFramework2023}
\begin{equation}
\label{eq:eq6-2}
\mu^A=\mu^B\;.
\end{equation}

\subsection{Interpretation of the agiture $\theta$}
\label{sect:agiture}

The agiture $\theta$ plays the role of temperature in the thermodynamics formalism that we are developing.  However, can we express it in terms of more familiar variables?  

We first note that in the formalism we have so far, there is one variable that is missing: pressure $P$. Intuitively, one would expect the agiture to be related to the pressure gradient $\nabla P$: a higher pressure gradient should indicate a higher agiture.    

\begin{figure}[htbp]
\begin{center}
\includegraphics[width=0.15\textwidth]{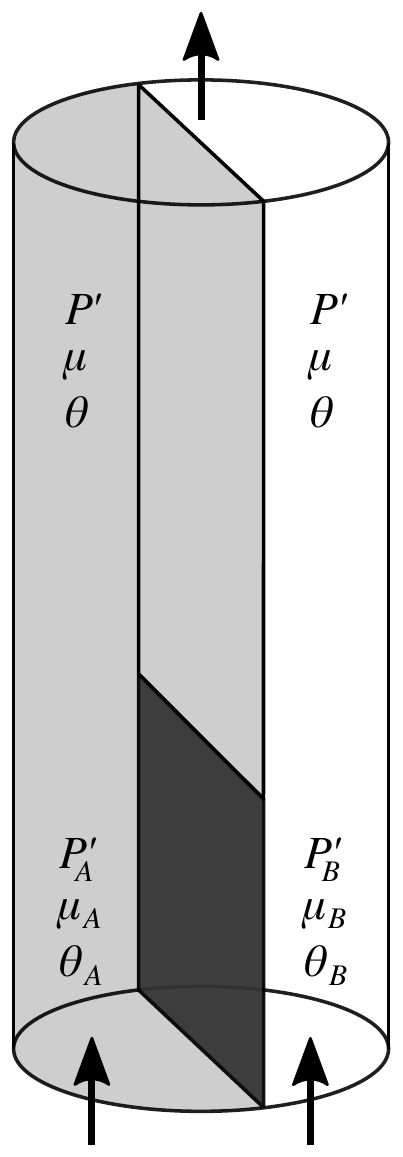}
\end{center}
\caption{{A} cylindrical porous media sample consisting of two halves, A and B, having different properties with respect to the two immiscible fluids injected at the bottom edge. There is an impenetrable wall separating the two halves at the bottom third of the boundary. This is indicated by the dark section. Above, the two halves are in direct contact, as shown in Fig.\ \ref{entropy-fig2}. A mixture of the two fluids is injected into the two halves at the lower edge. The pressure gradient is $p'_A$ on side A and $p'_B$ on side B, up to the upper edge of the impenetrable wall. Likewise, the flow derivatives are $\mu_A$ and $\mu_B$ and the agitures are $\theta_A$ and $\theta_B$.  Higher up, where the two halves communicate, the pressure gradients, the flow derivatives, and the agitures become pairwise equal.\label{entropy-fig3} (From \cite{sinha2025thermodynamics}.)}
\end{figure}

Figure \ref{entropy-fig3} shows a cylindrical porous medium consisting of two halves, A and B.  However, they exhibit different properties. At the bottom one-third of the porous cylinder, the boundary between the two halves is impenetrable. This is indicated by the dark section of the boundary.  Above, the two halves are in direct contact.  We inject the immiscible fluids into each of the halves A and B at the bottom.  This can be performed by injecting at different volumetric flow rates $Q_p^A=Q_w^A+Q_n^A$ and $Q_p^B=Q_w^B+Q_n^B$.  This gives rise to flow derivatives $\mu_A$ and $\mu_B$, and agitates $\theta_A$ and $\theta_B$ in the two halves sufficiently far from the inlet for the flow to be in a steady state.  The pressure gradient $p'$ then points along the axis of the cylinder. We find pressure gradients $p'_A$ and $p'_B$ in A and B, respectively. Further into the cylindrical sample above the impenetrable wall, the flow adjusts to a new steady state characterized by equal agiture $\theta$ and flow derivative $\mu$, and  in both halves according to Equations (\ref{eq:eq6-1}) and (\ref{eq:eq6-2}).  Likewise, the pressure gradient $p'$ must be equal in the two halves because there cannot be any net flow in either direction across the boundary between the two halves, apart from local fluctuations. 

Because the variables $\theta$ and $\mu$ on one side and $p'$ on the other side form alternate descriptions, we must assume that they are related.  The simplest relation between them we may write down is
\begin{equation}
\label{eq6-4}
|p'| = - c_\theta \theta +c_\mu \mu\;.
\end{equation}

The reason for the negative sign in front of the $\theta$ term is that the flow direction is opposite to the pressure gradient direction. We note that the unit of $c_\theta$ is viscosity and that of $c_\mu$ is viscosity times area.      

We note that Equation (\ref{eq:eq5-5}) defines the seepage velocity $v_p(\theta,S_w,\phi)$ where $S_w$ rather than the flow derivative $\mu$ is a control variable.  Simultaneously, the seepage velocity depends on the local pressure gradient $v_p(p',S_w,\phi)$.  Hence, we conjecture that $c_\mu=0$, so that
\begin{equation}
\label{eq:eq6-5}
|p'| = - c_\theta \theta\;.
\end{equation}

We note that $Q_p=Q_p(\theta,A_w,A_p)$, defined in Equation (\ref{eq:eq4-31}), may be written as
\begin{equation}
\label{eq:eq6-6}
Q_p=Q_p(p',A_w,A_p)=A_p v_p(p',S_w,\phi)\;,
\end{equation}
where we also used Equation (\ref{eq:eq5-5}).  We see that this is a constitutive equation relating the seepage velocity to the local pressure gradient, saturation, and porosity.              

Equation (\ref{eq:eq6-5}) has interesting consequences. We consider the following: The agiture is the conjugate variable to the entropy.  We may express it as
\begin{equation}
\label{eq:eq6-8}
\theta=\left(\frac{\partial Q_u}{\partial \Delta S_I}\right)_{A_w,A_p}\;,
\end{equation}

The entropy is given by 
\begin{equation}
\label{eq:eq6-9}
\Delta S_I=-\left(\frac{\partial Q_p}{\partial \theta}\right)_{A_w,A_p}=c_\theta\left(\frac{\partial Q_p}{\partial p'}\right)_{A_w,A_p}\;,
\end{equation} 
or, in terms of the entropy density,
\begin{equation}
\label{eq:eq6-10}
s=c_\theta\left(\frac{\partial v_p}{\partial p'}\right)_{S_w,\phi}\;.
\end{equation} 

We now define the {\it differential mobility\/} as 
\begin{equation}
\label{eq:eq6-11}
m(p',S_w,\phi)=-\left(\frac{\partial v_p}{\partial p'}\right)_{S_w,\phi}\;.
\end{equation}

Comparing Equations (\ref{eq:eq6-10}) and (\ref{eq:eq6-11}), we see that 
\begin{equation}
\label{eq:eq6-13}
s+s_0=-c_\theta m\;.
\end{equation}

In her MSc thesis \cite{hermundstad2025shannon}, Hermundstad compared the Shannon information entropy with differential mobility based on a dynamic pore network model \cite{sinha2021fluid} using wavelets to spatially de-correlate the signal.  The results indicate that Equation (\ref{eq:eq6-13}) is valid: The differential mobility is proportional to the configurational entropy.  

\subsection{Flow derivative}
\label{sect:mu}

Following Hansen et al.\ \cite{hansen2018relations}, we may define the {\it thermodynamic velocities\/} as
\begin{eqnarray}
{\hat v}_w=\left(\frac{\partial Q_p}{\partial A_w}\right)_{\theta,A_n}\;,\label{eq:eq6-14}\\
{\hat v}_n=\left(\frac{\partial Q_p}{\partial A_n}\right)_{\theta,A_w}\;,\label{eq:eq6-15}
\end{eqnarray}
where the control variables are $\theta$, $A_w$, and $A_n$, as defined in (\ref{eq:eq3-1}), making $A_p$ a dependent variable. Changing the variables $(\theta,A_w,A_n)\to(\theta,S_w,A_p)$, these two equations may be written as
\begin{eqnarray}
{\hat v}_w=v_p+S_n\left(\frac{\partial v_p}{\partial S_w}\right)_{\theta,\phi}\;,\label{eq:eq6-16}\\
{\hat v}_n=v_p-S_w\left(\frac{\partial v_p}{\partial S_w}\right)_{\theta,\phi}\;.\label{eq:eq6-17}
\end{eqnarray}

The flow derivative $\mu$, which is the conjugate of the wetting area $A_w$, and thereby also the saturation $S_w$, is given by
\begin{equation}
\label{eq:eq6-18}
\mu=-\left(\frac{\partial Q_p}{\partial A_w}\right)_{\theta,A_p}=\left(\frac{\partial v_p}{\partial S_w}\right)_{\theta,\phi}\;.
\end{equation}

Hence, Equation (\ref{eq:eq6-17}) may be recognized as a Legendre transformation substituting $S_w \to \mu$, 
\begin{equation}
\label{eq6-19}
{\hat v}_n(\theta,\mu,\phi)=v_p(\theta,S_w,\phi)-S_w(\theta,\mu,\phi)\mu\;.
\end{equation}
Hence, the non-wetting thermodynamic velocity is the Legendre transformation of the average seepage velocity with respect to saturation. 

\subsection{Thermodynamic velocities, seepage velocities and the co-moving velocity}
\label{sect:seepage}

We introduced in Section \ref{sect:mu} the {\it thermodynamic velocities,\/} Equations (\ref{eq:eq6-14}) and (\ref{eq:eq6-15}),
\begin{eqnarray}
{\hat v}_w=\left(\frac{\partial Q_p}{\partial A_w}\right)_{\theta,A_n}\;,\nonumber\\
{\hat v}_n=\left(\frac{\partial Q_p}{\partial A_n}\right)_{\theta,A_w}\;.\nonumber
\end{eqnarray}
These velocities are different from the {\it seepage velocities\/} defined as
\begin{eqnarray}
{\hat v}_w=\frac{Q_w}{A_w}\;,\label{eq:co-moving-10}\\
{\hat v}_n=\frac{Q_n}{A_n}\;.\label{eq:co-moving-11}
\end{eqnarray}
Both the thermodynamic velocities and the seepage velocities obey the relation
\begin{equation}
    v_p=S_w\hat{v}_w+S_n\hat{v}_n=S_wv_w+S_nv_n\;.
    \label{eq:co-moving-12}
\end{equation}
It is tempting to assume that this means that $\hat{v}_w=v_w$ and $\hat{v}_n=v_n$ because this equality is valid for all saturations $S_w$.  However, this is incorrect. It is straight forward to show that the most general relation between the thermodynamic velocities and the seepage velocities are given by the relation
\cite{hansen2018relations}
\begin{eqnarray}
\hat{v}_w&=&v_w+v_m S_n\;,\label{eq:co-moving-2}\\
\hat{v}_n&=&v_w-v_m S_w\;,\label{eq:co-moving-3}
\end{eqnarray}
where $v_m$ denotes the {\it co-moving velocity\/} \cite{royCoMovingVelocityImmiscible2022,alzubaidiImpactWettabilityComoving2024,hansenLinearityComovingVelocity2024,hansen2018relations,pedersen2023parameterizations,pedersen2025geometric,pedersen2025co}.

The thermodynamic velocities obey the two relations
\begin{equation}
\label{eq:co-moving-13}
0=S_w\frac{\partial \hat{v}_w}{\partial S_w}+S_n \frac{\partial \hat{v}_n}{\partial S_w}\;,
\end{equation}
and 
\begin{equation}
\label{eq:co-moving-14}
0=\frac{\partial v_p}{\partial S_w}-\hat{v}_w+\hat{v}_n\;.
\end{equation}
Combining these two equations with Equations (\ref{eq:co-moving-2}) and (\ref{eq:co-moving-3}) gives
\begin{equation}
\label{eq:co-moving-1}
v_m=S_w\frac{\partial v_w}{\partial S_w}+S_n \frac{\partial v_n}{\partial S_w}\;,
\end{equation}
and
\begin{equation}
\label{eq:co-moving-4}
v_m=\frac{\partial v_p}{\partial S_w}-v_w+v_n\;.
\end{equation}

We now combine the defining Equations (\ref{eq:co-moving-2}) and (\ref{eq:co-moving-3}) for the co-moving velocity with Equations (\ref{eq:eq6-16}) and (\ref{eq:eq6-17}) to find 
\begin{eqnarray}
v_w&=&v_p+S_n[\mu-v_m]\;,\label{eq:co-moving-20}\\
v_n&=&v_p-S_w[\mu-v_m]\;.\label{eq:co-moving-21}
\end{eqnarray}
Here, we have used the definition of the flow derivative in Equation (\ref{eq:eq6-18}). These equations imply that the natural variables for the co-moving velocity are $(\theta,\mu,\phi)$,
\begin{equation}
\label{eq:co-moving-22}
v_m=v_m(\theta,\mu,\phi)\;.
\end{equation}

Let us summarize.  We have the mapping $(v_w,v_n) \to (v_p,v_m)$ given by Equations (\ref{eq:co-moving-12}) and (\ref{eq:co-moving-1}),
\begin{eqnarray}
v_p&=&S_wv_w+S_nv_n\;,\nonumber\\
v_m&=&S_w\frac{\partial v_w}{\partial S_w}+S_n \frac{\partial v_n}{\partial S_w}\;.\nonumber
\end{eqnarray}
The opposite mapping $(v_p,v_m)\to (v_w,v_n)$ is given by Equations (\ref{eq:co-moving-20}) and (\ref{eq:co-moving-21}).  Such a mapping, which provides the seepage velocity of each fluid when the average seepage velocity $v_p$ is known together with $v_m$, is new.  In Section \ref{sec:non-linear} we describe flow regimes where the relation between flow rate and pressure gradient is a power law.  Without $v_m$, and therefore mapping $(v_p,v_m)\to (v_w,v_n)$, it would not be possible to deduce $v_w$ and $v_n$.  Instead, they would have to be measured directly, as in Yoitis et al.\ \cite{yiotis2019nonlinear}.  

As we will see in Section \ref{sect:comovingvelocity}, the functional form of $v_m=v_m(\theta, \mu, \phi)$ is surprisingly simple.  This makes the co-moving velocity a useful tool.      

\subsection{Fluctuations and higher-order statistical moments}
\label{sect:flow-area}

So far, we have only addressed average velocities. Can we generalize the thermodynamic formalism we have developed to address fluctuations and higher statistical moments?  The answer is yes \cite{roy2020flow}. The key concept here is the {\it differential transversal pore area\/} $a_p(S_w,v)$. We define $dA_p=a_p dv$ as the area covered by fluid with velocities in the range $[v,v+dv]$.  This means that 
\begin{equation}
\label{eq:ap-1}
A_p=\int_{-\infty}^\infty dv\ a_p(S_w,v)\;.
\end{equation}
We must also have that
\begin{equation}
\label{eq:ap-2}
Q_p=\int_{-\infty}^\infty dv\ v\ a_p(S_w,v)\;.
\end{equation}
The average seepage velocity is then given by
\begin{equation}
\label{eq:ap-3}
v_p=\frac{1}{A_p}\int_{-\infty}^\infty dv\ v\ a_p(S_w,v)\;.
\end{equation}

We may split the differential pore area into one for the wetting fluid, $a_w(S_w,v)$ and one for the non-wetting fluid $a_n(S_w,v)$, so that
\begin{equation}
\label{eq_ap-4}a_p(S_w,v)=a_w(S_w,v)+a_n(S_w,v)\;.
\end{equation}
We have that 
\begin{eqnarray}
A_w&=&\int_{-\infty}^\infty dv\ a_w(S_w,v)\;,\label{eq:ap-5}\\
v_w&=&\frac{1}{A_w}\int_{-\infty}^\infty dv\ v\ a_w(S_w,v)\;,\label{eq:ap-6}
\end{eqnarray}
and
\begin{eqnarray}
A_n&=&\int_{-\infty}^\infty dv\ a_n(S_w,v)\;,\label{eq:ap-7}\\
v_n&=&\frac{1}{A_n}\int_{-\infty}^\infty dv\ v\ a_n(S_w,v)\;.\label{eq:ap-8}
\end{eqnarray}
It is straightforward to derive the two expressions $A_p=A_w+A_n$ and $v_p=S_wv_w+S_nv_n$ from these equations.

The co-moving velocity, Equation (\ref{eq:co-moving-1}), may be combined with the defining equations in this section to form the {\it co-moving differential transversal pore area\/} $a_m$, 
\begin{equation}
\label{eq:ap-9}
a_m=S_w\frac{\partial}{\partial S_w}\left(\frac{a_w}{S_w}\right)+S_n\frac{\partial}{\partial S_w}\left(\frac{a_n}{S_n}\right)\;.
\end{equation}
We have by construction that
\begin{equation}
\label{eq:ap-10}
v_m=\int_{-\infty}^\infty dv\ v\ a_m\;.
\end{equation}

The area associated with the co-moving velocity is, however,
\begin{equation}
\label{eq:ap-11}
A_m=\int_{-\infty}^\infty dv\ a_m=0\;.
\end{equation}
This makes sense because $(v_p,v_m)$ and $(v_w,v_n)$ constitute different partitions of the velocity field. Hence, we must have that
\begin{equation}
\label{eq:ap-12}
A_p+A_m=A_w+A_n\;,
\end{equation}
forcing $A_m$ to be zero. As a further consequence, we find that there is no volumetric flow associated with the co-moving velocity, 
\begin{equation}
\label{eq:ap-13}
Q_m=A_m v_m=0\;.
\end{equation}
This result also makes sense when referring to $(v_p,v_m)$ and $(v_w,v_n)$ as separate partitions.  We must have that
\begin{equation}
\label{eq:ap-14}
Q_p+Q_m=Q_w+Q_n\;,
\end{equation}
forcing $Q_m$ to be zero.

We may now use Equations (\ref{eq:co-moving-2}) and (\ref{eq:co-moving-3}) to construct differential transversal area for the thermodynamic velocities $\hat{v}_w$ and $\hat{v}_n$, finding
\begin{eqnarray}
\hat{a}_w&=&a_w+a_m S_w S_n\;,\label{eq:ap-15}\\
\hat{a}_n&=&a_w-a_m S_w S_n\;.\label{eq:ap-16}
\end{eqnarray}
This implies that
\begin{eqnarray}
\hat{A}_w&=&\int_{-\infty}^\infty dv\ \hat{a}_w=A_w\;,\label{eq:ap-17}\\
\hat{A}_n&=&\int_{-\infty}^\infty dv\ \hat{a}_n=A_n\;.\label{eq:ap-18}
\end{eqnarray}
This result is not surprising.  This implies that the saturation is the same whether we consider the seepage velocities or the thermodynamic velocities, that is, $S_w=\hat{A}_w/A_p=A_w/A_p$.  The thermodynamic velocities are given by 
\begin{eqnarray}
\hat{v}_w&=&\frac{1}{A_w}\int_{-\infty}^\infty dv\ v\ \hat{a}_w\;,\label{eq:ap-19}\\
\hat{v}_n&=&\frac{1}{A_n}\int_{-\infty}^\infty dv\ v\ \hat{a}_n\;.\label{eq:ap-20}
\end{eqnarray}
Furthermore, we have that 
\begin{equation}
\label{eq:ap-21}
\hat{a}_w+\hat{a}_n=a_w+a_n=a_p\;.
\end{equation}

So far, we have looked at the zeroth and the first moment of the velocity distribution.  We generalize to the $q$th moment, 
\begin{equation}
\label{eq:ap-22}
v_p^q =\frac{1}{A_p} \int_{-\infty}^\infty dv\ v^q\ a_p(S_w,v)\;,
\end{equation}
In the same way, we define the $q$th moment of the seepage velocities, 
\begin{eqnarray}
v_w^q &=&\frac{1}{A_w} \int_{-\infty}^\infty dv\ v^q\ a_w(S_w,v)\;,\label{eq:ap-23}\\
v_n^q &=&\frac{1}{A_n} \int_{-\infty}^\infty dv\ v^q\ a_n(S_w,v)\;.\label{eq:ap-24}
\end{eqnarray}
We do the same for the thermodynamic velocities,
\begin{eqnarray}
\hat{v}_w^q &=&\frac{1}{A_w} \int_{-\infty}^\infty dv\ v^q\ \hat{a}_w(S_w,v)\;,\label{eq:ap-25}\\
\hat{v}_n^q &=&\frac{1}{A_n} \int_{-\infty}^\infty dv\ v^q\ \hat{a}_n(S_w,v)\;.\label{eq:ap-26}
\end{eqnarray}
It is straight forward to show that
\begin{equation}
v_p^q=v_w^qS_w+v_n^qS_n=\hat{v}_w^qS_w+\hat{v}_n^qS_n\;.\label{eq:ap-27}\\
\end{equation}

With the expressions for the higher moments in place, we my now construct the Fourier transforms of the differential transversal areas, 
\begin{eqnarray}
\tilde{a}_p(S_w,\omega)&=&\frac{1}{2\pi} \int_{-\infty}^\infty dv\ e^{iv\omega}\ a_p(S_w,v)\;,\label{eq:ap-22-1}\\
\tilde{a}_w(S_w,\omega)&=&\frac{1}{2\pi} \int_{-\infty}^\infty dv\ e^{iv\omega}\ a_w(S_w,v)\;,\label{eq:ap-23-1}\\
\tilde{a}_n(S_w,\omega)&=&\frac{1}{2\pi} \int_{-\infty}^\infty dv\ e^{iv\omega}\ a_n(S_w,v)\;,\label{eq:ap-24-1}
\end{eqnarray}
and we have that 
\begin{equation}
\label{eq:ap-25-1}
\tilde{a}_p(S_w,\omega)=\tilde{a}_w(S_w,\omega)+\tilde{a}_n(S_w,\omega)\;.
\end{equation}
We now do a moment expansion of $\tilde{a}_p$,
\begin{equation}
\label{eq:ap-26-1}
\tilde{a}_p=\sum_{m=0}^\infty \frac{(i\omega)^m}{m!} v_p^m\;.
\end{equation}
We also do a {\it cumulant\/} expansion,
\begin{equation}
\label{eq:ap-27-1}
\tilde{a}_p=\exp\left[\sum_{k=1}^\infty \frac{(i\omega)^k}{k!}\ C_p^k\right]\;,
\end{equation}
where $C_p^k$ denotes the $k$th cumulant.  The first four cumulants are in terms of the moments,
\begin{eqnarray}
C_p^1&=&v_p\;,\label{eq:ap-28}\\
C_p^2&=&v_p^2-(v_p)^2\;,\label{eq:ap-29}\\
C_p^3&=&v_p^3-3v_p^2 v_p+2(v_p)^3\;,\label{eq:ap-30}\\
C_p^4&=&v_p^4-4v_p^3v_p-3(v_p^2)^2+12v_p^2(v_p)^2-6(v_p)^4\;.\label{eq:ap-31}
\end{eqnarray}
Hence, the first cumulant is the average $v_p$ and the second cumulant is the variance $\Delta v_p^2=v_p^2-(v_p)^2$. If the underlying statistical distribution is Gaussian, all cumulants $C_p^k$  with $k$ larger than or equal to three are zero. 

By using Equations (\ref{eq:ap-25}) to (\ref{eq:ap-29}), we find that
\begin{equation}
\label{eq:ap-32}
\Delta v_p^2=\Delta v_w^2S_w+\Delta v_n^2S_n+S_wS_n(v_w-v_n)^2\;.
\end{equation}
We combine this expression with Equation (\ref{eq:co-moving-4}) to find
\begin{equation}
\label{eq:ap-33}
\Delta v_p^2=\Delta v_w^2S_w+\Delta v_n^2S_n+S_wS_n\left[\frac{\partial v_p}{\partial S_w}-v_m\right]^2\;.
\end{equation}
This remarkable equation shows that it is possible to determine the co-moving velocity $v_m$ from the velocity fluctuations $\Delta v_p^2$, $\Delta v_w^2$, and $\Delta v_n^2$ alone. It should be noted that this relation between the co-moving velocity and the velocity fluctuations is not related to the fluctuation-dissipation theorem. 

\subsection{Open questions}
\label{sec:open_questions_stat_mech}

We end this section on the statistical thermodynamics approach to immiscible two-phase flow in porous media, detailing some of the open questions that linger.

\begin{itemize}
\item We base our statistical mechanics on maximizing the configurational entropy given a knowledge of the averages of $Q_u$ and $A_w$, Equations (\ref{eq:eq4-4}) and (\ref{eq:eq4-5}).  This gives us the extensive variables $Q_p$ and $A_w$, and the intensive variables agiture $\theta$ and flow derivative $\mu$.  If more extensive Darcy-scale variables are identified, it is straightforward to include them in the analysis. One candidate is associated with the wetting angle.  

\item Central to the statistical mechanics approach is the partition function defined in Equation (\ref{eq:eq4-13}).  It is an integral over all possible physical realizations of fluid configurations that pass through a given REA.  It is important to find a {\it stylized model\/} for flow configurations.  By this, we mean a minimalistic, mechanism-oriented model capturing essential aspects of a complex system without claiming full realism \cite{bouchaud2008economics}.  An example of such a model is the Ising model discussed in Section \ref{subsubsec:Louge}. We need a stylized model for the fluid configurations in the spirit of the Ising model that is simple enough to make it possible to perform the partition function integral.

\item Given a stylized model for the fluid configurations on the pore scale, a natural next step after calculating the partition function would be to average over the geometric structure of the porous medium as described in Section \ref{sec:pi}. This is a very difficult calculation, but with the correct model, there is hope.  Such a model would need to treat geometric disorder in a stylized manner.
\item At the end of Section \ref{sec:pi} we sketched some of the tell-tale signs of a glassy flow state: hysteresis, broken ergodicity, and dynamics over a wide range of time scales.  For spin glasses, the Edwards-Anderson order parameter (\ref{eq:EA-definition}) signals whether the system is in the glassy phase or not. The EA order parameter is zero if the system is not in a glassy state and it is positive if it is.  A corresponding Darcy-scale quantity should be constructed to determine whether  the flow is glassy.

\item The introduction of the temperature-like emergent variable {\it agiture\/} is essential for the theory. As argued in Section \ref{sect:agiture}, a very likely candidate for the agiture is the pressure gradient, see Equation (\ref{eq:eq6-5}). It is difficult to test this hypothesis directly as it is to define temperature in ordinary thermodynamics. However, the consequence of Equation (\ref{eq:eq6-5}) is a relationship between the configurational entropy $\Delta S_I$ and differential mobility $dQ_p/dp'$, see Equation (\ref{eq:eq6-13}).  The difficulty here lies in measuring the configurational entropy.  Progress has been made \cite{hermundstad2025shannon}, but more work is needed.

\item Experimental verifications are needed.  The thermodynamic relations that ensue at the Darcy scale invite experimental investigations.  For example, the fluctuations
of the saturation within an REA is given by
\begin{equation}
\label{eq:summary-thermo}
\Delta S_w^2=-\frac{\theta}{A_p}\ \left(\frac{\partial^2 \hat{v}_n}{\partial \mu^2} \right)_\theta\;,
\end{equation}
and this can be tested experimentally. 

\item Even though we have discussed in Section \ref{sect:emergent} porous media samples that are heterogeneous in the direction orthogonal to the average flow direction, we have not attempted to move past one flow direction.  How would we need to modify the equations we have presented if they were to be generalized to three dimensional vectors, e.g., Equations (\ref{eq:co-moving-20}) and (\ref{eq:co-moving-21})?  This is mostly obvious, but less so the co-moving velocity.  Is Equation (\ref{eq:co-moving-4}) sufficient?  Clarity here is necessary for developing the theory into a tool. 

\item Generalizing the statistical thermodynamics approach to heterogeneous saturation distributions is an essential step in the direction of a theory capable of treating non-steady-state systems such as Darcy-scale viscous fingers. This means finding a way to incorporate the capillary forces at the Darcy scale.  A natural path is to develop an equivalent to non-equilibrium thermodynamics based in the statistical thermodynamics approach we have described here.

\end{itemize}

        \cleardoublepage
        \newpage

	\section{The co-moving velocity}\label{sect:comovingvelocity}

The co-moving velocity~\cite{royCoMovingVelocityImmiscible2022,alzubaidiImpactWettabilityComoving2024,hansenLinearityComovingVelocity2024} was introduced in section \ref{sect:seepage}, where it naturally appeared as part of the statistical thermodynamics approach.  This is a logical way to introduce the co-moving velocity, but it also drowns out some of its immediate practical value.  In order to highlight this aspect, we will discuss the co-moving velocity specifically in this section with an emphasis on its usefulness within the relative permeability framework.

\subsection{Co-moving velocity as a geometric concept}
\label{sec:geom}

We start this section by attempting to give an intuition for what it is. To do this, we need to a little bit of vector algebra in parameter space
\cite{pedersen2023parameterizations,pedersen2025geometric,pedersen2025co}.  With an REA in mind, we define an area vector ${\vec A}=(A_w,A_n)$, where $A_w$ and $A_n$ are the transversal wetting and non-wetting area respectively.  We also define a velocity vector ${\vec v}=(v_w,v_n)$. We may then express the volumetric flow rate $Q_p$ through the REA as
\begin{equation}
\label{eq:co-mov-1}
Q_p={\vec A}\cdot{\vec v}\;.
\end{equation}

We define another velocity vector based on the thermodynamic velocities instead, ${\vec {\hat v}}=({\hat v}_w,{\hat v}_n)$, and we have 
\begin{equation}
\label{eq:co-mov-2}
Q_p={\vec A}\cdot{\vec {\hat v}}\;.
\end{equation}
We subtract Equation (\ref{eq:co-mov-2}) from Equation (\ref{eq:co-mov-1}) to get
\begin{equation}
\label{eq:co-mov-3}
{\vec A}\cdot({\vec v}-{\vec{\hat v}})=0\;.
\end{equation}
The co-moving velocity is simply
\begin{equation}
\label{eq:co-mov-7}
{\vec v}_m={\vec v}-{\vec{\hat v}}\;,
\end{equation}
and from Equation (\ref{eq:co-mov-3}) we see that it is orthogonal to the area vector $\vec A$,
\begin{equation}
\label{eq:co-mov-4}
{\vec v}_m \perp  {\vec A}\;.
\end{equation}

What makes the thermodynamic velocities $\vec {\hat v}$ special?  If we make a small change in the volumetric flow rate, $Q_p\to Q_p+dQ_p$, we find that
\begin{equation}
\label{eq:co-mov-5}
dQ_p=d{\vec A}\cdot {\vec{\hat v}}\;.
\end{equation}
Hence, there is no $d{\vec{\hat v}}$ in this expression.  This is equivalent to Equation (\ref{eq:co-moving-13}). The seepage velocities $\vec v$, on the other hand, give rise to
\begin{equation}
\label{eq:co-mov-6}
dQ_p=d{\vec A}\cdot {\vec v}+{\vec A}\cdot d{\vec v}\;.
\end{equation}
By Equation (\ref{eq:co-mov-7}), we may express this equation as
\begin{equation}
\label{eq:co-mov-8}
dQ_p=d{\vec A}\cdot{\vec v} -d{\vec A}\cdot{\vec v}_m\;,
\end{equation}
which is equivalent to Equation (\ref{eq:co-moving-1}). 

What we have presented in these lines is barely the tip of an iceberg.  In fact, the entire thermodynamic formalism for immiscible two-phase flow in porous media developed in the previous Section boils down to parameterizing the space of variables.  This is the realm of differential geometry, a connection developed in the references given earlier, \cite{pedersen2023parameterizations,pedersen2025geometric,pedersen2025co}.  We will not pursue this direction further in spite of its elegance and beauty. However, we may summarize what we have learned from this approach: the {\it co-moving velocity is the seepage velocity vector relative to the thermodynamic velocity vector and it does not contribute to the volumetric flow rate .\/}  Hence, it is a rather abstract concept --- but, as we shall see, it can also be very useful in spite of this. 

\subsection{Co-moving velocity measured in a dynamic pore network model}
\label{sec:com-dyn-net}

\begin{figure}
\begin{center}
\includegraphics[width=1.0\textwidth]{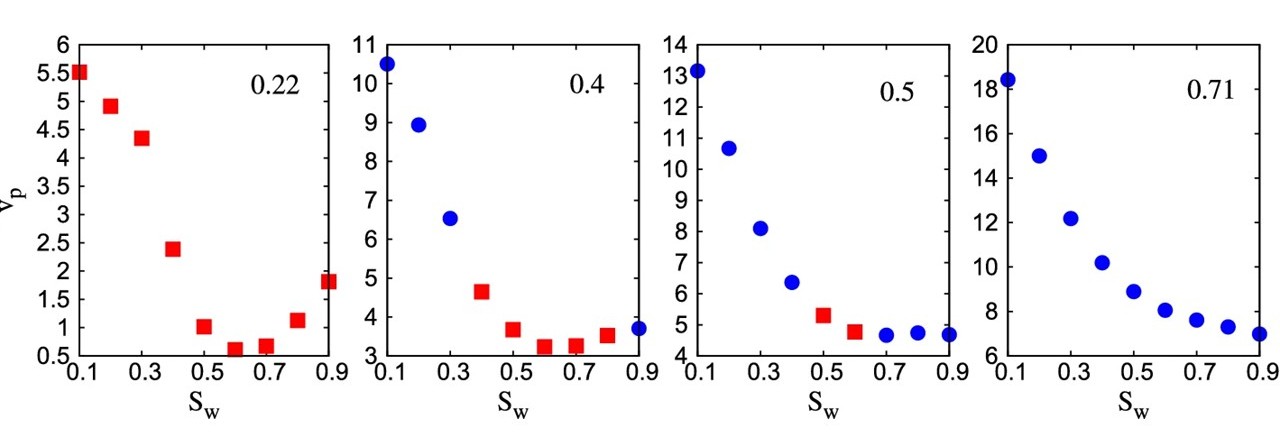}
\end{center}
\caption{\label{fig:vp-non-linear-co-moving} Average seepage velocity $v_p$ as a function of saturation for pressure gradient equal to $|\nabla p|=\Delta p/L=0.22$, 0.40, 0.50 and 0.71 MPa/m, respectively determined in a dynamic network model. The viscosities of the fluids were $\mu_w=0.03$ Pa s and $\mu_n=0.01$ Pa s. The red square indicates that the flow is in the nonlinear region II. A blue circle indicates that the flow is in the linear region I, see Section \ref{sec:non-linear} and Fig.~\ref{fig:flowregimeslinearnonlinear}. 
(From \cite{royCoMovingVelocityImmiscible2022}.)}
\end{figure}

Using the dynamic pore network simulator of Sinha et al.\ \cite{sinha2021fluid}, the co-moving velocity was measured over a wide range of capillary numbers \cite{royCoMovingVelocityImmiscible2022}.  The pore network model was set up as a square lattice in the form of the surface of a torus. This makes the average saturation $S_w$ a conserved control variable.  The second control variable is the average pressure gradient. The average seepage velocity $v_p$ is found using Equations  (\ref{eq:co-moving-2}) and (\ref{eq:co-moving-3}), whereas the co-moving velocity is found using Equation (\ref{eq:co-moving-14}). We show in Fig.\ 
\ref{fig:vp-non-linear-co-moving} the average seepage velocity $v_p$ as a function of saturation $S_w$ for four different pressure gradients.  The blue symbols indicate that the flow appears in the linear Darcy regime and the red symbols indicate that the flow appears in the non-linear regime characterized by $v_p\sim |\Delta p|^{1/\alpha}$, see Section \ref{sec:non-linear}.    

\begin{figure}
\begin{center}
\includegraphics[width=0.8\textwidth]{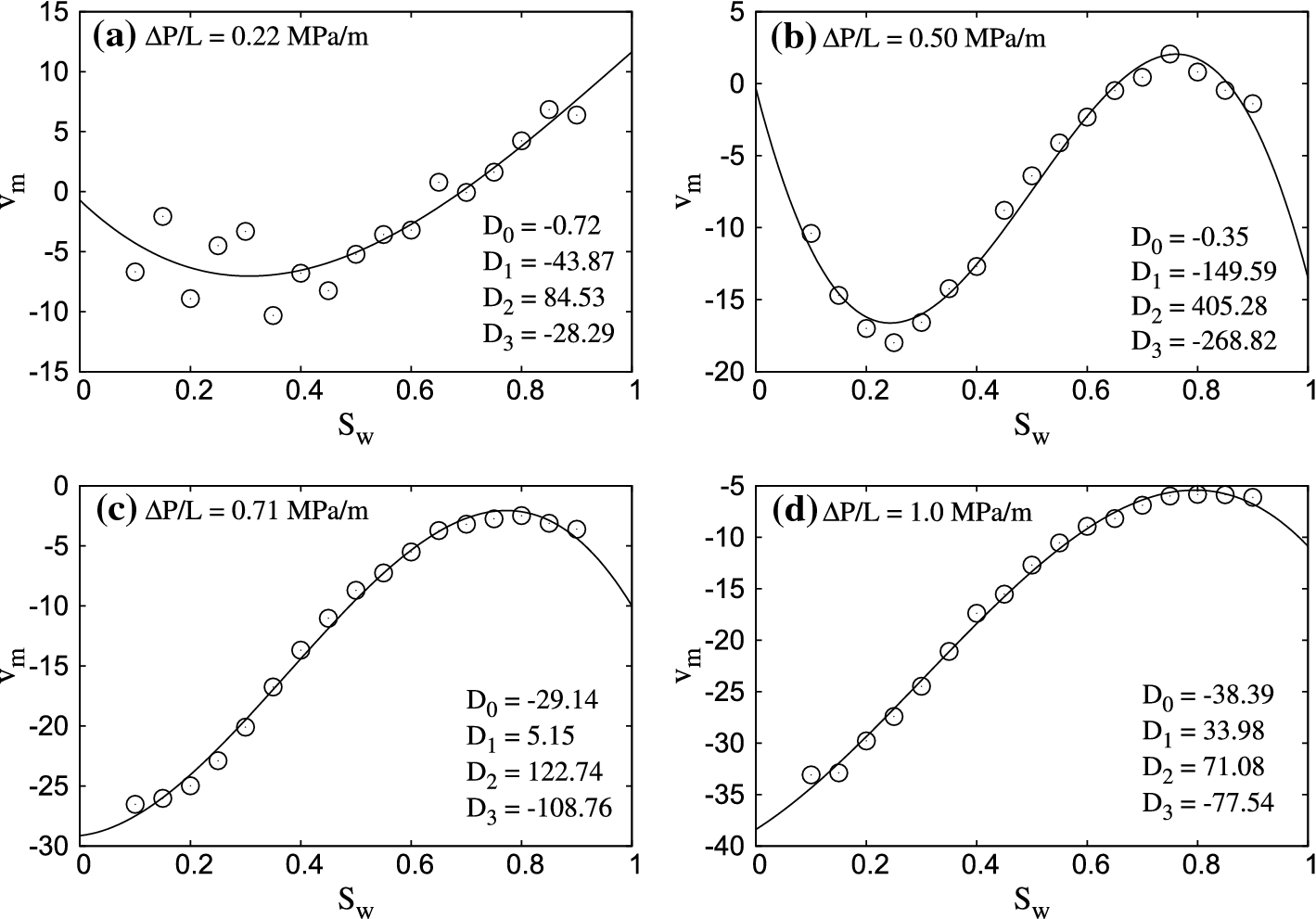}
\end{center}
\caption{\label{fig:vm-pore-network} $v_m$ calculated from Equation (\ref{eq:co-moving-14}) for the same pressure gradients as in Fig.\ \ref{fig:vp-non-linear-co-moving}. Each panel contains the values of the parameters $D_0$ to $D_3$ that produce the fitted lines.  
(From \cite{royCoMovingVelocityImmiscible2022}.)}
\end{figure}  

We show in Fig.\ \ref{fig:vm-pore-network} the resulting co-moving velocity $v_m$ as a function of the saturation $S_w$.  The fitted lines are based on third-order polynomials, 
\begin{equation}
\label{eq:co-mov-10}
v_m=\sum_{k=0}^3 D_k S_w^k\;.
\end{equation}

\begin{figure}
\begin{center}
\includegraphics[width=1.0\textwidth]{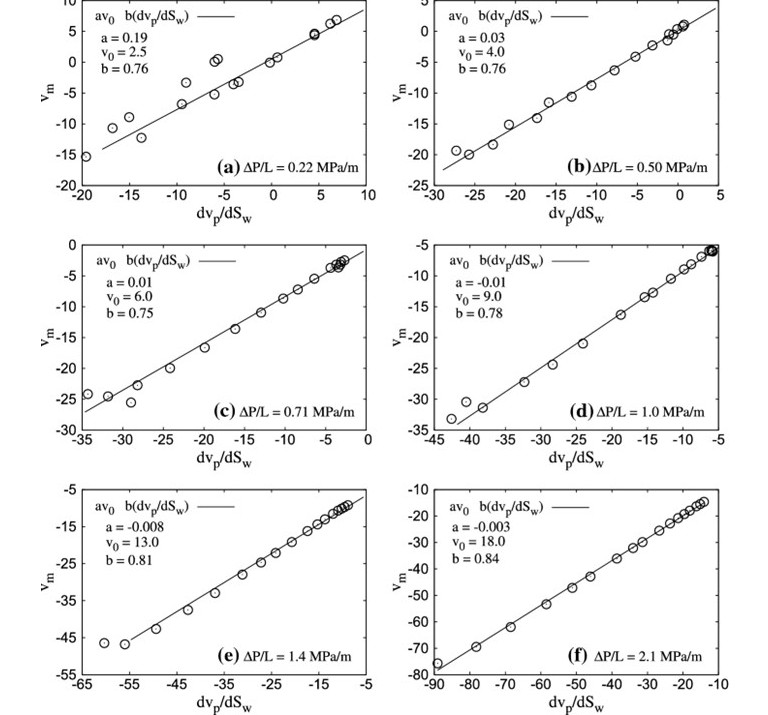}
\end{center}
\caption{\label{fig:co-moving-pore-model}   The co-moving velocity from Fig.\ \ref{fig:vm-pore-network} plotted against the flow derivative $\mu=dv_p/dS_w$.
(From \cite{royCoMovingVelocityImmiscible2022}.)}
\end{figure}  
\begin{figure}
\begin{center}
\includegraphics[width=1.0\textwidth]{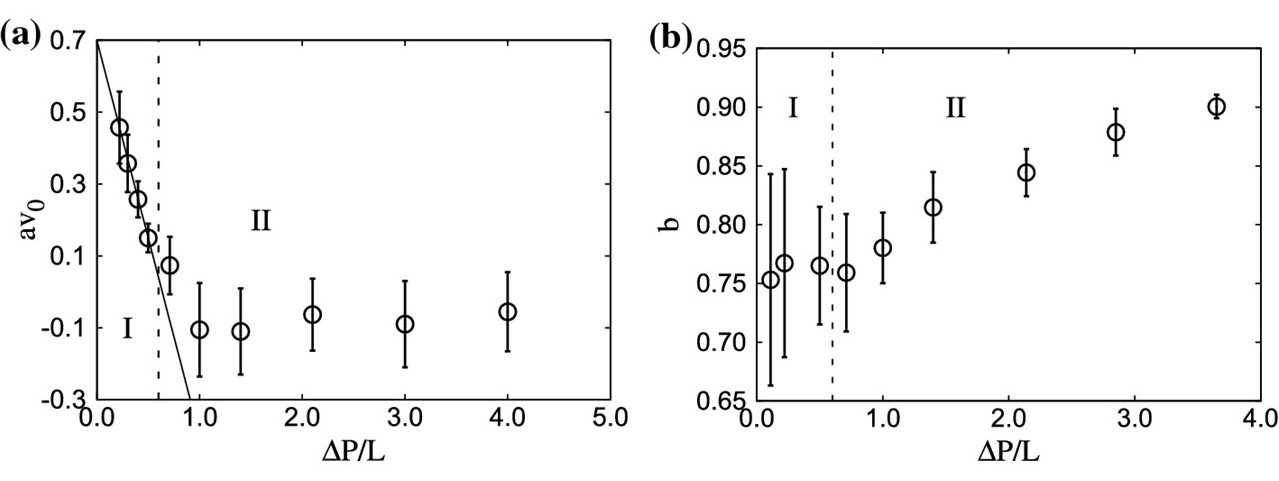}
\end{center}
\caption{\label{fig:a-b}  $av_0$ and $b$ as functions of the average pressure gradient $\Delta p/L$. (From \cite{royCoMovingVelocityImmiscible2022}.)}
\end{figure}  

However, in Section \ref{sect:seepage}, we demonstrated that the flow derivative $\mu=dv_p/dS_w$ is the natural variable for the co-moving velocity, see Equation (\ref{eq:co-moving-22}).  Hence, in Fig.\ \ref{fig:co-moving-pore-model} we plot the same $v_m$ data against $dv_p/dS_w$. The data indicate a linear relation between $v_m$ and $dv_p/dS_w$,
\begin{equation}
\label{eq:co-mov-11}
v_m=av_0+b\frac{dv_p}{dS_w}\;,
\end{equation}
where $a$ and $b$ are dimensionless constants and $v_0=v_p(S_w=1)$ acts as a velocity scale. Fig.\ \ref{fig:a-b} shows the measured values of $av_0$ and $b$ as functions of the average pressure gradient $\Delta p/L$.  We see two regimes marked I and II in these plots.  In regime I, the constant $av_0$ depends linearly on the average pressure gradient $\Delta p/L$, whereas $b$ remains constant.  In regime II it is opposite.  Here $av_0$ remains constant whereas $b$ increases in value. At very high values of the pressure gradient where the fluids are essentially miscible so that $v_w=v_n$, we expect from Equation (\ref{eq:co-moving-14}) we expect $v_m \to dv_p/dS_w$ so that $b\to 1$.     
\subsection{Co-moving velocity from relative permeability data}
\label{sec:com-rel-perm}

We write the generalized Darcy equations in terms of the seepage velocities $v_w$ and $v_n$ as
\begin{eqnarray}
v_w&=&-\frac{K}{\phi} \frac{k_{rw}}{\mu_w S_w} \frac{\partial p}{\partial x}\;,\label{eq:vwdarcy}\\
v_n&=&-\frac{K}{\phi} \frac{k_{rn}}{\mu_n S_n} \frac{\partial p}{\partial x}\;.
\label{eq:vndarcy}
\end{eqnarray}
We have here assumed that there saturation field $S_w$ is constant throughout the system
so that $\partial p_c(S_w)/\partial x =(dp_c/dS_w) \partial S_w/\partial x=0$. Since we have that $p_c=p_n-p_w$, we must here have $\partial p_w/\partial x=\partial (p_n-p_c)=\partial p_n$. Hence, we set $p_w=p_n=p$ under this assumption.  
We combine these two equations with the expression for $v_m$ given in Equation 
(\ref{eq:co-moving-1}) to find
\begin{equation}
\label{eq:co-mov-12}
v_m=v_0\left[S_w\frac{\partial}{\partial S_w} \left(\frac{k_{rw}}{S_w}\right)+MS_n\frac{\partial}{\partial S_w} \left(\frac{k_{rn}}{S_n}\right)\right]\;,
\end{equation}
where
\begin{equation}
\label{eq:co-mov-13}
v_0=-\frac{K}{\mu_w\phi}\frac{\partial p}{\partial x}
\end{equation}
sets a velocity scale and $M=\mu_w/\mu_n$ is the viscosity ratio between the two immiscible fluids. 

\begin{figure}[htbp]
\begin{center}
\includegraphics[width=1.0\textwidth]{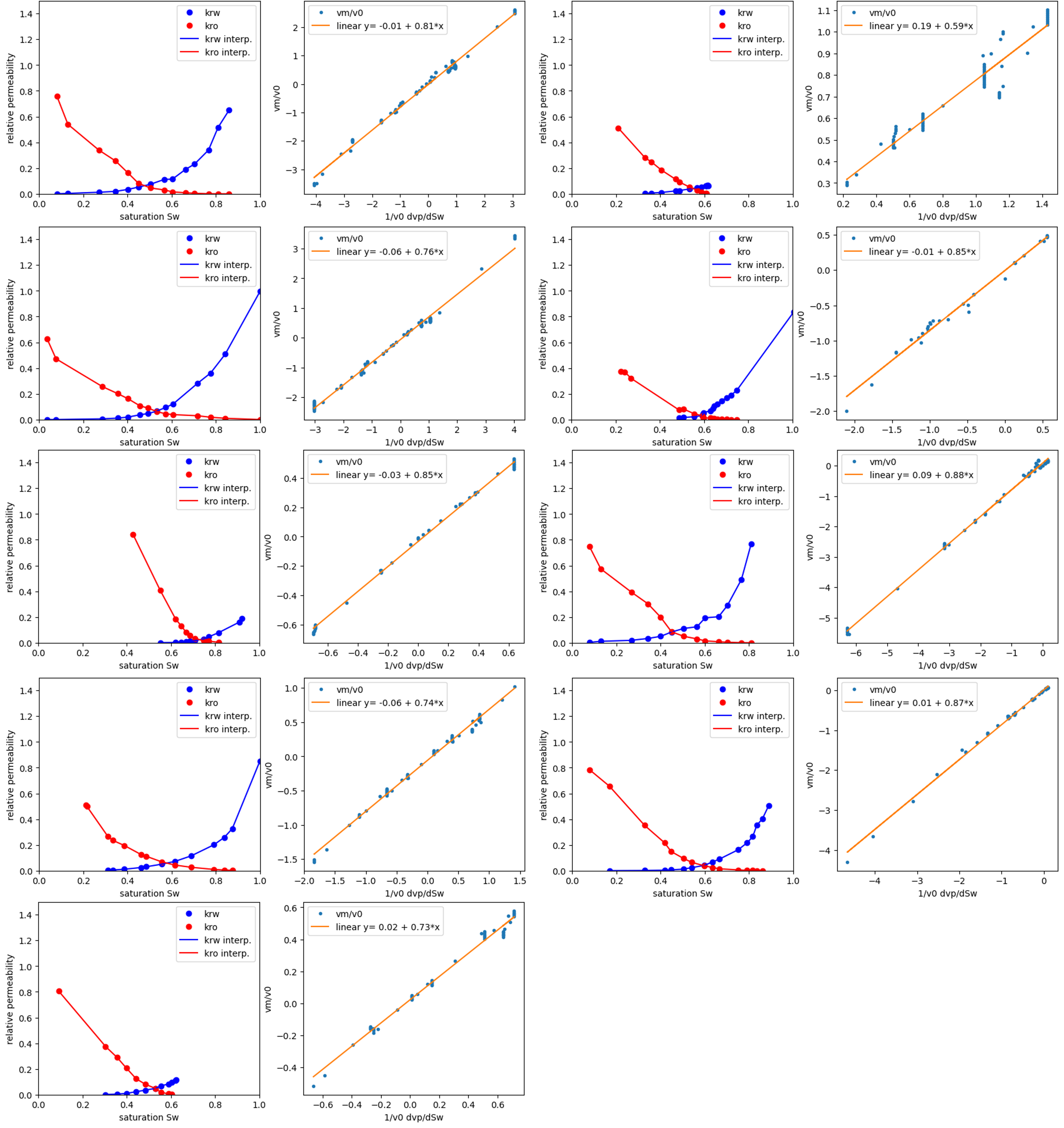}
\end{center}
\caption{Examples of selected drainage and imbibition steady-state and unsteady-state relative permeability and $v_m / v_0$ vs. $1/v_0 dv_p / dS_w$ which show a predominantly linear behavior \cite{bergDisplacementMassTransfer2013i,soropRelativePermeabilityMeasurements2015,bergConnectedPathwayRelative2016d,Liu2017,bergNonuniquenessUncertaintyQuantification2021a,Ruecker2021,gaoCapillarityPhasemobilityHydrocarbon2021,bergSensitivityUncertaintyAnalysis2021,anInverseModelingCore2023,maasViscousFingeringCCS2024}.\label{fig:comovingvelocityShell01}}
\end{figure}  

Fig.\ \ref{fig:comovingvelocityShell01} shows the experimentally measured relative permeability from steady-state and unsteady-state drainage and imbibition experiments \cite{bergDisplacementMassTransfer2013i,soropRelativePermeabilityMeasurements2015,bergConnectedPathwayRelative2016d,Liu2017,bergNonuniquenessUncertaintyQuantification2021a,Ruecker2021,gaoCapillarityPhasemobilityHydrocarbon2021,bergSensitivityUncertaintyAnalysis2021,anInverseModelingCore2023,maasViscousFingeringCCS2024}. When fitting $v_m / v_0$ vs. $1/v_0 dv_p / dS_w$ we obtain a predominantly linear relationship for all cases as in Equation (\ref{eq:co-mov-11}).  

\begin{figure}[htbp]
\begin{center}
\includegraphics[width=0.7\textwidth]{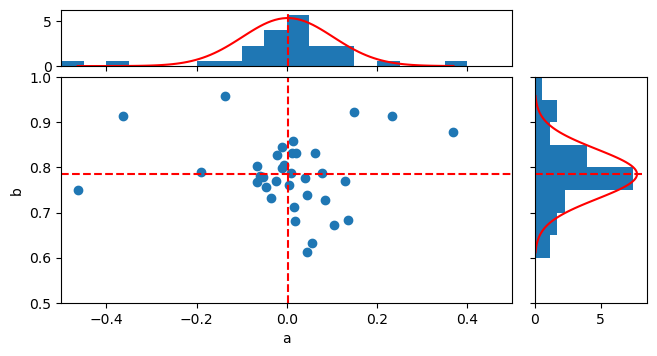}
\end{center}
\caption{Parameters $a$ and $b$ from fitting the $v_m / v_0$ vs. $1/v_0 dv_p / dS_w$ from Fig.~\ref{fig:comovingvelocityShell01} with a linear relationship $v_m/v_0=a+b\cdot 1/v_0 dv_p / dS_w$. The respective histograms were fitted using a Gaussian distribution (red lines).  \label{fig:comovingvelocityShell02}}
\end{figure}

In Fig.\ \ref{fig:comovingvelocityShell02} the offset $a$ and slope $b$ from the fits in Fig.\ \ref{fig:comovingvelocityShell01} are shown, suggesting that the offset $a \approx 0$.  We note, however, that when $v_p$ has a minimum when plotted against the saturation $S_w$, we must have from Equation (\ref{eq:co-moving-14}) that $a=\min_{v_p(S_w)}(v_n-v_w)/v_0$ since $\mu=dv_p/dS_w=0$ here. 

If we now express Equation (\ref{eq:co-mov-11}) in terms of the relative permeabilities, we find
\begin{equation}
\label{eq:co-mov-14}
v_m=av_0+bv_0\left[\frac{\partial k_{rw}}{\partial S_w}+ M\frac{\partial k_{rn}}{\partial S_w}\right]\;.
\end{equation}
Combining this equation with Equation (\ref{eq:co-mov-12}), we find
\begin{equation}
\label{eq:co-mov-15}
\left[(1-b)\frac{\partial k_{rw}}{\partial S_w}-\frac{k_{rw}}{S_w}\right]
-M\left[(1-b)\frac{\partial k_{rn}}{\partial S_n}-\frac{k_{rn}}{S_n}\right]=a\;,
\end{equation}
where we have used that $S_n=1-S_w$.
This equation means that knowing one of the relative permeabilities and the parameters $a$ and
$b$, will give the other relative permeability. 

Let us consider Equation (\ref{eq:co-mov-15}) in the limit where $M\gg 1$ as is the case when we are dealing with a wetting liquid and a non-wetting gas. The equation becomes
\begin{equation}
\label{eq:co-mov-16}
(1-b)\frac{\partial k_{rn}}{\partial S_n}-\frac{k_{rn}}{S_n}+\frac{a}{M}=0\;.
\end{equation}
We make the assumption that also $M/a\gg 1$, leading to the further simplification
\begin{equation}
\label{eq:co-mov-17}
(1-b)\frac{\partial k_{rn}}{\partial S_n} =\frac{k_{rn}}{S_n}\;.
\end{equation}
We solve this equation finding
\begin{equation}
\label{eq:co-mov-18}
k_{rn}\propto S_n^{1/(1-b)}\;,
\end{equation}
i.e., the Corey relative permeability with index $1/(1-b)$. With a typical range of 2.5 -- 3.0 for Corey indices for gases, we find a range $b\approx 0.6$ to 0.7, which is consistent with the $b$-values listed in Fig.\ \ref{fig:comovingvelocityShell02}.

We may summarize the findings so far in this Subsection as follows: The co-moving velocity makes it possible to address one of the big shortcomings of the empirically postulated equations in that its transport coefficients relative permeability are not constrained in any way, which leaves a wide range of possible choices of functional forms and resulting uncertainty, which is very difficult to constrain purely experimentally~\cite{bergSensitivityUncertaintyAnalysis2021}.  Intrinsic symmetries in the generalized Darcy equations are revealed by the co-moving velocity concept that constrains the possible choices of relative permeability and has the potential to reduce the experimental effort and simplify experimental protocols, based on Equation (\ref{eq:co-mov-15}). 

\begin{figure}[htbp]
\begin{center}
\includegraphics[width=1.0\textwidth]{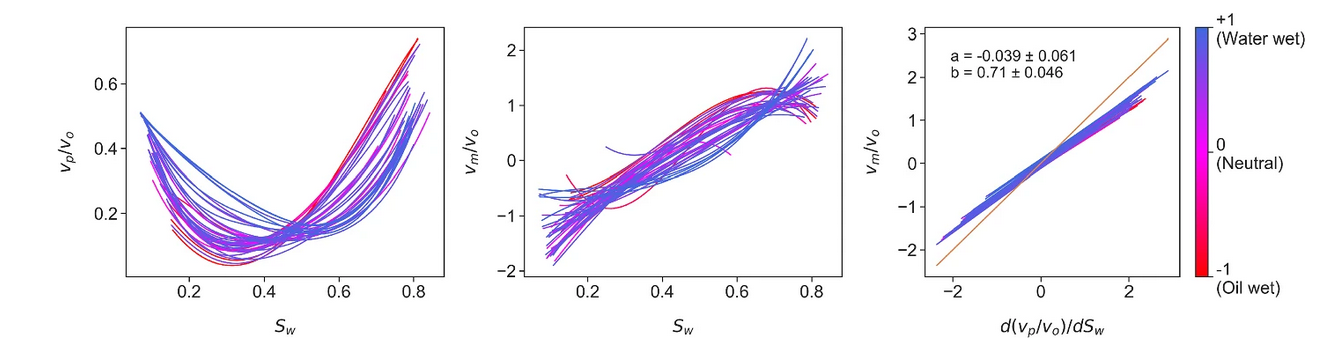}
\end{center}
\caption{Average seepage velocity $v_p$ and co-moving velocity $v_m$ for North Sea Sandstone based on 41 different wetting conditions. $v_0$ is a velocity scale.  Left panel: $v_p$ as a function of the saturation $S_w$. Middle panel: $v_m$ as a function of $S_w$. Right panel: $v_m$ as a function of $\mu=dv_p/dS_w$. The coefficients $a$ and $b$ defined in Equation (\ref{eq:co-mov-11}) averaged over all 41 wetting conditions are given in the last panel. (From \cite{alzubaidiImpactWettabilityComoving2024}.) \label{fig:alzubaidi-2}}
\end{figure}

Alzubaidi et al.\ \cite{alzubaidiImpactWettabilityComoving2024} posed the question: how sensitive is the co-moving velocity to the wettability properties of the immisicble fluids? They considered the reconstructed pore space of a North Sea sandstone and a Bentheimer sandstone, running the Lattice-Boltzmann algorithm simulating immiscible two-phase flow on them for different wetting properties.  For the Bentheimer sandstone they also used experimental data done for different wetting properties \cite{zou2018experimental}.  The results for the North Sea sandstone is shown in Fig.\ \ref{fig:alzubaidi-2}. The results, for the simulations of the Bentheimer sandstone gave similar results: The coefficients $a$ and $b$ defined in Equation (\ref{eq:co-mov-1}) showed almost no dependency on the wetting properties, but there was a difference between the two types of sandstones.  They summarized their findings in Table \ref{tab:alzubaidi-1}.    

\begin{table}[ht]
  \centering
  \caption{Averaged coefficients $a$ and $b$ defined in Equation (\ref{eq:co-mov-11}) based on the Lattice-Boltzmann simulations on the reconstructed pore spaces of a North Sea sandstone and a Bentheimer sandstone. (From \cite{alzubaidiImpactWettabilityComoving2024}.) }
  \label{tab:alzubaidi-1}
  \small
  \begin{tabular}{lcc}
    \toprule
     & $a$ & $b$ \\
    \midrule
    North Sea sandstone & $-0.039\pm 0.061$ & $0.71\pm 0.046$\\
    Bentheimer sandstone & $0.012\pm 0.078$ & $0.731\pm 0.050$\\
    \bottomrule
  \end{tabular}
\end{table}

\subsection{Away from steady-state flow}
\label{subsec:away-with-pc}

We will here discuss how to deal with the situation when the saturation field $S_w$ is {\it not\/} constant in the context of the statistical thermodynamics approach developed in Sections \ref{sect:statisticalmechanics} and \ref{sect:comovingvelocity}.  We start by a discussion of the Darcy scale capillary pressure function $p_c=p_c(S_w)$ and then go on to sketch how the non-steady flow --- e.g., imbibition or drainage processes --- may be included in the statistical thermodynamics approach.  As we shall argue, in this theory there is no need for an explicit $p_c$.    

In the relative permeability approach to this problem, the generalized Darcy equations, (\ref{eq:vwdarcy}) and (\ref{eq:vndarcy}) which were written under the assumption that gradients in the saturation $S_w$ leads to there being no difference between the phase pressures $p_w=p_n=p$ and $p_c=p_n-p_w=0$. Note that the assumption of $p_c=0$ is also made by Buckley \& Leverett~\cite{buckleyMechanismFluidDisplacement1942} in the derivation of the essential behavior of immiscible displacement in porous media while capillary pressure $p_c \neq 0$ provides in many situation a ``correction" of saturation profiles~\cite{bergSimultaneousDeterminationRelative2024a} but does not change the fundamental picture. Capillary pressure on the pore scale is, however, essential to explain Darcy-scale phenomena in particular in regime I (as per definition in Fig.~\ref{fig:flowregimeslinearnonlinear}) such as spontaneous imbibition~\cite{morrowRecoveryOilSpontaneous2001,masonDevelopmentsSpontaneousImbibition2013}, saturation gradients at capillary discontinuities~\cite{huangCapillaryEndEffects1998} and transition zones in hydrocarbon reservoirs~\cite{masalmehImprovedCharacterizationModeling2007}, and effects of wettability~\cite{al-futaisiImpactWettabilityAlteration2003,abdallahFundamentalsWettability2007,soropRelativePermeabilityMeasurements2015,liuInfluenceWettabilityPhase2018,armstrong2021multiscale,lei3DgeometrytriggeredTransitionMonotonic2025}.  But is the formulation of a Darcy-scale capillary pressure function really necessary to encompass the capillary effects at the Darcy scale?  We will propose an alternative further on in this Section.  

When the assumption of $p_c=0$ is not made, the generalized Darcy equations become in three dimensions
\begin{eqnarray}
{\bf v}_w&=&-\frac{K}{\phi} \frac{k_{rw}}{\mu_w S_w} \nabla p_w\;,\label{eq:vwdarcy-neq}\\
{\bf v}_n&=&-\frac{K}{\phi} \frac{k_{rn}}{\mu_n S_n} \nabla p_n\;,
\label{eq:vndarcy-neq}
\end{eqnarray}
where there now is a third constitutive equation defining the quasi-static capillary pressure function
\begin{equation}
\label{eq:pc-er-bare-tull}
p_c=p_n-p_w\;,
\end{equation}
The Laplace pressure from a hydrodynamic normal stress boundary condition~\cite{lealAdvancedTransportPhenomena} is used as a closure relationship for the set of governing equations (which may not be applicable for dynamics situations, see Fig.~\ref{fig:JoekarNiasarJFM2010}).

When expressed only as a function of saturation $S_w$ the capillary pressure function is highly hysteretic (see section~\ref{sect:Minkowskifunctionals}), depends on the geometry of the sample \cite{moura2015impact} and is from a pore-scale viewpoint rather problematic because the concept as such is non-local \cite{armstrongInterfacialVelocitiesCapillary2013}. When both fluid phases percolate, it makes sense to define a pressure $p_w$ in the percolating wetting phase and a pressure $p_n$ in the percolating non-wetting phase.  Clearly there is going to be an interface between the two fluids, and the capillary forces associated with it will be the capillary pressure curve $p_c$.  When, on the other hand, the fluids break up into clusters, as they do in regimes Ib, II and III (from Fig.~\ref{fig:flowregimeslinearnonlinear}), it becomes difficult to retain this intuitive picture of it. For a given static fluid cluster, the capillary pressure will balance the internal pressure in that cluster.  But, clusters come in different sizes and shapes and internal pressures.  How can one distill an equation as (\ref{eq:pc-er-bare-tull}) on the Darcy scale describing this pore-level situation?  It is furthermore inconsistent to define two extinct pressures $p_n$ and $p_w$ at the Darcy scale while simultaneously defining a {\it continuous\/} saturation function at this scale where the two fluids have lost their individuality; they present themselves as a single fluid possessing a continuous parameter, the saturation.           

We now consider how the capillary pressure-saturation function is used in Equations (\ref{eq:vwdarcy-neq}) and (\ref{eq:vndarcy-neq}).  Let us focus on the first of the two equations and write it as 

\begin{equation}
\label{eq:pc-not-gone-1}
{\bf v}_w=-\frac{K}{\phi} \frac{k_{rw}}{\mu_w S_w} \left[\nabla p_n-\nabla p_c\right]\;.
\end{equation}

If we now use the traditional capillary pressure-saturation relationship $p_c=p_c(S_w)$ then we arrive at
\begin{equation}
\label{eq:pc-gone-1}
{\bf v}_w=-\frac{K}{\phi} \frac{k_{rw}}{\mu_w S_w} \left[\nabla p_n-\frac{dp_c}{dS_w}\nabla S_w\right]\;.
\end{equation}
We have here made the standard assumption that $p_c$ is a function of the saturation $S_w$ alone. We introduce the mobilities $m_p=Kk_{rw}/\phi\mu_wS_w$ and $m_s=(Kk_{rw}/\phi\mu_wS_w) (dp_c/dS_w)$, so that Equation (\ref{eq:pc-gone-1}) becomes
\begin{equation}
\label{eq:pc-gone-2}
{\bf v}_w=-m_p\nabla p_n-m_s\nabla S_w\;.
\end{equation}
The first term on the right hand side of this equation is as it should be.  We have transport of fluid, i.e., transport of mass on the left hand side. The first term on the right hand side is proportional to the driving force $\nabla p$. However, the second term is not as it should be.  Newton taught us that masses are moved by forces,  not masses.  But the second term on the right hand says that the mass, ${\bf v}_w$ is moved by the mass gradient, $\nabla S_w$.  Why does it say so? Keep the saturation fixed.  The derivative $dp_c/dS_w$, which contains the information on the capillary pressure, is then fixed. Still, the velocity ${\bf v}_w$ changes when $\nabla S_w$ changes.  This is {\it not\/} what Newton taught us.  A similar situation has arisen when Hassanizadeh \& Gray~\cite{HassanizadehGray1993} introduced interfacial area $a_{nw}$ as state variable in capillary pressure. Following a similar derivation one can then construct that gradients in interfacial area $\nabla a_{nw}$ become driving forces for flow~\cite{niessnerComparisonTwoPhaseDarcys2011}. Gray \& Miller clarify that gradients in interfacial area $\nabla a_{nw}$ are {\it not\/} driving forces for flow~\cite{grayAnalysisInvestigatingExtended2024}. 

We emphasize that we are not questioning the existence of capillary pressure at the pore level, because we know that capillary pressure is the driving force for capillary rise in a single capillary~\cite{Gauss1830,Einstein_1901} and capillary imbibition dynamics~\cite{lucasUeberZeitgesetzKapillaren1918,washburnDynamicsCapillaryFlow1921}. Our critique concerns the {\it Darcy-scale\/} capillary pressure-saturation relationship $p_c=p_c(S_w)$ and how it is used as closure relationship for the Darcy-scale set of governing equations~\cite{whitakerFlowPorousMedia1986a}. 

Hence, we will do the radical step to simply get rid the Darcy-scale capillary pressure-saturation function all together in the context of the statistical thermodynamics approach of Sections \ref{sect:statisticalmechanics} and \ref{sect:comovingvelocity}.  Does this mean that we get rid of the capillary forces?  Absolutely not.  Whether there are deviations from steady-state flow or not at the Darcy scale, is not a question asked on the pore scale.  There is in fact no way to detect at the pore scale whether there are macroscopic Darcy-scale gradients present or not.  All the capillary forces at the pore scale will do their job no matter what. 

\begin{figure}[htbp]
\begin{center}
\includegraphics[width=0.7\textwidth]{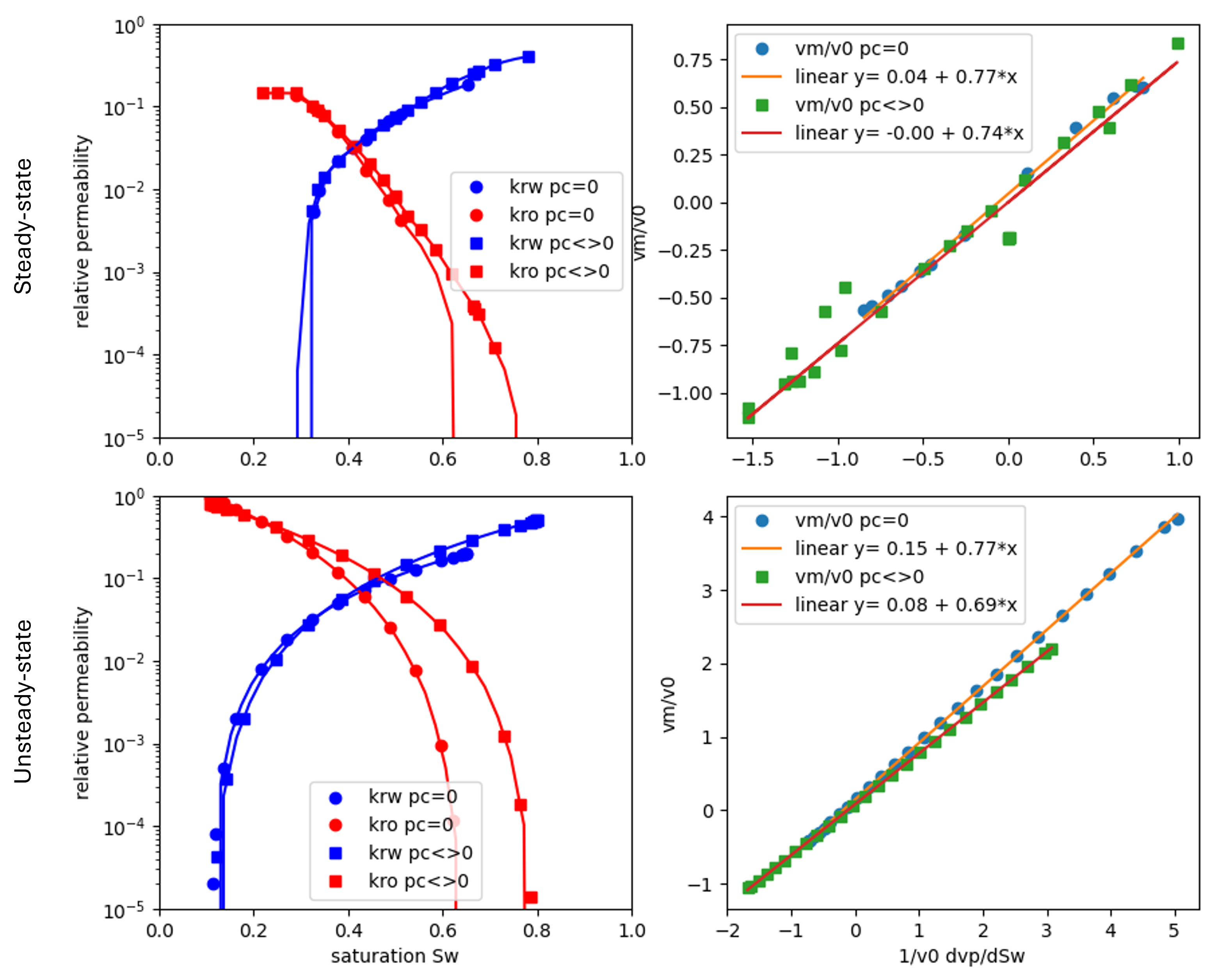}
\end{center}
\caption{Relative permeability from a steady-state (top) and an unsteady-state (bottom) experiment~\cite{bergNonuniquenessUncertaintyQuantification2021a} interpreted under the assumption of $p_c=0$ or $p_c \neq 0$ (left) and the impact on the co-moving velocity and parameters $a$ and $b$ (right).}  \label{fig:comovingvelocitypc0}
\end{figure}  

Before getting to how we will get rid of the Darcy-scale capillary pressure-saturation function, we address the important question of what happens to the co-moving velocity when the system is no longer homogeneous, i.e., $|\nabla S_w|\neq 0$?  We consider the relative permeability data for a preliminary answer.  It matters for the interpretation of the co-moving velocity from relative permeability data whether such relative permeability data has been interpreted under the assumption of $p_c=0$ or $p_c \neq 0$ as shown in Fig.~\ref{fig:comovingvelocitypc0}.  The $p_c=0$ interpretation is based on a single-flow rate interpretation, while the $p_c \neq 0$ interpretation utilizes multiple flow rates during which the capillary end-effect is integrated into the interpretation, which is caused by $p_c \neq 0$. 
In both cases we get a straight line in the $v_m/v_0$ vs. $1/v_0 dv_p/dS_w$ representation which is most caused by the fact that for the interpretation of unsteady-state experiments a functional form (here LET) was used, which similar as Corey functions naturally produces a straight line in the $v_m/v_0$ vs. $1/v_0 dv_p/dS_w$. 
There is, however, a clear impact on the slope of the $v_m/v_0$ vs. $1/v_0 dv_p/dS_w$ and respective parameters $a$ and $b$ (for the unsteady-state case the variation in $b$ is around $4 \cdot 10^{-2}$ while the uncertainty of the fit in $b$ is $9 \cdot 10^{-4}$). For the $p_c \neq 0$ interpretation of $k_r$ both $a$ and $b$ tend to decrease compared with the $p_c=0$ interpretation of $k_r$.  We note that this is consistent with the data shown in Fig.\ \ref{fig:a-b}, where the coefficients $a$ and $b$ are shown as a function of the the pressure gradient based on a dynamic pore network model: the steady and non-steady measurements may have effectively been made at different agitures. 

We note the following: The statistical mechanics approach described in Section \ref{sect:statisticalmechanics} develops a thermo\-dynamics-like formalism where all the variables have a counterpart in ordinary thermodynamics; e.g., the agiture has its counterpart in temperature and the flow derivative in chemical potential.  Except one variable: {\it the co-moving velocity.\/}  This statement, however, was recently shown not to be true.  Olsen et al.\ \cite{olsen2025new} have demonstrated that the partial molar volumes of fluid mixtures may be linked to the partial volumes found by Voronoi tesselation through a new thermodynamic function they named the {\it co-molar volume.\/}  They furthermore showed that this function is but one example of a new class of thermodynamic functions.  The co-moving velocity is another example, but in the context of flow in porous media. Hence, we are in a situation where the mathematical structure of the statistical thermodynamics framework for immiscible two-phase flow in porous media is formally identical to ordinary equilibrium thermodynamics of binary fluid mixtures, including the co-moving velocity. And here is the clue: We know how to move from equilibrium thermodynamics to non-equilibrium thermodynamics in this system, that is, introduce gradients in temperature and chemical potential.  What happens to the co-molar volume in this case?   

We can do the same for the immiscible two-phase flow problem in porous media. We will introduce gradients in the agiture $\nabla \theta$ and the flow derivative divided by the agiture $\nabla (\mu/\theta)$, where $\mu=dv_p/dS_w$.  These are now proper driving forces moving the different immisicble fluids around. This follows from analogy with ordinary equilibrium and non-equilibrium thermodynamics.    

The new concept in these considerations is that the non-equilibrium thermodynamics equivalent in the flow problem considers deviations from the steady-state flow state and not from a state with no flow.  We demonstrated in Section \ref{sect:agiture} that the agiture is proportional to the absolute value of the pressure gradient, see equation (\ref{eq:eq6-5}).  Hence, the driving forces are then given by $\nabla |\nabla p|$ and $\nabla (\mu/|\nabla p|)$.  Armed with this formalism, the statistical thermodynamics approach should be ready to deal with both imbibition and drainage.  It is then a complete story, where the input is the constitutive equation for the average seepage velocity ${\bf v}_p={\bf v}_p(\nabla p, S_w)$ and the coefficients $a=a(|\nabla p|)$ and $b=b(|\nabla p|)$, and the proper boundary and initial conditions, and the output is the saturation field and the velocity fields as a function of position and time.

\subsection{Open questions}

The most important of the open questions, how to deal with flow states away from steady state, we have already addressed in the previous Section \ref{subsec:away-with-pc}.  Here follows further open questions to consider.  

\begin{itemize}
\item In spite of the analysis of Hansen \cite{hansenLinearityComovingVelocity2024}, the question of linearity, see Equation (\ref{eq:co-mov-11}), lingers on.  The analysis in this paper was based on a small number of assumptions.  They were not general enough. We know this from the analysis of Olsen et al.\ \cite{olsen2025new} and Hansen and Sinha \cite{sinha2025thermodynamics} that it is possible to define other velocity pairs than the seepage velocities $(v_w,v_n)$ obeying Equation (\ref{eq:co-moving-12}) that do not obey linearity.  The linearity must therefore be a result of the definition used for the seepage velocities, Equations (\ref{eq:co-moving-10}) and (\ref{eq:co-moving-11}).  Why?

\item Is the co-moving velocity really linear? The data certainly support this, but they have finite accuracy. 

\item The co-moving velocity retains it linear character both in the linear Darcy regime and in the non-linear regime \cite{royCoMovingVelocityImmiscible2022}.  Why? 

\item Alzubaidi et al.\ \cite{alzubaidiImpactWettabilityComoving2024} found that the co-moving velocity is insensitive to changes in the wetting properties. Why?

\item What is the full range of consequences of assuming $p_n - p_w = 0$ as discussed in section~\ref{subsec:away-with-pc} ? In Fig.~\ref{fig:comovingvelocitypc0} we demonstrate that there is an impact but it would be good to check more examples and do a systematic variation in wettability (where from water-wet to oil-wet conditions the importance of capillary pressure for imbibition relative permeability interpretation systematically increases).

\item What about other variables such as viscosity ratio?  What is the sensitivity of the co-moving velocity to changes in these variables? 

\item What happens to the co-moving velocity when we are not in a steady-state flow situation?

\item Is it possible to measure the co-moving velocity without measuring the velocities of the individual fluids?  

\end{itemize}

		
		\cleardoublepage
		\newpage
		\section{Comparison of the new approaches}\label{sec:comparison}

        In the following, the four new approaches from sections~\ref{sec:Derivation2PhaseDarcyEquationsEntropyGeneration}, \ref{sect:spacetimeaveraging}, \ref{sect:statisticalmechanics} and \ref{sect:comovingvelocity} are compared with respect to their commonalities and differences. The main difference between the approaches is that they likely cover different flow regimes, \textit{cf.} Section~\ref{sec:non-linear}. Before discussing this in detail, the commonalities of the new approaches and the distinction from status quo in the literature are reviewed. 

        \subsection{Commonalities}
        \subsubsection{Space-time averaging}
        The first and perhaps most important commonality of the four new approaches in sections~\ref{sec:Derivation2PhaseDarcyEquationsEntropyGeneration}, \ref{sect:spacetimeaveraging}, \ref{sect:statisticalmechanics} and \ref{sect:comovingvelocity} is that they all consider thermodynamic upscaling without the use of a classical volume averaging of single variables; effectively taking the complex dynamics at pore and meso-scale into account. This is important because the phenomenological extension of Darcy's law from single to two-phase flow, and also most of the volume averaging-based upscaling approaches, do not take into account the complex spatio-temporal dynamics at the pore and cluster scale which ultimately leads to the picture of a complex energy landscape
        as depicted in Fig.~\ref{fig:energylandscape}. The energy landscape reflects the pore-scale flow regime (Fig. ~\ref{fig:AvraamPayatakesJFM1995}), \textit{i.e.} it may look different for connected pathway flow and ganglion dynamics. Furthermore, the dependency on pore level occupancy (with wetting and non-wetting phases) may introduce cooperative dynamics with topological changes at length and time scales, causing fluctuations in individual state variables not averaging out in simple volume averages alone. 
        
        In the space-time averaging approach, the NET approach, the statistical thermodynamics approach, the macro-scale variables are defined effectively as space-time averages. In the NET approach, the REV is constructed from time-averaged variables in a given space or REV \cite{Bedeaux2024_nano}. In the space-time averaging approach the representative elementary volume (REV) is explicitely defined as a space-time average~\cite{McClure2025}. In the statistical thermodynamics approach, the averaging is less evident because it utilizes a representative elementary area (REA) concept~\cite{fyhnLocalStatisticsImmiscible2023}. However, in the REA approach, the z-direction that would combine the REA to a REV is aligned with the flux, i.e., the trajectories of fluid-fluid displacements in time. Therefore, effectively, the statistical thermodynamics approach first makes a time-average over a single REA and then averages over multiple REAs. They also explicitly considered pore-scale dynamics and other fluctuations at the capillary energy scale. 

        \subsubsection{Reversibility}\label{sect:reversibility}
        Another commonality of the four new approaches is reversibility at the meso-scale. Reversibility means that there is symmetry in the Onsager coefficients. This is definitely the case on the molecular level. 
        While individual pore-scale displacement events are reversible only up to the point of topological change~\cite{Steijn2009}, reversibility can also originate from coalescence events, as depicted in Fig. ~\ref{fig:energylandscape} (see Fig. ~\ref{fig:RueckerGanglionDynamics}) and from cooperative dynamics. This is a key assumption in all approaches. Reversibility may not occur spontaneously, and energy is required to drive coalescence events, as is typical with a viscous driving force. 
        The last is somewhat analogous to processes at the molecular scale where irreversible processes can be associated with symmetry breaking~\cite{viscardyViscosityNewtonModern2010}. In multiphase flow in porous media, such symmetry breaking could be caused by pore scale displacement events accompanied by topological changes in the pore scale fluid configuration which are at the pore scale at an individual level irreversible ~\cite{mcclureGeometricEvolutionSource2019}. However, repeated coalescence and breakup can restore reversibility over longer time series.

        \subsubsection{Linear laws predicted with pressure gradient as only driving force}
        Comforting is that there are two routes  that can be used to arrive at the same permeabilities in NET. Both routes start with  the entropy production, one uses linear laws the other uses fluctuation dissipation theorems. The NET approach is limited to the linear law regime, and this is most likely present in the low-velocity limit. The high-velocity regimes have not been explored from this perspective. 
        
        All approaches differ from the volume averaging procedure used by Hassanizadeh \& Gray~\cite{HassanizadehGray1993} and the thermodynamically constrained averaging theory (TCAT)~\cite{grayIntroductionThermodynamicallyConstrained2014}. These earlier approaches contain different driving forces related to the gradients in the state variables. There is a vivid debate in the literature on which of the possible gradients can act as driving forces for the flow. The question has, for instance, been posed whether or not the interfacial area gradient can drive flow~\cite{grayAnalysisInvestigatingExtended2024,ebadi2024investigating}. The NET approach is positive about this possibility, as the REV variables contain area-dependent terms \cite{Bedeaux2024_nano}. 
        It was earlier hypothesized that such additional terms could arise, when moving from the volume averaging step to a REV, where all sub-scale temporal dynamics have been averaged out.         
        A consequence is that respective microscopic degrees of freedom are considered to be independent from one-another. In the derivation, this ultimately leads to additional terms in the transport equations~\cite{HassanizadehGray1993}. 
        
        However, as shown experimentally~\cite{Ruecker2021} and by LBM simulations~\cite{McClure2025} temporal fluctuations of microscopic degrees of freedom do not average out in space averaging alone and appear to be correlated in space and time~\cite{McClure2025}.                
        This is explicitly taken into account in the space-time averaging approach, which assumes that the collective energy dynamics of all fluctuations average out (in space-time averages), but not of individual state variables~\cite{McClure2025}. These findings give support also to the ideas of the REV-construction in the NET approach.
        The NET and statistical thermodynamics approaches also operate based on space and time averages. 
        This is important, as the averaging over space and time covers a wide range of flow regimes, including connected pathway flow and ganglion dynamics regimes, which all produce linear laws. This has not been obvious and is, therefore, an important finding. Many previous approaches, such as tube models, have covered only connected pathway flow. Now, it has been shown that linear laws are also obtained for other flow regimes.

        \subsection{Differences}
        However, there are distinct differences which  are summarized below. The key difference between the four new approaches is that they are applicable to different flow regimes. 

        \subsubsection{Flow regimes}\label{sect:differencesflowregimes}
\begin{figure}[htbp]
\begin{center}
\includegraphics[width=0.5\textwidth]{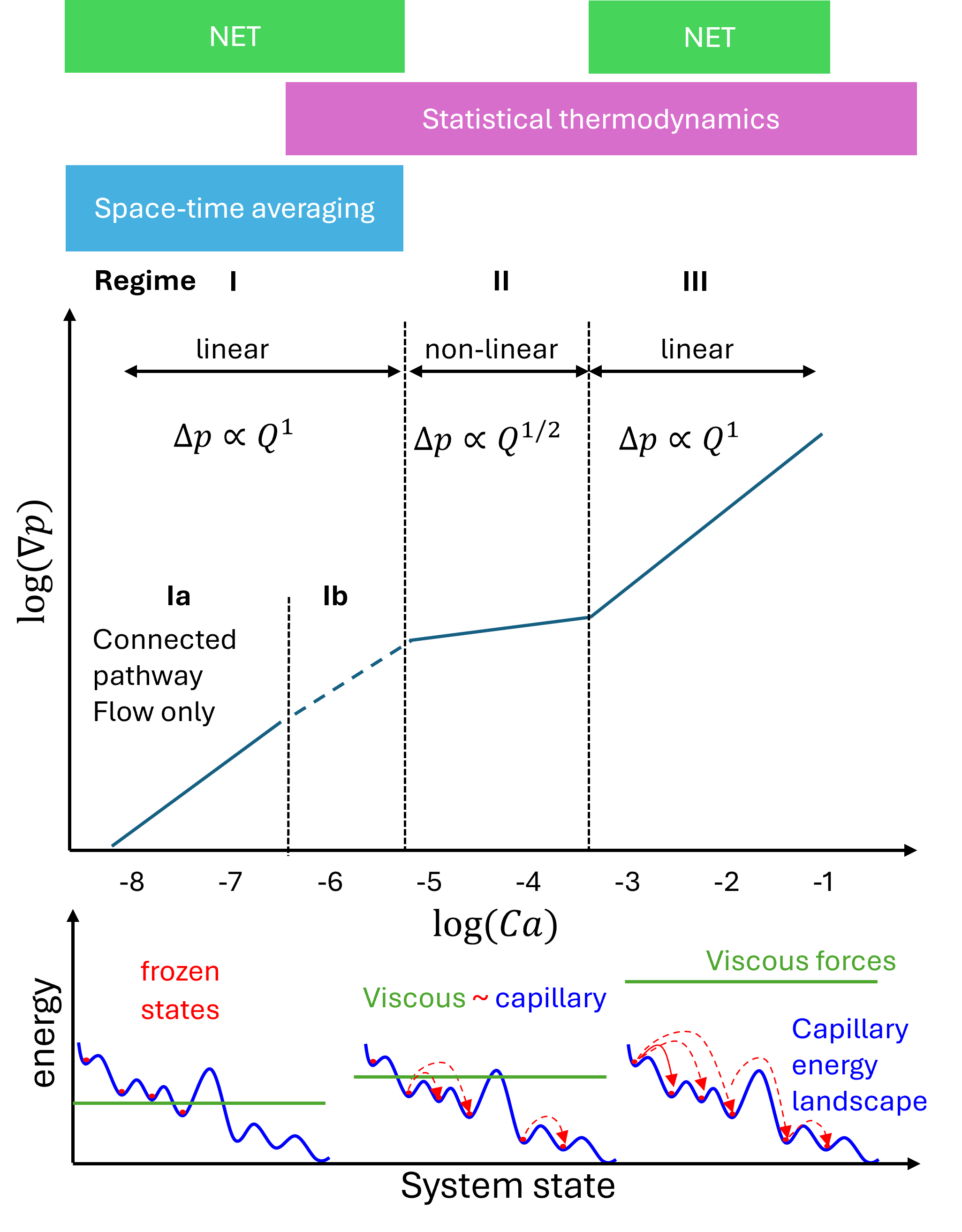}
\end{center}
\caption{Applicability and tentative mapping of the novel approaches (top) to the flow regimes from Fig.~\ref{fig:flowregimeslinearnonlinear} (see also Table~\ref{tab:flowregimes}) and cartoons of the respective energy level from viscous driving forces to capillary energy landscape (bottom). In regime III the externally applied energy level (viscous forces) are much larger than the capillary energy landscape (see also Fig.~\ref{fig:energylandscape}) and therefore, all states are accessible. On the other hand, in regime Ia the externally applied energy is less of the capillary energy barriers of the porous structure and therefore, the system falls into a ''frozen state'' of either static connected pathways or (capillary) trapping. Regime Ib is still largely in the frozen state but breakup and coalescence events release some level of energy which is sufficient to overcome the lower energy barriers which cause fluctuations around a still overall connected pathway flow dominance. Regime II is a transition regime where viscous and capillary forces become equal which leads to viscous mobilization and associated non-linearity. Note that the NET approach applies to regime III at the level of capillary fluctuations and also regime Ia at the level of molecular fluctuations. It also applies to the capillary fluctuations in regime Ib which the space-time averaging approach has proven to average out leaving a dominant connected pathway flow transport mechanism also in regime Ib, which is the main reason why the flux-force linearity still holds in Ib. 
} 
\label{fig:flowregimesapproaches}
\end{figure}

        The main difference between the different approaches is which flow regime is described. For the flow regimes, we mainly refer to the regimes outlined in Section ~\ref{sec:non-linear} (for example, Fig.~\ref{fig:gao2020} and Fig.~\ref{fig:talon-1}), where we found three distinct regimes: (I) a linear regime at small capillary numbers, (II) a non-linear regime at intermediate capillary numbers, and (III) a linear regime at high capillary numbers. 
        Table~\ref{tab:flowregimes} and Fig.~\ref{fig:flowregimesapproaches} hypothesize which approach covers which flow regime. 
        However, this mapping is tentative and hypothesized very much based on plausibility, such as whether the flux-force relationship is linear and whether viscous or capillary forces dominate. 

\begin{table}[ht]
  \centering
  \caption{Flow regimes from section~\ref{sec:non-linear} and how they are covered by the 3 new approaches. One of the goals of future research must be to show how the approaches can extend beyond their current ranges and cover all flow regimes.}
  \label{tab:flowregimes}
  \small
  \begin{tabular}{cllllc}
    \toprule
    \textbf{Regime} & \textbf{Flux-force} & \textbf{Capillary number} & \textbf{Force balance} & \textbf{Description Approach} & \textbf{Section}\\
    \midrule
    I & linear & $Ca<10^{-5}$ & capillary-dominated & NET, space-time averaging &  4, \ref{sect:spacetimeaveraging}\\
    II & non-linear & $10^{-5} < Ca < 10^{-2}$ & visco-capillary & statistical thermodynamics & \ref{sect:statisticalmechanics} \\
    III & linear & $Ca>10^{-2}$ & viscous-dominated & NET, statistical thermodynamics  &  \ref{sec:Derivation2PhaseDarcyEquationsEntropyGeneration}  \\
    \bottomrule
  \end{tabular}
\end{table}

        For the space-time averaging approach thus far, only the small-velocity limit has been investigated. In essence, the space-time averaging approach describes linear Darcy-type transport equations in the presence of capillary fluctuations. All non-wetting phase clusters are in a capillary-dominated regime, and the viscous forces are not sufficient for mobile flow. Topological changes and ganglion dynamics are only triggered by coalescence and breakup and their respective internal dynamics (which can be locally at much higher capillary numbers and even contain inertial contributions). Many capillary states are energetically inaccessible. Nevertheless, one may expect that these systems are ergodic \cite{Erpelding}, that the processes obey time reversal invariance for proper choices of integration times, and therefore that sufficient conditions apply for NET to be applicable. It is at low flow velocities that diffusion, thermal diffusion and pressure diffusion can add significantly  to the dissipation of energy, or entropy production. Tensors of different order do not couple, meaning \textit{e.g.} that vectorial diffusive forces do not couple to viscous forces, but they can couple to pressure diffusion \cite{Kjelstrup2019}.  Each flow regime has a particular slope.
        As Fig.~\ref{fig:gao2020} indicates, in regime I, there is already a transition to ganglion-dynamic-type flow patterns at higher velocities before the flux-pressure gradient relation becomes non-linear.         
        Therefore, from a flow pattern perspective, there is already some degree of viscous mobilization before formally entering nonlinear regime II. 
        However, over most of the mobile saturation range, their flux contribution is very minor (see Fig.~\ref{fig:RueckerGanglionDynamics}C \cite{Armstrong2016}), and the overall flux is dominated by the connected pathway flow.
        The space-time averaging approach demonstrates that the collective energy dynamics of the moving ganglia averages out in space-time averaging, which then leaves mainly the connected pathway flow component (with changing pathways due to the topological changes in regime Ib), which is the main reason why one single formulation of two-phase flow with the relative permeability concept holds over both sub-regimes Ia and Ib.  For higher velocities, the currently neglected higher-order terms could lead to a more complex description, which would also make the approach applicable to non-Darcy flow, i.e., the nonlinear ganglion dynamics regime (second flow regime in Fig.~\ref{fig:gao2020} and Fig.~\ref{fig:talon-1}). 

        Only the statistical thermodynamics approach has the potential to predict various flow regimes. NET and space-time averaging may describe constitutive relationships for a given flow regime, but cannot predict the flow regime itself. In the statistical thermodynamics approach, the transitions between flow regimes may appear as equivalent to ''phase transitions'' with respect to possible flow patterns. 
        The statistical thermodynamics approach covers the second and third regimes in the viscous limit very well, as shown in Fig.~\ref{fig:gao2020} and Fig.~\ref{fig:talon-1}, it struggles to describe the first regime, because that would represent essentially a largely ''frozen'' state (where many capillary states are energetically not accessible). Current progress extends the statistical thermodynamics approach into regime Ib. In the second regime, which is covered by the statistical thermodynamics approach, the lower part of the cluster size distribution is still capillary-dominated, but the upper part is subject to viscous mobilization, which allows more states to become accessible; the populations of states are then described by Boltzmann statistics with agiture as the equivalent to temperature.          

        Regime III, shown in Fig.~\ref{fig:gao2020} and Fig.~\ref{fig:talon-1}, is viscous-dominated and presumably contains a drop-traffic flow pattern (Fig.~\ref{fig:AvraamPayatakesJFM1995}) represents the viscous limit. All non-wetting phase clusters are now subject to viscous-dominated transport, and all states are available and accessible for each saturation value. This is in some sense equivalent to the dynamics at the molecular scale, and the fluctuation-dissipation theorem approach to Onsager reciprocal relationships may hold. 

        There are many open questions related to this mapping, and overall, this is considered an area for future research. Several types of mechanisms can be associated with time-reversal invariance, but any discrimination between them in this context has not been systematically investigated. True is that states can be accessed via trajectories in saturation space with the associated pore-scale mechanisms of coalescence and breakup (see ganglion dynamics in Fig.~\ref{fig:RueckerGanglionDynamics}), which is linked to the reversibility discussion in section~\ref{sect:reversibility}. The NET approach may, in principle, hold in this regime as well, including the possibility of transport coefficients with memory functions. 

        Nevertheless, from the perspective of transport dynamics, regime III is the most comprehensive and intuitive regime because viscous forces are much larger than capillary forces such that the multiphase flow does not really "feel" the capillary energy landscape. When successively approaching regime I the viscous forces decrease and become first comparable (regime II) and then smaller than the capillary forces (regime I) which leads to ''frozen states''. Understanding regime I as frozen state suggests that regime I is actually the most complicated regime to describe and understand from the perspective of upscaling theories while regime III is the most straight-forward one. Traditional approaches begin with regime I, while in this review paper we take the perspective of regime III as starting point. That opens the perspective about the true nature of regime I and challenges many implicit assumptions which appear natural like the percolation and connected pathway flows with the definition of capillary pressure (see section~\ref{subsec:away-with-pc}).

       \subsubsection{State variables}
        Some state variables, such as phase saturation, are used by all approaches, whereas other variables differ. For example, the space-time averaging approach uses Minkowski functionals directly as state variables. The Euler characteristic is then included among the variables. This is particular for this approach. In NET, the REV construction is particular. It contains lumped variables that contain mixed normal- and Minkowski-type variables, according to the recipy of Hill \cite{Bedeaux2024_nano}. The NET approach can be said to use macroscale REV variables obtained by coarse graining. The variables obey local equilibrium in the REV. The coarse-graining procedure assumes that the subsystems (the phases, interfaces and contact lines) are weakly coupled (energies are additive). The sum of these contributions will include the contributions of the Minkowski functionals. Minkowski functionals enter the formulations in a different way. 
        
        The statistical thermodynamics approach introduces the variable ''agiture'' which is not present in the other approaches. The statistical thermodynamics approach and the co-moving velocity concept start with an incomplete set of state variables to provide a proof-of-concept. This can be extended to more state variables. The space-time-averaging approach uses four Minkowski functionals, established in a capillary equation of state~\cite{McClure2018}.   

        The statistical thermodynamics approach i.e. the derivation of the partition function, is in some sense a logical next step after the identification of the state variables. This has been attempted, following an Euler-Lagrange formalism, which was unsuccessful because of discontinuities introduced by geometric evolution and non-differentiability of the Minkowski sum~\cite{McClure2020,mcclureGeometricEvolutionSource2019}; see also Fig.~\ref{fig:workflowpartitionfunction}. The approach by Hansen and co-workers has been successful by realizing that irrespective of that, the outcome will still be a Boltzmann type of statistics, and the complications are avoided by using Jaynes' maximum entropy principle.

        \subsubsection{Additional differences}
        A few additional differences can also be included in Table~\ref{tab:upscaling}:

        Following the recipy of Hill and Bedeaux and coworkers \cite{Hill1994,Bedeaux2024_nano}, the driving force for confined flows is given by the gradient in the integral pressure. The NET-, and only this approach, takes advantage of the grand potential to describe transport in porous media. In the absence of capillary forces, the gradient in the integral pressure is equal to the gradient in the hydrostatic pressure. This thermodynamic basis allow us to also include gradients in saturation and temperature in the entropy production \cite{Kjelstrup2018,Kjelstrup2019}. Gradients in surface tension and  line tension can be handled  \cite{Hill1994} and links can be made to regimes with diffusion, adsorption and thermal diffusion.   
           
        The NET approach considers fluctuations on all time- and spacial scales, \textit{i.e.} from the molecular scale to the viscous-dominated flow regime. The space-time averaging approach captures fluctuations also at the capillary energy scale. The molecular fluctuations are implicitly covered in continuum mechanics (Navier-Stokes equation). That is the reason why the NET approach is marked as also applicable in regime Ia as well as in III in Fig.~\ref{fig:flowregimesapproaches}.   


        Section~\ref{sect:comovingvelocity} distinguishes itself from other approaches in the sense that flow equations are not derived. The established two-phase Darcy equations are accepted as the starting point, and then symmetries within these flow equations are analyzed/identified and used to constrain relative permeability. 
        Using saturation $S_w$ as the only state variable might be too simplistic, given that it has been demonstrated in other approaches that there are more state variables. Nevertheless, the approach is already useful in itself, and instructive in a more general way, as it might be extended to a system with more state variables.

        An overview is provided in Tab. ~\ref{tab:upscaling} summarizes the four new approaches for describing multiphase flow in porous media/upscaling from the pore to the Darcy scale. The key concepts which had been discussed in the previous sections are briefly summarized, and the key differences are highlighted. 
        
        
        \begin{landscape}          
\begin{table}[ht]
  \centering
  \caption{Overview of the new approaches for upscaling 2-phase flow in porous medial, summarizing the key concepts, commonalities and key differences.}
  \label{tab:upscaling}
  \small
  \begin{tabularx}{\linewidth}{@{}l X X X@{}}
    \toprule
    \textbf{Section} & \textbf{Key Concepts} & \textbf{Upscaling Approach} & \textbf{Key Differences} \\
    \midrule
    \ref{sec:Derivation2PhaseDarcyEquationsEntropyGeneration}. NET &
      \begin{itemize}[itemsep=1pt, parsep=0pt, topsep=2pt, partopsep=0pt, leftmargin=*]
        \item Principles of classic NET applied to coarse-grained REV-variables 
        \item Hill’s nanothermodynamics is used to deal with fluid confinements in the REV
        \item Flux–force relations and experiments are defined by the REV entropy production
        \item Transport coefficients are obtained from fluctuation–dissipation theorems and from linear laws
      \end{itemize}
    &
      \textit{Gibbs equation, written for coarse-grained REV variables, identifies the entropy production}

      \medskip
      \textbf{Outcome:} First‐principles derivation procedure for linear laws
    &
      \begin{itemize}[itemsep=1pt, parsep=0pt, topsep=2pt, partopsep=0pt, leftmargin=*]
        \item Geometric variables appear from use of Hill's thermodynamics. 
        \item An effective pressure gradient drives two-phase flow of confined fluids 
      \end{itemize}
    \\[1em]

    \ref{sect:spacetimeaveraging}. Time‐and‐Space \\ Averaging &
      \begin{itemize}[itemsep=1pt, parsep=0pt, topsep=2pt, partopsep=0pt, leftmargin=*]
        \item Classic NET with time‐and‐space averaging of variables
        \item Steady-state over space‐and‐time required for bilinear entropy generation
        \item Two‐phase Darcy emerges at steady-state
      \end{itemize}
    &
      \textit{Entropy‐production including time‐and‐space averaged properties}

      \medskip
      \textbf{Outcome:} Proof of two‐phase Darcy under steady-state
    &
      \begin{itemize}[itemsep=1pt, parsep=0pt, topsep=2pt, partopsep=0pt, leftmargin=*]
        \item Minkowski functional implicit in fluctuation terms
        \item Total pressure gradient drives flow under steady-state
      \end{itemize}
    \\[1em]

    \ref{sect:statisticalmechanics}. Statistical \\ Thermodynamics &
      \begin{itemize}[itemsep=1pt, parsep=0pt, topsep=2pt, partopsep=0pt, leftmargin=*]
        \item Jaynes’ maximum‐entropy principle at capillary‐energy level
        \item Boltzmann‐like occupancy statistics for interface configurations
        \item Partition‐function framework incorporating Minkowski functionals
      \end{itemize}
    &
      \textit{Maximum Entropy}

      \medskip
      \textbf{Outcome:} Statistical‐mechanical framework for porous media flows
    &
      \begin{itemize}[itemsep=1pt, parsep=0pt, topsep=2pt, partopsep=0pt, leftmargin=*]
        \item Minkowski functionals are conjugate to capillary potentials
        \item Capillary ‘agiture’ and flow derivative drives flow
      \end{itemize}

    \ref{sect:comovingvelocity}. Co‐Moving \\ Velocity &
      \begin{itemize}[itemsep=1pt, parsep=0pt, topsep=2pt, partopsep=0pt, leftmargin=*]
        \item Starts at two‐phase Darcy
        \item Reveals symmetry relations between relative permeability functions
        \item and potentially also between more state variables
      \end{itemize}
    &
      \textit{Symmetry Analysis}

      \medskip
      \textbf{Outcome:} Reduced uncertainty in relative permeability determination
    &
      \begin{itemize}[itemsep=1pt, parsep=0pt, topsep=2pt, partopsep=0pt, leftmargin=*]
        \item No explicit use of Minkowski functionals
        \item Pressure gradient as defined in traditional Darcy formulation drives flow
      \end{itemize}
    \\
    \bottomrule
  \end{tabularx}
\end{table}
\end{landscape}

		
		\cleardoublepage
		\newpage
		\section{Applications, open questions and vision}\label{sec:Outlook}
        
        The classical 2-phase Darcy approach was developed to deal with engineering needs, mainly with 2-phase immiscible displacements. Although it has been used for more complex situations, the approach quickly reaches its limits (see Sections ~\ref{sect:massmomentumbalance} and \ref{sect:thermodynamicproblem}) or is implicitly extended beyond its original validity. However, many current scientific and technical challenges are accompanied by an increased level of complexity. 
        The key challenge is coupled processes, where the Darcy-scale consequences of  coupling explicitly depend on microscopic flow regimes. A simple pressure gradient is not sufficient as a driving force in  pore-scale flow regimes, where confinement, mixing, or phase transitions occur.
    
        \subsection{Immediate applications}
        \subsubsection{Confirmation of linear laws reduces model-based uncertainty}
        The immediate result of this review is that for both the capillary-dominated and viscous-dominated flow regimes, linear laws are obtained, which are consistent with the phenomenological 2-phase Darcy equations. The respective derivations do not make the limiting assumptions of historical volume-averaging approaches. 
        At least for steady-state conditions in the low velocity limit the 2-phase Darcy equations are correct, but this may not be the case for non-steady-state conditions (see section~\ref{sect:spacetimeaveraging2phaseflow}). 
        While this does, in principle, not change the status quo, it puts the commonly used 2-phase Darcy equations on a much more solid theoretical basis, and therefore reduces model-based uncertainties in many applications and provides practical guidance on meaningful domain sizes to determine relative permeability.       
        One concrete example is the Digital Rock method, where pore-scale numerical simulations are used to determine the relative permeability of rock (and potentially other porous materials). The new definition of the multiphase representative elementary volume (REV)~\cite{McClure2025} (section~\ref{sect:multiphaseREV}) based on the space-time averaging approach (section~\ref{sect:spacetimeaveraging}) provides the basis for the choice of domain sizes in Digital Rock simulations~\cite{ramstadSimulationTwoPhaseFlow2010,ramstadRelativePermeabilityCalculations2012a} that are practically achievable and enabling to optimize for limiting boundary conditions between field-of-view and resolution of micro-CT imaging, as well as such limiting conditions for computational costs.
        
        \subsubsection{Constraining the functional form of relative permeability}
        For the co-moving velocity, the most immediate application is to constrain the possible choices of relative permeability representations. In the traditional 2-phase Darcy description the "flexibility" in the relative permeability functions is perhaps one of the greatest weaknesses which allows a wide range of choices in representations, that ultimately leads to a higher degree of uncertainty also in their experimental determination. Furthermore, a wide range of effects, beginning with heterogeneity, flow instability, steady-state vs. non-steady-state and many more are all lumped into relative permeability functions, which makes it on the one hand very "flexible" but also much less "predictive". The relative permeability then becomes less of a material and more of a process parameter, which is highly process-dependent, such as the drainage vs. imbibition process. 
        In the current formulation, the co-moving velocity assumes only the saturation-pressure gradient relationship, i.e. saturation as the only state variable. In a more general formulation with more state variables, such as the Minkowski functionals, the underlying principle i.e. symmetries in the flow equations, might help establish more general relationships between state variables. 

        \subsubsection{Describing hysteresis}
        For geometric state characterization, the most immediate application is probably to describe hysteresis in multiphase flow~\cite{mukherjeeModelingRelativePermeability2025, AlZubaidi2023}. This has been a long-standing problem, and current hysteresis models are entirely phenomenological. 
        Since hysteresis is a key ingredient in the nonlinear dynamics description (see section~\ref{sec:non-linear} ), the correct description of hysteresis is not only a question of large practical relevance, but there is also the need from a theoretical perspective~\cite{adilaComparisonRelativePermeability2025}. 
        The next level of applications could be 3-phase. Given the significant experimental challenges, experimental data are scarce~\cite{oakThreePhaseRelativePermeability1990,scanzianiSituCharacterizationThreePhase2020} current 3-phase relative permeability models bear significant uncertainties and are overly simplistic~\cite{bluntEmpiricalModelThreePhase2000}. For instance, they do not cover cyclic hysteresis in 3-phase flow situations relevant for water-alternating-gas (WAG) enhanced hydrocarbon recovery and optimizing mechanisms for $\rm{CO_2}$ injection for CCS~\cite{masalmehGasCO2Mobility2025}. 

        \subsubsection{Coupled transport and phase behavior on a fundamental level}
        For problems such as gas diffusion layers where flow and phase transition take place at the same time~\cite{chapuisTwophaseFlowEvaporation2008}, a thermodynamically consistent formulation of flow and state variables is lacking. The NET formulation was extended to address coarse-grained variables. These variables may contain the effective pressure ( the integral pressure according to Hill), and the application of this may help find more consistent descriptions of coupling between transport and phase transitions. To date, coupling at this mesoscale has not been studied.

        \subsection{Open questions}
        There are significant questions that remain unanswered.
        While in the past decade some central questions such as flux contribution of ganglion dynamics~\cite{Armstrong2016}, multi-phase REV~\cite{McClure2025}, and non-linear flow regimes~\cite{zhang2021quantification} have been at least partially answered, this is rather the beginning of a new field. 
        Many central open questions must be addressed before a fully coherent framework for the upscaling of multiphase flow in porous media can be established:

        \begin{itemize}
            \item Classes of systems with respective flow regimes need be characterized for which the novel concepts are potentially applicable and add new perspectives / additional value.   
            We deal with a vast set of mechanisms with respect to flow regimes and flow patterns (Fig. ~\ref{fig:AvraamPayatakesJFM1995}). In addition, there are also stable vs. unstable displacements. 
            Does the tentative assignment of the approach to the flow regime outlined in Section ~\ref{sect:differencesflowregimes} really hold? We have a tentative understanding of which approach covers which flow regime, but this needs to be developed further and validated.      

            \item How can diffusion and thermal gradients~\cite{fernandoNonequilibriumComputerSimulations2017,burelbachUnifiedDescriptionColloidal2018} be implemented in a consistent manner into the description for multiphase flow in porous media? An increasing number of problems involve diffusive transport driven by temperature gradients~\cite{hafskjoldSoretSeparationThermoosmosis2022}.
            Examples range from transport phenomena in catalysis where chemical reactions generate temperature gradients~\cite{gladdenMRIOperandoMeasurements2010,zhengOperandoCharacterisationProducts2024} to temperature gradients in electrochemical devices~\cite{gunathilakaOperandoMagneticResonance2021,zhengOperandoCharacterisationProducts2024}, transport in insulation layers of cryogenic storage tanks~\cite{kimCryogenicCalorimeterTesting2023} and many more.
            Numerical simulators for multiphase flow in porous media face the challenge that in order to directly model molecular diffusion, the numerical dispersion needs to be suppressed below the level of molecular diffusion, which requires very fine grids and small time steps that come at a significantly increased computational cost. 

            \item Steady-state vs. unsteady-state - in the space-time averaging approach the 2-phase Darcy equations are only obtained for steady-state but not for unsteady state, however, the unsteady-state has not been investigated in sufficient detail. 
            So far, the experimental evidence about differences between steady-state and unsteady-state is inconclusive~\cite{alemnDifferenceSteadystateUnsteadystate1989,mainiComparisonSteadyStateUnsteadyState1990,ramstadRelativePermeabilityCalculations2012,ramstadPoreScaleSimulationsSingle2019} when the full range of uncertainties~\cite{bergSimultaneousDeterminationRelative2024a} is taken into account.
            
            \item Energy landscapes and connections between flow regimes and energy landscapes. How does the energy landscape which includes meta-stable states away from equilibrium ~\cite{einavHydrodynamicsNonEquilibriumSoil2023} look-like? Fig.\ \ref{fig:energylandscape} is only a cartoon, and how does it change with the flow regime, i.e. what is the difference between ganglion dynamics and drop traffic? A related question is how the balance between the capillary and viscous forces influences the energy landscape. On the one hand, a larger (viscous) driving force allows overcoming capillary energy barriers. On the other hand, the flow regime changes and the energy barriers may simply become smaller. One example is the drop-traffic regime, where non-wetting phase clusters are smaller than the capillary restrictions at pore throats, and therefore, the energy landscape becomes flatter. 


            \item Is there an underlying, deeper principle that allows us to determine constitutive and transport relationships? 
            There is an ongoing activity in the community such as those related to the principle of minimum power~\cite{valavanidesReviewSteadyStateTwoPhase2018a,gaoNewInsightsInterface2025}, the explicit or implicit thermodynamic arguments and assumptions related to (minimum or maximum) entropy production~\cite{HassanizadehGray1993} and TCAT~\cite{millerPedagogicalApproachThermodynamically2017}. 
            However, from a non-equilibrium thermodynamics perspective such principles hold only under strong limitations such as small driving forces~\cite{grootNonEquilibriumThermodynamics1984}.


            \item What are the state variables for characterizing flow regimes in multiphase flow? The geometric (capillary) state is conveniently parameterized by the four Minkowski functionals because they contain variables such as volume, interfacial area, and curvature, which also naturally occur in thermodynamic formulations. 
            This makes the respective state functions and thermodynamic formulations very intuitive. It has also been shown that relative permeability~\cite{Liu2017,AlZubaidi2023} and electrical resistivity~\cite{liuInfluenceWettabilityPhase2018} can be parameterized by Minkowski functionals.
            
            However, the Minkowski functionals are applicable for quasi-static situations as they do not contain time, and are also contain largely dependent variables as illustrated in Fig.~\ref{fig:flowregimestrajectoriesclustercapillarynumber}. 
            For dynamic situations new variables and parameters become important such as viscosity ratios,      
            capillary number~\cite{Armstrong2016, zhang2022nonlinear,suwandiRelativePermeabilityVariation2022} (visco-capillary balance),  fractional flow $f_w$,  Ohnesorge number $Oh$~\cite{chenInertialEffectsProcess2019,zacharoudiouPoreScaleModelingDrainage2020} (capturing inertial effects), etc. 

            \item Are relative permeability-saturation functions really described by single power laws such as the Corey model? Alternatively, do the different underlying flow regimes have all their own constitutive relationships, which are masked by the relatively large uncertainty range that generates a potentially misleading impression of one effective regime over the whole mobile saturation range? Or is that more a consequence of considering only saturation as a state variable, which is highly non-unique, that is, would flow regimes become more obvious by also considering Minkowski functionals ?
        

            \item What is the magnitude of (off-diagonal) relative permeability cross-coupling coefficients representing the viscous coupling (tangential stress boundary condition~\cite{lealAdvancedTransportPhenomena}) in comparison with the traditional (diagonal elements) relative permeability coefficients ? So far, experimental work has left this question unanswered~\cite{bentsenUseConventionalCocurrent1993,bentsenEffectMomentumTransfer1998,ayubInterfacialViscousCoupling1999,bentsenEffectNeglectingInterfacial2005}; however, recent work by Lasseux \textit{et al.}  has shown that viscous coupling may lead to additional terms in the upscaling from pore to Darcy scale~\cite{Lasseux2022}.

            \item Should relative and absolute permeability be functions of temperature ~\cite{standnesDerivationConventionalGeneralized2022}? 
            The traditional view is that absolute permeability is a function of the porous medium and (apart from thermal expansion effects) not a function of temperature (only fluid density and viscosity, and interfacial tensions would be a function of temperature). Derivation via Onsager coefficients opens up a temperature dependence.

        \end{itemize}

        \subsection{Experiments and pore scale modeling to drive theory forward}
        The four novel approaches are, to a large extent, motivated by experimental work conducted over the past decade with significant input from direct pore scale imaging by fast synchrotron beamline-based X-ray computed micro tomography. For instance, the space-time averaging approach in section~\ref{sect:spacetimeaveraging} was motivated by the experimental observation of fluctuations at any scale between the pore and Darcy scale~\cite{Ruecker2021}.        
        At the same time, a new theory will promote the development of experimental work. One historical example is the work by Hassanizadeh \& Gray~\cite{HassanizadehGray1993} which significantly influenced experimental and numerical modeling work and established the relationship between interfacial area and saturation~\cite{Culligan2004,Herring2013}, which in turn led to the discovery of the Euler characteristic~\cite{Herring2015} as the last missing capillary state variable~\cite{Ruecker2015}. 
        In other words, experimental work and theoretical development go hand-in-hand. This is complemented by numerical modeling ~\cite{McClure2018} which, after calibration/validation against experimental data~\cite{Armstrong2016}, complements experiments. 

        With the four novel concepts of Table~\ref{tab:upscaling}, the concrete need is to further test and validate the approaches. Experiments need to be conducted to test aspects that are specific/unique to novel concepts. The hope is to be able to differentiate between them and understand how they differ from older concepts. Clarifying exactly the driving force issues seems essential, and relations to relative permeability must also be made for novel displacement processes such as gas dropping out of solution and ripening~\cite{gaoCapillarityPhasemobilityHydrocarbon2021,dokhonPressureDeclineGas2024}. 

        The space-time averaging method clearly showed that there is a difference in dissipation between steady-state and non-steady-state conditions. This would imply that steady-state and unsteady-state relative permeability might be different, but the question is how much. The literature is inconclusive regarding this question~\cite{alemnDifferenceSteadystateUnsteadystate1989,mainiComparisonSteadyStateUnsteadyState1990,ramstadRelativePermeabilityCalculations2012,ramstadPoreScaleSimulationsSingle2019}.

        One of the general challenges is that theoretical concepts and respective predictions, such as phase diagrams, need to be validated by experimental data, which, however, has considerable uncertainty ranges when inadequate experimental and interpretation methodologies are used~       \cite{bergSimultaneousDeterminationRelative2024a}. Therefore, in general,
        a much more rigorous uncertainty assessment needs to be performed than is currently routinely done, that is, also considering the model-based uncertainties introduced by the interpretation methodology, which also captures how well the inverse model is constrained by the data~\cite{bergSimultaneousDeterminationRelative2024a}. 
        This may then allow us to address questions such as whether there are differences between steady-state and unsteady-state relative permeability, as predicted by the space-time averaging method~\cite{McClure2022}.
        There might be further complications, e.g. the mobile saturation ranges directly accessed in steady-state and unsteady-state techniques overlap only to some extent. In addition, inlet and outlet effects would need to be suppressed more, as the play out differently in steady-state and unsteady-state protocols. 
        
        Perhaps the classical steady-state vs. unsteady-state comparison is not suitable, and one needs to compare experiments where fractional flow is changed in steps as in the classical steady-state protocol vs. a continuous but gradual change in fractional flow. The NET approach would also allow us to revisit the question of viscous coupling coefficients, i.e. off-diagonal cross-coupling terms in relative permeability, which thus far has not been addressed in the literature in a satisfactory manner~\cite{bentsenUseConventionalCocurrent1993,bentsenEffectMomentumTransfer1998,ayubInterfacialViscousCoupling1999,bentsenEffectNeglectingInterfacial2005}.

        Considering the possibility of traveling waves and other periodic solutions (see Fig. ~\ref{fig:MitraOrbits}), it may be important to conduct experiments in a way that external triggers such as the capillary instability at the entry of the porous domain are suppressed. 
        While all solutions are intrinsic modes of porous media transport, depending on the magnitude of externally introduced perturbations,  experiments will provide solutions on the unstable branch, but not the stable base case.

        \subsection{The role of machine learning and AI to drive theory development forward}
 The obvious question is why to continue building theories and models while the world is moving away from model-driven to data-driven approaches, such as AI.
The main reason is that the parameter space is still unknown and might be too large for data-driven approaches to be tractable, keeping in mind that experimental data are very sparse. While there is moderate coverage for 2-phase flow (but with significant gaps, see e.g., discussion around viscous coupling parameters), it is already too sparse for 3-phase flow to train an AI. 
Solving the upscaling problem via model building will naturally provide the still unknown state variables (i.e., which parameters are written on the axis of the phase diagram for flow regimes, see Fig. ~\ref{fig:flowregimes}).
        
Nevertheless, AI has already supported theory development in many ways. AI has a significant supporting role in image processing~\cite{bergGenerationGroundTruth2018,tahmasebiMachineLearningGeo2020,wangDeepNeuralNetworks2021,alqahtaniSuperResolvedSegmentationXray2022,tangDeepLearningFullfeature2022,liangMultimineralSegmentationMicrotomographic2022,reedyHighresolutionMicroCT3D2022,mahdaviaraPoreSkelSkeletonizationGrayscale2023,waldnerDeeplearningbasedWorkflowReconstructing2024,strandbergAIassistedDeepLearning2024} leading to higher precision, accuracy, reduced user bias, and faster image processing, AI can also be used to overcome the hardware limits of current imaging techniques. Machine learning and AI methods provide techniques such as super-resolution access to a larger field of view at high resolution~\cite{wangLargescalePhysicallyAccurate2023} and factors of higher time resolution in tomographic imaging~\cite{goethalsDYRECTComputedTomography2025,niu2021geometrical,liu2020tomogan}. These tools are currently coming to fruition and will likely allow for a wealth of information on pore-scale multiphase flow and the measurement of pore-scale entities at time and spatial scales previously impossible. Linking these tools to direct theoretical questions is currently underdeveloped, with the current work focusing on methodology development. A tighter coupling between theory and AI tools is required. 

With AI methods, a new form of empiricism is emerging that provides potentially greater generalization and predictability of a larger range of physical properties. AI has been used to estimate effective properties. Examples include the prediction of stress distribution during finite element analysis \cite{liang2018deep}, identification of Koopman eigenfunctions~\cite{lusch2018deep}, and estimation of relative permeability from the geometric state of fluids \cite{AlZubaidi2023}. The work of Spurin and co-workers on dynamic mode decomposition has made the first steps in this direction~\cite{spurinDynamicModeDecomposition2023b,raizadaDynamicModeDecomposition2025}.
Methods have also proposed the prediction of Darcy scale properties, such as absolute permeability, directly from micro-CT images \cite{alqahtani2021flow}. Perhaps unexpected trends could be discovered with such AI methods, such as the identification of key state variables through reduced-order modeling and/or studying how the transformed network makes a prediction, such as using the integrated gradients method \cite{sundararajan2017axiomatic}. However, in general, current usage has focused on predicting the effective properties of porous materials rather than as a discovery tool.

        An interesting opportunity arises from the dynamical systems community, where AI is used to discover the governing equations. \cite{schmidt2009distilling} propose symbolic regression to extract free-form natural laws, identifying conservation principles without prior knowledge of the domain by using physical invariants from experimental time-series data. 
        \cite{imDatadrivenDiscoveryGoverning2023} follow a similar approach using symbolic regression and stochastic optimization for transport in heterogeneous porous media. 
        \cite{bongard2007automated} extend this by introducing an active learning framework that alternates between modeling and assessment, allowing for the reverse engineering of nonlinear dynamical systems through targeted perturbations. \cite{brunton2016discovering} used a sparse regression model to extract parsimonious dynamic models from noisy data, which is scalable and interpretable for large systems. Lastly, \cite{bapst2020unveiling} applied graph neural networks to predict the long-term dynamics of glassy systems using only static structural information, offering new insights into structural predictors of dynamical arrest. These methods are particularly interesting for the behavior of multiphase flow systems at low capillary numbers, which are believed to have behavior analogous to glassy systems and/or high capillary numbers, where fluctuation dynamics could result in significant energy expenditure under nonsteady-state conditions.


        \subsection{Long term vision}
        The current state is that we have theoretical derivations of the transport equations or equivalent descriptions for each of the three flow regimes.      
        The long-range goal is to obtain a thermodynamically consistent description of the multiphase flow in porous media that covers all flow regimes. In essence, so far we have been a "passive consumer" of porous media. Now, we are moving to a more active design role where we optimize the function of the porous medium towards a specific application, which requires a better conceptual understanding of porous media flow physics.
        An example of feasibility is the favorable relative permeability for mixed-wet conditions caused by a bi-continuous surface (see Fig. ~\ref{fig:minimalsurfaces}), which may be a consequence of the principle of minimum power (see discussion towards the end of section~\ref{sect:thermodynamicproblem}). 
              
        Such a description must cover all aspects of the method in a consistent manner, as described by the entropy balance. So far, the approaches presented are not complete. To varying degrees, they have been tested for consistency, and we have a tentative understanding of which flow regimes are covered. Currently, the different approaches focus on different aspects for which individual choices are better suited than others. To achieve this goal, it may be necessary to make the new approaches consistent in terms of state variables and REV. 

        In principle, a statistical thermodynamics approach would cover everything from the pore scale to the Darcy scale, including state variables, constitutive relationships, and transport equations, by considering the rugged energy landscape~\cite{bhandarEnergyLandscapesBistability2004,cueto-felguerosoDiscretedomainDescriptionMultiphase2016} of the system.  
        However, this is not tractable because of the complexity and difficulty in dealing with frozen states found in the (for most applications, most relevant) capillary-dominated flow regime. 
        The partition function is very complicated to compute to honor the exact dynamics at the pore scale because the pore space is not homogeneous. However, the level at which it is presented here provides the correct framework, but then requires inputs, such as constitutive relationships. These are provided by the NET and space-time averaging approaches. While the ultimate goal and long-term vision is to provide one consistent framework, the immediate steps are to harmonize the different approaches and set them up to complement each other, i.e. that the NET and space-time averaging provide the required input such as constitutive relationships, i.e. provide the specific physics of the system as input to the framework.

		
		\section*{Acknowledgment}\label{sec:Acknowledgment}
		
		J.E.M. is gratefully acknowledged for his key contributions to the development of the theory behind his work. This work was partly supported by the Research Council of Norway through its Centers of Excellence funding scheme (project number 262644). AH, SK, DB and SB are grateful for this. AH also acknowledges funding from the European Research Council (Grant Agreement 101141323 AGIPORE). R.T.A. acknowledges his Australian Research Council Future Fellowship (FT210100165).


		
						

		
		
		
		\newpage
		\cleardoublepage
		\bibliographystyle{elsarticle-num} 
        \bibliography{references.bib}

\begin{thebibliography}{100}
\expandafter\ifx\csname url\endcsname\relax
  \def\url#1{\texttt{#1}}\fi
\expandafter\ifx\csname urlprefix\endcsname\relax\def\urlprefix{URL }\fi
\expandafter\ifx\csname href\endcsname\relax
  \def\href#1#2{#2} \def\path#1{#1}\fi

\bibitem{hassanizadehUpscalingMultiphaseFlow2005}
S.~M. Hassanizadeh, Upscaling {{Multiphase Flow}} in {{Porous Media}}: {{From
  Pore}} to {{Core}} and {{Beyond}}, Springer Netherlands, Dordrecht, 2005.

\bibitem{tullerHydraulicConductivityVariably2001}
M.~Tuller, D.~Or, Hydraulic conductivity of variably saturated porous media:
  {{Film}} and corner flow in angular pore space, Water Resources Research
  37~(5) (2001) 1257--1276.
\newblock \href {https://doi.org/10.1029/2000WR900328}
  {\path{doi:10.1029/2000WR900328}}.

\bibitem{avraamFlowRegimesRelative1995}
D.~G. Avraam, A.~C. Payatakes, Flow regimes and relative permeabilities during
  steady-state two-phase flow in porous media, Journal of Fluid Mechanics 293
  (1995) 207--236.
\newblock \href {https://doi.org/10.1017/S0022112095001698}
  {\path{doi:10.1017/S0022112095001698}}.

\bibitem{gaoPoreOccupancyRelative2019}
Y.~Gao, A.~Qaseminejad~Raeini, M.~J. Blunt, B.~Bijeljic, Pore occupancy,
  relative permeability and flow intermittency measurements using {{X-ray}}
  micro-tomography in a complex carbonate, Advances in Water Resources 129
  (2019) 56--69.
\newblock \href {https://doi.org/10.1016/j.advwatres.2019.04.007}
  {\path{doi:10.1016/j.advwatres.2019.04.007}}.

\bibitem{raizadaDynamicModeDecomposition2025}
A.~Raizada, S.~Berg, S.~M. Benson, H.~A. Tchelepi, C.~Spurin, Dynamic {{Mode
  Decomposition}} of {{4D}} imaging data to explore intermittent fluid
  connectivity in subsurface flows, Advances in Water Resources 203 (2025)
  105013.
\newblock \href {https://doi.org/10.1016/j.advwatres.2025.105013}
  {\path{doi:10.1016/j.advwatres.2025.105013}}.

\bibitem{ramstadClusterEvolutionSteadystate2006}
T.~Ramstad, A.~Hansen, Cluster evolution in steady-state two-phase flow in
  porous media, Physical Review E 73~(2) (2006) 026306.
\newblock \href {https://doi.org/10.1103/PhysRevE.73.026306}
  {\path{doi:10.1103/PhysRevE.73.026306}}.

\bibitem{Armstrong2016}
R.~T. Armstrong, J.~E. McClure, M.~A. Berrill, M.~R{\"u}cker, S.~Schl{\"u}ter,
  S.~Berg, Beyond {{Darcy}}'s law: {{The}} role of phase topology and ganglion
  dynamics for two-fluid flow, Physical Review E 94~(4) (2016) 043113.
\newblock \href {https://doi.org/10.1103/PhysRevE.94.043113}
  {\path{doi:10.1103/PhysRevE.94.043113}}.

\bibitem{battiatoTheoryApplicationsMacroscale2019}
I.~Battiato, P.~T. Ferrero~V, D.~O'~Malley, C.~T. Miller, P.~S. Takhar, F.~J.
  {Vald{\'e}s-Parada}, B.~D. Wood, Theory and {{Applications}} of {{Macroscale
  Models}} in {{Porous Media}}, Transport in Porous Media 130~(1) (2019) 5--76.
\newblock \href {https://doi.org/10.1007/s11242-019-01282-2}
  {\path{doi:10.1007/s11242-019-01282-2}}.

\bibitem{Haines1930}
W.~Haines, Studies in the physical properties of soils. v. the hysteresis
  effect in capillary properties, and the modes of water distribution
  associated therewith, J Agric Sci 20~(1) (1930) 97--116.

\bibitem{Armstrong2019}
R.~T. Armstrong, J.~E. McClure, V.~Robins, Z.~Liu, C.~H. Arns, S.~Schl{\"u}ter,
  S.~Berg, Porous {{Media Characterization Using Minkowski Functionals}}:
  {{Theories}}, {{Applications}} and {{Future Directions}}, Transport in Porous
  Media 130~(1) (2019) 305--335.
\newblock \href {https://doi.org/10.1007/s11242-018-1201-4}
  {\path{doi:10.1007/s11242-018-1201-4}}.

\bibitem{Bedeaux2024_nano}
D.~Bedeaux, S.~Kjelstrup, S.~S. Schnell, Nanothermodynamics. Theory and
  applications, World Scientific, Singapore, 2024.

\bibitem{Bedeaux2021}
D.~Bedeaux, S.~Kjelstrup, Fluctuation-{{Dissipation Theorems}} for {{Multiphase
  Flow}} in {{Porous Media}}, Entropy 24~(1) (2021) 46.
\newblock \href {https://doi.org/10.3390/e24010046}
  {\path{doi:10.3390/e24010046}}.

\bibitem{Bedeaux2022}
D.~Bedeaux, S.~Kjelstrup, Fluctuation-mdissipation theorems of flow in porous
  media, Entropy 24 (2022) 46.
\newblock \href {https://doi.org/10.3390/e24010046}
  {\path{doi:10.3390/e24010046}}.

\bibitem{bearDynamicsofFluidsinPorousMedia1972}
J.~Bear (Ed.), Dynamics of Fluids in Porous Media, Elsevier, New York, 1982.

\bibitem{wangPredictionsEffectivePhysical2008}
M.~Wang, N.~Pan, Predictions of effective physical properties of complex
  multiphase materials, Materials Science and Engineering: R: Reports 63~(1)
  (2008) 1--30.
\newblock \href {https://doi.org/10.1016/j.mser.2008.07.001}
  {\path{doi:10.1016/j.mser.2008.07.001}}.

\bibitem{sahimiFlowTransportPorous2011}
M.~Sahimi, Flow and Transport in Porous Media and Fractured Rock: From
  Classical Methods to Modern Approaches, John Wiley \& Sons, 2011.

\bibitem{feder2022physics}
J.~Feder, E.~G. Flekk{\o}y, A.~Hansen, Physics of Flow in Porous Media,
  Cambridge University Press, 2022.

\bibitem{katsanouSurfaceWaterGroundwater2017}
K.~Katsanou, H.~K. Karapanagioti, Surface {{Water}} and {{Groundwater Sources}}
  for {{Drinking Water}}, in: A.~Gil, L.~A. Galeano, M.~{\'A}. Vicente (Eds.),
  Applications of {{Advanced Oxidation Processes}} ({{AOPs}}) in {{Drinking
  Water Treatment}}, Vol.~67, Springer International Publishing, Cham, 2017,
  pp. 1--19.
\newblock \href {https://doi.org/10.1007/698.2017.140}
  {\path{doi:10.1007/698.2017.140}}.

\bibitem{helmigMultiphaseFlowTransport1997}
R.~Helmig, Multiphase {{Flow}} and {{Transport Processes}} in the
  {{Subsurface}}, 1st Edition, Environmental {{Science}} and {{Engineering}},
  Springer, Berlin, Heidelberg, 1997.

\bibitem{schijvenRemovalVirusesSoil2000}
J.~F. Schijven, S.~M. Hassanizadeh, Removal of {{Viruses}} by {{Soil Passage}}:
  {{Overview}} of {{Modeling}}, {{Processes}}, and {{Parameters}}, Critical
  Reviews in Environmental Science and Technology 30~(1) (2000) 49--127.
\newblock \href {https://doi.org/10.1080/10643380091184174}
  {\path{doi:10.1080/10643380091184174}}.

\bibitem{torkzabanVirusTransportSaturated2006}
S.~Torkzaban, S.~M. Hassanizadeh, J.~F. Schijven, H.~A.~M. De~Bruin, A.~M.
  De~Roda~Husman, Virus {{Transport}} in {{Saturated}} and {{Unsaturated Sand
  Columns}}, Vadose Zone Journal 5~(3) (2006) 877--885.
\newblock \href {https://doi.org/10.2136/vzj2005.0086}
  {\path{doi:10.2136/vzj2005.0086}}.

\bibitem{kuhlmannInfluenceSoilStructure2012}
A.~Kuhlmann, I.~Neuweiler, S.~E. A. T.~M. Van Der~Zee, R.~Helmig, Influence of
  soil structure and root water uptake strategy on unsaturated flow in
  heterogeneous media, Water Resources Research 48~(2) (2012) 2011WR010651.
\newblock \href {https://doi.org/10.1029/2011WR010651}
  {\path{doi:10.1029/2011WR010651}}.

\bibitem{luGeneralizedSoilWater2016}
N.~Lu, Generalized {{Soil Water Retention Equation}} for {{Adsorption}} and
  {{Capillarity}}, Journal of Geotechnical and Geoenvironmental Engineering
  142~(10) (2016) 04016051.
\newblock \href {https://doi.org/10.1061/(ASCE)GT.1943-5606.0001524}
  {\path{doi:10.1061/(ASCE)GT.1943-5606.0001524}}.

\bibitem{vereeckenModelingSoilProcesses2016}
H.~Vereecken, A.~Schnepf, J.~Hopmans, M.~Javaux, D.~Or, T.~Roose,
  J.~Vanderborght, M.~Young, W.~Amelung, M.~Aitkenhead, S.~Allison,
  S.~Assouline, P.~Baveye, M.~Berli, N.~Br{\"u}ggemann, P.~Finke, M.~Flury,
  T.~Gaiser, G.~Govers, T.~Ghezzehei, P.~Hallett, H.~Hendricks~Franssen,
  J.~Heppell, R.~Horn, J.~Huisman, D.~Jacques, F.~Jonard, S.~Kollet,
  F.~Lafolie, K.~Lamorski, D.~Leitner, A.~McBratney, B.~Minasny, C.~Montzka,
  W.~Nowak, Y.~Pachepsky, J.~Padarian, N.~Romano, K.~Roth, Y.~Rothfuss,
  E.~Rowe, A.~Schwen, J.~{\v S}im{\r u}nek, A.~Tiktak, J.~Van~Dam, S.~Van
  Der~Zee, H.~Vogel, J.~Vrugt, T.~W{\"o}hling, I.~Young, Modeling {{Soil
  Processes}}: {{Review}}, {{Key Challenges}}, and {{New Perspectives}}, Vadose
  Zone Journal 15~(5) (2016) 1--57.
\newblock \href {https://doi.org/10.2136/vzj2015.09.0131}
  {\path{doi:10.2136/vzj2015.09.0131}}.

\bibitem{colbeckPhysicalAspectsWater1978}
S.~C. Colbeck, The {{Physical Aspects}} of {{Water Flow Through Snow}}, in:
  V.~T. Chow (Ed.), Advances in {{Hydroscience}}, Vol.~11, Elsevier, 1978, pp.
  165--206.
\newblock \href {https://doi.org/10.1016/B978-0-12-021811-0.50008-5}
  {\path{doi:10.1016/B978-0-12-021811-0.50008-5}}.

\bibitem{illangasekareModelingMeltwaterInfiltration1990}
T.~H. Illangasekare, R.~J. Walter~Jr., M.~F. Meier, W.~T. Pfeffer, Modeling of
  meltwater infiltration in subfreezing snow, Water Resources Research 26~(5)
  (1990) 1001--1012.
\newblock \href {https://doi.org/10.1029/WR026i005p01001}
  {\path{doi:10.1029/WR026i005p01001}}.

\bibitem{waldnerEffectSnowStructure2004}
P.~A. Waldner, M.~Schneebeli, U.~Schultze-Zimmermann, H.~Fl{\"u}hler, Effect of
  snow structure on water flow and solute transport, Hydrological Processes
  18~(7) (2004) 1271--1290.
\newblock \href {https://doi.org/10.1002/hyp.1401}
  {\path{doi:10.1002/hyp.1401}}.

\bibitem{krolRapidMRIProfiling2025}
Q.~Krol, M.~Skuntz, S.~L. Codd, J.~D. Seymour, Rapid {{MRI Profiling}} of
  {{Two-Phase Flow}} in {{Porous Media}} (Jul. 2025).
\newblock \href {http://arxiv.org/abs/5335548} {\path{arXiv:5335548}}, \href
  {https://doi.org/10.2139/ssrn.5335548} {\path{doi:10.2139/ssrn.5335548}}.

\bibitem{Goldsztein2005}
G.~H. Goldsztein, Transport of nutrients in bones, SIAM Journal on Applied
  Mathematics 65~(6) (2005) 2128--2140.
\newblock \href {https://doi.org/10.1137/040616632}
  {\path{doi:10.1137/040616632}}.

\bibitem{vaughanPoroelasticModelDescribing2013}
B.~L. Vaughan, P.~A. Galie, J.~P. Stegemann, J.~B. Grotberg, A {{Poroelastic
  Model Describing Nutrient Transport}} and {{Cell Stresses Within}} a
  {{Cyclically Strained Collagen Hydrogel}}, Biophysical Journal 105~(9) (2013)
  2188--2198.
\newblock \href {https://doi.org/10.1016/j.bpj.2013.08.048}
  {\path{doi:10.1016/j.bpj.2013.08.048}}.

\bibitem{classMultiphaseProcessesPorous2006}
H.~Class, R.~Helmig, J.~Niessner, U.~{\"O}lmann, Multiphase {{Processes}} in
  {{Porous Media}}, in: R.~Helmig, A.~Mielke, B.~I. Wohlmuth (Eds.), Multifield
  {{Problems}} in {{Solid}} and {{Fluid Mechanics}}, Vol.~28, Springer Berlin
  Heidelberg, 2006, pp. 45--82.
\newblock \href {https://doi.org/10.1007/978-3-540-34961-7.2}
  {\path{doi:10.1007/978-3-540-34961-7.2}}.

\bibitem{geeRecentStudiesFlow1991}
G.~W. Gee, C.~T. Kincaid, R.~J. Lenhard, C.~S. Simmons, Recent {{Studies}} of
  {{Flow}} and {{Transport}} in the {{Vadose Zone}}, Reviews of Geophysics
  29~(S1) (1991) 227--239.
\newblock \href {https://doi.org/10.1002/rog.1991.29.s1.227}
  {\path{doi:10.1002/rog.1991.29.s1.227}}.

\bibitem{rybakMultirateTimeIntegration2015}
I.~Rybak, J.~Magiera, R.~Helmig, C.~Rohde, Multirate time integration for
  coupled saturated/unsaturated porous medium and free flow systems,
  Computational Geosciences 19~(2) (2015) 299--309.
\newblock \href {https://doi.org/10.1007/s10596-015-9469-8}
  {\path{doi:10.1007/s10596-015-9469-8}}.

\bibitem{wankmullerGlobalInfluenceSoil2024}
F.~J.~P. Wankm{\"u}ller, L.~Delval, P.~Lehmann, M.~J. Baur, A.~Cecere, S.~Wolf,
  D.~Or, M.~Javaux, A.~Carminati, Global influence of soil texture on ecosystem
  water limitation, Nature 635~(8039) (2024) 631--638.
\newblock \href {https://doi.org/10.1038/s41586-024-08089-2}
  {\path{doi:10.1038/s41586-024-08089-2}}.

\bibitem{dakeFundamentalsReservoirEngineering2010}
L.~P. Dake, Fundamentals of Reservoir Engineering, no.~8 in Developments in
  Petroleum Science, Elsevier, Amsterdam Boston, 2010.

\bibitem{lakeFundamentalsEnhancedOil2014}
L.~Lake, R.~T. Johns, W.~R. Rossen, G.~A. Pope, Fundamentals of {{Enhanced Oil
  Recovery}}, Society of Petroleum Engineers, 2014.
\newblock \href {https://doi.org/10.2118/9781613993286}
  {\path{doi:10.2118/9781613993286}}.

\bibitem{mohantyPhysicsOilEntrapment1987}
K.~K. Mohanty, H.~T. Davis, L.~E. Scriven, Physics of {{Oil Entrapment}} in
  {{Water-Wet Rock}}, SPE Reservoir Engineering 2~(01) (1987) 113--128.
\newblock \href {https://doi.org/10.2118/9406-PA} {\path{doi:10.2118/9406-PA}}.

\bibitem{stegemeierRelationshipTrappedOil1974}
G.~L. Stegemeier, Relationship of {{Trapped Oil Saturation}} to {{Petrophysical
  Properties}} of {{Porous Media}}, in: {{SPE Improved Oil Recovery
  Symposium}}, SPE, Tulsa, Oklahoma, 1974, pp. SPE--4754--MS.
\newblock \href {https://doi.org/10.2118/4754-MS} {\path{doi:10.2118/4754-MS}}.

\bibitem{morrowWettabilityItsEffect1990}
N.~R. Morrow, Wettability and {{Its Effect}} on {{Oil Recovery}}, Journal of
  Petroleum Technology 42~(12) (1990) 1476--1484.
\newblock \href {https://doi.org/10.2118/21621-PA}
  {\path{doi:10.2118/21621-PA}}.

\bibitem{tangInfluenceBrineComposition1999}
G.-Q. Tang, N.~R. Morrow, Influence of brine composition and fines migration on
  crude oil/brine/rock interactions and oil recovery, Journal of Petroleum
  Science and Engineering 24~(2-4) (1999) 99--111.
\newblock \href {https://doi.org/10.1016/S0920-4105(99)00034-0}
  {\path{doi:10.1016/S0920-4105(99)00034-0}}.

\bibitem{sorbiePolymerImprovedOilRecovery1991}
K.~S. Sorbie, Polymer-{{Improved Oil Recovery}}, 1st Edition, {Blackie and Son
  Ltd.}, 1991.

\bibitem{hirasakiRecentAdvancesSurfactant2011}
G.~J. Hirasaki, C.~A. Miller, M.~Puerto, Recent {{Advances}} in {{Surfactant
  EOR}}, SPE Journal 16~(04) (2011) 889--907.
\newblock \href {https://doi.org/10.2118/115386-PA}
  {\path{doi:10.2118/115386-PA}}.

\bibitem{morrowRecoveryOilSpontaneous2001}
N.~R. Morrow, G.~Mason, Recovery of oil by spontaneous imbibition, Current
  Opinion in Colloid \& Interface Science 6~(4) (2001) 321--337.
\newblock \href {https://doi.org/10.1016/S1359-0294(01)00100-5}
  {\path{doi:10.1016/S1359-0294(01)00100-5}}.

\bibitem{kumarReservoirSimulationCO22005}
A.~Kumar, R.~Ozah, M.~Noh, G.~A. Pope, S.~Bryant, K.~Sepehrnoori, L.~W. Lake,
  Reservoir {{Simulation}} of {{CO2 Storage}} in {{Deep Saline Aquifers}}, SPE
  Journal 10~(03) (2005) 336--348.
\newblock \href {https://doi.org/10.2118/89343-PA}
  {\path{doi:10.2118/89343-PA}}.

\bibitem{nordbottenInjectionStorageCO22005}
J.~M. Nordbotten, M.~A. Celia, S.~Bachu, Injection and {{Storage}} of {{CO2}}
  in {{Deep Saline Aquifers}}: {{Analytical Solution}} for {{CO2 Plume
  Evolution During Injection}}, Transport in Porous Media 58~(3) (2005)
  339--360.
\newblock \href {https://doi.org/10.1007/s11242-004-0670-9}
  {\path{doi:10.1007/s11242-004-0670-9}}.

\bibitem{juanesImpactRelativePermeability2006}
R.~Juanes, E.~J. Spiteri, F.~M. Orr, M.~J. Blunt, Impact of relative
  permeability hysteresis on geological {{CO}}{\textsubscript{2}} storage,
  Water Resources Research 42~(12) (2006) 2005WR004806.
\newblock \href {https://doi.org/10.1029/2005WR004806}
  {\path{doi:10.1029/2005WR004806}}.

\bibitem{koppInvestigationsCO2Storage2009}
A.~Kopp, H.~Class, R.~Helmig, Investigations on {{CO2}} storage capacity in
  saline aquifers, International Journal of Greenhouse Gas Control 3~(3) (2009)
  263--276.
\newblock \href {https://doi.org/10.1016/j.ijggc.2008.10.002}
  {\path{doi:10.1016/j.ijggc.2008.10.002}}.

\bibitem{classBenchmarkStudyProblems2009}
H.~Class, A.~Ebigbo, R.~Helmig, H.~K. Dahle, J.~M. Nordbotten, M.~A. Celia,
  P.~Audigane, M.~Darcis, J.~{Ennis-King}, Y.~Fan, B.~Flemisch, S.~E. Gasda,
  M.~Jin, S.~Krug, D.~Labregere, A.~Naderi~Beni, R.~J. Pawar, A.~Sbai, S.~G.
  Thomas, L.~Trenty, L.~Wei, A benchmark study on problems related to {{CO2}}
  storage in geologic formations: {{Summary}} and discussion of the results,
  Computational Geosciences 13~(4) (2009) 409--434.
\newblock \href {https://doi.org/10.1007/s10596-009-9146-x}
  {\path{doi:10.1007/s10596-009-9146-x}}.

\bibitem{Perrin2010}
J.-C. Perrin, S.~Benson, An {{Experimental Study}} on the {{Influence}} of
  {{Sub-Core Scale Heterogeneities}} on {{CO2 Distribution}} in {{Reservoir
  Rocks}}, Transport in Porous Media 82~(1) (2010) 93--109.
\newblock \href {https://doi.org/10.1007/s11242-009-9426-x}
  {\path{doi:10.1007/s11242-009-9426-x}}.

\bibitem{szulczewskiLifetimeCarbonCapture2012}
M.~L. Szulczewski, C.~W. MacMinn, H.~J. Herzog, R.~Juanes, Lifetime of carbon
  capture and storage as a climate-change mitigation technology, Proceedings of
  the National Academy of Sciences 109~(14) (2012) 5185--5189.
\newblock \href {https://doi.org/10.1073/pnas.1115347109}
  {\path{doi:10.1073/pnas.1115347109}}.

\bibitem{tuckerCarbonCaptureStorage2018}
O.~Tucker, Carbon {{Capture}} and {{Storage}}, IOP Publishing, 2018.
\newblock \href {https://doi.org/10.1088/978-0-7503-1581-4}
  {\path{doi:10.1088/978-0-7503-1581-4}}.

\bibitem{buiCarbonCaptureStorage2018}
M.~Bui, C.~S. Adjiman, A.~Bardow, E.~J. Anthony, A.~Boston, S.~Brown, P.~S.
  Fennell, S.~Fuss, A.~Galindo, L.~A. Hackett, J.~P. Hallett, H.~J. Herzog,
  G.~Jackson, J.~Kemper, S.~Krevor, G.~C. Maitland, M.~Matuszewski, I.~S.
  Metcalfe, C.~Petit, G.~Puxty, J.~Reimer, D.~M. Reiner, E.~S. Rubin, S.~A.
  Scott, N.~Shah, B.~Smit, J.~P.~M. Trusler, P.~Webley, J.~Wilcox,
  N.~Mac~Dowell, Carbon capture and storage ({{CCS}}): The way forward, Energy
  \& Environmental Science 11~(5) (2018) 1062--1176.
\newblock \href {https://doi.org/10.1039/C7EE02342A}
  {\path{doi:10.1039/C7EE02342A}}.

\bibitem{wangPoreScaleModelingMultiphase2025}
J.~Wang, Y.~Yang, G.~Imani, J.~Liu, H.~Song, H.~Sun, L.~Zhang, J.~Zhong,
  K.~Zhang, J.~Yao, Pore-{{Scale Modeling}} of {{Multiphase Reactive
  Transport}} in {{Porous Media}} during {{Geological Carbon Storage}} in
  {{Saline Aquifers}}: {{Mechanisms}}, {{Progress}}, and {{Challenges}}, Gas
  Science and Engineering (2025) 205784\href
  {https://doi.org/10.1016/j.jgsce.2025.205784}
  {\path{doi:10.1016/j.jgsce.2025.205784}}.

\bibitem{heinemannEnablingLargescaleHydrogen2021}
N.~Heinemann, J.~Alcalde, J.~M. Miocic, S.~J.~T. Hangx, J.~Kallmeyer,
  C.~{Ostertag-Henning}, A.~Hassanpouryouzband, E.~M. Thaysen, G.~J. Strobel,
  C.~{Schmidt-Hattenberger}, K.~Edlmann, M.~Wilkinson, M.~Bentham,
  R.~Stuart~Haszeldine, R.~Carbonell, A.~Rudloff, Enabling large-scale hydrogen
  storage in porous media -- the scientific challenges, Energy \& Environmental
  Science 14~(2) (2021) 853--864.
\newblock \href {https://doi.org/10.1039/D0EE03536J}
  {\path{doi:10.1039/D0EE03536J}}.

\bibitem{Higgs2022}
S.~Higgs, Y.~Da~Wang, C.~Sun, J.~{Ennis-King}, S.~J. Jackson, R.~T. Armstrong,
  P.~Mostaghimi, In-situ hydrogen wettability characterisation for underground
  hydrogen storage, International Journal of Hydrogen Energy 47~(26) (2022)
  13062--13075.
\newblock \href {https://doi.org/10.1016/j.ijhydene.2022.02.022}
  {\path{doi:10.1016/j.ijhydene.2022.02.022}}.

\bibitem{Jha2021}
N.~K. Jha, A.~{Al-Yaseri}, M.~Ghasemi, D.~{Al-Bayati}, M.~Lebedev,
  M.~Sarmadivaleh, Pore scale investigation of hydrogen injection in sandstone
  via {{X-ray}} micro-tomography, International Journal of Hydrogen Energy
  46~(70) (2021) 34822--34829.
\newblock \href {https://doi.org/10.1016/j.ijhydene.2021.08.042}
  {\path{doi:10.1016/j.ijhydene.2021.08.042}}.

\bibitem{lysyyHydrogenRelativePermeability2022}
M.~Lysyy, T.~F{\o}yen, E.~B. Johannesen, M.~Fern{\o}, G.~Ersland, Hydrogen
  {{Relative Permeability Hysteresis}} in {{Underground Storage}}, Geophysical
  Research Letters 49~(17) (2022) e2022GL100364.
\newblock \href {https://doi.org/10.1029/2022GL100364}
  {\path{doi:10.1029/2022GL100364}}.

\bibitem{Gao2023}
Y.~Gao, T.~Sorop, N.~Brussee, H.~{Van Der Linde}, A.~Coorn, M.~Appel, S.~Berg,
  Advanced {{Digital-SCAL Measurements}} of {{Gas Trapped}} in {{Sandstone}},
  Petrophysics -- The SPWLA Journal of Formation Evaluation and Reservoir
  Description 64~(3) (2023) 368--383.
\newblock \href {https://doi.org/10.30632/PJV64N3-2023a4}
  {\path{doi:10.30632/PJV64N3-2023a4}}.

\bibitem{alzaabiWettabilityPoreOccupancy2025}
A.~AlZaabi, H.~M. Alzahrani, A.~Alhosani, B.~Bijeljic, M.~J. Blunt,
  Wettability, pore occupancy, connectivity and {{Ostwald}} ripening of
  nitrogen, carbon dioxide, and hydrogen in carbonate rocks: {{A}} comparative
  study, International Journal of Hydrogen Energy 135 (2025) 596--608.
\newblock \href {https://doi.org/10.1016/j.ijhydene.2025.04.399}
  {\path{doi:10.1016/j.ijhydene.2025.04.399}}.

\bibitem{gomezmendezInsightsUndergroundHydrogen2024}
I.~Gomez~Mendez, W.~M.~M. {El-Sayed}, A.~H. Menefee, Z.~T. Karpyn, Insights
  into {{Underground Hydrogen Storage Challenges}}: {{A Review}} on
  {{Hydrodynamic}} and {{Biogeochemical Experiments}} in {{Porous Media}},
  Energy \& Fuels 38~(21) (2024) 20015--20032.
\newblock \href {https://doi.org/10.1021/acs.energyfuels.4c03142}
  {\path{doi:10.1021/acs.energyfuels.4c03142}}.

\bibitem{dokhonPressureDeclineGas2024}
W.~Dokhon, S.~Goodarzi, H.~M. Alzahrani, M.~J. Blunt, B.~Bijeljic, Pressure
  decline and gas expansion in underground hydrogen storage: {{A}} pore-scale
  percolation study, International Journal of Hydrogen Energy 86 (2024)
  261--274.
\newblock \href {https://doi.org/10.1016/j.ijhydene.2024.08.139}
  {\path{doi:10.1016/j.ijhydene.2024.08.139}}.

\bibitem{rooseMathematicalModelPlant2001}
T.~Roose, A.~Fowler, P.~Darrah, A mathematical model of plant nutrient uptake,
  Journal of Mathematical Biology 42~(4) (2001) 347--360.
\newblock \href {https://doi.org/10.1007/s002850000075}
  {\path{doi:10.1007/s002850000075}}.

\bibitem{Hochberg2019}
U.~Hochberg, A.~Ponomarenko, Y.-J. Zhang, F.~E. Rockwell, N.~M. Holbrook,
  Visualizing {{Embolism Propagation}} in {{Gas-Injected Leaves}}, Plant
  Physiology 180~(2) (2019) 874--881.
\newblock \href {https://doi.org/10.1104/pp.18.01284}
  {\path{doi:10.1104/pp.18.01284}}.

\bibitem{brodersenDynamicsEmbolismRepair2010}
C.~R. Brodersen, A.~J. McElrone, B.~Choat, M.~A. Matthews, K.~A. Shackel, The
  {{Dynamics}} of {{Embolism Repair}} in {{Xylem}}: {{In Vivo Visualizations
  Using High-Resolution Computed Tomography}}, Plant Physiology 154~(3) (2010)
  1088--1095.
\newblock \href {https://doi.org/10.1104/pp.110.162396}
  {\path{doi:10.1104/pp.110.162396}}.

\bibitem{ottMicroscaleSoluteTransport2014}
H.~Ott, M.~Andrew, J.~Snippe, M.~J. Blunt, Microscale solute transport and
  precipitation in complex rock during drying, Geophysical Research Letters
  41~(23) (2014) 8369--8376.
\newblock \href {https://doi.org/10.1002/2014GL062266}
  {\path{doi:10.1002/2014GL062266}}.

\bibitem{ottSaltPrecipitationDue2015}
H.~Ott, S.~Roels, K.~De~Kloe, Salt precipitation due to supercritical gas
  injection: {{I}}. {{Capillary-driven}} flow in unimodal sandstone,
  International Journal of Greenhouse Gas Control 43 (2015) 247--255.
\newblock \href {https://doi.org/10.1016/j.ijggc.2015.01.005}
  {\path{doi:10.1016/j.ijggc.2015.01.005}}.

\bibitem{ottSaltPrecipitationDue2021}
H.~Ott, J.~Snippe, K.~De~Kloe, Salt precipitation due to supercritical gas
  injection: {{II}}. {{Capillary}} transport in multi porosity rocks,
  International Journal of Greenhouse Gas Control 105 (2021) 103233.
\newblock \href {https://doi.org/10.1016/j.ijggc.2020.103233}
  {\path{doi:10.1016/j.ijggc.2020.103233}}.

\bibitem{pratRecentAdvancesPorescale2002}
M.~Prat, Recent advances in pore-scale models for drying of porous media,
  Chemical Engineering Journal 86~(1) (2002) 153--164.
\newblock \href {https://doi.org/10.1016/S1385-8947(01)00283-2}
  {\path{doi:10.1016/S1385-8947(01)00283-2}}.

\bibitem{orAdvancesSoilEvaporation2013}
D.~Or, P.~Lehmann, E.~Shahraeeni, N.~Shokri, Advances in {{Soil Evaporation
  Physics}}---{{A Review}}, Vadose Zone Journal 12~(4) (Nov. 2013).
\newblock \href {https://doi.org/10.2136/vzj2012.0163}
  {\path{doi:10.2136/vzj2012.0163}}.

\bibitem{lehmannCharacteristicLengthsAffecting2008}
P.~Lehmann, S.~Assouline, D.~Or, Characteristic lengths affecting evaporative
  drying of porous media, Physical Review E 77~(5) (2008) 056309.
\newblock \href {https://doi.org/10.1103/PhysRevE.77.056309}
  {\path{doi:10.1103/PhysRevE.77.056309}}.

\bibitem{zhangNumericalStudyEvaporationinduced2014}
C.~Zhang, L.~Li, D.~Lockington, Numerical study of evaporation-induced salt
  accumulation and precipitation in bare saline soils: {{Mechanism}} and
  feedback, Water Resources Research 50~(10) (2014) 8084--8106.
\newblock \href {https://doi.org/10.1002/2013WR015127}
  {\path{doi:10.1002/2013WR015127}}.

\bibitem{hassaniPredictingLongtermDynamics2020}
A.~Hassani, A.~Azapagic, N.~Shokri, Predicting long-term dynamics of soil
  salinity and sodicity on a global scale, Proceedings of the National Academy
  of Sciences 117~(52) (2020) 33017--33027.
\newblock \href {https://doi.org/10.1073/pnas.2013771117}
  {\path{doi:10.1073/pnas.2013771117}}.

\bibitem{hassaniGlobalPredictionsPrimary2021}
A.~Hassani, A.~Azapagic, N.~Shokri, Global predictions of primary soil
  salinization under changing climate in the 21st century, Nature
  Communications 12~(1) (2021) 1--17.
\newblock \href {https://doi.org/10.1038/s41467-021-26907-3}
  {\path{doi:10.1038/s41467-021-26907-3}}.

\bibitem{Desarnaud2015}
J.~Desarnaud, H.~Derluyn, L.~Molari, S.~{de Miranda}, V.~Cnudde,
  N.~Shahidzadeh, Drying of salt contaminated porous media: {{Effect}} of
  primary and secondary nucleation, Journal of Applied Physics 118~(11) (Sep.
  2015).
\newblock \href {https://doi.org/10.1063/1.4930292}
  {\path{doi:10.1063/1.4930292}}.

\bibitem{lavegliaCircularDesignMaterial2024}
A.~Laveglia, D.~V. Madrid, N.~Ukrainczyk, V.~Cnudde, N.~De~Belie, E.~Koenders,
  Circular design, material properties, service life and cradle-to-cradle
  carbon footprint of lime-based building materials, Science of The Total
  Environment 948 (2024) 174875.
\newblock \href {https://doi.org/10.1016/j.scitotenv.2024.174875}
  {\path{doi:10.1016/j.scitotenv.2024.174875}}.

\bibitem{grossogiordanoSurfaceCouplingWater2024}
F.~Grosso~Giordano, D.~Valdez~Madrid, L.~Schr{\"o}er, N.~Boon, V.~Cnudde,
  N.~De~Belie, Beyond the surface: {{Coupling}} water permeability assessments
  to {{X-ray}} micro-computed tomography for evaluation of self-healing on
  lime-based mortars, Construction and Building Materials 431 (2024) 136603.
\newblock \href {https://doi.org/10.1016/j.conbuildmat.2024.136603}
  {\path{doi:10.1016/j.conbuildmat.2024.136603}}.

\bibitem{madridControllingSaltWeathering2025}
D.~E.~V. Madrid, C.~{Rodriguez-Navarro}, C.~W. Winardhi, N.~De~Belie,
  V.~Cnudde, Controlling salt weathering of lime mortars by mixed-in additives:
  {{A}} multi-scale analysis, Construction and Building Materials 490 (2025)
  142603.
\newblock \href {https://doi.org/10.1016/j.conbuildmat.2025.142603}
  {\path{doi:10.1016/j.conbuildmat.2025.142603}}.

\bibitem{madridEffectCementContent2025}
D.~E. Madrid, N.~Alderete, V.~Boterberg, N.~De~Belie, V.~Cnudde, Effect of
  cement content on the pore structure and imbibition rate of lime-cement
  mortars, Cement and Concrete Composites (2025) 106352\href
  {https://doi.org/10.1016/j.cemconcomp.2025.106352}
  {\path{doi:10.1016/j.cemconcomp.2025.106352}}.

\bibitem{abriolaMultiphaseApproachModeling1985}
L.~M. Abriola, G.~F. Pinder, A {{Multiphase Approach}} to the {{Modeling}} of
  {{Porous Media Contamination}} by {{Organic Compounds}}: 1. {{Equation
  Development}}, Water Resources Research 21~(1) (1985) 11--18.
\newblock \href {https://doi.org/10.1029/WR021i001p00011}
  {\path{doi:10.1029/WR021i001p00011}}.

\bibitem{pennellInfluenceViscousBuoyancy1996}
K.~D. Pennell, G.~A. Pope, L.~M. Abriola, Influence of {{Viscous}} and
  {{Buoyancy Forces}} on the {{Mobilization}} of {{Residual
  Tetrachloroethylene}} during {{Surfactant Flushing}}, Environmental Science
  \& Technology 30~(4) (1996) 1328--1335.
\newblock \href {https://doi.org/10.1021/es9505311}
  {\path{doi:10.1021/es9505311}}.

\bibitem{ahmadiLargescalePropertiesTwophase1996}
A.~Ahmadi, M.~Quintard, Large-scale properties for two-phase flow in random
  porous media, Journal of Hydrology 183~(1-2) (1996) 69--99.
\newblock \href {https://doi.org/10.1016/S0022-1694(96)80035-7}
  {\path{doi:10.1016/S0022-1694(96)80035-7}}.

\bibitem{sogaReviewNAPLSource2004}
K.~Soga, A review of {{NAPL}} source zone remediation efficiency and the mass
  flux approach, Journal of Hazardous Materials 110~(1-3) (2004) 13--27.
\newblock \href {https://doi.org/10.1016/j.jhazmat.2004.02.034}
  {\path{doi:10.1016/j.jhazmat.2004.02.034}}.

\bibitem{palmerPrinciplesContaminantHydrogeology2019}
C.~M. Palmer, Principles of {{Contaminant Hydrogeology}}, 2nd Edition, CRC
  Press, 2019.
\newblock \href {https://doi.org/10.1201/9780138742150}
  {\path{doi:10.1201/9780138742150}}.

\bibitem{Pak2020}
T.~Pak, L.~F. d.~L. Luz, T.~Tosco, G.~S.~R. Costa, P.~R.~R. Rosa, N.~L.
  Archilha, Pore-scale investigation of the use of reactive nanoparticles for
  in situ remediation of contaminated groundwater source, Proceedings of the
  National Academy of Sciences 117~(24) (2020) 13366--13373.
\newblock \href {https://doi.org/10.1073/pnas.1918683117}
  {\path{doi:10.1073/pnas.1918683117}}.

\bibitem{dudukovicMultiphaseReactorsRevisited1999}
M.~P. Dudukovic, F.~Larachi, P.~L. Mills, Multiphase reactors -- revisited,
  Chemical Engineering Science 54~(13-14) (1999) 1975--1995.
\newblock \href {https://doi.org/10.1016/S0009-2509(98)00367-4}
  {\path{doi:10.1016/S0009-2509(98)00367-4}}.

\bibitem{gladdenRecentAdvancesMRI2003}
L.~F. Gladden, Recent {{Advances}} in {{MRI Studies}} of {{Chemical Reactors}}:
  {{Ultrafast Imaging}} of {{Multiphase Flows}}, Topics in Catalysis 24~(1-4)
  (2003) 19--28.
\newblock \href {https://doi.org/10.1023/B:TOCA.0000003072.56070.2e}
  {\path{doi:10.1023/B:TOCA.0000003072.56070.2e}}.

\bibitem{sauerschellMethanationPilotPlant2022}
S.~Sauerschell, S.~Bajohr, T.~Kolb, Methanation {{Pilot Plant}} with a {{Slurry
  Bubble Column Reactor}}: {{Setup}} and {{First Experimental Results}}, Energy
  \& Fuels 36~(13) (2022) 7166--7176.
\newblock \href {https://doi.org/10.1021/acs.energyfuels.2c00655}
  {\path{doi:10.1021/acs.energyfuels.2c00655}}.

\bibitem{Owejan2009}
J.~P. Owejan, J.~J. Gagliardo, J.~M. Sergi, S.~G. Kandlikar, T.~A. Trabold,
  Water management studies in {{PEM}} fuel cells, {{Part I}}: {{Fuel}} cell
  design and in situ water distributions, International Journal of Hydrogen
  Energy 34~(8) (2009) 3436--3444.
\newblock \href {https://doi.org/10.1016/j.ijhydene.2008.12.100}
  {\path{doi:10.1016/j.ijhydene.2008.12.100}}.

\bibitem{Alink2011}
R.~Alink, D.~Gerteisen, W.~M{\'e}rida, Investigating the {{Water Transport}} in
  {{Porous Media}} for {{PEMFCs}} by {{Liquid Water Visualization}} in
  {{ESEM}}, Fuel Cells 11~(4) (2011) 481--488.
\newblock \href {https://doi.org/10.1002/fuce.201000110}
  {\path{doi:10.1002/fuce.201000110}}.

\bibitem{Doerenkamp2021}
T.~Doerenkamp, M.~Sabharwal, J.~Eller, Insights into the {{Stability}} and
  {{Formation}} of {{Water Droplets Using Operando X-Ray Tomographic
  Microscopy}}, ECS Meeting Abstracts MA2021-02~(36) (2021) 1026.
\newblock \href {https://doi.org/10.1149/MA2021-02361026mtgabs}
  {\path{doi:10.1149/MA2021-02361026mtgabs}}.

\bibitem{Shrestha2020}
P.~Shrestha, {\relax Ch}.~Lee, K.~F. Fahy, M.~Balakrishnan, N.~Ge, A.~Bazylak,
  Formation of {{Liquid Water Pathways}} in {{PEM Fuel Cells}}: {{A}} 3-{{D
  Pore-Scale Perspective}}, Journal of The Electrochemical Society 167~(5)
  (2020) 054516.
\newblock \href {https://doi.org/10.1149/1945-7111/ab7a0b}
  {\path{doi:10.1149/1945-7111/ab7a0b}}.

\bibitem{Eller2017}
J.~Eller, J.~Roth, F.~Marone, M.~Stampanoni, F.~N. B{\"u}chi, Operando
  {{Properties}} of {{Gas Diffusion Layers}}: {{Saturation}} and {{Liquid
  Permeability}}, Journal of The Electrochemical Society 164~(2) (2017)
  F115--F126.
\newblock \href {https://doi.org/10.1149/2.0881702jes}
  {\path{doi:10.1149/2.0881702jes}}.

\bibitem{Bazylak2009}
A.~Bazylak, Liquid water visualization in {{PEM}} fuel cells: {{A}} review,
  International Journal of Hydrogen Energy 34~(9) (2009) 3845--3857.
\newblock \href {https://doi.org/10.1016/j.ijhydene.2009.02.084}
  {\path{doi:10.1016/j.ijhydene.2009.02.084}}.

\bibitem{Shafaque2020}
H.~W. Shafaque, C.~Lee, K.~F. Fahy, J.~K. Lee, J.~M. LaManna, E.~Baltic, D.~S.
  Hussey, D.~L. Jacobson, A.~Bazylak, Boosting {{Membrane Hydration}} for
  {{High Current Densities}} in {{Membrane Electrode Assembly
  CO}}{\textsubscript{2}} {{Electrolysis}}, ACS Applied Materials \& Interfaces
  12~(49) (2020) 54585--54595.
\newblock \href {https://doi.org/10.1021/acsami.0c14832}
  {\path{doi:10.1021/acsami.0c14832}}.

\bibitem{Lee2020}
C.~Lee, B.~Zhao, J.~K. Lee, K.~F. Fahy, K.~Krause, A.~Bazylak, Bubble
  {{Formation}} in the {{Electrolyte Triggers Voltage Instability}} in {{CO2
  Electrolyzers}}, iScience 23~(5) (2020) 101094.
\newblock \href {https://doi.org/10.1016/j.isci.2020.101094}
  {\path{doi:10.1016/j.isci.2020.101094}}.

\bibitem{maasViscousFingeringCCS2024}
J.~G. Maas, N.~Springer, A.~Hebing, J.~Snippe, S.~Berg, Viscous fingering in
  {{CCS}} - {{A}} general criterion for viscous fingering in porous media,
  International Journal of Greenhouse Gas Control 132 (2024) 104074.
\newblock \href {https://doi.org/10.1016/j.ijggc.2024.104074}
  {\path{doi:10.1016/j.ijggc.2024.104074}}.

\bibitem{simonInfluenceGasDiffusion2017}
C.~Simon, F.~Hasch{\'e}, H.~A. Gasteiger, Influence of the {{Gas Diffusion
  Layer Compression}} on the {{Oxygen Transport}} in {{PEM Fuel Cells}} at
  {{High Water Saturation Levels}}, Journal of The Electrochemical Society
  164~(6) (2017) F591--F599.
\newblock \href {https://doi.org/10.1149/2.0691706jes}
  {\path{doi:10.1149/2.0691706jes}}.

\bibitem{zhangSimilaritiesDifferencesGas2024}
T.~Zhang, L.~Meng, C.~Chen, L.~Du, N.~Wang, L.~Xing, C.~Tang, J.~Hu, S.~Ye,
  Similarities and {{Differences}} between {{Gas Diffusion Layers Used}} in
  {{Proton Exchange Membrane Fuel Cell}} and {{Water Electrolysis}} for
  {{Material}} and {{Mass Transport}}, Advanced Science 11~(32) (2024) 2309440.
\newblock \href {https://doi.org/10.1002/advs.202309440}
  {\path{doi:10.1002/advs.202309440}}.

\bibitem{yueMultiphaseFlowProcessing2018}
J.~Yue, Multiphase flow processing in microreactors combined with heterogeneous
  catalysis for efficient and sustainable chemical synthesis, Catalysis Today
  308 (2018) 3--19.
\newblock \href {https://doi.org/10.1016/j.cattod.2017.09.041}
  {\path{doi:10.1016/j.cattod.2017.09.041}}.

\bibitem{miriNewInsightsPhysics2015}
R.~Miri, R.~Van~Noort, P.~Aagaard, H.~Hellevang, New insights on the physics of
  salt precipitation during injection of {{CO2}} into saline aquifers,
  International Journal of Greenhouse Gas Control 43 (2015) 10--21.
\newblock \href {https://doi.org/10.1016/j.ijggc.2015.10.004}
  {\path{doi:10.1016/j.ijggc.2015.10.004}}.

\bibitem{adlerMultiphaseFlowPorous1988}
P.~M. Adler, H.~Brenner, Multiphase {{Flow}} in {{Porous Media}}, Annual Review
  of Fluid Mechanics 20~(1) (1988) 35--59.
\newblock \href {https://doi.org/10.1146/annurev.fl.20.010188.000343}
  {\path{doi:10.1146/annurev.fl.20.010188.000343}}.

\bibitem{zhaoComprehensiveComparisonPorescale2019}
B.~Zhao, C.~W. MacMinn, B.~K. Primkulov, Y.~Chen, A.~J. Valocchi, J.~Zhao,
  Q.~Kang, K.~Bruning, J.~E. McClure, C.~T. Miller, A.~Fakhari, D.~Bolster,
  T.~Hiller, M.~Brinkmann, L.~{Cueto-Felgueroso}, D.~A. Cogswell, R.~Verma,
  M.~Prodanovi{\'c}, J.~Maes, S.~Geiger, M.~Vassvik, A.~Hansen, E.~Segre,
  R.~Holtzman, Z.~Yang, C.~Yuan, B.~Chareyre, R.~Juanes, Comprehensive
  comparison of pore-scale models for multiphase flow in porous media,
  Proceedings of the National Academy of Sciences 116~(28) (2019) 13799--13806.
\newblock \href {https://doi.org/10.1073/pnas.1901619116}
  {\path{doi:10.1073/pnas.1901619116}}.

\bibitem{ramstadPoreScaleSimulationsSingle2019a}
T.~Ramstad, C.~F. Berg, K.~Thompson, Pore-{{Scale Simulations}} of {{Single-}}
  and {{Two-Phase Flow}} in {{Porous Media}}: {{Approaches}} and
  {{Applications}}, Transport in Porous Media 130~(1) (2019) 77--104.
\newblock \href {https://doi.org/10.1007/s11242-019-01289-9}
  {\path{doi:10.1007/s11242-019-01289-9}}.

\bibitem{Ruecker2021}
M.~R{\"u}cker, A.~Georgiadis, R.~T. Armstrong, H.~Ott, N.~Brussee, H.~Van
  Der~Linde, L.~Simon, F.~Enzmann, M.~Kersten, S.~Berg, The {{Origin}} of
  {{Non-thermal Fluctuations}} in {{Multiphase Flow}} in {{Porous Media}},
  Frontiers in Water 3 (2021) 671399.
\newblock \href {https://doi.org/10.3389/frwa.2021.671399}
  {\path{doi:10.3389/frwa.2021.671399}}.

\bibitem{armstrongSubsecondPorescaleDisplacement2014}
R.~T. Armstrong, H.~Ott, A.~Georgiadis, M.~R{\"u}cker, A.~Schwing, S.~Berg,
  Subsecond pore-scale displacement processes and relaxation dynamics in
  multiphase flow, Water Resources Research 50~(12) (2014) 9162--9176.
\newblock \href {https://doi.org/10.1002/2014WR015858}
  {\path{doi:10.1002/2014WR015858}}.

\bibitem{bluntPorescaleImagingModelling2013}
M.~J. Blunt, B.~Bijeljic, H.~Dong, O.~Gharbi, S.~Iglauer, P.~Mostaghimi,
  A.~Paluszny, C.~Pentland, Pore-scale imaging and modelling, Advances in Water
  Resources 51 (2013) 197--216.
\newblock \href {https://doi.org/10.1016/j.advwatres.2012.03.003}
  {\path{doi:10.1016/j.advwatres.2012.03.003}}.

\bibitem{nordahlIdentifyingRepresentativeElementary2008}
K.~Nordahl, P.~S. Ringrose, Identifying the {{Representative Elementary
  Volume}} for {{Permeability}} in {{Heterolithic Deposits Using Numerical Rock
  Models}}, Mathematical Geosciences 40~(7) (2008) 753--771.
\newblock \href {https://doi.org/10.1007/s11004-008-9182-4}
  {\path{doi:10.1007/s11004-008-9182-4}}.

\bibitem{dengHilbertsSixthProblem2025}
Y.~Deng, Z.~Hani, X.~Ma, Hilbert's sixth problem: Derivation of fluid equations
  via {{Boltzmann}}'s kinetic theory (2025).
\newblock \href {https://doi.org/10.48550/ARXIV.2503.01800}
  {\path{doi:10.48550/ARXIV.2503.01800}}.

\bibitem{neumanTheoreticalDerivationDarcys1977}
S.~P. Neuman, Theoretical derivation of {{Darcy}}'s law, Acta Mechanica
  25~(3-4) (1977) 153--170.
\newblock \href {https://doi.org/10.1007/BF01376989}
  {\path{doi:10.1007/BF01376989}}.

\bibitem{whitakerFlowPorousMedia1986}
S.~Whitaker, Flow in porous media {{I}}: {{A}} theoretical derivation of
  {{Darcy}}'s law, Transport in Porous Media 1~(1) (1986) 3--25.
\newblock \href {https://doi.org/10.1007/BF01036523}
  {\path{doi:10.1007/BF01036523}}.

\bibitem{AdamsonGast1997}
A.~W. Adamson, A.~P. Gast, Physical Chemistry of Surfaces, John Wiley \& Sons,
  1997.

\bibitem{Hassanizadeh2024}
S.~M. Hassanizadeh, The {{Origin}} of {{Surface Tension}}, InterPore Journal
  1~(1) (2024) ipj260424--3.
\newblock \href {https://doi.org/10.69631/ipj.v1i1nr21}
  {\path{doi:10.69631/ipj.v1i1nr21}}.

\bibitem{Gauss1830}
C.~F. Gauss, Principia generalia theoriae figurae fluidorum in statu
  aequilibrii, Commentarii Societatis Regiae Scientiarum Gottingensis. Classes
  Mathematica 7 (1830) 39--88.

\bibitem{Einstein_1901}
A.~Einstein, Folgerungen aus den capillaritätserscheinungen, Annalen der
  Physik 309~(3) (1901) 513–523.

\bibitem{Buckingham1907}
E.~Buckingham, Studies on the movement of soil moisture., Tech. rep., Bull. 38.
  USDA, Bureau of Soils, Washington D.C. (1907).

\bibitem{narasimhanBuckingham1907Appreciation2005}
T.~N. Narasimhan, Buckingham, 1907: {{An Appreciation}}, Vadose Zone Journal
  4~(2) (2005) 434--441.
\newblock \href {https://doi.org/10.2136/vzj2004.0126}
  {\path{doi:10.2136/vzj2004.0126}}.

\bibitem{richardsCAPILLARYCONDUCTIONLIQUIDS1931}
L.~A. Richards, {{CAPILLARY CONDUCTION OF LIQUIDS THROUGH POROUS MEDIUMS}},
  Physics 1~(5) (1931) 318--333.
\newblock \href {https://doi.org/10.1063/1.1745010}
  {\path{doi:10.1063/1.1745010}}.

\bibitem{muskatFlowHeterogeneousFluids1936}
M.~Muskat, M.~W. Meres, The {{Flow}} of {{Heterogeneous Fluids Through Porous
  Media}}, Physics 7~(9) (1936) 346--363.
\newblock \href {https://doi.org/10.1063/1.1745403}
  {\path{doi:10.1063/1.1745403}}.

\bibitem{wyckoffFlowGasLiquidMixtures1936}
R.~D. Wyckoff, H.~G. Botset, The {{Flow}} of {{Gas-Liquid Mixtures Through
  Unconsolidated Sands}}, Physics 7~(9) (1936) 325--345.
\newblock \href {https://doi.org/10.1063/1.1745402}
  {\path{doi:10.1063/1.1745402}}.

\bibitem{leverettCapillaryBehaviorPorous1941}
M.~Leverett, Capillary {{Behavior}} in {{Porous Solids}}, Transactions of the
  AIME 142~(01) (1941) 152--169.
\newblock \href {https://doi.org/10.2118/941152-G}
  {\path{doi:10.2118/941152-G}}.

\bibitem{hammondSpontaneousImbibitionSurfactant2011}
P.~S. Hammond, E.~Unsal, Spontaneous {{Imbibition}} of {{Surfactant Solution}}
  into an {{Oil-Wet Capillary}}: {{Wettability Restoration}} by
  {{Surfactant}}-{{Contaminant Complexation}}, Langmuir 27~(8) (2011)
  4412--4429.
\newblock \href {https://doi.org/10.1021/la1048503}
  {\path{doi:10.1021/la1048503}}.

\bibitem{caiDiscussionEffectTortuosity2011}
J.~Cai, B.~Yu, A {{Discussion}} of the {{Effect}} of {{Tortuosity}} on the
  {{Capillary Imbibition}} in {{Porous Media}}, Transport in Porous Media
  89~(2) (2011) 251--263.
\newblock \href {https://doi.org/10.1007/s11242-011-9767-0}
  {\path{doi:10.1007/s11242-011-9767-0}}.

\bibitem{masonDevelopmentsSpontaneousImbibition2013}
G.~Mason, N.~R. Morrow, Developments in spontaneous imbibition and
  possibilities for future work, Journal of Petroleum Science and Engineering
  110 (2013) 268--293.
\newblock \href {https://doi.org/10.1016/j.petrol.2013.08.018}
  {\path{doi:10.1016/j.petrol.2013.08.018}}.

\bibitem{HassanizadehGray1993}
S.~Majid~Hassanizadeh, W.~G. Gray, Toward an improved description of the
  physics of two-phase flow, Advances in Water Resources 16~(1) (1993) 53--67.
\newblock \href {https://doi.org/10.1016/0309-1708(93)90029-F}
  {\path{doi:10.1016/0309-1708(93)90029-F}}.

\bibitem{armstrongModelingVelocityField2015b}
R.~T. Armstrong, N.~Evseev, D.~Koroteev, S.~Berg, Modeling the velocity field
  during {{Haines}} jumps in porous media, Advances in Water Resources 77
  (2015) 57--68.
\newblock \href {https://doi.org/10.1016/j.advwatres.2015.01.008}
  {\path{doi:10.1016/j.advwatres.2015.01.008}}.

\bibitem{hubbertDarcysLawField1956}
M.~K. Hubbert, Darcy's {{Law}} and the {{Field Equations}} of the {{Flow}} of
  {{Underground Fluids}}, Transactions of the AIME 207~(01) (1956) 222--239.
\newblock \href {https://doi.org/10.2118/749-G} {\path{doi:10.2118/749-G}}.

\bibitem{bearFundamentalsTransportPhenomena1984}
J.~Bear, M.~Y. Corapcioglu (Eds.), Fundamentals of {{Transport Phenomena}} in
  {{Porous Media}}, Springer Netherlands, Dordrecht, 1984.
\newblock \href {https://doi.org/10.1007/978-94-009-6175-3}
  {\path{doi:10.1007/978-94-009-6175-3}}.

\bibitem{kingEffectivePropertiesFlow1995}
M.~J. King, P.~R. King, C.~A. McGill, J.~K. Williams, Effective {{Properties}}
  for {{Flow Calculations}}, in: P.~M. Adler (Ed.), Multiphase {{Flow}} in
  {{Porous Media}}, Springer Netherlands, Dordrecht, 1995, pp. 169--196.
\newblock \href {https://doi.org/10.1007/978-94-017-2372-5-7}
  {\path{doi:10.1007/978-94-017-2372-5-7}}.

\bibitem{killoughReservoirSimulationHistoryDependent1976}
J.~Killough, Reservoir {{Simulation With History-Dependent Saturation
  Functions}}, Society of Petroleum Engineers Journal 16~(01) (1976) 37--48.
\newblock \href {https://doi.org/10.2118/5106-pa} {\path{doi:10.2118/5106-pa}}.

\bibitem{kjosavikRelativePermeabilityCorrelation2002}
A.~Kjosavik, J.~K. Ringen, S.~M. Skjaeveland, Relative {{Permeability
  Correlation}} for {{Mixed-Wet Reservoirs}}, SPE Journal 7~(01) (2002) 49--58.
\newblock \href {https://doi.org/10.2118/77328-PA}
  {\path{doi:10.2118/77328-PA}}.

\bibitem{masalmehImprovedCharacterizationModeling2007}
S.~K. Masalmeh, I.~A. Shiekah, X.~D. Jing, Improved {{Characterization}} and
  {{Modeling}} of {{Capillary Transition Zones}} in {{Carbonate Reservoirs}},
  SPE Reservoir Evaluation \& Engineering 10~(02) (2007) 191--204.
\newblock \href {https://doi.org/10.2118/109094-PA}
  {\path{doi:10.2118/109094-PA}}.

\bibitem{landCalculationImbibitionRelative1968}
C.~S. Land, Calculation of {{Imbibition Relative Permeability}} for {{Two-}}
  and {{Three-Phase Flow From Rock Properties}}, Society of Petroleum Engineers
  Journal 8~(02) (1968) 149--156.
\newblock \href {https://doi.org/10.2118/1942-pa} {\path{doi:10.2118/1942-pa}}.

\bibitem{dullienPorousMediaFluid1979}
F.~A.~L. Dullien, Porous {{Media}}: {{Fluid Transport}} and {{Pore Structure}},
  Academic Press Inc, New York, 1979.

\bibitem{armstrongCriticalCapillaryNumber2014a}
R.~T. Armstrong, A.~Georgiadis, H.~Ott, D.~Klemin, S.~Berg, Critical capillary
  number: {{Desaturation}} studied with fast {{X}}-ray computed
  microtomography, Geophysical Research Letters 41~(1) (2014) 55--60.
\newblock \href {https://doi.org/10.1002/2013GL058075}
  {\path{doi:10.1002/2013GL058075}}.

\bibitem{hilferCapillarySaturationDesaturation2015}
R.~Hilfer, R.~T. Armstrong, S.~Berg, A.~Georgiadis, H.~Ott, Capillary
  saturation and desaturation, Physical Review E 92~(6) (2015) 063023.
\newblock \href {https://doi.org/10.1103/PhysRevE.92.063023}
  {\path{doi:10.1103/PhysRevE.92.063023}}.

\bibitem{niessnerComparisonTwoPhaseDarcys2011}
J.~Niessner, S.~Berg, S.~M. Hassanizadeh, Comparison of {{Two-Phase Darcy}}'s
  {{Law}} with a {{Thermodynamically Consistent Approach}}, Transport in Porous
  Media 88~(1) (2011) 133--148.
\newblock \href {https://doi.org/10.1007/s11242-011-9730-0}
  {\path{doi:10.1007/s11242-011-9730-0}}.

\bibitem{Popper34}
K.~R. Popper, The Logic of Scientific Discovery, Hutchinson, London, 1934.

\bibitem{krolLocalHydraulicResistance2021}
Q.~Krol, I.~Fouxon, P.~Corso, M.~Holzner, Local {{Hydraulic Resistance}} in
  {{Heterogeneous Porous Media}}, Geophysical Research Letters 48~(22) (2021)
  e2021GL094694.
\newblock \href {https://doi.org/10.1029/2021GL094694}
  {\path{doi:10.1029/2021GL094694}}.

\bibitem{andersonWettabilityLiteratureSurvey1987a}
W.~G. Anderson, Wettability {{Literature Survey Part}} 5: {{The Effects}} of
  {{Wettability}} on {{Relative Permeability}}, Journal of Petroleum Technology
  39~(11) (1987) 1453--1468.
\newblock \href {https://doi.org/10.2118/16323-PA}
  {\path{doi:10.2118/16323-PA}}.

\bibitem{mccaughanMolecularDynamicsSimulation2013}
J.~McCaughan, S.~Iglauer, F.~Bresme, Molecular {{Dynamics Simulation}} of
  {{Water}}/{{CO2-quartz Interfacial Properties}}: {{Application}} to
  {{Subsurface Gas Injection}}, Energy Procedia 37 (2013) 5387--5402.
\newblock \href {https://doi.org/10.1016/j.egypro.2013.06.457}
  {\path{doi:10.1016/j.egypro.2013.06.457}}.

\bibitem{zhengSurrogateModelsStudying2020}
L.~Zheng, M.~R{\"u}cker, T.~Bultreys, A.~Georgiadis, M.~M. {Mooijer-van den
  Heuvel}, F.~Bresme, J.~P.~M. Trusler, E.~A. M{\"u}ller, Surrogate {{Models}}
  for {{Studying}} the {{Wettability}} of {{Nanoscale Natural Rough Surfaces
  Using Molecular Dynamics}}, Energies 13~(11) (2020) 2770.
\newblock \href {https://doi.org/10.3390/en13112770}
  {\path{doi:10.3390/en13112770}}.

\bibitem{lenhardModelHystereticConstitutive1987}
R.~J. Lenhard, J.~C. Parker, A model for hysteretic constitutive relations
  governing multiphase flow: 2. {{Permeability}}-saturation relations, Water
  Resources Research 23~(12) (1987) 2197--2206.
\newblock \href {https://doi.org/10.1029/WR023i012p02197}
  {\path{doi:10.1029/WR023i012p02197}}.

\bibitem{jerauldEffectPorestructureHysteresis1990}
G.~R. Jerauld, S.~J. Salter, The effect of pore-structure on hysteresis in
  relative permeability and capillary pressure: {{Pore-level}} modeling,
  Transport in Porous Media 5~(2) (1990) 103--151.
\newblock \href {https://doi.org/10.1007/BF00144600}
  {\path{doi:10.1007/BF00144600}}.

\bibitem{marleMacroscopicEquationsGoverning1982}
C.~Marle, On macroscopic equations governing multiphase flow with diffusion and
  chemical reactions in porous media, International Journal of Engineering
  Science 20~(5) (1982) 643--662.
\newblock \href {https://doi.org/10.1016/0020-7225(82)90118-5}
  {\path{doi:10.1016/0020-7225(82)90118-5}}.

\bibitem{whitakerFlowPorousMedia1986a}
S.~Whitaker, Flow in porous media {{II}}: {{The}} governing equations for
  immiscible, two-phase flow, Transport in Porous Media 1~(2) (1986) 105--125.
\newblock \href {https://doi.org/10.1007/BF00714688}
  {\path{doi:10.1007/BF00714688}}.

\bibitem{quintardTwophaseFlowHeterogeneous1988}
M.~Quintard, S.~Whitaker, Two-phase flow in heterogeneous porous media: {{The}}
  method of large-scale averaging, Transport in Porous Media 3~(4) (1988)
  357--413.
\newblock \href {https://doi.org/10.1007/BF00233177}
  {\path{doi:10.1007/BF00233177}}.

\bibitem{quintardTransportOrderedDisordered1993}
M.~Quintard, S.~Whitaker, Transport in ordered and disordered porous media:
  Volume-averaged equations, closure problems, and comparison with experiment,
  Chemical Engineering Science 48~(14) (1993) 2537--2564.
\newblock \href {https://doi.org/10.1016/0009-2509(93)80266-S}
  {\path{doi:10.1016/0009-2509(93)80266-S}}.

\bibitem{quintardTransportOrderedDisordered1994}
M.~Quintard, S.~Whitaker, Transport in ordered and disordered porous media
  {{II}}: {{Generalized}} volume averaging, Transport in Porous Media 14~(2)
  (1994) 179--206.
\newblock \href {https://doi.org/10.1007/BF00615200}
  {\path{doi:10.1007/BF00615200}}.

\bibitem{latva-kokkoScalingDynamicContact2007}
M.~{Latva-Kokko}, D.~H. Rothman, Scaling of {{Dynamic Contact Angles}} in a
  {{Lattice-Boltzmann Model}}, Physical Review Letters 98~(25) (2007) 254503.
\newblock \href {https://doi.org/10.1103/PhysRevLett.98.254503}
  {\path{doi:10.1103/PhysRevLett.98.254503}}.

\bibitem{snoeijerMovingContactLines2013}
J.~H. Snoeijer, B.~Andreotti, Moving {{Contact Lines}}: {{Scales}},
  {{Regimes}}, and {{Dynamical Transitions}}, Annual Review of Fluid Mechanics
  45~(1) (2013) 269--292.
\newblock \href {https://doi.org/10.1146/annurev-fluid-011212-140734}
  {\path{doi:10.1146/annurev-fluid-011212-140734}}.

\bibitem{mcclureTrackingInterfaceCommon2016}
J.~E. McClure, M.~A. Berrill, W.~G. Gray, C.~T. Miller, Tracking interface and
  common curve dynamics for two-fluid flow in porous media, Journal of Fluid
  Mechanics 796 (2016) 211--232.
\newblock \href {https://doi.org/10.1017/jfm.2016.212}
  {\path{doi:10.1017/jfm.2016.212}}.

\bibitem{akaiWettingBoundaryCondition2018}
T.~Akai, B.~Bijeljic, M.~J. Blunt, Wetting boundary condition for the
  color-gradient lattice {{Boltzmann}} method: {{Validation}} with analytical
  and experimental data, Advances in Water Resources 116 (2018) 56--66.
\newblock \href {https://doi.org/10.1016/j.advwatres.2018.03.014}
  {\path{doi:10.1016/j.advwatres.2018.03.014}}.

\bibitem{qiuPhasefieldModelingTwophase2025}
Y.~Qiu, L.~{Cueto-Felgueroso}, A.~A. Pahlavan, B.~K. Primkulov, R.~Juanes,
  Phase-field modeling of two-phase displacement in a capillary tube, Physical
  Review Fluids 10~(9) (2025) 094004.
\newblock \href {https://doi.org/10.1103/km7k-rmvb}
  {\path{doi:10.1103/km7k-rmvb}}.

\bibitem{arceArtScienceUpscaling2005}
P.~E. Arce, M.~Quintard, S.~Whitaker, The {{Art}} and {{Science}} of
  {{Upscaling}}, in: Chemical {{Engineering}}, John Wiley \& Sons, Ltd, 2005,
  Ch.~1, pp. 1--39.
\newblock \href {https://doi.org/10.1002/0470025018.ch1}
  {\path{doi:10.1002/0470025018.ch1}}.

\bibitem{Fadili2004}
A.~Fadili, R.~Ababou, Dual homogenization of immiscible steady two-phase flows
  in random porous media, Water Resources Research 40 (2004) W01513.
\newblock \href {https://doi.org/10.1029/2003WR002465}
  {\path{doi:10.1029/2003WR002465}}.

\bibitem{Lasseux2022}
D.~Lasseux, F.~J. {Vald{\'e}s-Parada}, A macroscopic model for immiscible
  two-phase flow in porous media, Journal of Fluid Mechanics 944 (2022) A43.
\newblock \href {https://doi.org/10.1017/jfm.2022.487}
  {\path{doi:10.1017/jfm.2022.487}}.

\bibitem{lasseux2023upscaled}
D.~Lasseux, F.~J. Vald{\'e}s-Parada, Upscaled dynamic capillary pressure for
  two-phase flow in porous media, Journal of Fluid Mechanics 959 (2023) R2.

\bibitem{zhangUpscalingNavierStokesCahnHilliardModel2025}
C.~Zhang, P.~Liu, C.~Peng, L.-P. Wang, Z.~Guo, Upscaling the
  {{Navier-Stokes-Cahn-Hilliard}} model for incompressible multiphase flow in
  inhomogeneous porous media (Apr. 2025).
\newblock \href {http://arxiv.org/abs/2504.16009} {\path{arXiv:2504.16009}},
  \href {https://doi.org/10.48550/arXiv.2504.16009}
  {\path{doi:10.48550/arXiv.2504.16009}}.

\bibitem{briones-carrilloUpscaledCoefficientsImmiscible2025}
J.~A. {Briones-Carrillo}, C.~G. {Aguilar-Madera}, G.~{Espinosa-Paredes},
  A.~{P{\'e}rez-Valseca}, E.~C. {Herrera-Hern{\'a}ndez},
  V.~{Mat{\'i}as-P{\'e}rez}, I.~{Navarro-de Le{\'o}n}, A.~T.
  {Finol-Gonz{\'a}lez}, Upscaled {{Coefficients}} for {{Immiscible Two-Phase
  Flow}} in {{Porous Media}} for {{Drainage}} and {{Imbibition Processes}},
  Transport in Porous Media 152~(10) (2025) 83.
\newblock \href {https://doi.org/10.1007/s11242-025-02222-z}
  {\path{doi:10.1007/s11242-025-02222-z}}.

\bibitem{PermeabilityPorousMaterials1950}
E.~C. Childs, N.~Collis-George, The permeability of porous materials,
  Proceedings of the Royal Society of London. Series A. Mathematical and
  Physical Sciences (Apr. 1950).
\newblock \href {https://doi.org/10.1098/rspa.1950.0068}
  {\path{doi:10.1098/rspa.1950.0068}}.

\bibitem{burdineRelativePermeabilityCalculations1953}
N.~T. Burdine, Relative {{Permeability Calculations From Pore Size Distribution
  Data}}, Journal of Petroleum Technology 5~(03) (1953) 71--78.
\newblock \href {https://doi.org/10.2118/225-G} {\path{doi:10.2118/225-G}}.

\bibitem{mualemNewModelPredicting1976}
Y.~Mualem, A new model for predicting the hydraulic conductivity of unsaturated
  porous media, Water Resources Research 12~(3) (1976) 513--522.
\newblock \href {https://doi.org/10.1029/WR012i003p00513}
  {\path{doi:10.1029/WR012i003p00513}}.

\bibitem{lenhardParametricModelPredicting}
R.~J. Lenhard, M.~Oostrom, A {{Parametric Model}} for {{Predicting Relative
  Permeability-Saturation-Capillary Pressure Relationships}} of
  {{Oil}}--{{Water Systems}} in {{Porous Media}} with {{Mixed Wettability}},
  Transport in Porous Media 31 (1998) 109--–131.

\bibitem{genuchtenClosedformEquationPredicting1980}
M.~T. van Genuchten, A {{Closed-form Equation}} for {{Predicting}} the
  {{Hydraulic Conductivity}} of {{Unsaturated Soils}}, Soil Science Society of
  America Journal 44~(5) (1980) 892--898.
\newblock \href {https://doi.org/10.2136/sssaj1980.03615995004400050002x}
  {\path{doi:10.2136/sssaj1980.03615995004400050002x}}.

\bibitem{petersPredictionAbsoluteUnsaturated2023}
A.~Peters, S.~C. Iden, W.~Durner, Prediction of absolute unsaturated hydraulic
  conductivity -- comparison of four different capillary bundle models,
  Hydrology and Earth System Sciences 27~(24) (2023) 4579--4593.
\newblock \href {https://doi.org/10.5194/hess-27-4579-2023}
  {\path{doi:10.5194/hess-27-4579-2023}}.

\bibitem{patzekShapeFactorHydraulic2001}
T.~W. Patzek, D.~B. Silin, Shape {{Factor}} and {{Hydraulic Conductance}} in
  {{Noncircular Capillaries}}: {{I}}. {{One-Phase Creeping Flow}}, Journal of
  Colloid and Interface Science 236~(2) (2001) 295--304.
\newblock \href {https://doi.org/10.1006/jcis.2000.7413}
  {\path{doi:10.1006/jcis.2000.7413}}.

\bibitem{patzekShapeFactorCorrelations2001}
T.~W. Patzek, J.~G. Kristensen, Shape {{Factor Correlations}} of {{Hydraulic
  Conductance}} in {{Noncircular Capillaries}}: {{II}}. {{Two-Phase Creeping
  Flow}}, Journal of Colloid and Interface Science 236~(2) (2001) 305--317.
\newblock \href {https://doi.org/10.1006/jcis.2000.7414}
  {\path{doi:10.1006/jcis.2000.7414}}.

\bibitem{silinMicrotomographyPoreScaleModeling2011}
D.~Silin, L.~Tomutsa, S.~M. Benson, T.~W. Patzek, Microtomography and
  {{Pore-Scale Modeling}} of {{Two-Phase Fluid Distribution}}, Transport in
  Porous Media 86~(2) (2011) 495--515.
\newblock \href {https://doi.org/10.1007/s11242-010-9636-2}
  {\path{doi:10.1007/s11242-010-9636-2}}.

\bibitem{heibaPercolationTheoryTwoPhase1992}
A.~A. Heiba, M.~Sahimi, L.~E. Scriven, H.~T. Davis, Percolation {{Theory}} of
  {{Two-Phase Relative Permeability}}, SPE Reservoir Engineering 7~(01) (1992)
  123--132.
\newblock \href {https://doi.org/10.2118/11015-pa}
  {\path{doi:10.2118/11015-pa}}.

\bibitem{ramstadFluxdependentPercolationTransition2009}
T.~Ramstad, A.~Hansen, P.-E. {\O}ren, Flux-dependent percolation transition in
  immiscible two-phase flows in porous media, Physical Review E 79~(3) (2009)
  036310.
\newblock \href {https://doi.org/10.1103/PhysRevE.79.036310}
  {\path{doi:10.1103/PhysRevE.79.036310}}.

\bibitem{raoofPoreFlowComplexPorenetwork2013}
A.~Raoof, H.~M. Nick, S.~M. Hassanizadeh, C.~J. Spiers, {{PoreFlow}}: {{A}}
  complex pore-network model for simulation of reactive transport in variably
  saturated porous media, Computers \& Geosciences 61 (2013) 160--174.
\newblock \href {https://doi.org/10.1016/j.cageo.2013.08.005}
  {\path{doi:10.1016/j.cageo.2013.08.005}}.

\bibitem{brooksPropertiesPorousMedia1966}
R.~H. Brooks, A.~T. Corey, Properties of {{Porous Media Affecting Fluid Flow}},
  Journal of the Irrigation and Drainage Division 92~(2) (1966) 61--88.
\newblock \href {https://doi.org/10.1061/jrcea4.0000425}
  {\path{doi:10.1061/jrcea4.0000425}}.

\bibitem{sahimiFlowPhenomenaRocks1993}
M.~Sahimi, Flow phenomena in rocks: From continuum models to fractals,
  percolation, cellular automata, and simulated annealing, Reviews of Modern
  Physics 65~(4) (1993) 1393--1534.
\newblock \href {https://doi.org/10.1103/revmodphys.65.1393}
  {\path{doi:10.1103/revmodphys.65.1393}}.

\bibitem{hilferMacroscopicEquationsMotion1998}
R.~Hilfer, Macroscopic equations of motion for two-phase flow in porous media,
  Physical Review E 58~(2) (1998) 2090--2096.
\newblock \href {https://doi.org/10.1103/PhysRevE.58.2090}
  {\path{doi:10.1103/PhysRevE.58.2090}}.

\bibitem{hilferMacroscopicTwophaseFlow2000}
R.~Hilfer, H.~Besserer, Macroscopic two-phase flow in porous media, Physica B:
  Condensed Matter 279~(1-3) (2000) 125--129.
\newblock \href {https://doi.org/10.1016/S0921-4526(99)00694-8}
  {\path{doi:10.1016/S0921-4526(99)00694-8}}.

\bibitem{hilferPercolationBasicConcept2010}
R.~Hilfer, F.~Doster, Percolation as a {{Basic Concept}} for {{Macroscopic
  Capillarity}}, Transport in Porous Media 82~(3) (2010) 507--519.
\newblock \href {https://doi.org/10.1007/s11242-009-9395-0}
  {\path{doi:10.1007/s11242-009-9395-0}}.

\bibitem{ferryKuboFormulaLinear1992}
D.~K. Ferry, The {{Kubo Formula}} and {{Linear Response}}, in: D.~K. Ferry,
  C.~Jacoboni (Eds.), Quantum {{Transport}} in {{Semiconductors}}, Springer US,
  Boston, MA, 1992, pp. 17--36.
\newblock \href {https://doi.org/$10.1007/978-1-4899-2359-2_2$}
  {\path{doi:$10.1007/978-1-4899-2359-2_2$}}.

\bibitem{morrow1970physics}
N.~R. Morrow, Physics and thermodynamics of capillary action in porous media,
  Industrial \& Engineering Chemistry 62~(6) (1970) 32--56.

\bibitem{hassanizadehMechanicsThermodynamicsMultiphase1990}
S.~Hassanizadeh, W.~G. Gray, Mechanics and thermodynamics of multiphase flow in
  porous media including interphase boundaries, Advances in Water Resources
  13~(4) (1990) 169--186.
\newblock \href {https://doi.org/10.1016/0309-1708(90)90040-B}
  {\path{doi:10.1016/0309-1708(90)90040-B}}.

\bibitem{grayThermodynamicallyConstrainedAveraging2005}
W.~G. Gray, C.~T. Miller, Thermodynamically constrained averaging theory
  approach for modeling flow and transport phenomena in porous medium systems:
  1. {{Motivation}} and overview, Advances in Water Resources 28~(2) (2005)
  161--180.
\newblock \href {https://doi.org/10.1016/j.advwatres.2004.09.005}
  {\path{doi:10.1016/j.advwatres.2004.09.005}}.

\bibitem{grayClosureConditionsTwoFluid2002}
W.~G. Gray, A.~F.~B. Tompson, W.~E. Soll, Closure {{Conditions}} for
  {{Two-Fluid Flow}} in {{Porous Media}}, Transport in Porous Media 47 (2002)
  29--65.

\bibitem{jongschaapEquilibriumThermodynamicsCallens2001}
R.~J. Jongschaap, H.~C. {\"O}ttinger, Equilibrium thermodynamics ---
  {{Callen}}'s postulational approach, Journal of Non-Newtonian Fluid Mechanics
  96~(1-2) (2001) 5--17.
\newblock \href {https://doi.org/10.1016/S0377-0257(00)00137-3}
  {\path{doi:10.1016/S0377-0257(00)00137-3}}.

\bibitem{veveakisReviewExtremumPostulates2015}
E.~Veveakis, K.~{Regenauer-Lieb}, Review of extremum postulates, Current
  Opinion in Chemical Engineering 7 (2015) 40--46.
\newblock \href {https://doi.org/10.1016/j.coche.2014.10.006}
  {\path{doi:10.1016/j.coche.2014.10.006}}.

\bibitem{millerThermodynamicallyConstrainedAveraging2005a}
C.~T. Miller, W.~G. Gray, Thermodynamically constrained averaging theory
  approach for modeling flow and transport phenomena in porous medium systems:
  2. {{Foundation}}, Advances in Water Resources 28~(2) (2005) 181--202.
\newblock \href {https://doi.org/10.1016/j.advwatres.2004.09.006}
  {\path{doi:10.1016/j.advwatres.2004.09.006}}.

\bibitem{grayThermodynamicallyConstrainedAveraging2006}
W.~G. Gray, C.~T. Miller, Thermodynamically constrained averaging theory
  approach for modeling flow and transport phenomena in porous medium systems:
  3. {{Single-fluid-phase}} flow, Advances in Water Resources 29~(11) (2006)
  1745--1765.
\newblock \href {https://doi.org/10.1016/j.advwatres.2006.03.010}
  {\path{doi:10.1016/j.advwatres.2006.03.010}}.

\bibitem{millerThermodynamicallyConstrainedAveraging2008}
C.~T. Miller, W.~G. Gray, Thermodynamically constrained averaging theory
  approach for modeling flow and transport phenomena in porous medium systems:
  4. {{Species}} transport fundamentals, Advances in Water Resources 31~(3)
  (2008) 577--597.
\newblock \href {https://doi.org/10.1016/j.advwatres.2007.11.004}
  {\path{doi:10.1016/j.advwatres.2007.11.004}}.

\bibitem{grayThermodynamicallyConstrainedAveraging2009}
W.~G. Gray, C.~T. Miller, Thermodynamically constrained averaging theory
  approach for modeling flow and transport phenomena in porous medium systems:
  5. {{Single-fluid-phase}} transport, Advances in Water Resources 32~(5)
  (2009) 681--711.
\newblock \href {https://doi.org/10.1016/j.advwatres.2008.10.013}
  {\path{doi:10.1016/j.advwatres.2008.10.013}}.

\bibitem{jacksonThermodynamicallyConstrainedAveraging2009}
A.~S. Jackson, C.~T. Miller, W.~G. Gray, Thermodynamically constrained
  averaging theory approach for modeling flow and transport phenomena in porous
  medium systems: 6. {{Two-fluid-phase}} flow, Advances in Water Resources
  32~(6) (2009) 779--795.
\newblock \href {https://doi.org/10.1016/j.advwatres.2008.11.010}
  {\path{doi:10.1016/j.advwatres.2008.11.010}}.

\bibitem{grayThermodynamicallyConstrainedAveraging2009a}
W.~G. Gray, C.~T. Miller, Thermodynamically constrained averaging theory
  approach for modeling flow and transport phenomena in porous medium systems:
  7. {{Single-phase}} megascale flow models, Advances in Water Resources 32~(8)
  (2009) 1121--1142.
\newblock \href {https://doi.org/10.1016/j.advwatres.2009.05.010}
  {\path{doi:10.1016/j.advwatres.2009.05.010}}.

\bibitem{grayThermodynamicallyConstrainedAveraging2010}
W.~G. Gray, C.~T. Miller, Thermodynamically constrained averaging theory
  approach for modeling flow and transport phenomena in porous medium systems:
  8. {{Interface}} and common curve dynamics, Advances in Water Resources
  33~(12) (2010) 1427--1443.
\newblock \href {https://doi.org/10.1016/j.advwatres.2010.07.002}
  {\path{doi:10.1016/j.advwatres.2010.07.002}}.

\bibitem{jacksonThermodynamicallyConstrainedAveraging2012a}
A.~Jackson, I.~Rybak, R.~Helmig, W.~Gray, C.~Miller, Thermodynamically
  constrained averaging theory approach for modeling flow and transport
  phenomena in porous medium systems: 9. {{Transition}} region models, Advances
  in Water Resources 42 (2012) 71--90.
\newblock \href {https://doi.org/10.1016/j.advwatres.2012.01.006}
  {\path{doi:10.1016/j.advwatres.2012.01.006}}.

\bibitem{grayIntroductionThermodynamicallyConstrained2014}
W.~G. Gray, C.~T. Miller, Introduction to the {{Thermodynamically Constrained
  Averaging Theory}} for {{Porous Medium Systems}}, Advances in {{Geophysical}}
  and {{Environmental Mechanics}} and {{Mathematics}}, Springer International
  Publishing, Cham, 2014.
\newblock \href {https://doi.org/10.1007/978-3-319-04010-3}
  {\path{doi:10.1007/978-3-319-04010-3}}.

\bibitem{millerPedagogicalApproachThermodynamically2017}
C.~T. Miller, F.~J. {Vald{\'e}s-Parada}, B.~D. Wood, A {{Pedagogical Approach}}
  to the {{Thermodynamically Constrained Averaging Theory}}, Transport in
  Porous Media 119~(3) (2017) 585--609.
\newblock \href {https://doi.org/10.1007/s11242-017-0900-6}
  {\path{doi:10.1007/s11242-017-0900-6}}.

\bibitem{millerThermodynamicallyConstrainedAveraging2018}
C.~T. Miller, W.~G. Gray, C.~E. Kees, Thermodynamically {{Constrained Averaging
  Theory}}: {{Principles}}, {{Model Hierarchies}}, and {{Deviation Kinetic
  Energy Extensions}}, Entropy 20~(4) (2018) 253.
\newblock \href {https://doi.org/10.3390/e20040253}
  {\path{doi:10.3390/e20040253}}.

\bibitem{Culligan2004}
K.~A. Culligan, D.~Wildenschild, B.~S.~B. Christensen, W.~G. Gray, M.~L.
  Rivers, A.~F.~B. Tompson, Interfacial area measurements for unsaturated flow
  through a porous medium, Water Resources Research 40~(12) (Dec. 2004).
\newblock \href {https://doi.org/10.1029/2004WR003278}
  {\path{doi:10.1029/2004WR003278}}.

\bibitem{porterLatticeBoltzmannSimulationsCapillary2009}
M.~L. Porter, M.~G. Schaap, D.~Wildenschild, Lattice-{{Boltzmann}} simulations
  of the capillary pressure--saturation--interfacial area relationship for
  porous media, Advances in Water Resources 32~(11) (2009) 1632--1640.
\newblock \href {https://doi.org/10.1016/j.advwatres.2009.08.009}
  {\path{doi:10.1016/j.advwatres.2009.08.009}}.

\bibitem{meisenheimer2020exploring}
D.~E. Meisenheimer, J.~E. McClure, M.~L. Rivers, D.~Wildenschild, Exploring the
  effect of flow condition on the constitutive relationships for two-phase
  flow, Advances in Water Resources 137 (2020) 103506.

\bibitem{Herring2013}
A.~L. Herring, E.~J. Harper, L.~Andersson, A.~Sheppard, B.~K. Bay,
  D.~Wildenschild, Effect of fluid topology on residual nonwetting phase
  trapping: {{Implications}} for geologic {{CO2}} sequestration, Advances in
  Water Resources 62 (2013) 47--58.
\newblock \href {https://doi.org/10.1016/j.advwatres.2013.09.015}
  {\path{doi:10.1016/j.advwatres.2013.09.015}}.

\bibitem{Schlueter2016}
S.~Schl{\"u}ter, S.~Berg, M.~R{\"u}cker, R.~T. Armstrong, H.-J. Vogel,
  R.~Hilfer, D.~Wildenschild, Pore-scale displacement mechanisms as a source of
  hysteresis for two-phase flow in porous media, Water Resources Research
  52~(3) (2016) 2194--2205.
\newblock \href {https://doi.org/10.1002/2015WR018254}
  {\path{doi:10.1002/2015WR018254}}.

\bibitem{dyeCapillaryPressureDynamics2013}
Capillary {{Pressure Dynamics}} in a {{Two-Fluid-Phase Porous Medium System}},
  Vol. H53M-05.

\bibitem{McClure2018}
J.~E. McClure, R.~T. Armstrong, M.~A. Berrill, S.~Schl{\"u}ter, S.~Berg, W.~G.
  Gray, C.~T. Miller, Geometric state function for two-fluid flow in porous
  media, Physical Review Fluids 3~(8) (2018) 084306.
\newblock \href {https://doi.org/10.1103/PhysRevFluids.3.084306}
  {\path{doi:10.1103/PhysRevFluids.3.084306}}.

\bibitem{constantinidesNetworkSimulationSteadystate1996}
G.~N. Constantinides, A.~C. Payatakes, Network simulation of steady-state
  two-phase flow in consolidated porous media, AIChE Journal 42~(2) (1996)
  369--382.
\newblock \href {https://doi.org/10.1002/aic.690420207}
  {\path{doi:10.1002/aic.690420207}}.

\bibitem{valavanidesSteadyStateTwoPhaseFlow2012}
M.~Valavanides, Steady-{{State Two-Phase Flow}} in {{Porous Media}}: {{Review}}
  of {{Progress}} in the {{Development}} of the {{{\emph{DeProF}}}} {{Theory
  Bridging Pore}} to {{Statistical Thermodynamics Scales}}, Oil \& Gas Science
  and Technology -- Revue d'IFP Energies nouvelles 67~(5) (2012) 787--804.
\newblock \href {https://doi.org/10.2516/ogst/2012056}
  {\path{doi:10.2516/ogst/2012056}}.

\bibitem{valavanidesTruetomechanismModelSteadystate}
M.~S. Valavanides, A.~C. Payatakes, True-to-mechanism model of steady-state
  two-phase flow in porous media, using decomposition into prototype flows,
  Advances in Water Resources 24 (2001) 385--407.

\bibitem{valavanidesReviewSteadyStateTwoPhase2018a}
M.~S. Valavanides, Review of {{Steady-State Two-Phase Flow}} in {{Porous
  Media}}: {{Independent Variables}}, {{Universal Energy Efficiency Map}},
  {{Critical Flow Conditions}}, {{Effective Characterization}} of {{Flow}} and
  {{Pore Network}}, Transport in Porous Media 123~(1) (2018) 45--99.
\newblock \href {https://doi.org/10.1007/s11242-018-1026-1}
  {\path{doi:10.1007/s11242-018-1026-1}}.

\bibitem{valavanidesFlowRateDependency2023}
M.~S. Valavanides, Flow {{Rate Dependency}} of {{Steady-State Two-Phase Flows}}
  in {{Pore Networks}}: {{Universal}}, {{Relative Permeability Scaling
  Function}} and {{System-Characteristic Invariants}}, Transport in Porous
  Media 150~(3) (2023) 521--557.
\newblock \href {https://doi.org/10.1007/s11242-023-02012-5}
  {\path{doi:10.1007/s11242-023-02012-5}}.

\bibitem{gaoNewInsightsInterface2025}
Z.~Gao, S.~Chen, Q.~Shao, P.~Cui, X.~Guo, B.~Wang, T.~Lv, Y.~Han, New insights
  into the interface reconstruction event: {{The}} principle of minimum
  operating power during two-phase displacement, Physics of Fluids 37~(7) (Jul.
  2025).
\newblock \href {https://doi.org/10.1063/5.0274455}
  {\path{doi:10.1063/5.0274455}}.

\bibitem{standnesImplicationsMolecularThermal2018}
D.~C. Standnes, Implications of {{Molecular Thermal Fluctuations}} on {{Fluid
  Flow}} in {{Porous Media}} and {{Its Relevance}} to {{Absolute
  Permeability}}, Energy \& Fuels 32~(8) (2018) 8024--8039.
\newblock \href {https://doi.org/10.1021/acs.energyfuels.8b00478}
  {\path{doi:10.1021/acs.energyfuels.8b00478}}.

\bibitem{standnesDissipationMechanismsFluids2021}
D.~C. Standnes, Dissipation {{Mechanisms}} for {{Fluids}} and {{Objects}} in
  {{Relative Motion Described}} by the {{Navier}}--{{Stokes Equation}}, ACS
  Omega 6~(29) (2021) 18598--18609.
\newblock \href {https://doi.org/10.1021/acsomega.1c01033}
  {\path{doi:10.1021/acsomega.1c01033}}.

\bibitem{standnesDerivationConventionalGeneralized2022}
D.~C. Standnes, Derivation of the {{Conventional}} and a {{Generalized Form}}
  of {{Darcy}}'s {{Law}} from the {{Langevin Equation}}, Transport in Porous
  Media 141~(1) (2022) 1--15.
\newblock \href {https://doi.org/10.1007/s11242-021-01707-x}
  {\path{doi:10.1007/s11242-021-01707-x}}.

\bibitem{Berg2013}
S.~Berg, H.~Ott, S.~A. Klapp, A.~Schwing, R.~Neiteler, N.~Brussee, A.~Makurat,
  L.~Leu, F.~Enzmann, J.-O. Schwarz, M.~Kersten, S.~Irvine, M.~Stampanoni,
  Real-time {{3D}} imaging of {{Haines}} jumps in porous media flow,
  Proceedings of the National Academy of Sciences 110~(10) (2013) 3755--3759.
\newblock \href {https://doi.org/10.1073/pnas.1221373110}
  {\path{doi:10.1073/pnas.1221373110}}.

\bibitem{standnesUsingTotalChemical2024}
D.~C. Standnes, E.~Ebeltoft, {\AA}.~Haugen, A.~Kristoffersen, Using the total
  chemical potential to generalize the capillary pressure concept and therefrom
  derive a governing equation for two-phase flow in porous media, International
  Journal of Multiphase Flow 181 (2024) 105024.
\newblock \href {https://doi.org/10.1016/j.ijmultiphaseflow.2024.105024}
  {\path{doi:10.1016/j.ijmultiphaseflow.2024.105024}}.

\bibitem{bresmeTopicalIssueThermal2019}
F.~Bresme, V.~Vesovic, H.~Bataller, F.~Croccolo, Topical {{Issue}} on {{Thermal
  Non-Equilibrium Phenomena}} in {{Soft Matter}}, The European Physical Journal
  E 42~(11) (2019) 148.
\newblock \href {https://doi.org/10.1140/epje/i2019-11918-4}
  {\path{doi:10.1140/epje/i2019-11918-4}}.

\bibitem{hassanizadehThermodynamicBasisCapillary1993a}
S.~M. Hassanizadeh, W.~G. Gray, Thermodynamic basis of capillary pressure in
  porous media, Water Resources Research 29~(10) (1993) 3389--3405.
\newblock \href {https://doi.org/10.1029/93WR01495}
  {\path{doi:10.1029/93WR01495}}.

\bibitem{nieberDynamicCapillaryPressure2005}
J.~L. Nieber, R.~Z. Dautov, A.~G. Egorov, A.~Y. Sheshukov, Dynamic {{Capillary
  Pressure Mechanism}} for {{Instability}} in {{Gravity-Driven Flows}};
  {{Review}} and {{Extension}} to {{Very Dry Conditions}}, Transport in Porous
  Media 58~(1) (2005) 147--172.
\newblock \href {https://doi.org/10.1007/s11242-004-5473-5}
  {\path{doi:10.1007/s11242-004-5473-5}}.

\bibitem{botteroNonequilibriumCapillarityEffects2011}
S.~Bottero, S.~M. Hassanizadeh, P.~J. Kleingeld, T.~J. Heimovaara,
  Nonequilibrium capillarity effects in two-phase flow through porous media at
  different scales, Water Resources Research 47~(10) (2011) 2011WR010887.
\newblock \href {https://doi.org/10.1029/2011WR010887}
  {\path{doi:10.1029/2011WR010887}}.

\bibitem{barenblattMathematicalModelNonequilibrium2003}
G.~I. Barenblatt, T.~W. Patzek, D.~B. Silin, The {{Mathematical Model}} of
  {{Nonequilibrium Effects}} in {{Water-Oil Displacement}}, SPE Journal 8~(04)
  (2003) 409--416.
\newblock \href {https://doi.org/10.2118/87329-PA}
  {\path{doi:10.2118/87329-PA}}.

\bibitem{dattaSpatialFluctuationsFluid2013}
S.~S. Datta, H.~Chiang, T.~S. Ramakrishnan, D.~A. Weitz, Spatial
  {{Fluctuations}} of {{Fluid Velocities}} in {{Flow}} through a
  {{Three-Dimensional Porous Medium}}, Physical Review Letters 111~(6) (Aug.
  2013).
\newblock \href {https://doi.org/10.1103/physrevlett.111.064501}
  {\path{doi:10.1103/physrevlett.111.064501}}.

\bibitem{dattaFluidBreakupSimultaneous2014}
S.~S. Datta, J.-B. Dupin, D.~A. Weitz, Fluid breakup during simultaneous
  two-phase flow through a three-dimensional porous medium, Physics of Fluids
  26~(6) (Jun. 2014).
\newblock \href {https://doi.org/10.1063/1.4884955}
  {\path{doi:10.1063/1.4884955}}.

\bibitem{masalmehLowSalinityFlooding2014}
S.~K. Masalmeh, T.~Sorop, B.~M. J.~M. Suijkerbuijk, E.~C.~M. Vermolen,
  S.~Douma, H.~A. {van del Linde}, S.~G.~J. Pieterse, Low {{Salinity
  Flooding}}: {{Experimental Evaluation}} and {{Numerical Interpretation}}, in:
  International {{Petroleum Technology Conference}}, OnePetro, 2014, pp. 1--13.
\newblock \href {https://doi.org/10.2523/IPTC-17558-MS}
  {\path{doi:10.2523/IPTC-17558-MS}}.

\bibitem{reynoldsCharacterizingFlowBehavior2015}
C.~A. Reynolds, S.~Krevor, Characterizing flow behavior for gas injection:
  {{Relative}} permeability of {{CO2-brine}} and {{N2-water}} in heterogeneous
  rocks, Water Resources Research 51~(12) (2015) 9464--9489.
\newblock \href {https://doi.org/10.1002/2015WR018046}
  {\path{doi:10.1002/2015WR018046}}.

\bibitem{soropRelativePermeabilityMeasurements2015}
T.~G. Sorop, S.~K. Masalmeh, B.~M. Suijkerbuijk, H.~A. Van Der~Linde,
  H.~Mahani, N.~J. Brussee, F.~A. Marcelis, A.~Coorn, Relative {{Permeability
  Measurements}} to {{Quantify}} the {{Low Salinity Flooding Effect}} at
  {{Field Scale}}, in: Abu {{Dhabi International Petroleum Exhibition}} and
  {{Conference}}, SPE, Abu Dhabi, UAE, 2015, p. D021S024R003.
\newblock \href {https://doi.org/10.2118/177856-MS}
  {\path{doi:10.2118/177856-MS}}.

\bibitem{gaoXrayMicrotomographyIntermittency2017}
Y.~Gao, Q.~Lin, B.~Bijeljic, M.~J. Blunt, X-ray {{Microtomography}} of
  {{Intermittency}} in {{Multiphase Flow}} at {{Steady State Using}} a
  {{Differential Imaging Method}}, Water Resources Research 53~(12) (2017)
  10274--10292.
\newblock \href {https://doi.org/10.1002/2017WR021736}
  {\path{doi:10.1002/2017WR021736}}.

\bibitem{gao2020pore}
Y.~Gao, Q.~Lin, B.~Bijeljic, M.~J. Blunt, Pore-scale dynamics and the
  multiphase darcy law, Physical Review Fluids 5~(1) (2020) 013801.

\bibitem{linImagingMeasurementPoreScale2018}
Q.~Lin, B.~Bijeljic, R.~Pini, M.~J. Blunt, S.~Krevor, Imaging and
  {{Measurement}} of {{Pore-Scale Interfacial Curvature}} to {{Determine
  Capillary Pressure Simultaneously With Relative Permeability}}, Water
  Resources Research 54~(9) (2018) 7046--7060.
\newblock \href {https://doi.org/10.1029/2018WR023214}
  {\path{doi:10.1029/2018WR023214}}.

\bibitem{alcornCorescaleSensitivityStudy2020}
Z.~P. Alcorn, S.~B. Fredriksen, M.~Sharma, T.~F{\o}yen, C.~Wergeland, M.~A.
  Fern{\o}, A.~Graue, G.~Ersland, Core-scale sensitivity study of {{CO2}} foam
  injection strategies for mobility control, enhanced oil recovery, and {{CO2}}
  storage, E3S Web of Conferences 146 (2020) 02002.
\newblock \href {https://doi.org/10.1051/e3sconf/202014602002}
  {\path{doi:10.1051/e3sconf/202014602002}}.

\bibitem{spurinIntermittentFluidConnectivity2019}
C.~Spurin, T.~Bultreys, B.~Bijeljic, M.~J. Blunt, S.~Krevor, Intermittent fluid
  connectivity during two-phase flow in a heterogeneous carbonate rock,
  Physical Review E 100~(4) (2019) 043103.
\newblock \href {https://doi.org/10.1103/PhysRevE.100.043103}
  {\path{doi:10.1103/PhysRevE.100.043103}}.

\bibitem{wangObtainingHighQuality2019}
Y.~Wang, S.~K. Masalmeh, Obtaining {{High Quality SCAL Data}}: {{Combining
  Different Measurement Techniques}}, {{Saturation Monitoring}}, {{Numerical
  Interpretation}} and {{Continuous Monitoring}} of {{Experimental Data}}, E3S
  Web of Conferences 89 (2019) 02007.
\newblock \href {https://doi.org/10.1051/e3sconf/20198902007}
  {\path{doi:10.1051/e3sconf/20198902007}}.

\bibitem{menkeUsingNanoXRMHighContrast2022}
H.~P. Menke, Y.~Gao, S.~Linden, M.~G. Andrew, Using {{Nano-XRM}} and
  {{High-Contrast Imaging}} to {{Inform Micro-Porosity Permeability During
  Stokes}}--{{Brinkman Single}} and {{Two-Phase Flow Simulations}} on
  {{Micro-CT Images}}, Frontiers in Water 4 (2022) 935035.
\newblock \href {https://doi.org/10.3389/frwa.2022.935035}
  {\path{doi:10.3389/frwa.2022.935035}}.

\bibitem{Mecke_98}
K.~Mecke, {Integral geometry in statistical physics}, International Journal of
  Modern Physics B {12}~({9}) ({1998}) {861--899}.
\newblock \href {https://doi.org/{10.1142/S0217979298000491}}
  {\path{doi:{10.1142/S0217979298000491}}}.

\bibitem{meckeEulerCharacteristicRelated1991}
K.~R. Mecke, H.~Wagner, Euler characteristic and related measures for random
  geometric sets, Journal of Statistical Physics 64~(3-4) (1991) 843--850.
\newblock \href {https://doi.org/10.1007/BF01048319}
  {\path{doi:10.1007/BF01048319}}.

\bibitem{Arns2001}
C.~H. Arns, M.~A. Knackstedt, W.~V. Pinczewski, K.~R. Mecke,
  Euler-{{Poincar{\'e}}} characteristics of classes of disordered media,
  Physical Review E 63~(3) (2001) 031112.
\newblock \href {https://doi.org/10.1103/PhysRevE.63.031112}
  {\path{doi:10.1103/PhysRevE.63.031112}}.

\bibitem{hilferMacroscopicCapillarityHysteresis2006}
R.~Hilfer, Macroscopic capillarity and hysteresis for flow in porous media,
  Physical Review E 73~(1) (2006) 016307.
\newblock \href {https://doi.org/10.1103/PhysRevE.73.016307}
  {\path{doi:10.1103/PhysRevE.73.016307}}.

\bibitem{Vogel2010}
H.-J. Vogel, U.~Weller, S.~Schl{\"u}ter, Quantification of soil structure based
  on {{Minkowski}} functions, Computers \& Geosciences 36~(10) (2010)
  1236--1245.
\newblock \href {https://doi.org/10.1016/j.cageo.2010.03.007}
  {\path{doi:10.1016/j.cageo.2010.03.007}}.

\bibitem{hadwiger1957vorlesungen}
H.~Hadwiger, Vorlesungen tiber inhalt, oberfl{\"a}che und isoperirnetrie
  (1957).

\bibitem{Klain_95}
D.~A. Klain, A short proof of hadwiger’s characterization theorem,
  Mathematika 42 (1995) 329--339.
\newblock \href {https://doi.org/10.1112/S0025579300014625}
  {\path{doi:10.1112/S0025579300014625}}.

\bibitem{chernyavskiyQuantitativeDescriptionInternal2025}
M.~Chernyavskiy, V.~Timoshenko, A.~Morkovkin, D.~Manukovskaya, A.~Zernyuk,
  P.~Grishin, A.~Kalashnikov, E.~Grachev, Quantitative description of internal
  {{3D}} structure of a geological sample using algebraic topology methods,
  Scientific Reports 15~(1) (Jul. 2025).
\newblock \href {https://doi.org/10.1038/s41598-025-05692-9}
  {\path{doi:10.1038/s41598-025-05692-9}}.

\bibitem{sadeghnejadMinkowskiFunctionalEvaluation2023}
S.~Sadeghnejad, M.~Reinhardt, F.~Enzmann, P.~Arnold, B.~Brandst{\"a}tter,
  H.~Ott, F.~Wilde, S.~Hupfer, T.~Sch{\"a}fer, M.~Kersten, Minkowski functional
  evaluation of representative elementary volume of rock microtomography images
  at multiple resolutions, Advances in Water Resources 179 (2023) 104501.
\newblock \href {https://doi.org/10.1016/j.advwatres.2023.104501}
  {\path{doi:10.1016/j.advwatres.2023.104501}}.

\bibitem{Herring2015}
A.~L. Herring, L.~Andersson, S.~Schl{\"u}ter, A.~Sheppard, D.~Wildenschild,
  Efficiently engineering pore-scale processes: {{The}} role of force dominance
  and topology during nonwetting phase trapping in porous media, Advances in
  Water Resources 79 (2015) 91--102.
\newblock \href {https://doi.org/10.1016/j.advwatres.2015.02.005}
  {\path{doi:10.1016/j.advwatres.2015.02.005}}.

\bibitem{porter2010image}
M.~L. Porter, D.~Wildenschild, Image analysis algorithms for estimating porous
  media multiphase flow variables from computed microtomography data: a
  validation study, Computational Geosciences 14 (2010) 15--30.

\bibitem{Pak2015}
T.~Pak, I.~B. Butler, S.~Geiger, M.~I.~J. {van Dijke}, K.~S. Sorbie, Droplet
  fragmentation: {{3D}} imaging of a previously unidentified pore-scale process
  during multiphase flow in porous media, Proceedings of the National Academy
  of Sciences 112~(7) (2015) 1947--1952.
\newblock \href {https://doi.org/10.1073/pnas.1420202112}
  {\path{doi:10.1073/pnas.1420202112}}.

\bibitem{Bultreys2024}
T.~Bultreys, S.~Ellman, C.~M. Schlep{\"u}tz, M.~N. Boone, G.~K. Pakkaner,
  S.~Wang, M.~Borji, S.~Van~Offenwert, N.~Moazami~Goudarzi, W.~Goethals, C.~W.
  Winardhi, V.~Cnudde, {{4D}} microvelocimetry reveals multiphase flow field
  perturbations in porous media, Proceedings of the National Academy of
  Sciences 121~(12) (2024) e2316723121.
\newblock \href {https://doi.org/10.1073/pnas.2316723121}
  {\path{doi:10.1073/pnas.2316723121}}.

\bibitem{Zenyuk2016}
I.~V. Zenyuk, A.~Lamibrac, J.~Eller, D.~Y. Parkinson, F.~Marone, F.~N.
  B{\"u}chi, A.~Z. Weber, Investigating {{Evaporation}} in {{Gas Diffusion
  Layers}} for {{Fuel Cells}} with {{X-ray Computed Tomography}}, The Journal
  of Physical Chemistry C 120~(50) (2016) 28701--28711.
\newblock \href {https://doi.org/10.1021/acs.jpcc.6b10658}
  {\path{doi:10.1021/acs.jpcc.6b10658}}.

\bibitem{holtzmanOriginHysteresisMemory2020}
R.~Holtzman, M.~Dentz, R.~Planet, J.~Ort{\'i}n, The origin of hysteresis and
  memory of two-phase flow in disordered media, Communications Physics 3~(1)
  (2020) 222.
\newblock \href {https://doi.org/10.1038/s42005-020-00492-1}
  {\path{doi:10.1038/s42005-020-00492-1}}.

\bibitem{nepalMechanismsInterfaceJumps2025}
A.~Nepal, J.~J. Hidalgo, J.~Ort{\'i}n, I.~Lunati, M.~Dentz, Mechanisms of
  interface jumps, pinning and hysteresis during cyclic fluid displacements in
  an isolated pore, Journal of Colloid and Interface Science (2025) 137767\href
  {https://doi.org/10.1016/j.jcis.2025.137767}
  {\path{doi:10.1016/j.jcis.2025.137767}}.

\bibitem{Ruecker2015}
M.~R{\"u}cker, S.~Berg, R.~Armstrong, A.~Georgiadis, H.~Ott, L.~Simon,
  F.~Enzmann, M.~Kersten, The fate of oil clusters during fractional flow:
  Trajectories in the saturationcapillary number space, in: International
  Symposium of the Society of Core Analysts held in St. John’s Newfoundland
  and Labrador, Canada, 16-21 August, 2015, 2015, pp. 1--12.

\bibitem{McClure2020}
J.~E. McClure, T.~Ramstad, Z.~Li, R.~T. Armstrong, S.~Berg, Modeling
  {{Geometric State}} for {{Fluids}} in {{Porous Media}}: {{Evolution}} of the
  {{Euler Characteristic}}, Transport in Porous Media 133~(2) (2020) 229--250.
\newblock \href {https://doi.org/10.1007/s11242-020-01420-1}
  {\path{doi:10.1007/s11242-020-01420-1}}.

\bibitem{Liu2017}
Z.~Liu, A.~Herring, C.~Arns, S.~Berg, R.~T. Armstrong, Pore-{{Scale
  Characterization}} of {{Two-Phase Flow Using Integral Geometry}}, Transport
  in Porous Media 118~(1) (2017) 99--117.
\newblock \href {https://doi.org/10.1007/s11242-017-0849-5}
  {\path{doi:10.1007/s11242-017-0849-5}}.

\bibitem{Khorsandi2017}
S.~Khorsandi, L.~Li, R.~T. Johns, Equation of {{State}} for {{Relative
  Permeability}}, {{Including Hysteresis}} and {{Wettability Alteration}}, SPE
  Journal 22~(06) (2017) 1915--1928.
\newblock \href {https://doi.org/10.2118/182655-PA}
  {\path{doi:10.2118/182655-PA}}.

\bibitem{Purswani2021}
P.~Purswani, R.~T. Johns, Z.~T. Karpyn, M.~Blunt, Predictive {{Modeling}} of
  {{Relative Permeability Using}} a {{Generalized Equation}} of {{State}}, SPE
  Journal 26~(01) (2021) 191--205.

\bibitem{AlZubaidi2023}
F.~{Al-Zubaidi}, P.~Mostaghimi, Y.~Niu, R.~T. Armstrong, G.~Mohammadi, J.~E.
  McClure, S.~Berg, Effective permeability of an immiscible fluid in porous
  media determined from its geometric state, Physical Review Fluids 8~(6)
  (2023) 064004.
\newblock \href {https://doi.org/10.1103/PhysRevFluids.8.064004}
  {\path{doi:10.1103/PhysRevFluids.8.064004}}.

\bibitem{armstrong2021multiscale}
R.~T. Armstrong, C.~Sun, P.~Mostaghimi, S.~Berg, M.~R{\"u}cker, P.~Luckham,
  A.~Georgiadis, J.~E. McClure, Multiscale characterization of wettability in
  porous media, Transport in Porous Media 140~(1) (2021) 215--240.

\bibitem{sethEfficiencyConversionWork2007}
S.~Seth, N.~R. Morrow, Efficiency of the {{Conversion}} of {{Work}} of
  {{Drainage}} to {{Surface Energy}} for {{Sandstone}} and {{Carbonate}}, SPE
  Reservoir Evaluation \& Engineering 10~(04) (2007) 338--347.
\newblock \href {https://doi.org/10.2118/102490-PA}
  {\path{doi:10.2118/102490-PA}}.

\bibitem{McClure2021}
J.~E. McClure, S.~Berg, R.~T. Armstrong, Capillary fluctuations and energy
  dynamics for flow in porous media, Physics of Fluids 33~(8) (2021) 083323.
\newblock \href {https://doi.org/10.1063/5.0057428}
  {\path{doi:10.1063/5.0057428}}.

\bibitem{mcclureThermodynamicsFluctuationsBased2021}
J.~E. McClure, S.~Berg, R.~T. Armstrong, Thermodynamics of fluctuations based
  on time-and-space averages, Physical Review E 104~(3) (2021) 035106.
\newblock \href {https://doi.org/10.1103/PhysRevE.104.035106}
  {\path{doi:10.1103/PhysRevE.104.035106}}.

\bibitem{McClure2022}
J.~E. McClure, M.~Fan, S.~Berg, R.~T. Armstrong, C.~F. Berg, Z.~Li, T.~Ramstad,
  Relative permeability as a stationary process: {{Energy}} fluctuations in
  immiscible displacement, Physics of Fluids 34~(9) (Sep. 2022).
\newblock \href {https://doi.org/10.1063/5.0107149}
  {\path{doi:10.1063/5.0107149}}.

\bibitem{arbogastDerivationDoublePorosity1990}
T.~Arbogast, J.~Douglas, Jr., U.~Hornung, Derivation of the {{Double Porosity
  Model}} of {{Single Phase Flow}} via {{Homogenization Theory}}, SIAM Journal
  on Mathematical Analysis 21~(4) (1990) 823--836.
\newblock \href {https://doi.org/10.1137/0521046} {\path{doi:10.1137/0521046}}.

\bibitem{McClure2025}
J.~E. McClure, M.~Fan, S.~Berg, R.~T. Armstrong, C.~F. Berg, Z.~Li, T.~Ramstad,
  Derivation of {{A Representative Elementary Volume}} ({{REV}}) for {{Upscaled
  Two-Phase Flow}} in {{Porous Media}}, Petrophysics -- The SPWLA Journal of
  Formation Evaluation and Reservoir Description 66~(1) (2025) 68--79.
\newblock \href {https://doi.org/10.30632/PJV66N1-2025a5}
  {\path{doi:10.30632/PJV66N1-2025a5}}.

\bibitem{fyhn2021rheology}
H.~Fyhn, S.~Sinha, S.~Roy, A.~Hansen, Rheology of immiscible two-phase flow in
  mixed wet porous media: Dynamic pore network model and capillary fiber bundle
  model results, Transport in Porous Media 139~(3) (2021) 491--512.

\bibitem{fyhn2023effective}
H.~Fyhn, S.~Sinha, A.~Hansen, Effective rheology of immiscible two-phase flow
  in porous media consisting of random mixtures of grains having two types of
  wetting properties, Frontiers in Physics 11 (2023) 1175426.
\newblock \href {https://doi.org/10.3389/fphy.2023.1175426}
  {\path{doi:10.3389/fphy.2023.1175426}}.

\bibitem{lealAdvancedTransportPhenomena}
L.~G. Leal, Advanced {{Transport Phenomena}}: {{Fluid Mechanics}} and
  {{Convective Transport Processes}}, Cambridge University Press, 2007.

\bibitem{royCoMovingVelocityImmiscible2022}
S.~Roy, H.~Pedersen, S.~Sinha, A.~Hansen, The {{Co-Moving Velocity}} in
  {{Immiscible Two-Phase Flow}} in {{Porous Media}}, Transport in Porous Media
  143~(1) (2022) 69--102.
\newblock \href {https://doi.org/10.1007/s11242-022-01783-7}
  {\path{doi:10.1007/s11242-022-01783-7}}.

\bibitem{alzubaidiImpactWettabilityComoving2024}
F.~Alzubaidi, J.~E. McClure, H.~Pedersen, A.~Hansen, C.~F. Berg, P.~Mostaghimi,
  R.~T. Armstrong, The {{Impact}} of {{Wettability}} on the {{Co-moving
  Velocity}} of {{Two-Fluid Flow}} in {{Porous Media}}, Transport in Porous
  Media 151~(10-11) (2024) 1967--1982.
\newblock \href {https://doi.org/10.1007/s11242-024-02102-y}
  {\path{doi:10.1007/s11242-024-02102-y}}.

\bibitem{hansenLinearityComovingVelocity2024}
A.~Hansen, Linearity of the {{Co-moving Velocity}}, Transport in Porous Media
  151~(13) (2024) 2477--2489.
\newblock \href {https://doi.org/10.1007/s11242-024-02121-9}
  {\path{doi:10.1007/s11242-024-02121-9}}.

\bibitem{hansenStatisticalMechanicsFramework2023}
A.~Hansen, E.~G. Flekk{\o}y, S.~Sinha, P.~A. Slotte, A statistical mechanics
  framework for immiscible and incompressible two-phase flow in porous media,
  Advances in Water Resources 171 (2023) 104336.
\newblock \href {https://doi.org/10.1016/j.advwatres.2022.104336}
  {\path{doi:10.1016/j.advwatres.2022.104336}}.

\bibitem{picchiRelativePermeabilityScaling2019}
D.~Picchi, I.~Battiato, Relative {{Permeability Scaling From Pore}}-{{Scale
  Flow Regimes}}, Water Resources Research 55~(4) (2019) 3215--3233.
\newblock \href {https://doi.org/10.1029/2018WR024251}
  {\path{doi:10.1029/2018WR024251}}.

\bibitem{bergSensitivityUncertaintyAnalysis2021}
S.~Berg, E.~Unsal, H.~Dijk, Sensitivity and {{Uncertainty Analysis}} for
  {{Parameterization}} of {{Multiphase Flow Models}}, Transport in Porous Media
  140~(1) (2021) 27--57.
\newblock \href {https://doi.org/10.1007/s11242-021-01576-4}
  {\path{doi:10.1007/s11242-021-01576-4}}.

\bibitem{Kjelstrup2018}
S.~Kjelstrup, D.~Bedeaux, A.~Hansen, B.~Hafskjold, O.~Galteland, Non-isothermal
  transport of multi-phase fluids in porous media. {I. The} entropy production,
  Frontiers in Physics 6 (2019) 126.
\newblock \href {https://doi.org/10.3389/fphy.2018.00126}
  {\path{doi:10.3389/fphy.2018.00126}}.

\bibitem{Kjelstrup2019}
S.~Kjelstrup, D.~Bedeaux, A.~Hansen, B.~Hafskjold, O.~Galteland, Non-isothermal
  transport of multi-phase fluids in porous media. {I. The} constitutive
  equations, Frontiers in Physics 6 (2019) 150.
\newblock \href {https://doi.org/10.3389/fphy.2018.00150}
  {\path{doi:10.3389/fphy.2018.00150}}.

\bibitem{Bedeaux2025}
D.~Bedeaux, S.~Kjelstrup, S.~Berg, U.~Alfazazi, R.~T. Armstrong,
  Fluctuation-dissipation theorems for multi-phase flow with memory in poous
  media, phys.fluid.-dyn (2025) {arXive}2502.11539.v1.

\bibitem{bedeauxFluctuationdissipationTheoremsMultiphase2025}
D.~Bedeaux, S.~Kjelstrup, S.~Berg, U.~Alfazazi, R.~T. Armstrong,
  Fluctuation-dissipation theorems for multi-phase flow with memory in porous
  media (Feb. 2025).
\newblock \href {http://arxiv.org/abs/2502.11539} {\path{arXiv:2502.11539}},
  \href {https://doi.org/10.48550/arXiv.2502.11539}
  {\path{doi:10.48550/arXiv.2502.11539}}.

\bibitem{hansenThermodynamicsImmiscibleTwophase2009}
A.~Hansen, T.~Ramstad, Towards a thermodynamics of immiscible two-phase
  steady-state flow in porous media, Computational Geosciences 13~(2) (2009)
  227--234.
\newblock \href {https://doi.org/10.1007/s10596-008-9109-7}
  {\path{doi:10.1007/s10596-008-9109-7}}.

\bibitem{mcclureGeometricEvolutionSource2019}
J.~E. McClure, S.~Berg, R.~T. Armstrong, Geometric evolution as a source of
  discontinuous behavior in soft condensed matter (2019).
\newblock \href {https://doi.org/10.48550/ARXIV.1906.04073}
  {\path{doi:10.48550/ARXIV.1906.04073}}.

\bibitem{carlsonSimulationRelativePermeability1981}
F.~M. Carlson, Simulation of {{Relative Permeability Hysteresis}} to the
  {{Nonwetting Phase}}, in: {{SPE Annual Technical Conference}} and
  {{Exhibition}}, SPE, San Antonio, Texas, 1981, pp. SPE--10157--MS.
\newblock \href {https://doi.org/10.2118/10157-MS}
  {\path{doi:10.2118/10157-MS}}.

\bibitem{al-futaisiImpactWettabilityAlteration2003}
A.~Al-Futaisi, T.~W. Patzek, Impact of wettability alteration on two-phase flow
  characteristics of sandstones: {{A}} quasi-static description, Water
  Resources Research 39~(2) (Feb. 2003).
\newblock \href {https://doi.org/10.1029/2002wr001366}
  {\path{doi:10.1029/2002wr001366}}.

\bibitem{abdallahFundamentalsWettability2007}
W.~Abdallah, J.~S. Buckley, A.~Carnegie, J.~Edwards, B.~Herold, E.~Fordham,
  A.~Graue, T.~Habashy, N.~Seleznev, C.~Signer, H.~Hussain, B.~Montaron,
  M.~Ziauddin, Fundamentals of {{Wettability}}, Oilfield Review (2007) 44--61.

\bibitem{masalmehImpactRelativePermeability2010a}
S.~K. Masalmeh, L.~Wei, Impact of {{Relative Permeability Hysteresis}}, {{IFT}}
  dependent and {{Three Phase Models}} on the {{Performance}} of {{Gas Based
  EOR Processes}}, in: Abu {{Dhabi International Petroleum Exhibition}} and
  {{Conference}}, SPE, Abu Dhabi, UAE, 2010, pp. SPE--138203--MS.
\newblock \href {https://doi.org/10.2118/138203-MS}
  {\path{doi:10.2118/138203-MS}}.

\bibitem{ghanbarianTortuosityPorousMedia2013}
B.~Ghanbarian, A.~G. Hunt, R.~P. Ewing, M.~Sahimi, Tortuosity in {{Porous
  Media}}: {{A Critical Review}}, Soil Science Society of America Journal
  77~(5) (2013) 1461--1477.
\newblock \href {https://doi.org/10.2136/sssaj2012.0435}
  {\path{doi:10.2136/sssaj2012.0435}}.

\bibitem{caiGeneralizedModelingSpontaneous2014}
J.~Cai, E.~Perfect, C.-L. Cheng, X.~Hu, Generalized {{Modeling}} of
  {{Spontaneous Imbibition Based}} on {{Hagen}}--{{Poiseuille Flow}} in
  {{Tortuous Capillaries}} with {{Variably Shaped Apertures}}, Langmuir 30~(18)
  (2014) 5142--5151.
\newblock \href {https://doi.org/10.1021/la5007204}
  {\path{doi:10.1021/la5007204}}.

\bibitem{joekar-niasarTrappingHysteresisTwophase2013}
V.~Joekar-Niasar, F.~Doster, R.~T. Armstrong, D.~Wildenschild, M.~A. Celia,
  Trapping and hysteresis in two-phase flow in porous media: {{A}} pore-network
  study, Water Resources Research 49~(7) (2013) 4244--4256.
\newblock \href {https://doi.org/10.1002/wrcr.20313}
  {\path{doi:10.1002/wrcr.20313}}.

\bibitem{joekar-niasarNonequilibriumEffectsCapillarity2010a}
V.~{Joekar-Niasar}, S.~M. Hassanizadeh, H.~K. Dahle, Non-equilibrium effects in
  capillarity and interfacial area in two-phase flow: Dynamic pore-network
  modelling, Journal of Fluid Mechanics 655 (2010) 38--71.
\newblock \href {https://doi.org/10.1017/S0022112010000704}
  {\path{doi:10.1017/S0022112010000704}}.

\bibitem{linMinimalSurfacesPorous2019}
Q.~Lin, B.~Bijeljic, S.~Berg, R.~Pini, M.~J. Blunt, S.~Krevor, Minimal surfaces
  in porous media: {{Pore-scale}} imaging of multiphase flow in an
  altered-wettability {{Bentheimer}} sandstone, Physical Review E 99~(6) (2019)
  063105.
\newblock \href {https://doi.org/10.1103/PhysRevE.99.063105}
  {\path{doi:10.1103/PhysRevE.99.063105}}.

\bibitem{shojaeiMinimalSurfacesPorous2022}
M.~J. Shojaei, B.~Bijeljic, Y.~Zhang, M.~J. Blunt, Minimal {{Surfaces}} in
  {{Porous Materials}}: {{X-Ray Image-Based Measurement}} of the {{Contact
  Angle}} and {{Curvature}} in {{Gas Diffusion Layers}} to {{Design Optimal
  Performance}} of {{Fuel Cells}}, ACS Applied Energy Materials 5~(4) (2022)
  4613--4621.
\newblock \href {https://doi.org/10.1021/acsaem.2c00023}
  {\path{doi:10.1021/acsaem.2c00023}}.

\bibitem{silinPoreSpaceMorphology2006}
D.~Silin, T.~Patzek, Pore space morphology analysis using maximal inscribed
  spheres, Physica A: Statistical Mechanics and its Applications 371~(2) (2006)
  336--360.
\newblock \href {https://doi.org/10.1016/j.physa.2006.04.048}
  {\path{doi:10.1016/j.physa.2006.04.048}}.

\bibitem{bejanadrianShapeStructureEngineering2000}
A.~Bejan, {Shape and Structure, from Engineering to Nature}, Cambridge
  University Press, 2000.

\bibitem{Schlueter_Vogel_11}
S.~Schlueter, H.-J. Vogel, {On the reconstruction of structural and functional
  properties in random heterogeneous media}, Advances in Water Resources
  {34}~({2}) ({2011}) {314--325}.
\newblock \href {https://doi.org/{10.1016/j.advwatres.2010.12.004}}
  {\path{doi:{10.1016/j.advwatres.2010.12.004}}}.

\bibitem{yanWettabilitydrivenPorefillingInstabilities2025}
L.~Yan, J.~C. M{\"u}ller, T.~L. Van~Noorden, B.~Weigand, A.~Raoof,
  Wettability-driven pore-filling instabilities: {{Microfluidic}} and numerical
  insights, Journal of Colloid and Interface Science 696 (2025) 137884.
\newblock \href {https://doi.org/10.1016/j.jcis.2025.137884}
  {\path{doi:10.1016/j.jcis.2025.137884}}.

\bibitem{lei3DgeometrytriggeredTransitionMonotonic2025}
W.~Lei, W.~Gong, X.~Lu, Y.~Yang, H.-e. Yang, M.~Wang,
  3-{{D-geometry-triggered}} transition from monotonic to non-monotonic effects
  of wettability on multiphase displacements in homogeneous porous media,
  Journal of Fluid Mechanics 1014 (2025) R2.
\newblock \href {https://doi.org/10.1017/jfm.2025.10196}
  {\path{doi:10.1017/jfm.2025.10196}}.

\bibitem{Wang2024}
Y.~D. Wang, L.~M. Kearney, M.~J. Blunt, C.~Sun, K.~Tang, P.~Mostaghimi, R.~T.
  Armstrong, {\emph{In Situ}} characterization of heterogeneous surface wetting
  in porous materials, Advances in Colloid and Interface Science 326 (2024)
  103122.
\newblock \href {https://doi.org/10.1016/j.cis.2024.103122}
  {\path{doi:10.1016/j.cis.2024.103122}}.

\bibitem{ederRoleMinkowskiFunctionals2018}
G.~Eder, The role of {{Minkowski}} functionals in the thermodynamics of
  two-phase systems, AIP Advances 8~(1) (2018) 015127.
\newblock \href {https://doi.org/10.1063/1.5017592}
  {\path{doi:10.1063/1.5017592}}.

\bibitem{simeskiModelingAdsorptionSilica2020}
F.~Simeski, A.~M.~P. Boelens, M.~Ihme, Modeling {{Adsorption}} in {{Silica
  Pores}} via {{Minkowski Functionals}} and {{Molecular Electrostatic
  Moments}}, Energies 13~(22) (2020) 5976.
\newblock \href {https://doi.org/10.3390/en13225976}
  {\path{doi:10.3390/en13225976}}.

\bibitem{Hill1994}
T.~L. Hill, Thermodynamics of small systems, 2nd Edition, Dover, 1994.

\bibitem{Galteland2019}
O.~Galteland, S.~Kjelstrup, D.~Bedeaux, B.~Hafsjold, Pressure inside a
  nano-porous medium. the case of a single-phase fluid, Frontiers in Physics 7
  (2019) 150.

\bibitem{ebadiTopologicalEvolutionUnexplored2024}
M.~Ebadi, D.~Meisenheimer, D.~Wildenschild, J.~E. McClure, P.~Mostaghimi,
  R.~Armstrong, Topological evolution: An unexplored aspect of hysteresis for
  multiphase flow in porous media (Nov. 2024).
\newblock \href {https://doi.org/10.22541/au.173169343.37921907/v1}
  {\path{doi:10.22541/au.173169343.37921907/v1}}.

\bibitem{noether1918}
E.~Noether, Invariante variationsprobleme, Nachr. Ges. Wiss. Göttingen
  Math.-Phys. Kl. 23 (1918) 235--257.

\bibitem{noetherInvariantVariationProblems1971}
E.~Noether, Invariant variation problems, Transport Theory and Statistical
  Physics 1~(3) (1971) 186--207.
\newblock \href {https://doi.org/10.1080/00411457108231446}
  {\path{doi:10.1080/00411457108231446}}.

\bibitem{viscardyViscosityNewtonModern2010}
S.~Viscardy, Viscosity from {{Newton}} to {{Modern Non-equilibrium Statistical
  Mechanics}} (Nov. 2010).
\newblock \href {http://arxiv.org/abs/cond-mat/0601210}
  {\path{arXiv:cond-mat/0601210}}, \href
  {https://doi.org/10.48550/arXiv.cond-mat/0601210}
  {\path{doi:10.48550/arXiv.cond-mat/0601210}}.

\bibitem{liuInfluenceWettabilityPhase2018}
Z.~Liu, J.~E. McClure, R.~T. Armstrong, Influence of wettability on phase
  connectivity and electrical resistivity, Physical Review E 98~(4) (Oct.
  2018).
\newblock \href {https://doi.org/10.1103/physreve.98.043102}
  {\path{doi:10.1103/physreve.98.043102}}.

\bibitem{gloverConnectednessTheoryRelative2025}
P.~W.~J. Glover, W.~Wei, P.~Lorinczi, Connectedness theory of relative
  permeability, Transport in Porous Media accepted for publication (2025).

\bibitem{mukherjeeModelingRelativePermeability2025}
S.~Mukherjee, R.~T. Johns, Modeling of {{Relative Permeability Hysteresis Using
  Limited Experimental Data}} and {{Physically Constrained ANN}}, Transport in
  Porous Media 152~(6) (2025) 39.
\newblock \href {https://doi.org/10.1007/s11242-025-02178-0}
  {\path{doi:10.1007/s11242-025-02178-0}}.

\bibitem{lawLineTensionIts2017}
B.~M. Law, S.~P. McBride, J.~Y. Wang, H.~S. Wi, G.~Paneru, S.~Betelu,
  B.~Ushijima, Y.~Takata, B.~Flanders, F.~Bresme, H.~Matsubara, T.~Takiue,
  M.~Aratono, Line tension and its influence on droplets and particles at
  surfaces, Progress in Surface Science 92~(1) (2017) 1--39.
\newblock \href {https://doi.org/10.1016/j.progsurf.2016.12.002}
  {\path{doi:10.1016/j.progsurf.2016.12.002}}.

\bibitem{sunProbingEffectiveWetting2020i}
C.~Sun, J.~E. McClure, P.~Mostaghimi, A.~L. Herring, S.~Berg, R.~T. Armstrong,
  Probing {{Effective Wetting}} in {{Subsurface Systems}}, Geophysical Research
  Letters 47~(5) (2020) e2019GL086151.
\newblock \href {https://doi.org/10.1029/2019GL086151}
  {\path{doi:10.1029/2019GL086151}}.

\bibitem{morrowEffectsSurfaceRoughness1975}
N.~R. Morrow, The {{Effects}} of {{Surface Roughness On Contact}}: {{Angle With
  Special Reference}} to {{Petroleum Recovery}}, Journal of Canadian Petroleum
  Technology 14~(04) (Oct. 1975).
\newblock \href {https://doi.org/10.2118/75-04-04}
  {\path{doi:10.2118/75-04-04}}.

\bibitem{sunCharacterizationWettingUsing2020}
C.~Sun, J.~E. McClure, P.~Mostaghimi, A.~L. Herring, D.~E. Meisenheimer,
  D.~Wildenschild, S.~Berg, R.~T. Armstrong, Characterization of wetting using
  topological principles, Journal of Colloid and Interface Science 578 (2020)
  106--115.
\newblock \href {https://doi.org/10.1016/j.jcis.2020.05.076}
  {\path{doi:10.1016/j.jcis.2020.05.076}}.

\bibitem{sunUniversalDescriptionWetting2022a}
C.~Sun, J.~McClure, S.~Berg, P.~Mostaghimi, R.~T. Armstrong, Universal
  description of wetting on multiscale surfaces using integral geometry,
  Journal of Colloid and Interface Science 608 (2022) 2330--2338.
\newblock \href {https://doi.org/10.1016/j.jcis.2021.10.152}
  {\path{doi:10.1016/j.jcis.2021.10.152}}.

\bibitem{haghaniReviewWettabilityCharacterization2025}
R.~Haghani, C.~F. Berg, A {{Review}} on {{Wettability Characterization}} from
  {{3D Pore-Scale Images}}, Transport in Porous Media 152~(11) (2025) 93.
\newblock \href {https://doi.org/10.1007/s11242-025-02228-7}
  {\path{doi:10.1007/s11242-025-02228-7}}.

\bibitem{lenormandNumericalModelsExperiments1988}
R.~Lenormand, E.~Touboul, C.~Zarcone, Numerical models and experiments on
  immiscible displacements in porous media, Journal of Fluid Mechanics 189
  (1988) 165--187.
\newblock \href {https://doi.org/10.1017/S0022112088000953}
  {\path{doi:10.1017/S0022112088000953}}.

\bibitem{bluntMultiphaseFlowPermeable2017}
M.~J. Blunt, Multiphase {{Flow}} in {{Permeable Media}}: {{A Pore-Scale
  Perspective}}, Cambridge University Press, 2017.

\bibitem{payatakesDynamicsOilGanglia1982}
A.~C. Payatakes, Dynamics of {{Oil Ganglia During Immiscible Displacement}} in
  {{Water-Wet Porous Media}}, Annual Review of Fluid Mechanics 14~(1) (1982)
  365--393.
\newblock \href {https://doi.org/10.1146/annurev.fl.14.010182.002053}
  {\path{doi:10.1146/annurev.fl.14.010182.002053}}.

\bibitem{avraamGeneralizedRelativePermeability1995}
D.~G. Avraam, A.~C. Payatakes, Generalized relative permeability coefficients
  during steady-state two-phase flow in porous media, and correlation with the
  flow mechanisms, Transport in Porous Media 20~(1) (1995) 135--168.
\newblock \href {https://doi.org/10.1007/BF00616928}
  {\path{doi:10.1007/BF00616928}}.

\bibitem{bergMultiphaseFlowPorous2014a}
S.~Berg, R.~Armstrong, H.~Ott, A.~Georgiadis, S.~A. Klapp, A.~Schwing,
  R.~Neiteler, N.~Brussee, A.~Makurat, L.~Leu, F.~Enzmann, M.~Wolf, F.~Khan,
  M.~Kersten, S.~Irvine, M.~Stampanoni, Multiphase flow in porous rock imaged
  under dynamic flow conditions with fast x-ray computed microtomography,
  Petrophysics 55 (2014) 304--312.

\bibitem{andrewImagingDynamicMultiphase2015}
M.~Andrew, H.~Menke, M.~J. Blunt, B.~Bijeljic, The {{Imaging}} of {{Dynamic
  Multiphase Fluid Flow Using Synchrotron-Based X-ray Microtomography}} at
  {{Reservoir Conditions}}, Transport in Porous Media 110~(1) (2015) 1--24.
\newblock \href {https://doi.org/10.1007/s11242-015-0553-2}
  {\path{doi:10.1007/s11242-015-0553-2}}.

\bibitem{Steijn2009}
V.~van Steijn, C.~R. Kleijn, M.~T. Kreutzer, Flows around confined bubbles and
  their importance in triggering pinch-off, Physics Review Letters 103 (2009)
  214501.
\newblock \href {https://doi.org/10.1103/PhysRevLett.103.214501}
  {\path{doi:10.1103/PhysRevLett.103.214501}}.

\bibitem{zhang2022nonlinear}
Y.~Zhang, B.~Bijeljic, M.~J. Blunt, Nonlinear multiphase flow in hydrophobic
  porous media, Journal of Fluid Mechanics 934 (2022) R3.

\bibitem{suwandiRelativePermeabilityVariation2022}
N.~Suwandi, F.~Jiang, T.~Tsuji, Relative {{Permeability Variation Depending}}
  on {{Viscosity Ratio}} and {{Capillary Number}}, Water Resources Research
  58~(6) (Jun. 2022).
\newblock \href {https://doi.org/10.1029/2021wr031501}
  {\path{doi:10.1029/2021wr031501}}.

\bibitem{chenInertialEffectsProcess2019}
Y.~Chen, A.~J. Valocchi, Q.~Kang, H.~S. Viswanathan, Inertial {{Effects
  During}} the {{Process}} of {{Supercritical CO2 Displacing Brine}} in a
  {{Sandstone}}: {{Lattice Boltzmann Simulations Based}} on the
  {{Continuum-Surface-Force}} and {{Geometrical Wetting Models}}, Water
  Resources Research 55~(12) (2019) 11144--11165.
\newblock \href {https://doi.org/10.1029/2019WR025746}
  {\path{doi:10.1029/2019WR025746}}.

\bibitem{zacharoudiouPoreScaleModelingDrainage2020}
I.~Zacharoudiou, E.~S. Boek, J.~Crawshaw, Pore-{{Scale Modeling}} of {{Drainage
  Displacement Patterns}} in {{Association With Geological Sequestration}} of
  {{CO2}}, Water Resources Research 56~(11) (2020) e2019WR026332.
\newblock \href {https://doi.org/10.1029/2019WR026332}
  {\path{doi:10.1029/2019WR026332}}.

\bibitem{dicarloAcousticMeasurementsPorescale2003}
D.~A. DiCarlo, J.~I.~G. Cidoncha, C.~Hickey, Acoustic measurements of
  pore-scale displacements, Geophysical Research Letters 30~(17) (2003)
  2003GL017811.
\newblock \href {https://doi.org/10.1029/2003GL017811}
  {\path{doi:10.1029/2003GL017811}}.

\bibitem{armstrongInterfacialVelocitiesCapillary2013}
R.~T. Armstrong, S.~Berg, Interfacial velocities and capillary pressure
  gradients during {{Haines}} jumps, Physical Review E 88~(4) (2013) 043010.
\newblock \href {https://doi.org/10.1103/PhysRevE.88.043010}
  {\path{doi:10.1103/PhysRevE.88.043010}}.

\bibitem{ruckerConnectedPathwayFlow2015}
M.~R{\"u}cker, S.~Berg, R.~T. Armstrong, A.~Georgiadis, H.~Ott, A.~Schwing,
  R.~Neiteler, N.~Brussee, A.~Makurat, L.~Leu, M.~Wolf, F.~Khan, F.~Enzmann,
  M.~Kersten, From connected pathway flow to ganglion dynamics, Geophysical
  Research Letters 42~(10) (2015) 3888--3894.
\newblock \href {https://doi.org/10.1002/2015GL064007}
  {\path{doi:10.1002/2015GL064007}}.

\bibitem{spurinRedNoiseSteadyState2022}
C.~Spurin, M.~R{\"u}cker, M.~Moura, T.~Bultreys, G.~Garfi, S.~Berg, M.~J.
  Blunt, S.~Krevor, Red {{Noise}} in {{Steady}}-{{State Multiphase Flow}} in
  {{Porous Media}}, Water Resources Research 58~(7) (2022) e2022WR031947.
\newblock \href {https://doi.org/10.1029/2022WR031947}
  {\path{doi:10.1029/2022WR031947}}.

\bibitem{spurin2023pore}
C.~Spurin, G.~G. Roberts, C.~P. O’Malley, T.~Kurotori, S.~Krevor, M.~J.
  Blunt, H.~Tchelepi, Pore-scale fluid dynamics resolved in pressure
  fluctuations at the darcy scale, Geophysical Research Letters 50~(18) (2023)
  e2023GL104473.

\bibitem{onuchicTHEORYPROTEINFOLDING1997}
J.~N. Onuchic, Z.~{Luthey-Schulten}, P.~G. Wolynes, {{THEORY OF PROTEIN
  FOLDING}}: {{The Energy Landscape Perspective}}, Annual Review of Physical
  Chemistry 48~(1) (1997) 545--600.
\newblock \href {https://doi.org/10.1146/annurev.physchem.48.1.545}
  {\path{doi:10.1146/annurev.physchem.48.1.545}}.

\bibitem{bhandarEnergyLandscapesBistability2004}
A.~S. Bhandar, M.~J. Vogel, P.~H. Steen, Energy landscapes and bistability to
  finite-amplitude disturbances for the capillary bridge, Physics of Fluids
  16~(8) (2004) 3063--3069.
\newblock \href {https://doi.org/10.1063/1.1761510}
  {\path{doi:10.1063/1.1761510}}.

\bibitem{debenedettiSupercooledLiquidsGlass2001}
P.~G. Debenedetti, F.~H. Stillinger, Supercooled liquids and the glass
  transition, Nature 410~(6825) (2001) 259--267.
\newblock \href {https://doi.org/10.1038/35065704}
  {\path{doi:10.1038/35065704}}.

\bibitem{debenedettiEquationStateEnergy1999}
P.~G. Debenedetti, F.~H. Stillinger, T.~M. Truskett, C.~J. Roberts, The
  {{Equation}} of {{State}} of an {{Energy Landscape}}, The Journal of Physical
  Chemistry B 103~(35) (1999) 7390--7397.
\newblock \href {https://doi.org/10.1021/jp991384m}
  {\path{doi:10.1021/jp991384m}}.

\bibitem{cueto-felguerosoDiscretedomainDescriptionMultiphase2016}
L.~Cueto-Felgueroso, R.~Juanes, A discrete-domain description of multiphase
  flow in porous media: {{Rugged}} energy landscapes and the origin of
  hysteresis, Geophysical Research Letters 43~(4) (2016) 1615--1622.
\newblock \href {https://doi.org/10.1002/2015GL067015}
  {\path{doi:10.1002/2015GL067015}}.

\bibitem{wangTrappedLiquidDrop2013}
Z.~Wang, C.-C. Chang, S.-J. Hong, Y.-J. Sheng, H.-K. Tsao, Trapped liquid drop
  in a microchannel: {{Multiple}} stable states, Physical Review E 87~(6) (Jun.
  2013).
\newblock \href {https://doi.org/10.1103/physreve.87.062401}
  {\path{doi:10.1103/physreve.87.062401}}.

\bibitem{crandallDistributionOccurrenceLocalizedbursts2009}
D.~Crandall, G.~Ahmadi, M.~Ferer, D.~H. Smith, Distribution and occurrence of
  localized-bursts in two-phase flow through porous media, Physica A:
  Statistical Mechanics and its Applications 388~(5) (2009) 574--584.
\newblock \href {https://doi.org/10.1016/j.physa.2008.11.010}
  {\path{doi:10.1016/j.physa.2008.11.010}}.

\bibitem{meisenheimerPredictingEffectRelaxation2021}
D.~E. Meisenheimer, D.~Wildenschild, Predicting the {{Effect}} of
  {{Relaxation}} on {{Interfacial Area Development}} in {{Multiphase Flow}},
  Water Resources Research 57~(3) (2021) e2020WR028770.
\newblock \href {https://doi.org/10.1029/2020WR028770}
  {\path{doi:10.1029/2020WR028770}}.

\bibitem{hassanizadehDynamicEffectCapillary2002}
S.~M. Hassanizadeh, M.~A. Celia, H.~K. Dahle, Dynamic {{Effect}} in the
  {{Capillary Pressure}}--{{Saturation Relationship}} and its {{Impacts}} on
  {{Unsaturated Flow}}, Vadose Zone Journal 1~(1) (2002) 38--57.
\newblock \href {https://doi.org/10.2136/vzj2002.3800}
  {\path{doi:10.2136/vzj2002.3800}}.

\bibitem{schluterTimeScalesRelaxation2017}
S.~Schl{\"u}ter, S.~Berg, T.~Li, H.-J. Vogel, D.~Wildenschild, Time scales of
  relaxation dynamics during transient conditions in two-phase flow, Water
  Resources Research 53~(6) (2017) 4709--4724.
\newblock \href {https://doi.org/10.1002/2016WR019815}
  {\path{doi:10.1002/2016WR019815}}.

\bibitem{shiozawaUnexpectedWaterContent2004}
S.~Shiozawa, H.~Fujimaki, Unexpected water content profiles under flux-limited
  one-dimensional downward infiltration in initially dry granular media, Water
  Resources Research 40~(7) (2004) 2003WR002197.
\newblock \href {https://doi.org/10.1029/2003WR002197}
  {\path{doi:10.1029/2003WR002197}}.

\bibitem{dicarloExperimentalMeasurementsSaturation2004}
D.~A. DiCarlo, Experimental measurements of saturation overshoot on
  infiltration, Water Resources Research 40~(4) (2004) 2003WR002670.
\newblock \href {https://doi.org/10.1029/2003WR002670}
  {\path{doi:10.1029/2003WR002670}}.

\bibitem{cueto-felguerosoNonlocalInterfaceDynamics2008}
L.~{Cueto-Felgueroso}, R.~Juanes, Nonlocal {{Interface Dynamics}} and {{Pattern
  Formation}} in {{Gravity-Driven Unsaturated Flow}} through {{Porous Media}},
  Physical Review Letters 101~(24) (2008) 244504.
\newblock \href {https://doi.org/10.1103/PhysRevLett.101.244504}
  {\path{doi:10.1103/PhysRevLett.101.244504}}.

\bibitem{steinleInfluenceInitialConditions2016}
R.~Steinle, R.~Hilfer, Influence of {{Initial Conditions}} on {{Propagation}},
  {{Growth}} and {{Decay}} of {{Saturation Overshoot}}, Transport in Porous
  Media 111~(2) (2016) 369--380.
\newblock \href {https://doi.org/10.1007/s11242-015-0598-2}
  {\path{doi:10.1007/s11242-015-0598-2}}.

\bibitem{einavHydrodynamicsNonEquilibriumSoil2023}
I.~Einav, M.~Liu, Hydrodynamics of {{Non}}-{{Equilibrium Soil Water
  Retention}}, Water Resources Research 59~(1) (2023) e2022WR033409.
\newblock \href {https://doi.org/10.1029/2022WR033409}
  {\path{doi:10.1029/2022WR033409}}.

\bibitem{vanduijnTravellingWaveSolutions2018}
C.~Van~Duijn, K.~Mitra, I.~Pop, Travelling wave solutions for the {{Richards}}
  equation incorporating non-equilibrium effects in the capillarity pressure,
  Nonlinear Analysis: Real World Applications 41 (2018) 232--268.
\newblock \href {https://doi.org/10.1016/j.nonrwa.2017.10.015}
  {\path{doi:10.1016/j.nonrwa.2017.10.015}}.

\bibitem{mitraWettingFrontsUnsaturated2019}
K.~Mitra, C.~Van~Duijn, Wetting fronts in unsaturated porous media: {{The}}
  combined case of hysteresis and dynamic capillary pressure, Nonlinear
  Analysis: Real World Applications 50 (2019) 316--341.
\newblock \href {https://doi.org/10.1016/j.nonrwa.2019.05.005}
  {\path{doi:10.1016/j.nonrwa.2019.05.005}}.

\bibitem{mitraFrontsTwophasePorous2020a}
K.~Mitra, T.~K{\"o}ppl, I.~S. Pop, C.~J. Van~Duijn, R.~Helmig, Fronts in
  two-phase porous media flow problems: {{The}} effects of hysteresis and
  dynamic capillarity, Studies in Applied Mathematics 144~(4) (2020) 449--492.
\newblock \href {https://doi.org/10.1111/sapm.12304}
  {\path{doi:10.1111/sapm.12304}}.

\bibitem{unsalBubbleSnapoffCapillaryBack2009}
E.~Unsal, G.~Mason, N.~R. Morrow, D.~W. Ruth, Bubble {{Snap-off}} and
  {{Capillary-Back Pressure}} during {{Counter-Current Spontaneous Imbibition}}
  into {{Model Pores}}, Langmuir 25~(6) (2009) 3387--3395.
\newblock \href {https://doi.org/10.1021/la803568a}
  {\path{doi:10.1021/la803568a}}.

\bibitem{unsalCoCountercurrentSpontaneous2007}
E.~Unsal, G.~Mason, D.~Ruth, N.~Morrow, Co- and counter-current spontaneous
  imbibition into groups of capillary tubes with lateral connections permitting
  cross-flow, Journal of Colloid and Interface Science 315~(1) (2007) 200--209.
\newblock \href {https://doi.org/10.1016/j.jcis.2007.06.070}
  {\path{doi:10.1016/j.jcis.2007.06.070}}.

\bibitem{unsalCocurrentCountercurrentImbibition2007}
E.~Unsal, G.~Mason, N.~Morrow, D.~Ruth, Co-current and counter-current
  imbibition in independent tubes of non-axisymmetric geometry, Journal of
  Colloid and Interface Science 306~(1) (2007) 105--117.
\newblock \href {https://doi.org/10.1016/j.jcis.2006.10.042}
  {\path{doi:10.1016/j.jcis.2006.10.042}}.

\bibitem{wangInherentErrorsCurrent2025}
G.~Wang, A.~Beteta, K.~S. Sorbie, E.~J. Mackay, Inherent {{Errors}} in
  {{Current Core-Flooding Relative Permeability Data}} for {{Modelling
  Underground Hydrogen Storage}}, InterPore Journal 2~(1) (2025) IPJ260225--6.
\newblock \href {https://doi.org/10.69631/ipj.v2i1nr42}
  {\path{doi:10.69631/ipj.v2i1nr42}}.

\bibitem{betetaRoleImmiscibleFingering2022}
A.~Beteta, K.~S. Sorbie, K.~McIver, G.~Johnson, R.~Gasimov, W.~Van~Zeil, The
  {{Role}} of {{Immiscible Fingering}} on the {{Mechanism}} of {{Secondary}}
  and {{Tertiary Polymer Flooding}} of {{Viscous Oil}}, Transport in Porous
  Media 143~(2) (2022) 343--372.
\newblock \href {https://doi.org/10.1007/s11242-022-01774-8}
  {\path{doi:10.1007/s11242-022-01774-8}}.

\bibitem{larkiViscousFingeringDynamics2025}
S.~A. Larki, A.~Skauge, K.~Sorbie, E.~Mackay, Viscous {{Fingering Dynamics}} in
  {{Carbon Capture}}: {{Impact}} of {{CO2 Solubility}} and {{Relative
  Permeability Selection}} in {{CCS}}, in: {{IOR}}+ 2025 - 23rd {{European
  Symposium}} on {{IOR}}, European Association of Geoscientists \& Engineers,
  Edinburgh, Scotland, United Kingdom,, 2025, pp. 1--13.
\newblock \href {https://doi.org/10.3997/2214-4609.202531064}
  {\path{doi:10.3997/2214-4609.202531064}}.

\bibitem{wangHowUsefulAre2025}
G.~Wang, A.~Beteta, K.~Sorbie, E.~Mackay, How {{Useful Are}} the {{Existing
  Core-Flooding Derived Hydrogen-Water Relative Permeabilities}} for
  {{Modelling Underground Hydrogen Storage}}?, in: 86th {{EAGE Annual
  Conference}} \& {{Exhibition}}, Vol. 2025, European Association of
  Geoscientists \& Engineers, 2025, pp. 1--5.
\newblock \href {https://doi.org/10.3997/2214-4609.202510777}
  {\path{doi:10.3997/2214-4609.202510777}}.

\bibitem{scotsonXrayComputedTomography2021}
C.~P. Scotson, S.~J. Duncan, K.~A. Williams, S.~A. Ruiz, T.~Roose, X-ray
  computed tomography imaging of solute movement through ridged and flat plant
  systems, European Journal of Soil Science 72~(1) (2021) 198--214.
\newblock \href {https://doi.org/10.1111/ejss.12985}
  {\path{doi:10.1111/ejss.12985}}.

\bibitem{ottWavelengthViscousUnstableCO2Brine2025}
H.~Ott, O.~Amrollahinasab, S.~Berg, The {{Wavelength}} of {{Viscous-Unstable
  CO2-Brine Displacement}}, SSRN Electronic Journal (2025).
\newblock \href {https://doi.org/10.2139/ssrn.5068154}
  {\path{doi:10.2139/ssrn.5068154}}.

\bibitem{bergStabilityCO2brineImmiscible2012}
S.~Berg, H.~Ott, Stability of {{CO2}}--brine immiscible displacement,
  International Journal of Greenhouse Gas Control 11 (2012) 188--203.
\newblock \href {https://doi.org/10.1016/j.ijggc.2012.07.001}
  {\path{doi:10.1016/j.ijggc.2012.07.001}}.

\bibitem{aryanaNonequilibriumEffectsMultiphase2013}
S.~A. Aryana, A.~R. Kovscek, Nonequilibrium {{Effects}} and {{Multiphase Flow}}
  in {{Porous Media}}, Transport in Porous Media 97~(3) (2013) 373--394.
\newblock \href {https://doi.org/10.1007/s11242-013-0129-y}
  {\path{doi:10.1007/s11242-013-0129-y}}.

\bibitem{wangExtensionDarcysLaw2019}
Y.~Wang, S.~A. Aryana, M.~B. Allen, An extension of {{Darcy}}'s law
  incorporating dynamic length scales, Advances in Water Resources 129 (2019)
  70--79.
\newblock \href {https://doi.org/10.1016/j.advwatres.2019.05.010}
  {\path{doi:10.1016/j.advwatres.2019.05.010}}.

\bibitem{renBayesianModelSelection2017}
G.~Ren, J.~Rafiee, S.~A. Aryana, R.~M. Younis, A {{Bayesian}} model selection
  analysis of equilibrium and nonequilibrium models for multiphase flow in
  porous media, International Journal of Multiphase Flow 89 (2017) 313--320.
\newblock \href {https://doi.org/10.1016/j.ijmultiphaseflow.2016.11.006}
  {\path{doi:10.1016/j.ijmultiphaseflow.2016.11.006}}.

\bibitem{tallakstad2009steady}
K.~T. Tallakstad, H.~A. Knudsen, T.~Ramstad, G.~L{\o}voll, K.~J. M{\aa}l{\o}y,
  R.~Toussaint, E.~G. Flekk{\o}y, Steady-state two-phase flow in porous media:
  statistics and transport properties, Physical Review Letters 102~(7) (2009)
  074502.

\bibitem{tallakstad2009steadyb}
K.~T. Tallakstad, G.~L{\o}voll, H.~A. Knudsen, T.~Ramstad, E.~G. Flekk{\o}y,
  K.~J. M{\aa}l{\o}y, Steady-state, simultaneous two-phase flow in porous
  media: An experimental study, Physical Review E 80~(3) (2009) 036308.

\bibitem{rassi2011nuclear}
E.~M. Rassi, S.~L. Codd, J.~D. Seymour, Nuclear magnetic resonance
  characterization of the stationary dynamics of partially saturated media
  during steady-state infiltration flow, New Journal of Physics 13~(1) (2011)
  015007.

\bibitem{sinha2012effective}
S.~Sinha, A.~Hansen, Effective rheology of immiscible two-phase flow in porous
  media, Europhysics Letters 99~(4) (2012) 44004.

\bibitem{sinha2021fluid}
S.~Sinha, M.~A. Gjennestad, M.~Vassvik, A.~Hansen, Fluid meniscus algorithms
  for dynamic pore-network modeling of immiscible two-phase flow in porous
  media, Frontiers in Physics 8 (2021) 548497.
\newblock \href {https://doi.org/10.3389/fphy.2020.548497}
  {\path{doi:10.3389/fphy.2020.548497}}.

\bibitem{roy2024immiscible}
S.~Roy, S.~Sinha, A.~Hansen, Immiscible two-phase flow in porous media:
  Effective rheology in the continuum limit, Transport in Porous Media 151~(6)
  (2024) 1295--1311.

\bibitem{sinhaEffectiveRheologyTwoPhase2017}
S.~Sinha, A.~T. Bender, M.~Danczyk, K.~Keepseagle, C.~A. Prather, J.~M. Bray,
  L.~W. Thrane, J.~D. Seymour, S.~L. Codd, A.~Hansen, Effective {{Rheology}} of
  {{Two-Phase Flow}} in {{Three-Dimensional Porous Media}}: {{Experiment}} and
  {{Simulation}}, Transport in Porous Media 119~(1) (2017) 77--94.
\newblock \href {https://doi.org/10.1007/s11242-017-0874-4}
  {\path{doi:10.1007/s11242-017-0874-4}}.

\bibitem{sinhaEffectiveRheologyBubbles2013}
S.~Sinha, A.~Hansen, D.~Bedeaux, S.~Kjelstrup, Effective rheology of bubbles
  moving in a capillary tube, Physical Review E 87~(2) (2013) 025001.
\newblock \href {https://doi.org/10.1103/PhysRevE.87.025001}
  {\path{doi:10.1103/PhysRevE.87.025001}}.

\bibitem{xu2014non}
X.~Xu, X.~Wang, Non-darcy behavior of two-phase channel flow, Physical Review E
  90~(2) (2014) 023010.

\bibitem{roy2019effectiveb}
S.~Roy, A.~Hansen, S.~Sinha, Effective rheology of two-phase flow in a
  capillary fiber bundle model, Frontiers in Physics 7 (2019) 92.

\bibitem{lanza2022non}
F.~Lanza, A.~Rosso, L.~Talon, A.~Hansen, Non-newtonian rheology in a capillary
  tube with varying radius, Transport in Porous Media 145~(1) (2022) 245--269.

\bibitem{cheon2023steady}
H.~L. Cheon, H.~Fyhn, A.~Hansen, {\O}.~Wilhelmsen, S.~Sinha, Steady-state
  two-phase flow of compressible and incompressible fluids in a capillary tube
  of varying radius, Transport in Porous Media 147~(1) (2023) 15--33.

\bibitem{aursjo2014film}
O.~Aursj{\o}, M.~Erpelding, K.~T. Tallakstad, E.~G. Flekk{\o}y, A.~Hansen,
  K.~J. M{\aa}l{\o}y, Film flow dominated simultaneous flow of two viscous
  incompressible fluids through a porous medium, Frontiers in physics 2 (2014)
  63.

\bibitem{zhang2021quantification}
Y.~Zhang, B.~Bijeljic, Y.~Gao, Q.~Lin, M.~J. Blunt, Quantification of nonlinear
  multiphase flow in porous media, Geophysical Research Letters 48~(5) (2021)
  e2020GL090477.

\bibitem{yiotis2013blob}
A.~Yiotis, L.~Talon, D.~Salin, Blob population dynamics during immiscible
  two-phase flows in reconstructed porous media, Physical Review
  E—Statistical, Nonlinear, and Soft Matter Physics 87~(3) (2013) 033001.
\newblock \href {https://doi.org/10.1103/PhysRevE.87.033001}
  {\path{doi:10.1103/PhysRevE.87.033001}}.

\bibitem{yiotis2019nonlinear}
A.~Yiotis, A.~Dollari, M.~Kainourgiakis, D.~Salin, L.~Talon, Nonlinear darcy
  flow dynamics during ganglia stranding and mobilization in heterogeneous
  porous domains, Physical Review Fluids 4~(11) (2019) 114302.
\newblock \href {https://doi.org/10.1103/PhysRevFluids.4.114302}
  {\path{doi:10.1103/PhysRevFluids.4.114302}}.

\bibitem{botticiniOriginPressureflowNonlinearity2025}
P.~Botticini, D.~Picchi, S.~Sinha, A.~Hansen, Origin of pressure-flow
  non-linearity in two-phase intermittent flow in porous media (Oct. 2025).
\newblock \href {http://arxiv.org/abs/2510.12588} {\path{arXiv:2510.12588}},
  \href {https://doi.org/10.48550/arXiv.2510.12588}
  {\path{doi:10.48550/arXiv.2510.12588}}.

\bibitem{lovoll2004growth}
G.~L{\o}voll, Y.~M{\'e}heust, R.~Toussaint, J.~Schmittbuhl, K.~J. M{\aa}l{\o}y,
  Growth activity during fingering in a porous hele-shaw cell, Physical Review
  E—Statistical, Nonlinear, and Soft Matter Physics 70~(2) (2004) 026301.
\newblock \href {https://doi.org/10.1103/PhysRevE.70.026301}
  {\path{doi:10.1103/PhysRevE.70.026301}}.

\bibitem{toussaint2005influence}
R.~Toussaint, G.~L{\o}voll, Y.~M{\'e}heust, K.~J. M{\aa}l{\o}y, J.~Schmittbuhl,
  Influence of pore-scale disorder on viscous fingering during drainage,
  Europhysics Letters 71~(4) (2005) 583.
\newblock \href {https://doi.org/10.1209/epl/i2005-10136-9}
  {\path{doi:10.1209/epl/i2005-10136-9}}.

\bibitem{barabasi1995fractal}
A.-L. Barab{\'a}si, H.~E. Stanley, Fractal concepts in surface growth,
  Cambridge university press, 1995.

\bibitem{sinha2024disorder}
S.~Sinha, Y.~M{\'e}heust, H.~Fyhn, S.~Roy, A.~Hansen, Disorder-induced
  non-linear growth of fingers in immiscible two-phase flow in porous media,
  Physics of Fluids 36~(3) (2024).
\newblock \href {https://doi.org/10.1063/5.0193570}
  {\path{doi:10.1063/5.0193570}}.

\bibitem{burelbachUnifiedDescriptionColloidal2018}
J.~Burelbach, D.~Frenkel, I.~Pagonabarraga, E.~Eiser, A unified description of
  colloidal thermophoresis, The European Physical Journal E 41~(1) (2018) 7.
\newblock \href {https://doi.org/10.1140/epje/i2018-11610-3}
  {\path{doi:10.1140/epje/i2018-11610-3}}.

\bibitem{Kjelstrup2020}
S.~Kjelstrup, D.~Bedeaux, Non-equilibrium thermodynamics for heterogeneous
  systems, 2nd Edition, World Scientific, 2020.

\bibitem{Green1954}
M.~S. Green, Markoff random processes and the statistical mechanics of
  time-dependent phenomena. {II. I}rreversible processes in fluids, J. Chem.
  Phys. 22 (1954) 398--413.

\bibitem{Kubo1966}
R.~Kubo, The fluctuation-dissipation theorem, Rep. Prog. Phys. 29 (1966) 255.

\bibitem{Rauter2020}
M.~Rauter, O.~Galteland, M.~Erdös, M.~O. A., T.~Vlugt, S.~K. Schnell,
  D.~Bedeaux, S.~Kjelstrup, Two-phase equilibrium conditions in nanopores,
  Nanomaterials 10 (2020) 608.
\newblock \href {https://doi.org/10.3390/nano10040608}
  {\path{doi:10.3390/nano10040608}}.

\bibitem{Gray1998}
W.~Gray, S.~Hassanizadeh, Macroscale continuum mechanics for multiphase
  porous-media flow including phases, interfaces, common lines and commpon
  points, Adv. Water Resources 21 (1998) 261.
\newblock \href {https://doi.org/10.1016/S0309- 1708(96)00063-2}
  {\path{doi:10.1016/S0309- 1708(96)00063-2}}.

\bibitem{Callen1951}
H.~B. Callen, T.~A. Welton, Irreversibility and generalized noise, Phys. Rev.
  83 (1951) 34.

\bibitem{Liu2011}
X.~Liu, T.~Vlugt, A.~Bardow, {Predictive Darken equation for Maxwell-Stefan
  diffusivities in multicomponent mixtures}, Ind. Eng. Chem. Res. 50 (2011)
  10350--10358.

\bibitem{Bresme2014}
F.~Bresme, J.~Armstrong, Note: Local thermal conductivities from boundary
  driven non-equilibrium molecular dynamics simulations, J. Chem. Phys. (2014)
  016102.

\bibitem{Winkler2020}
M.~Winkler, M.~A. Gjennestad, D.~Bedeaux, S.~Kjelstrup, R.~Cabriolu, A.~Hansen,
  Onsager-symmetry obeyed in athermal mesoscopic systems: Two-phase flow in
  porous media, Frontiers in Physics 8 (2020) 60.
\newblock \href {https://doi.org/10.3389/fphy.2020.00060}
  {\path{doi:10.3389/fphy.2020.00060}}.

\bibitem{Alfazazi2024}
U.~Alfazazi, D.~Bedeaux, S.~Kjelstrup, M.~Moura, M.~Ebadi, P.~Mostaghimi, J.~E.
  McClure, R.~T. Armstrong, Interpreting pore-scale fluctuations{: P}redicting
  transport coefficients in multiphase flow through porous media using the
  {Green-Kubo} formulation - an experimental investigation, Physics of Fluids
  (2024).

\bibitem{Umar2025}
U.~Alfazazi, M.~Moura, Y.~D. Wang, D.~Bedeaux, S.~Kjelstrup, P.~Mostaghimi,
  R.~T. Armstrong1, Using the fluctuation-dissipation theorem to measure total
  phase mobility during fractional flow experiments, The 38th International
  Symposium of the Society of Core Analysts (2025) 1040.

\bibitem{Onsager1931a}
L.~Onsager, Reciprocal relations in irreversible processes {I}, Phys.Rev. 37
  (1931) 405--426.
\newblock \href {https://doi.org/10.1103/PhysRev.37.405}
  {\path{doi:10.1103/PhysRev.37.405}}.

\bibitem{Onsager1931b}
L.~Onsager, Reciprocal relations in irreversible processes {II}, Phys.Rev. 38
  (1931) 2265--2279.
\newblock \href {https://doi.org/10.1103/PhysRev.38.2265}
  {\path{doi:10.1103/PhysRev.38.2265}}.

\bibitem{Moura2024}
M.~Moura, D.~Bedeaux, R.~T. Armstrong, S.~Kjelstrup, Fluctuation-dissipation
  theorems and the measurement of the onsager coefficients for two-phase flow
  in poous media, phys.fluid.-dyn (2024) {arXive}2410.03661v1.

\bibitem{Rubi}
J.~Rubi, S.~Kjelstrup, Mesoscopic nonequilibrium thermodynamics gives the same
  thermodynamic basis to {Butler-Volmer and Nernst} equations, J. Phys. Chem. B
  107 (2003) 13471--13477.

\bibitem{hafskjoldSoretSeparationThermoosmosis2022}
B.~Hafskjold, D.~Bedeaux, S.~Kjelstrup, {\O}.~Wilhelmsen, Soret separation and
  thermo-osmosis in porous media, The European Physical Journal E 45~(5) (2022)
  41.
\newblock \href {https://doi.org/10.1140/epje/s10189-022-00194-2}
  {\path{doi:10.1140/epje/s10189-022-00194-2}}.

\bibitem{dechalendarPorescaleModellingOstwald2018}
J.~A. De~Chalendar, C.~Garing, S.~M. Benson, Pore-scale modelling of
  {{Ostwald}} ripening, Journal of Fluid Mechanics 835 (2018) 363--392.
\newblock \href {https://doi.org/10.1017/jfm.2017.720}
  {\path{doi:10.1017/jfm.2017.720}}.

\bibitem{xuEgalitarianismBubblesPorous2017}
K.~Xu, R.~Bonnecaze, M.~Balhoff, Egalitarianism among {{Bubbles}} in {{Porous
  Media}}: {{An Ostwald Ripening Derived Anticoarsening Phenomenon}}, Physical
  Review Letters 119~(26) (2017) 264502.
\newblock \href {https://doi.org/10.1103/PhysRevLett.119.264502}
  {\path{doi:10.1103/PhysRevLett.119.264502}}.

\bibitem{BergDeterminationCriticalGas2020}
S.~Berg, Y.~Gao, A.~Georgiadis, N.~Brussee, A.~Coorn, H.~{van Dder Linde},
  J.~Dietderich, F.~O. Alpak, D.~Eriksen, M.~{Mooijer-van Den Heuvel},
  J.~Southwick, M.~Appel, O.~B. Wilson, Determination of {{Critical Gas
  Saturation}} by {{Micro-CT}}, Petrophysics -- The SPWLA Journal of Formation
  Evaluation and Reservoir Description 61~(2) (2020) 133--150.
\newblock \href {https://doi.org/10.30632/PJV61N2-2020a1}
  {\path{doi:10.30632/PJV61N2-2020a1}}.

\bibitem{gaoCapillarityPhasemobilityHydrocarbon2021}
Y.~Gao, A.~Georgiadis, N.~Brussee, A.~Coorn, H.~Van Der~Linde, J.~Dietderich,
  F.~O. Alpak, D.~Eriksen, M.~{Mooijer-van Den Heuvel}, M.~Appel, T.~Sorop,
  O.~B. Wilson, S.~Berg, Capillarity and phase-mobility of a hydrocarbon
  gas--liquid system, Oil \& Gas Science and Technology -- Revue d'IFP Energies
  nouvelles 76 (2021) 43.
\newblock \href {https://doi.org/10.2516/ogst/2021025}
  {\path{doi:10.2516/ogst/2021025}}.

\bibitem{wangCapillaryEquilibriumBubbles2021}
C.~Wang, Y.~Mehmani, K.~Xu, Capillary equilibrium of bubbles in porous media,
  Proceedings of the National Academy of Sciences 118~(17) (2021) e2024069118.
\newblock \href {https://doi.org/10.1073/pnas.2024069118}
  {\path{doi:10.1073/pnas.2024069118}}.

\bibitem{goodarziTrappingHysteresisOstwald2024}
S.~Goodarzi, Y.~Zhang, S.~Foroughi, B.~Bijeljic, M.~J. Blunt, Trapping,
  hysteresis and {{Ostwald}} ripening in hydrogen storage: {{A}} pore-scale
  imaging study, International Journal of Hydrogen Energy 56 (2024) 1139--1151.
\newblock \href {https://doi.org/10.1016/j.ijhydene.2023.12.029}
  {\path{doi:10.1016/j.ijhydene.2023.12.029}}.

\bibitem{wangTimeAndSpaceAveragingApplied2024b}
T.~Wang, J.~E. McClure, Y.~Da~Wang, S.~Berg, C.~Chen, P.~Mostaghimi, R.~T.
  Armstrong, Time-{{And}}-{{Space Averaging Applied}} to {{Intermittent
  Multiphase Flow Experiments}}, Water Resources Research 60~(6) (2024)
  e2023WR036577.
\newblock \href {https://doi.org/10.1029/2023WR036577}
  {\path{doi:10.1029/2023WR036577}}.

\bibitem{lenormandMechanismsDisplacementOne1983}
R.~Lenormand, C.~Zarcone, A.~Sarr, Mechanisms of the displacement of one fluid
  by another in a network of capillary ducts, Journal of Fluid Mechanics
  135~(-1) (1983) 337.
\newblock \href {https://doi.org/10.1017/S0022112083003110}
  {\path{doi:10.1017/S0022112083003110}}.

\bibitem{reynolds2017dynamic}
C.~A. Reynolds, H.~Menke, M.~Andrew, M.~J. Blunt, S.~Krevor, Dynamic fluid
  connectivity during steady-state multiphase flow in a sandstone, Proceedings
  of the National Academy of Sciences 114~(31) (2017) 8187--8192.

\bibitem{standnesNovelRelativePermeability2017}
D.~C. Standnes, S.~Evje, P.~{\O}. Andersen, A {{Novel Relative Permeability
  Model Based}} on {{Mixture Theory Approach Accounting}} for
  {{Solid}}--{{Fluid}} and {{Fluid}}--{{Fluid Interactions}}, Transport in
  Porous Media 119~(3) (2017) 707--738.
\newblock \href {https://doi.org/10.1007/s11242-017-0907-z}
  {\path{doi:10.1007/s11242-017-0907-z}}.

\bibitem{mcclureLBPMSoftwarePackage2021}
J.~E. McClure, Z.~Li, M.~Berrill, T.~Ramstad, The {{LBPM}} software package for
  simulating multiphase flow on digital images of porous rocks, Computational
  Geosciences 25~(3) (2021) 871--895.
\newblock \href {https://doi.org/10.1007/s10596-020-10028-9}
  {\path{doi:10.1007/s10596-020-10028-9}}.

\bibitem{anderson1972more}
P.~W. Anderson, More is different: Broken symmetry and the nature of the
  hierarchical structure of science., Science 177~(4047) (1972) 393--396.

\bibitem{shannon1948mathematical}
C.~E. Shannon, A mathematical theory of communication, The Bell system
  technical journal 27~(3) (1948) 379--423.
\newblock \href {https://doi.org/10.1002/j.1538-7305.1948.tb01338.x}
  {\path{doi:10.1002/j.1538-7305.1948.tb01338.x}}.

\bibitem{jaynes1957information}
E.~T. Jaynes, Information theory and statistical mechanics, Physical review
  106~(4) (1957) 620.
\newblock \href {https://doi.org/10.1103/PhysRev.106.620}
  {\path{doi:10.1103/PhysRev.106.620}}.

\bibitem{de1995philosophical}
M.~De~Laplace, A philosophical essay on probabilities, Courier Corporation,
  1995.

\bibitem{sinha2025thermodynamics}
A.~Hansen, S.~Sinha, Thermodynamics-like formalism for immiscible and
  incompressible two-phase flow in porous media, Entropy 27~(2) (2025) 121.
\newblock \href {https://doi.org/10.3390/e27020121}
  {\path{doi:10.3390/e27020121}}.

\bibitem{bear2012introduction}
J.~Bear, Y.~Bachmat, Introduction to modeling of transport phenomena in porous
  media, Vol.~4, Springer Science \& Business Media, 2012.

\bibitem{fyhnLocalStatisticsImmiscible2023}
H.~Fyhn, S.~Sinha, A.~Hansen, Local statistics of immiscible and incompressible
  two-phase flow in porous media, Physica A: Statistical Mechanics and its
  Applications 616 (2023) 128626.
\newblock \href {https://doi.org/10.1016/j.physa.2023.128626}
  {\path{doi:10.1016/j.physa.2023.128626}}.

\bibitem{baxter1985exactly}
R.~J. Baxter, Exactly solved models in statistical mechanics, in: Integrable
  systems in statistical mechanics, World Scientific, 1985, pp. 5--63.

\bibitem{binder1986spin}
K.~Binder, A.~P. Young, Spin glasses: Experimental facts, theoretical concepts,
  and open questions, Reviews of Modern physics 58~(4) (1986) 801.
\newblock \href {https://doi.org/10.1103/RevModPhys.58.801}
  {\path{doi:10.1103/RevModPhys.58.801}}.

\bibitem{sherrington1975solvable}
D.~Sherrington, S.~Kirkpatrick, Solvable model of a spin-glass, Physical review
  letters 35~(26) (1975) 1792.
\newblock \href {https://doi.org/10.1103/PhysRevLett.35.1792}
  {\path{doi:10.1103/PhysRevLett.35.1792}}.

\bibitem{sinha2025sk}
S.~Sinha, H.~Carmona, J.~S. Andrade~Jr., A.~Hansen, Mapping immiscible
  two-phase flow in porous media onto a spin glass (2025).

\bibitem{xuStatisticalMechanicsUnsaturated2015}
J.~Xu, M.~Y. Louge, Statistical mechanics of unsaturated porous media, Physical
  Review E 92~(6) (2015) 062405.
\newblock \href {https://doi.org/10.1103/PhysRevE.92.062405}
  {\path{doi:10.1103/PhysRevE.92.062405}}.

\bibitem{ising1925beitrag}
E.~Ising, Beitrag zur theorie des ferromagnetismus, Zeitschrift f{\"u}r Physik
  31~(1) (1925) 253--258.

\bibitem{onsager1944crystal}
L.~Onsager, Crystal statistics. i. a two-dimensional model with an
  order-disorder transition, Physical review 65~(3-4) (1944) 117.

\bibitem{hauge2005lars}
E.~H. Hauge, Lars onsager: The nth student who became one of the greatest
  scientists of the 20th century, Energy 30~(6) (2005) 787--793.
\newblock \href {https://doi.org/10.1016/j.energy.2004.04.009}
  {\path{doi:10.1016/j.energy.2004.04.009}}.

\bibitem{hermundstad2025shannon}
A.~Hermundstad, Shannon entropy of immiscible two-phase flow in porous media
  (2025).

\bibitem{hansen2018relations}
A.~Hansen, S.~Sinha, D.~Bedeaux, S.~Kjelstrup, M.~A. Gjennestad, M.~Vassvik,
  Relations between seepage velocities in immiscible, incompressible two-phase
  flow in porous media, Transport in Porous Media 125 (2018) 565--587.
\newblock \href {https://doi.org/10.1007/s11242-018-1139-6}
  {\path{doi:10.1007/s11242-018-1139-6}}.

\bibitem{pedersen2023parameterizations}
H.~Pedersen, A.~Hansen, Parameterizations of immiscible two-phase flow in
  porous media, Frontiers in Physics 11 (2023) 1127345.
\newblock \href {https://doi.org/10.3389/fphy.2023.1127345}
  {\path{doi:10.3389/fphy.2023.1127345}}.

\bibitem{pedersen2025geometric}
H.~Pedersen, A.~Hansen, Geometric structure of parameter space in immiscible
  two-phase flow in porous media, Frontiers in Physics 13 (2025) 1571054.
\newblock \href {https://doi.org/10.3389/fphy.2025.1571054}
  {\path{doi:10.3389/fphy.2025.1571054}}.

\bibitem{pedersen2025co}
H.~Pedersen, The co-moving velocity and projective transformations, arXiv
  preprint arXiv:2502.04810 (2025).
\newblock \href {https://doi.org/arXiv.2502.04810}
  {\path{doi:arXiv.2502.04810}}.

\bibitem{roy2020flow}
S.~Roy, S.~Sinha, A.~Hansen, Flow-area relations in immiscible two-phase flow
  in porous media, Frontiers in Physics 8 (2020) 4.
\newblock \href {https://doi.org/10.3389/fphy.2020.00004}
  {\path{doi:10.3389/fphy.2020.00004}}.

\bibitem{bouchaud2008economics}
J.-P. Bouchaud, Economics needs a scientific revolution, Nature 455~(7217)
  (2008) 1181--1181.
\newblock \href {https://doi.org/10.1038/4551181a}
  {\path{doi:10.1038/4551181a}}.

\bibitem{bergDisplacementMassTransfer2013i}
S.~Berg, S.~Oedai, H.~Ott, Displacement and mass transfer between saturated and
  unsaturated {{CO2}}--brine systems in sandstone, International Journal of
  Greenhouse Gas Control 12 (2013) 478--492.
\newblock \href {https://doi.org/10.1016/j.ijggc.2011.04.005}
  {\path{doi:10.1016/j.ijggc.2011.04.005}}.

\bibitem{bergConnectedPathwayRelative2016d}
S.~Berg, M.~R{\"u}cker, H.~Ott, A.~Georgiadis, H.~Van Der~Linde, F.~Enzmann,
  M.~Kersten, R.~Armstrong, S.~De~With, J.~Becker, A.~Wiegmann, Connected
  pathway relative permeability from pore-scale imaging of imbibition, Advances
  in Water Resources 90 (2016) 24--35.
\newblock \href {https://doi.org/10.1016/j.advwatres.2016.01.010}
  {\path{doi:10.1016/j.advwatres.2016.01.010}}.

\bibitem{bergNonuniquenessUncertaintyQuantification2021a}
S.~Berg, E.~Unsal, H.~Dijk, Non-uniqueness and uncertainty quantification of
  relative permeability measurements by inverse modelling, Computers and
  Geotechnics 132 (2021) 103964.
\newblock \href {https://doi.org/10.1016/j.compgeo.2020.103964}
  {\path{doi:10.1016/j.compgeo.2020.103964}}.

\bibitem{anInverseModelingCore2023}
S.~An, N.~Wenck, S.~Manoorkar, S.~Berg, C.~Taberner, R.~Pini, S.~Krevor,
  Inverse {{Modeling}} of {{Core Flood Experiments}} for {{Predictive Models}}
  of {{Sandstone}} and {{Carbonate Rocks}}, Water Resources Research 59~(12)
  (2023) e2023WR035526.
\newblock \href {https://doi.org/10.1029/2023WR035526}
  {\path{doi:10.1029/2023WR035526}}.

\bibitem{zou2018experimental}
S.~Zou, R.~T. Armstrong, J.-Y. Arns, C.~H. Arns, F.~Hussain, Experimental and
  theoretical evidence for increased ganglion dynamics during fractional flow
  in mixed-wet porous media, Water Resources Research 54~(5) (2018) 3277--3289.

\bibitem{buckleyMechanismFluidDisplacement1942}
S.~Buckley, M.~Leverett, Mechanism of {{Fluid Displacement}} in {{Sands}},
  Transactions of the AIME 146~(01) (1942) 107--116.
\newblock \href {https://doi.org/10.2118/942107-G}
  {\path{doi:10.2118/942107-G}}.

\bibitem{bergSimultaneousDeterminationRelative2024a}
S.~Berg, H.~Dijk, E.~Unsal, R.~Hofmann, B.~Zhao, V.~Raju~Ahuja, Simultaneous
  determination of relative permeability and capillary pressure from an
  unsteady-state core flooding experiment?, Computers and Geotechnics 168
  (2024) 106091.
\newblock \href {https://doi.org/10.1016/j.compgeo.2024.106091}
  {\path{doi:10.1016/j.compgeo.2024.106091}}.

\bibitem{huangCapillaryEndEffects1998}
D.~D. Huang, M.~M. Honarpour, Capillary end effects in coreflood calculations,
  Journal of Petroleum Science and Engineering 19~(1-2) (1998) 103--117.
\newblock \href {https://doi.org/10.1016/S0920-4105(97)00040-5}
  {\path{doi:10.1016/S0920-4105(97)00040-5}}.

\bibitem{moura2015impact}
M.~Moura, E.-A. Fiorentino, K.~J. M{\aa}l{\o}y, G.~Sch{\"a}fer, R.~Toussaint,
  Impact of sample geometry on the measurement of pressure-saturation curves:
  Experiments and simulations, Water Resources Research 51~(11) (2015)
  8900--8926.

\bibitem{grayAnalysisInvestigatingExtended2024}
W.~G. Gray, C.~T. Miller, Analysis of `{{Investigating}} an extended multiphase
  flow model that includes specific interfacial area', {{Computer Methods}} in
  {{Applied Mechanics}} and {{Engineering}}, 418:116594, 2024, Computer Methods
  in Applied Mechanics and Engineering 426 (2024) 116984.
\newblock \href {https://doi.org/10.1016/j.cma.2024.116984}
  {\path{doi:10.1016/j.cma.2024.116984}}.

\bibitem{lucasUeberZeitgesetzKapillaren1918}
R.~Lucas, {Ueber das Zeitgesetz des kapillaren Aufstiegs von
  Fl{\"u}ssigkeiten}, Kolloid-Zeitschrift 23~(1) (1918) 15--22.
\newblock \href {https://doi.org/10.1007/BF01461107}
  {\path{doi:10.1007/BF01461107}}.

\bibitem{washburnDynamicsCapillaryFlow1921}
E.~W. Washburn, The {{Dynamics}} of {{Capillary Flow}}, Physical Review 17~(3)
  (1921) 273--283.
\newblock \href {https://doi.org/10.1103/PhysRev.17.273}
  {\path{doi:10.1103/PhysRev.17.273}}.

\bibitem{olsen2025new}
K.~P. Olsen, B.~Hafskjold, A.~Lervik, A.~Hansen, A new thermodynamic function
  for binary mixtures: The co-molar volume, arXiv preprint arXiv:2504.13602
  (2025).

\bibitem{ebadi2024investigating}
M.~Ebadi, J.~McClure, P.~Mostaghimi, R.~T. Armstrong, Investigating an extended
  multiphase flow model that includes specific interfacial area, Computer
  Methods in Applied Mechanics and Engineering 418 (2024) 116594.

\bibitem{Erpelding}
M.~Erpelding, S.~Sinha, K.~T. Tallakstad, A.~Hansen, E.~G. Flekkøy, K.~J.
  Måløy, History independence of steady state in simultaneous two-phase flow
  through two-dimensional porous media, Phys. Rev. E. 80 (2013) 053004.

\bibitem{ramstadSimulationTwoPhaseFlow2010}
T.~Ramstad, P.-E. {\O}ren, S.~Bakke, Simulation of {{Two-Phase Flow}} in
  {{Reservoir Rocks Using}} a {{Lattice Boltzmann Method}}, SPE Journal 15~(04)
  (2010) 917--927.
\newblock \href {https://doi.org/10.2118/124617-PA}
  {\path{doi:10.2118/124617-PA}}.

\bibitem{ramstadRelativePermeabilityCalculations2012a}
T.~Ramstad, N.~Idowu, C.~Nardi, P.-E. {\O}ren, Relative {{Permeability
  Calculations}} from {{Two-Phase Flow Simulations Directly}} on {{Digital
  Images}} of {{Porous Rocks}}, Transport in Porous Media 94~(2) (2012)
  487--504.
\newblock \href {https://doi.org/10.1007/s11242-011-9877-8}
  {\path{doi:10.1007/s11242-011-9877-8}}.

\bibitem{adilaComparisonRelativePermeability2025}
A.~Adila, M.~Ebadi, W.~Xi, X.~Qi, Y.~Jing, P.~Mostaghimi, R.~T. Armstrong,
  Comparison of relative permeability hysteresis in oil-brine and gas-brine
  systems: {{A}} pore-scale investigation, International Journal of Multiphase
  Flow (2025) 105483\href
  {https://doi.org/10.1016/j.ijmultiphaseflow.2025.105483}
  {\path{doi:10.1016/j.ijmultiphaseflow.2025.105483}}.

\bibitem{oakThreePhaseRelativePermeability1990}
M.~J. Oak, Three-{{Phase Relative Permeability}} of {{Water-Wet Berea}}, in:
  {{SPE}}/{{DOE Enhanced Oil Recovery Symposium}}, SPE, Tulsa, Oklahoma, 1990,
  pp. SPE--20183--MS.
\newblock \href {https://doi.org/10.2118/20183-MS}
  {\path{doi:10.2118/20183-MS}}.

\bibitem{scanzianiSituCharacterizationThreePhase2020}
A.~Scanziani, A.~Alhosani, Q.~Lin, C.~Spurin, G.~Garfi, M.~J. Blunt,
  B.~Bijeljic, In {{Situ Characterization}} of {{Three}}-{{Phase Flow}} in
  {{Mixed}}-{{Wet Porous Media Using Synchrotron Imaging}}, Water Resources
  Research 56~(9) (2020) e2020WR027873.
\newblock \href {https://doi.org/10.1029/2020WR027873}
  {\path{doi:10.1029/2020WR027873}}.

\bibitem{bluntEmpiricalModelThreePhase2000}
M.~J. Blunt, An {{Empirical Model}} for {{Three-Phase Relative Permeability}},
  SPE Journal 5~(04) (2000) 435--445.
\newblock \href {https://doi.org/10.2118/67950-PA}
  {\path{doi:10.2118/67950-PA}}.

\bibitem{masalmehGasCO2Mobility2025}
S.~Masalmeh, A.~Farzaneh, M.~Sohrabi, Gas/{{CO2 Mobility}} under 2-{{Phase}}
  and 3-{{Phase Flow}} in {{Carbonate}} for {{CO2-EOR}} and {{CCS Projects}},
  in: {{IOR}}+ 2025 - 23rd {{European Symposium}} on {{IOR}}, European
  Association of Geoscientists \& Engineers, Edinburgh, Scotland, United
  Kingdom,, 2025, pp. 1--23.
\newblock \href {https://doi.org/10.3997/2214-4609.202531034}
  {\path{doi:10.3997/2214-4609.202531034}}.

\bibitem{chapuisTwophaseFlowEvaporation2008}
O.~Chapuis, M.~Prat, M.~Quintard, E.~{Chane-Kane}, O.~Guillot, N.~Mayer,
  Two-phase flow and evaporation in model fibrous media: {{Application}} to the
  gas diffusion layer of {{PEM}} fuel cells, Journal of Power Sources 178~(1)
  (2008) 258--268.
\newblock \href {https://doi.org/10.1016/j.jpowsour.2007.12.011}
  {\path{doi:10.1016/j.jpowsour.2007.12.011}}.

\bibitem{fernandoNonequilibriumComputerSimulations2017}
B.~Fernando, Non-equilibrium {{Computer Simulations}} of {{Coupling Effects}}
  under {{Thermal Gradients}}, CMST 23~(3) (2017) 165--173.
\newblock \href {https://doi.org/10.12921/cmst.2017.0000018}
  {\path{doi:10.12921/cmst.2017.0000018}}.

\bibitem{gladdenMRIOperandoMeasurements2010}
L.~F. Gladden, F.~J. Abeg{\~a}o, C.~P. Dunckley, D.~J. Holland, M.~H. Sankey,
  A.~J. Sederman, {{MRI}}: {{Operando}} measurements of temperature,
  hydrodynamics and local reaction rate in a heterogeneous catalytic reactor,
  Catalysis Today 155~(3-4) (2010) 157--163.
\newblock \href {https://doi.org/10.1016/j.cattod.2009.10.012}
  {\path{doi:10.1016/j.cattod.2009.10.012}}.

\bibitem{zhengOperandoCharacterisationProducts2024}
Q.~Zheng, J.~H. Williams, S.~V. Elgersma, M.~D. Mantle, A.~J. Sederman, G.~L.
  Bezemer, C.~M. Gu{\'e}don, L.~F. Gladden, Operando characterisation of the
  products of {{Fischer-Tropsch}} synthesis in a fixed-bed reactor studied by
  magnetic resonance, Catalysis Today 428 (2024) 114416.
\newblock \href {https://doi.org/10.1016/j.cattod.2023.114416}
  {\path{doi:10.1016/j.cattod.2023.114416}}.

\bibitem{gunathilakaOperandoMagneticResonance2021}
I.~E. Gunathilaka, J.~M. Pringle, L.~A. O'Dell, Operando magnetic resonance
  imaging for mapping of temperature and redox species in
  thermo-electrochemical cells, Nature Communications 12~(1) (Nov. 2021).
\newblock \href {https://doi.org/10.1038/s41467-021-26813-8}
  {\path{doi:10.1038/s41467-021-26813-8}}.

\bibitem{kimCryogenicCalorimeterTesting2023}
J.-D. Kim, H.-T. Kim, H.-J. Jeon, H.-J. Tak, D.-H. Park, J.-H. Kim, B.-H. Min,
  J.-M. Lee, Cryogenic {{Calorimeter Testing}} of {{Insulation Materials}} for
  {{Development}} of {{Liquid Hydrogen Storage System}}, in: Proceedings of the
  {{Thirty-third}} (2023) {{International Ocean}} and {{Polar Engineering
  Conference}}, {International Society of Offshore and Polar Engineers
  (ISOPE)}, Ottawa, Canada, 2023, pp. 3494--3498.

\bibitem{alemnDifferenceSteadystateUnsteadystate1989}
M.~A. Aleman, T.~R. Ramahohan, J.~C. Slattery, The difference between
  steady-state and unsteady-state relative permeabilities, Transport in Porous
  Media 4~(5) (Oct. 1989).
\newblock \href {https://doi.org/10.1007/BF00179531}
  {\path{doi:10.1007/BF00179531}}.

\bibitem{mainiComparisonSteadyStateUnsteadyState1990}
B.~Maini, G.~Coskuner, K.~Jha, A {{Comparison Of Steady-State And
  Unsteady-State Relative Permeabilities Of Viscocities Oil And Water In Ottawa
  Sand}}, Journal of Canadian Petroleum Technology 29~(02) (Mar. 1990).
\newblock \href {https://doi.org/10.2118/90-02-02}
  {\path{doi:10.2118/90-02-02}}.

\bibitem{ramstadRelativePermeabilityCalculations2012}
T.~Ramstad, N.~Idowu, C.~Nardi, P.-E. {\O}ren, Relative {{Permeability
  Calculations}} from {{Two-Phase Flow Simulations Directly}} on {{Digital
  Images}} of {{Porous Rocks}}, Transport in Porous Media 94~(2) (2012)
  487--504.
\newblock \href {https://doi.org/10.1007/s11242-011-9877-8}
  {\path{doi:10.1007/s11242-011-9877-8}}.

\bibitem{ramstadPoreScaleSimulationsSingle2019}
T.~Ramstad, C.~F. Berg, K.~Thompson, Pore-{{Scale Simulations}} of {{Single-}}
  and {{Two-Phase Flow}} in {{Porous Media}}: {{Approaches}} and
  {{Applications}}, Transport in Porous Media 130~(1) (2019) 77--104.
\newblock \href {https://doi.org/10.1007/s11242-019-01289-9}
  {\path{doi:10.1007/s11242-019-01289-9}}.

\bibitem{grootNonEquilibriumThermodynamics1984}
S.~R.~D. Groot, P.~Mazur, Non-{{Equilibrium Thermodynamics}}, Dover
  Publications, New York, 1984.

\bibitem{bentsenUseConventionalCocurrent1993}
R.~G. Bentsen, A.~A. Manai, On the use of conventional cocurrent and
  countercurrent effective permeabilities to estimate the four generalized
  permeability coefficients which arise in coupled, two-phase flow, Transport
  in Porous Media 11~(3) (1993) 243--262.
\newblock \href {https://doi.org/10.1007/BF00614814}
  {\path{doi:10.1007/BF00614814}}.

\bibitem{bentsenEffectMomentumTransfer1998}
R.~G. Bentsen, Effect of momentum transfer between fluid phases on effective
  mobility, Journal of Petroleum Science and Engineering 21~(1-2) (1998)
  27--42.
\newblock \href {https://doi.org/10.1016/S0920-4105(98)00035-7}
  {\path{doi:10.1016/S0920-4105(98)00035-7}}.

\bibitem{ayubInterfacialViscousCoupling1999}
M.~Ayub, R.~G. Bentsen, Interfacial viscous coupling: A myth or reality?,
  Journal of Petroleum Science and Engineering 23~(1) (1999) 13--26.
\newblock \href {https://doi.org/10.1016/S0920-4105(99)00003-0}
  {\path{doi:10.1016/S0920-4105(99)00003-0}}.

\bibitem{bentsenEffectNeglectingInterfacial2005}
R.~G. Bentsen, Effect of neglecting interfacial coupling when using vertical
  flow experiments to determine relative permeability, Journal of Petroleum
  Science and Engineering 48~(1-2) (2005) 81--93.
\newblock \href {https://doi.org/10.1016/j.petrol.2005.03.010}
  {\path{doi:10.1016/j.petrol.2005.03.010}}.

\bibitem{bergGenerationGroundTruth2018}
S.~Berg, N.~Saxena, M.~Shaik, C.~Pradhan, Generation of ground truth images to
  validate micro-{{CT}} image-processing pipelines, The Leading Edge 37~(6)
  (2018) 412--420.
\newblock \href {https://doi.org/10.1190/tle37060412.1}
  {\path{doi:10.1190/tle37060412.1}}.

\bibitem{tahmasebiMachineLearningGeo2020}
P.~Tahmasebi, S.~Kamrava, T.~Bai, M.~Sahimi, Machine learning in geo- and
  environmental sciences: {{From}} small to large scale, Advances in Water
  Resources 142 (2020) 103619.
\newblock \href {https://doi.org/10.1016/j.advwatres.2020.103619}
  {\path{doi:10.1016/j.advwatres.2020.103619}}.

\bibitem{wangDeepNeuralNetworks2021}
Y.~D. Wang, M.~Shabaninejad, R.~T. Armstrong, P.~Mostaghimi, Deep neural
  networks for improving physical accuracy of {{2D}} and {{3D}} multi-mineral
  segmentation of rock micro-{{CT}} images, Applied Soft Computing 104 (2021)
  107185.
\newblock \href {https://doi.org/10.1016/j.asoc.2021.107185}
  {\path{doi:10.1016/j.asoc.2021.107185}}.

\bibitem{alqahtaniSuperResolvedSegmentationXray2022}
N.~J. Alqahtani, Y.~Niu, Y.~D. Wang, T.~Chung, Z.~Lanetc, A.~Zhuravljov, R.~T.
  Armstrong, P.~Mostaghimi, Super-{{Resolved Segmentation}} of {{X-ray Images}}
  of {{Carbonate Rocks Using Deep Learning}}, Transport in Porous Media 143~(2)
  (2022) 497--525.
\newblock \href {https://doi.org/10.1007/s11242-022-01781-9}
  {\path{doi:10.1007/s11242-022-01781-9}}.

\bibitem{tangDeepLearningFullfeature2022}
K.~Tang, Q.~Meyer, R.~White, R.~T. Armstrong, P.~Mostaghimi, Y.~Da~Wang,
  S.~Liu, C.~Zhao, K.~{Regenauer-Lieb}, P.~K.~M. Tung, Deep learning for
  full-feature {{X-ray}} microcomputed tomography segmentation of proton
  electron membrane fuel cells, Computers \& Chemical Engineering 161 (2022)
  107768.
\newblock \href {https://doi.org/10.1016/j.compchemeng.2022.107768}
  {\path{doi:10.1016/j.compchemeng.2022.107768}}.

\bibitem{liangMultimineralSegmentationMicrotomographic2022}
J.~Liang, Y.~Sun, M.~Lebedev, B.~Gurevich, M.~Nzikou, S.~Vialle,
  S.~Glubokovskikh, Multi-mineral segmentation of micro-tomographic images
  using a convolutional neural network, Computers \& Geosciences 168 (2022)
  105217.
\newblock \href {https://doi.org/10.1016/j.cageo.2022.105217}
  {\path{doi:10.1016/j.cageo.2022.105217}}.

\bibitem{reedyHighresolutionMicroCT3D2022}
C.~L. Reedy, C.~L. Reedy, High-resolution micro-{{CT}} with {{3D}} image
  analysis for porosity characterization of historic bricks, Heritage Science
  10~(1) (2022) 83.
\newblock \href {https://doi.org/10.1186/s40494-022-00723-4}
  {\path{doi:10.1186/s40494-022-00723-4}}.

\bibitem{mahdaviaraPoreSkelSkeletonizationGrayscale2023}
M.~Mahdaviara, M.~Sharifi, A.~Raoof, {{PoreSkel}}: {{Skeletonization}} of
  grayscale micro-{{CT}} images of porous media using deep learning techniques,
  Advances in Water Resources 180 (2023) 104544.
\newblock \href {https://doi.org/10.1016/j.advwatres.2023.104544}
  {\path{doi:10.1016/j.advwatres.2023.104544}}.

\bibitem{waldnerDeeplearningbasedWorkflowReconstructing2024}
S.~Waldner, J.~Huwyler, M.~Puchkov, A deep-learning-based workflow for
  reconstructing and segmenting challenging sets of time-resolved {{X-ray}}
  micro-computed tomography data, SoftwareX 27 (2024) 101796.
\newblock \href {https://doi.org/10.1016/j.softx.2024.101796}
  {\path{doi:10.1016/j.softx.2024.101796}}.

\bibitem{strandbergAIassistedDeepLearning2024}
A.~Strandberg, H.~Chevreau, N.~Skoglund, {{AI-assisted}} deep learning
  segmentation and quantitative analysis of {{X-ray}} microtomography data from
  biomass ashes, MethodsX 13 (2024) 102812.
\newblock \href {https://doi.org/10.1016/j.mex.2024.102812}
  {\path{doi:10.1016/j.mex.2024.102812}}.

\bibitem{wangLargescalePhysicallyAccurate2023}
Y.~D. Wang, Q.~Meyer, K.~Tang, J.~E. McClure, R.~T. White, S.~T. Kelly, M.~M.
  Crawford, F.~Iacoviello, D.~J.~L. Brett, P.~R. Shearing, P.~Mostaghimi,
  C.~Zhao, R.~T. Armstrong, Large-scale physically accurate modelling of real
  proton exchange membrane fuel cell with deep learning, Nature Communications
  14~(1) (2023) 745.
\newblock \href {https://doi.org/10.1038/s41467-023-35973-8}
  {\path{doi:10.1038/s41467-023-35973-8}}.

\bibitem{goethalsDYRECTComputedTomography2025}
W.~Goethals, T.~Bultreys, S.~Berg, M.~N. Boone, J.~Aelterman, {{DYRECT Computed
  Tomography}}: {{DYnamic Reconstruction}} of {{Events}} on a {{Continuous
  Timescale}}, IEEE Transactions on Computational Imaging (2025) 1--13\href
  {https://doi.org/10.1109/TCI.2025.3566241}
  {\path{doi:10.1109/TCI.2025.3566241}}.

\bibitem{niu2021geometrical}
Y.~Niu, Y.~Da~Wang, P.~Mostaghimi, J.~E. McClure, J.~Yin, R.~T. Armstrong,
  Geometrical-based generative adversarial network to enhance digital rock
  image quality, Physical Review Applied 15~(6) (2021) 064033.

\bibitem{liu2020tomogan}
Z.~Liu, T.~Bicer, R.~Kettimuthu, D.~Gursoy, F.~De~Carlo, I.~Foster, Tomogan:
  low-dose synchrotron x-ray tomography with generative adversarial networks:
  discussion, Journal of the Optical Society of America A 37~(3) (2020)
  422--434.

\bibitem{liang2018deep}
L.~Liang, M.~Liu, C.~Martin, W.~Sun, A deep learning approach to estimate
  stress distribution: a fast and accurate surrogate of finite-element
  analysis, Journal of The Royal Society Interface 15~(138) (2018) 20170844.

\bibitem{lusch2018deep}
B.~Lusch, J.~N. Kutz, S.~L. Brunton, Deep learning for universal linear
  embeddings of nonlinear dynamics, Nature communications 9~(1) (2018) 4950.

\bibitem{spurinDynamicModeDecomposition2023b}
C.~Spurin, R.~T. Armstrong, J.~McClure, S.~Berg, Dynamic mode decomposition for
  analysing multi-phase flow in porous media, Advances in Water Resources 175
  (2023) 104423.
\newblock \href {https://doi.org/10.1016/j.advwatres.2023.104423}
  {\path{doi:10.1016/j.advwatres.2023.104423}}.

\bibitem{alqahtani2021flow}
N.~J. Alqahtani, T.~Chung, Y.~D. Wang, R.~T. Armstrong, P.~Swietojanski,
  P.~Mostaghimi, Flow-based characterization of digital rock images using deep
  learning, SPE Journal 26~(04) (2021) 1800--1811.

\bibitem{sundararajan2017axiomatic}
M.~Sundararajan, A.~Taly, Q.~Yan, Axiomatic attribution for deep networks, in:
  International conference on machine learning, PMLR, 2017, pp. 3319--3328.

\bibitem{schmidt2009distilling}
M.~Schmidt, H.~Lipson, Distilling free-form natural laws from experimental
  data, science 324~(5923) (2009) 81--85.

\bibitem{imDatadrivenDiscoveryGoverning2023}
J.~Im, F.~P.~J. De~Barros, S.~Masri, M.~Sahimi, R.~M. Ziff, Data-driven
  discovery of the governing equations for transport in heterogeneous media by
  symbolic regression and stochastic optimization, Physical Review E 107~(1)
  (2023) L013301.
\newblock \href {https://doi.org/10.1103/PhysRevE.107.L013301}
  {\path{doi:10.1103/PhysRevE.107.L013301}}.

\bibitem{bongard2007automated}
J.~Bongard, H.~Lipson, Automated reverse engineering of nonlinear dynamical
  systems, Proceedings of the National Academy of Sciences 104~(24) (2007)
  9943--9948.

\bibitem{brunton2016discovering}
S.~L. Brunton, J.~L. Proctor, J.~N. Kutz, Discovering governing equations from
  data by sparse identification of nonlinear dynamical systems, Proceedings of
  the national academy of sciences 113~(15) (2016) 3932--3937.

\bibitem{bapst2020unveiling}
V.~Bapst, T.~Keck, A.~Grabska-Barwi{\'n}ska, C.~Donner, E.~D. Cubuk, S.~S.
  Schoenholz, A.~Obika, A.~W. Nelson, T.~Back, D.~Hassabis, et~al., Unveiling
  the predictive power of static structure in glassy systems, Nature physics
  16~(4) (2020) 448--454.

\end{thebibliography}
		
		
		
		
		
		%
		
	\end{document}